\RequirePackage{ifpdf}
\documentclass[11pt,a4paper]{article}
\pdfoutput=1 

\usepackage{amsmath} 
\usepackage{jheppub}
\usepackage{pstricks}
\usepackage{longtable}
\usepackage[final]{pdfpages}
\usepackage{ifpdf}
\usepackage{flexisym}
\usepackage{breqn}

\def\eps{\epsilon}
 
\def\beq{\begin{equation}} 
\def\eeq{\end{equation}}

\def\b0{\beta_0}
\def\ca{\text{\tiny C}_\text{\tiny A}}
\def\cf{\text{\tiny C}_\text{\tiny F}}
\def\unM{\hat{\cal M}}
\def\unas{ \left( \frac{\hat{a}_s}{\mu_0^{\epsilon}} S_{\epsilon} \right) }
\def\rnM{{\cal M}}
\def\rnas{ \left( a_s  \right) }
\def\cD{{\cal D}}

\def\ca{\text{\tiny C}_\text{\tiny A}}
\def\cf{\text{\tiny C}_\text{\tiny F}}

\title{Two-Loop QCD Correction to massive spin-2 resonance $\rightarrow$ 3 gluons }

\author[a]{Taushif Ahmed,}
\author[a]{Maguni Mahakhud,}
\author[b]{Prakash Mathews,}
\author[a]{Narayan Rana}
\author[c]{and V. Ravindran}

\affiliation[a]{Harish-Chandra Research Institute,\\Chhatnag Road, Jhunsi, Allahabad 211 019, Uttar Pradesh, India}
\affiliation[b]{Saha Institute of Nuclear Physics,\\1/AF Bidhan Nagar, Kolkata 700 064, India}
\affiliation[c]{The Institute of Mathematical Sciences,\\C.I.T Campus, 4th Cross St, Tharamani, Chennai 600 113, Tamil Nadu, India }

\emailAdd{taushif@hri.res.in}
\emailAdd{maguni@hri.res.in}
\emailAdd{prakash.mathews@saha.ac.in}
\emailAdd{narayan@hri.res.in}
\emailAdd{ravindra@imsc.res.in}

\abstract{We present the ${\cal O}(\alpha_s^3)$ virtual QCD corrections to  
the process $h \rightarrow g+g+g$ due to interference of born and two-loop amplitudes, where $h$ is a massive spin-2 particle
and $g$ is the gluon.  We assume that the SM fields couple to $h$ through
the SM energy momentum tensor. Our result constitutes one of the ingredients to full NNLO QCD contribution to
production of a massive spin-2 particle along with a jet  in the scattering process at the
LHC.  In particular, this massive spin-2 could be a KK mode of a ADD graviton in
large extra dimensional model or a RS KK mode in warped extra dimensional model or
a generic massive spin-2.
In addition, it provides an opportunity to
study the ultraviolet and infrared structures of QCD amplitudes involving tensorial coupling resulting from 
energy momentum operator.
Using dimensional regularization, we find that infrared poles of this amplitude 
are in agreement with the proposal by Catani confirming the factorization property of 
QCD amplitudes with tensorial insertion.} 

\preprint{}


\allowdisplaybreaks[1]

\begin{document}
\maketitle
\flushbottom

\section{Introduction}

Theoretical studies on production and decay of a spin $J^P$ particle in the hadron
colliders in the Standard Model (SM) and its extensions are underway 
due to a wealth of information available from ATLAS \cite{ATLAS:2011ab}
and CMS \cite{Chatrchyan:2011fq} collaborations 
at the Large Hadron Collider (LHC).   In particular,
investigations involving particles with higher spin, namely $J^P=2^+$ bosons gained
attention \cite{Ellis:2012xd,Ellis:2012jv} due to the discovery of 
a new boson at the LHC with a mass $m \approx 125$ GeV as 
its spin is yet to be determined with no doubt.  
Searches of spin-2 particles earlier in Tevatron
and recently in LHC were motivated due to their presence 
in theories with large extra dimensions,
such as ADD \cite{ADD} and warped extra dimensions
such as RS \cite{RS1} models. These beyond the SM (BSM) theories provide 
an alternate scenario to explain the hierarchy problem of the SM through the introduction
of compact higher dimensions with factorisable (ADD) or non-factorisable (RS) geometries. 
The impact of these higher dimensions can be understood in four dimensional framework
through the existence of a tower of massive spin-2 Kaluza-Klein (KK) gravitons.  Their signatures
at the colliders can provide valuable information on these models. 
For the production of scalar $J^P=0^+$ and vector $J^P=1^+$ particles 
in the SM and BSM, there are extensive studies
available in literature. They include inclusive and semi-inclusive rates taking into
account quantum chromodynamics (QCD) and electroweak (EW) radiative corrections to impressively high orders in perturbative expansion. 
Thanks to universal features of some of the perturbative corrections in QCD, one can
safely use those computed in the SM for BSM processes.  For spin-2 particles, searches
in Drell-Yan (DY), di-photon and jet+missing energy at the LHC are already underway. Also the bounds on the
scale of the new physics and the number of extra-dimensions are severely constrained from these
searches.  These studies use theoretical results computed in QCD at NLO level in perturbation
theory.  Such computations are inevitable due to various uncertainties resulting from
renormalization and factorization scales as well as from the parton distribution functions.  
Needless to say that the corrections are also big giving large K-factors which are
often sensitive to the observables and their physical scales.    
In ADD and RS models spin-2 KK gravitons contribute to physical
processes through either exchange of
virtual graviton state or production of a real graviton state. In the virtual mode, 
KK gravitons get exchanged between the SM particles and the summation over their high multiplicity
leads to a compensation of small coupling strength.
Hence, the cross section will
be appreciable at collider energies, giving rise to non-resonant enhancement of
the high invariant mass regions of a di-final state production
\cite{HLZ,GRW,di-final} or final
states involving more particles \cite{tri-final}.
Next to leading order QCD corrections were computed for the di-final
state processes in the ADD model {\em viz.}\ $\ell^+ \ell^-$ \cite{di-ll1,di-ll2,di-ll3},
$\gamma \gamma$ \cite{di-ph1,di-ph2} and $ZZ$ \cite{di-ZZ1,di-ZZ2} and $W^+ W^-$
\cite{di-WW1,di-WW2},
in addition they have been extended to NLO+PS accuracy
\cite{di-ph+ps, Frederix:2013lga}.
On-shell spin-2 particle production results in missing energy signal.  Again
large multiplicity of final states enhances the observable effects.
Productions of on-shell spin-2 gravitons in association with (a) jet \cite{jEt}, (b) photon
\cite{Gao:2009pn} and (c) electroweak gauge boson \cite{ZWEt} have been studied to
NLO in QCD.  
Similar results taking into account NLO QCD effects for processes involving 
the resonant production of spin-2 graviton are available for 
di-lepton production \cite{di-ll2,di-ll3},
di-photon production \cite{di-ph2}, di-neutral electroweak gauge boson production
\cite{di-ZZ2} and charged electroweak gauge boson production $W^+ W^-$ \cite{di-WW2}.
Since the NLO corrections and the associated scale uncertainties are 
not completely under control in most of the observable, an attempt 
\cite{deFlorian:2013wpa} was undertaken to include soft and
collinear contribution to NNLO accuracy
for processes involving KK gravitons 
in both ADD and RS models.
This was possible as the full two loop QCD matrix element of energy momentum tensor 
is now available \cite{deFlorian:2013sza}.
The phenomenological results show significant reduction
in dependence on renormalization and factorization scales and these results improve the stability of the perturbative predictions.  

Observables with jet + missing energy are sensitive to new physics
from many BSM scenarios.  This missing energy could be due to a heavy
resonance produced in the final state escaping the detector.  
The NLO QCD effects to this process in large extra dimensional model, namely ADD 
have been available for sometime and the importance of K-factor
and scale uncertainties are documented in \cite{jEt}.  In this article,
we compute ${\cal O}(\alpha_s^3)$ virtual correction in massless QCD to  
the process $h \rightarrow g+g+g$ due to interference of born and two-loop amplitudes, where $h$ is a massive spin-2 particle.
The full NNLO analysis also requires square of one loop amplitudes,  
real emission processes and appropriate mass counter terms which we reserve for future work.  
We have assumed a minimal coupling between massive spin-2 field and the fields of the SM.   
Hence our results are applicable to scattering processes involving a massive spin-2 particle and three gluons 
such as a massive graviton production with a jet
in ADD and RS models or production of a massive spin-2 Higgs like boson along with a jet after appropriate analytical continuation \cite{Gehrmann:2002zr} of kinematical variables to the respective
physical regions. 
Similar computations with massive spin-0 and spin-1 boson productions 
at two loop level in QCD exist in the literature for the processes 
$Higgs \rightarrow g+g+g$ \cite{Gehrmann:2011aa} 
and $g + g \rightarrow V + jet, V=Z,\gamma$ \cite{Gehrmann:2013vga} respectively. 

Spin-2 field being a rank-2 tensor, we encounter for the first
time the two loop amplitudes with higher tensorial integrals resulting from
rank-2 derivative couplings of spin-2 fields with the SM ones.  In addition, we encounter 
more than 2000 two loop Feynman amplitudes contributing due to the universal coupling of
spin-2 field with all the SM particles.  While these increase 
technical complexities at the intermediate stages of computation, the results confirm
the universal infra-red structure of QCD amplitudes.  In other words,
we find that soft and collinear divergences not only factorise but also agree with
the predictions from Catani's work \cite{catani1} (see also \cite{sterman}) on two loop QCD amplitudes for multi-leg 
processes.  We also observe that there are no additional
UV divergences as the interaction is through energy momentum tensor of the SM which
is conserved.  Hence, our present work is also useful to study the field theoretical structure of 
QCD amplitudes with tensor operator insertions, in particular with the energy momentum tensor of the 
SM.  

In the next section, we describe the generic effective Lagrangian that describes coupling of
spin-2 fields with those of the SM.  Section 3 is devoted to the computational
details.  Section 4 and Appendix contain our final results.  In section 5, we conclude with
our findings.

\section{Theory}

\subsection{The effective Lagrangian}
We consider the SM with an additional spin-2 field $h^{\mu\nu}$.  We assume that the spin-2 field couples minimally with the SM ones
through the SM energy momentum tensor $T_{\mu\nu}^{SM}$.  Since we are interested only in the
QCD effects of the process under study, we restrict ourselves to the QCD part of $T_{\mu\nu}^{SM}$ and 
hence the action reads \cite{ADD,RS1} as
\begin{eqnarray}\label{intlag}
{\cal S} = {\cal S}_{SM} +{\cal S}_h -  \frac{\kappa}{2} \int d^4 x ~T^{QCD}_{\mu\nu}
(x)~ h^{\mu\nu} (x) \, ,
\end{eqnarray}
where $\kappa$ is a dimensionful coupling and $T^{QCD}_{\mu\nu}$ is the energy momentum tensor of QCD\\ given by
\begin{eqnarray}\label{emT}
T^{QCD}_{\mu\nu} &=& -g_{\mu\nu} {\cal L}_{QCD} - F_{\mu\rho}^a F^{a\rho}_\nu
- \frac{1}{\xi} g_{\mu\nu} \partial^\rho(A_\rho^a\partial^\sigma A_\sigma^a)
+ \frac{1}{\xi}(A_\nu^a \partial_\mu(\partial^\sigma A_\sigma^a) + A_\mu^a\partial_\nu
(\partial^\sigma A_\sigma^a))
\nonumber\\[1ex]
&&+\frac{i}{4} \Big[ \overline \psi \gamma_\mu (\overrightarrow{\partial}_\nu -i g_s T^a A^a_\nu)\psi
-\overline \psi (\overleftarrow{\partial}_\nu + i g_s T^a A^a_\nu) \gamma_\mu \psi
+\overline \psi \gamma_\nu (\overrightarrow{\partial}_\mu -i g_s T^a A^a_\mu)\psi
\nonumber\\[1ex]
&&-\overline \psi (\overleftarrow{\partial}_\mu + i g_s T^a A^a_\mu) \gamma_\nu \psi\Big]
+\partial_\mu \overline \omega^a (\partial_\nu \omega^a - g_s f^{abc} A_\nu^c \omega^b)
\nonumber\\[1ex]
&&+\partial_\nu \overline \omega^a (\partial_\mu \omega^a- g_s f^{abc} A_\mu^c \omega^b).
\end{eqnarray}
$g_s$ is the strong coupling constant and $\xi$ is the gauge fixing parameter.  The $T^a$ are generators and $f^{abc}$
are the structure constants of $SU(3)$.   
Note that spin-2 fields couple to ghost fields ($\omega^a$) \cite{Mathews:2004pi} as well in order to cancel unphysical
degrees of freedom of gluon fields ($A^a_{\mu}$).

\subsection{Notation}
We consider the decay of a massive spin-2 field into three gluons 
\begin{equation}\label{ggggr}
 h(q) \rightarrow g(p_1) + g(p_2) + g(p_3).
\end{equation}
The associated Mandelstam variables are defined as
\begin{equation}
 s \equiv (p_1 + p_2)^2, \hspace{1cm} t \equiv (p_2 + p_3)^2, \hspace{1cm} u \equiv (p_1 + p_3)^2
\end{equation}
which satisfy 
\begin{equation}
 s + t + u = M_h^2 \equiv Q^2
\end{equation}
where $M_h$ is the mass of the spin-2 field.  We also define the following
dimensionless invariants which appear in harmonic polylogarithms (HPL) \cite{remiddi} and 2dHPL \cite{gr1,gr2} as 
\begin{equation}
 x \equiv s/Q^2, \hspace{1cm} y \equiv u/Q^2, \hspace{1cm} z \equiv t/Q^2
\end{equation}
satisfying  
\begin{equation}
 x + y + z = 1.
\end{equation}

\subsection{Ultraviolet renormalization}
We describe here the ultraviolet (UV) renormalization of the matrix elements of the decay of a spin-2 resonance with
minimal coupling up to second order in QCD perturbation theory.  We regularize the theory in $d=4+\epsilon$ dimensions and the dimensionful
strong coupling constant in $d$ dimensions is made dimensionless one ($\hat g_s$) by introducing the scale $\mu_0$.  We expand the unrenormalized
amplitude in powers of $\hat a_s=\hat g_s^2/16 \pi^2$ as 
\begin{equation} \label{unm}
 |{\cal M} \rangle = \unas^{\frac{1}{2}} |\unM^{(0)} \rangle + \unas^{\frac{3}{2}} |\unM^{(1)} \rangle + \unas^{\frac{5}{2}} |\unM^{(2)} \rangle + {\cal O}(\hat{a}_s^3) \; ,
\end{equation}
where $S_{\eps} = \exp[\frac{\eps}{2} (\gamma_E - \ln 4\pi)]$ with Euler constant $\gamma_E = 0.5772 \ldots$ , results from loop integrals beyond leading order.
$|\unM^{(i)} \rangle$ is the unrenormalized color-space vector representing the $i^{th}$ loop amplitude.
In $\overline {MS}$ scheme, the renormalized coupling constant $a_s \equiv a_s (\mu_R^2)$ at
renormalization scale $\mu_R$ is related to unrenormalized coupling constant $\hat{a}_s$ by 
\begin{eqnarray} \label{renas}
 \frac{\hat{a}_s}{\mu_0^{\epsilon}} S_{\epsilon} &=& \frac{a_s}{\mu_R^{\epsilon}} {Z}(\mu_R^2)
\nonumber\\[1ex]
 &=&\frac{a_s}{\mu_R^{\epsilon}} \left[   1 + a_s \frac{2 \b0}{\eps} + a_s^2 \left( \frac{4\b0^2}{\eps^2} + \frac{\beta_1}{\eps} \right) + {\cal O}(a_s^3)  \right] \; ,
\end{eqnarray}
where  
\begin{equation}
 \b0 = \left( \frac{11}{3} C_A - \frac{4}{3} T_F n_f \right), \hspace{0.5cm}
 \beta_1 = \left( \frac{34}{3} C_A^2 - \frac{20}{3} C_A T_F n_f - 4 C_F T_F n_f \right)
\end{equation}
with $C_A = N$, $C_F = (N^2 -1)/{2 N} $, $T_F = 1/2$ and $n_f$ is the number of active quark flavors.  Since, the spin-2 resonance couples
to the SM particles through energy momentum tensor (eqn.(\ref{intlag})) which is conserved, the coupling constant $\kappa$ is protected 
from any UV renormalization.  Hence, there will be no additional UV renormalization required other than the strong coupling constant renormalization.
Using the eqn.(\ref{renas}), we now can express $|{\cal M } \rangle$ (eqn.(\ref{unm})) in powers  
of renormalized $a_s$ with UV finite matrix elements $|\rnM^{(i)} \rangle$  
\begin{equation}
 |\rnM \rangle = \rnas^{\frac{1}{2}} \Bigg( |\rnM^{(0)} \rangle + a_s |\rnM^{(1)} \rangle + a_s^{2} |\rnM^{(2)} \rangle + {\cal O}({a}_s^3) \Bigg)
\end{equation}
where
\begin{eqnarray}
 |\rnM^{(0)} \rangle &=& \left(  \frac{1}{\mu^{\eps}_R} \right)^{\frac{1}{2}}    |\unM^{(0)} \rangle \; ,
\nonumber\\[1ex]
|\rnM^{(1)} \rangle &=& \left(  \frac{1}{\mu^{\eps}_R} \right)^{\frac{3}{2}} \left[ |\unM^{(1)} \rangle  + \mu^{\eps}_R  \frac{r_1}{2}  |\unM^{(0)} \rangle \right] \; ,
\nonumber\\[1ex]
|\rnM^{(2)} \rangle &=& \left(  \frac{1}{\mu^{\eps}_R} \right)^{\frac{5}{2}} \left[ |\unM^{(2)} \rangle  + \mu^{\eps}_R  \frac{3r_1}{2}  |\unM^{(1)} \rangle 
+ \mu^{2\eps}_R \left( \frac{r_2}{2} - \frac{r_1^2}{8}  \right)  |\unM^{(0)} \rangle \right]
\end{eqnarray}
with
\begin{equation}
 r_1 = \frac{2 \b0}{\eps} \:, \hspace{1cm} r_2 = \left( \frac{4\b0^2}{\eps^2} + \frac{\beta_1}{\eps} \right).
\end{equation}

\subsection{Infrared factorization}
Beyond leading order, the UV renormalized matrix elements $|{\cal M}^{(i)}\rangle$, $i>0$ contain divergences arising from the infrared sector of massless QCD. 
They result from soft gluons and collinear massless partons present in the loops.  They will cancel against similar divergences coming from real
emission contributions in the infrared safe observables order by order in $a_s$, thanks to KLN theorem \cite{Kinoshita:1962ur, Lee:1964is}.  
The infrared divergence structure and their factorization property in QCD amplitudes have been well studied for long time.  In \cite{catani1}, Catani predicted the infrared 
divergences of multi-parton QCD amplitudes precisely in dimensional regularization up to two loops excluding two loop single pole in $\epsilon$.          
In \cite{sterman}, Sterman and Tejeda-Yeomans provided a systematic way of understanding the structure of infrared divergences using factorization properties of the scattering amplitudes
along with infrared evolution equations.  They demonstrated the connection of single pole in $\epsilon$ to a soft anomalous dimension matrix, later computed in \cite{Aybat:2006wq, Aybat:2006mz}.  
The structure of single pole term for the electromagnetic and Higgs form factors was first shown in \cite{Ravindran:2003um, Ravindran:2004mb}.  
Using soft collinear effective field theory, Becher and Neubert \cite{Becher:2009cu} derived the exact formula for the infra-red divergences of scattering amplitudes with an arbitrary number of loops and legs
in massless QCD including single pole in dimensional regularization.  Using Wilson lines for hard partons and soft and eikonal jet functions in dimensional regularization,
Gardi and Magnea also arrived at a similar all order result \cite{Gardi:2009qi}. 

According to Catani's prediction, the renormalized amplitudes $|{\cal M}^{(i)}\rangle$ for the process (eqn.(\ref{ggggr})) 
can be expressed in terms of the universal subtraction operators ${\bf I}^{(i)}_g(\epsilon)$ as
follows\footnote{\tiny{The numerical coefficients 2 and 4 with ${\bf I}^{(i)}$ come due to the different definition of $a_s$ between ours and Catani.}}
\begin{eqnarray}\label{catani1}
 |\rnM^{(1)}  \rangle &=& 2 \hspace{0.1cm} {\bf{I}}_{g}^{(1)} (\eps) \hspace{0.1cm} |\rnM^{(0)} \rangle
+ |\rnM^{(1)fin}  \rangle 
\nonumber\\[2ex] 
|\rnM^{(2)} \rangle &=& 2 \hspace{0.1cm} {\bf{I}}_{g}^{(1)} (\eps) \hspace{0.1cm} |\rnM^{(1)} \rangle
+ 4 \hspace{0.1cm} {\bf{I}}^{(2)}_{g} (\eps) \hspace{0.1cm} |\rnM^{(0)}  \rangle
+ |\rnM^{(2)fin} \rangle 
\end{eqnarray} 
where,
\begin{eqnarray}
 {\bf{I}}^{(1)}_{g} (\eps) &=& \frac{1}{2}   \frac{e^{- \frac{\eps}{2} \gamma_E}}{\Gamma(1+\frac{\eps}{2})}  {\cal V}_g^{sing}(\eps) 
\left[ \Big( - \frac{s}{\mu^2_R} \Big)^{\frac{\eps}{2}} + \Big( - \frac{t}{\mu^2_R} \Big)^{\frac{\eps}{2}} + \Big( - \frac{u}{\mu^2_R} \Big)^{\frac{\eps}{2}}   \right]
\nonumber\\[1ex]
{\bf{I}}^{(2)}_{g} (\eps) &=& - \hspace{0.1cm} \frac{1}{2} {\bf{I}}^{(1)}_{g} (\eps) \Big[ {\bf{I}}^{(1)}_{g} (\eps) - \frac{2 \b0}{\eps}  \Big]
 +\hspace{0.1cm} \frac{e^{\frac{\eps}{2} \gamma_E} \hspace{0.1cm} \Gamma(1+\eps)}{\Gamma(1+\frac{\eps}{2})} \Big[ -\frac{\b0}{\eps} + K \Big] \hspace{0.1cm} {\bf{I}}^{(1)}_{g} (2\eps)
\nonumber\\[1ex]
&& +\hspace{0.1cm} {\bf{H}}^{(2)}_{g} (\eps)
\end{eqnarray}
and  
\begin{equation}
 {\cal V}_g^{sing}(\eps) = C_A \frac{4}{\eps^2} - \frac{\b0}{\eps}, \hspace{0.5cm} K = \left(\frac{67}{18}-\frac{\pi^2}{6} \right) C_A
 - \frac{10}{9} T_F n_f
\end{equation}

\begin{equation}
 {\bf{H}}^{(2)}_g (\eps) = \frac{3}{2 \epsilon } \Bigg\{ C_A^2 \Big( - \frac{5}{12} - \frac{11}{24} \zeta_2 - \frac{1}{2} \zeta_3 \Big) + C_A n_f \Big(\frac{29}{27} +  \frac{1}{12} \zeta_2 \Big) 
- \frac{1}{2} C_F n_f -\frac{5}{27} n_f^2  \Bigg\}
\end{equation}

\section{Calculation of two-loop amplitude}
We now describe the computation of $\langle {\cal M}^{(0)} |{\cal M}^{(1)} \rangle$ \& $\langle {\cal M}^{(0)} |{\cal M}^{(2)} \rangle$ matrix elements where all the
Lorentz and color indices of external particles are summed over.
The computation involves large number of Feynman diagrams.  We need to perform various algebraic simplifications with Dirac, Lorentz and color indices before
the loop integrals are evaluated.  
Due to tensorial coupling of spin-2 resonance with the SM fields, the loops contain higher rank tensor integrals as compared to the ones normally encountered
in the SM processes. We have systematically automated this computation using various symbolic manipulation programs developed in house and 
few publicly available packages that use FORM \cite{Vermaseren:2000nd} and Mathematica.  In the following, we describe the method in detail.

\subsection{Feynman diagrams and simplification}
We use QGRAF \cite{qgraf} to generate the Feynman diagrams.  We find that there are 4 diagrams at tree level, 108 at one loop and 2362 at two loops, leaving
out tadpole and self energy corrections to the external legs.  The output of the QGRAF is then converted to the format that is suitable
for further symbolic manipulation using FORM and Mathematica.  A set of FORM routines is used to perform simplification of the squared 
matrix elements involving gluon and spin-2 resonance polarization and color sums.     
We have used Feynman gauge throughout and for the external on-shell gluon legs, physical polarizations are summed using
\begin{equation}
 \sum_s \varepsilon^{\mu}(p_i,s) \varepsilon^{\nu *}(p_i,s) = -g^{\mu\nu} + \frac{p_i^{\mu} q_i^{\nu} + q_i^{\mu} p_i^{\nu}}{p.q}
\end{equation}
where, $p_i$ is the $i^{th}$-gluon momentum and $q_i$ is the corresponding reference momentum which is an arbitrary light-like 4-vector. 
We choose $q_1 = p_2$, $q_2 = p_1$ and $q_3 = p_1$ for simplicity.  
The spin-2 polarization sum in $d$ dimensions is given by \cite{di-ll1}
\begin{eqnarray}
 B^{\mu \nu ; \rho \sigma}(q) &=& \left( g^{\mu\rho} - \frac{\, \, q^{\mu} q^{\rho} }{q.q} \right)  \left(g^{\nu\sigma} - \frac{\, \,q^{\nu} q^{\sigma} }{q.q} \right) 
                           + \left( g^{\mu\sigma} - \frac{\, \,q^{\mu} q^{\sigma} }{q.q} \right)  \left(g^{\nu\rho} - \frac{\, \,q^{\nu} q^{\rho} }{q.q} \right) 
\nonumber\\[1ex]
&& -\, \, \frac{2}{d-1} \left( g^{\mu\nu} - \frac{\, \,q^{\mu} q^{\nu} }{q.q} \right) \left( g^{\rho\sigma} - \frac{\, \,q^{\rho} q^{\sigma} }{q.q} \right)
\end{eqnarray}
\subsection{Reduction of tensor integrals}
Beyond leading order, the resulting expressions involve tensorial one and two loop integrals which need to be reduced to a set of scalar 
integrals using a convenient tensorial reduction procedure.  Tensorial reduction is quite straightforward at one loop level but not so
at two loop level and beyond.  In addition, finding a minimal set of integrals after the tensorial reduction
is important to achieve the task with large number of Feynman integrals.  A systematic approach to deal with higher rank tensor integrals
and large number of scalar integrals is to use Integration by parts (IBP) \cite{chet} and Lorentz invariant (LI) \cite{gr} identities.   

The IBP identities follow from the fact that 
in the dimensional regularization, the integral of the total derivative with respect to any loop momenta vanishes, that is
\begin{equation}
 \int \frac{d^d k_1}{(2 \pi)^d} \cdot \cdot \cdot \int \frac{d^d k_L}{(2 \pi)^d} \frac{\partial}{\partial k_i} \cdot \left( v_j \frac{1}{\prod_l \cD_l^{n_l} } \right) = 0
\end{equation}
where $n_l$ is an element of $\vec n = (n_1,\cdot \cdot \cdot,n_N)$ with $n_l \in Z$, $L$ is the number of loops and $\cD_l$s are propagators which depend on the loop and external momenta and also masses.
The four vector $v_j^\mu$ can be both loop and external momenta.
Performing the differentiation on the left hand side and expressing the scalar products of $k_i$ and $p_j$ linearly in terms of
$\cD_l$'s, one obtains the IBP identities given by
\begin{equation}
\sum_i a_i J(b_{i,1} + n_1, ... , b_{i,N} + n_N ) = 0 
\end{equation}
where
\begin{equation}
 J(\vec m) = J(m_1,\cdot \cdot \cdot ,m_N)=
 \int \frac{d^d k_1}{(2 \pi)^d} \cdot \cdot \cdot \frac{d^d k_L}{(2 \pi)^d} \frac{1}{\prod_l \cD_l^{m_l}}   
\end{equation}
with $b_{i,j} \in \{-1,0,1\}$ and $a_i$ are polynomial in $n_j$.
The LI identities follow from the fact that the loop integrals are invariant under Lorentz transformations of the external momenta, that is
\begin{equation}
 p_i^{\mu} p_j^{\nu} \left(  \sum_k p_{k [\nu} \frac{\partial}{\partial p_k^{\mu]}}  \right) J(\vec n) = 0.
\end{equation}
While these identities are useful to express the tensorial integrals in terms of a set of master integrals,
in practice, the computation becomes tedious due to the appearance of large variety of Feynman integrals involving different set of propagators
each requiring a set of IBP and LI identities independently.  We have reduced such varieties to a few by shifting the loop momenta suitably using
an in-house algorithm which uses FORM.  For one-loop diagrams, we can express each Feynman integral to contain terms from one of the following
three sets:
\begin{eqnarray}\label{onebasis}
\{ \cD, \hspace{0.1cm} \cD_{1}, \hspace{0.1cm} \cD_{12}, \hspace{0.1cm} \cD_{123} \} \, ,
\{ \cD, \hspace{0.1cm} \cD_{2}, \hspace{0.1cm} \cD_{23}, \hspace{0.1cm} \cD_{123} \} \, ,\{ \cD, \hspace{0.1cm} \cD_{3}, \hspace{0.1cm} \cD_{31}, \hspace{0.1cm} \cD_{123} \}
\end{eqnarray}
where, 
\begin{eqnarray}
 \cD = k_1^2, \hspace{0.1cm} \cD_{i} = (k_1 - p_i)^2, 
\hspace{0.1cm} \cD_{ij} = (k_1 - p_i - p_j)^2, 
\hspace{0.1cm} \cD_{ijk} = (k_1 - p_i - p_j - p_k)^2 
\end{eqnarray}
In each set in eqn.(\ref{onebasis}), $\cD$'s are linearly independent and form a complete basis in the sense that 
any Lorentz invariant $k_1\cdot p_i$ can be expressed in terms of $\cD$'s.
At two loops, there are nine independent Lorentz invariants involving loop momenta $k_1$ and $k_2$, namely $\{ ( k_\alpha \cdot k_\beta ), 
(k_\alpha \cdot p_i) \}, \alpha,\beta = 1, 2;\hspace{0.1cm} i = 1,...,3 $.  After appropriate shifting of loop momenta, we can
express each two loop Feynman integral to contain terms belonging to one of the following six sets:  

\begin{eqnarray}\label{twobasis}
 &&\{ \cD_0, \hspace{0.1cm} \cD_1, \hspace{0.1cm} \cD_2, \hspace{0.1cm} \cD_{1;1}, \hspace{0.1cm} \cD_{2;1}, \hspace{0.1cm}
 \cD_{1;12}, \hspace{0.1cm} \cD_{2;12}, \hspace{0.1cm} \cD_{1;123}, \hspace{0.1cm} \cD_{2;123} \}
\nonumber\\[1ex]
 &&\{ \cD_0, \hspace{0.1cm} \cD_1, \hspace{0.1cm} \cD_2, \hspace{0.1cm} \cD_{1;2}, \hspace{0.1cm} \cD_{2;2}, \hspace{0.1cm}
 \cD_{1;23}, \hspace{0.1cm} \cD_{2;23}, \hspace{0.1cm} \cD_{1;123}, \hspace{0.1cm} \cD_{2;123} \}
\nonumber\\[1ex]
 &&\{ \cD_0, \hspace{0.1cm} \cD_1, \hspace{0.1cm} \cD_2, \hspace{0.1cm} \cD_{1;3}, \hspace{0.1cm} \cD_{2;3}, \hspace{0.1cm}
 \cD_{1;31}, \hspace{0.1cm} \cD_{2;31}, \hspace{0.1cm} \cD_{1;123}, \hspace{0.1cm} \cD_{2;123} \}
\nonumber\\[1ex]
 &&\{ \cD_0, \hspace{0.1cm} \cD_1, \hspace{0.1cm} \cD_2, \hspace{0.1cm} \cD_{1;1}, \hspace{0.1cm} \cD_{2;1}, \hspace{0.1cm}
 \cD_{0;3}, \hspace{0.1cm} \cD_{1;12}, \hspace{0.1cm} \cD_{2;12}, \hspace{0.1cm} \cD_{1;123} \}
\nonumber\\[1ex]
 &&\{ \cD_0, \hspace{0.1cm} \cD_1, \hspace{0.1cm} \cD_2, \hspace{0.1cm} \cD_{1;2}, \hspace{0.1cm} \cD_{2;2}, \hspace{0.1cm}
 \cD_{0;1}, \hspace{0.1cm} \cD_{1;23}, \hspace{0.1cm} \cD_{2;23}, \hspace{0.1cm} \cD_{1;123} \}
\nonumber\\[1ex]
 &&\{ \cD_0, \hspace{0.1cm} \cD_1, \hspace{0.1cm} \cD_2, \hspace{0.1cm} \cD_{1;3}, \hspace{0.1cm} \cD_{2;3}, \hspace{0.1cm}
 \cD_{0;2}, \hspace{0.1cm} \cD_{1;31}, \hspace{0.1cm} \cD_{2;31}, \hspace{0.1cm} \cD_{1;123} \}
\end{eqnarray}
where,
\begin{eqnarray}
 && \cD_0 = (k_1 - k_2)^2, \hspace{0.1cm} \cD_{\alpha} = k_{\alpha}^2, \hspace{0.1cm} \cD_{\alpha;i} = (k_{\alpha} - p_i)^2, 
\hspace{0.1cm} \cD_{\alpha; ij} = (k_{\alpha} - p_i - p_j)^2, 
\nonumber\\[1ex]
&& \cD_{0;i} = (k_1 - k_2 - p_i)^2, \hspace{0.1cm} \cD_{\alpha;ijk} = (k_{\alpha} - p_i - p_j - p_k)^2
\end{eqnarray}

Given the fewer number of sets (eqns.(\ref{onebasis}) \& (\ref{twobasis})), it is easier to use IBP and LI identities using Laporta algorithm \cite{Laporta:2001dd}.  These identities can be
generated using publicly available packages such as AIR \cite{Anastasiou:2004vj}, FIRE \cite{Smirnov:2008iw}, REDUZE \cite{Studerus:2009ye, vonManteuffel:2012np}, LiteRed \cite{Lee:2012cn, Lee:2013mka} etc.
For our computation, we use a Mathematica based package 
LiteRedV1.51 along with MintV1.1 \cite{Nason:2007vt}. This package has the option to exploit symmetry relations within each set and 
also among different sets.

\subsection{Master integrals}
\begin{figure}
 \centering 
 \includegraphics[width=0.82\textwidth]{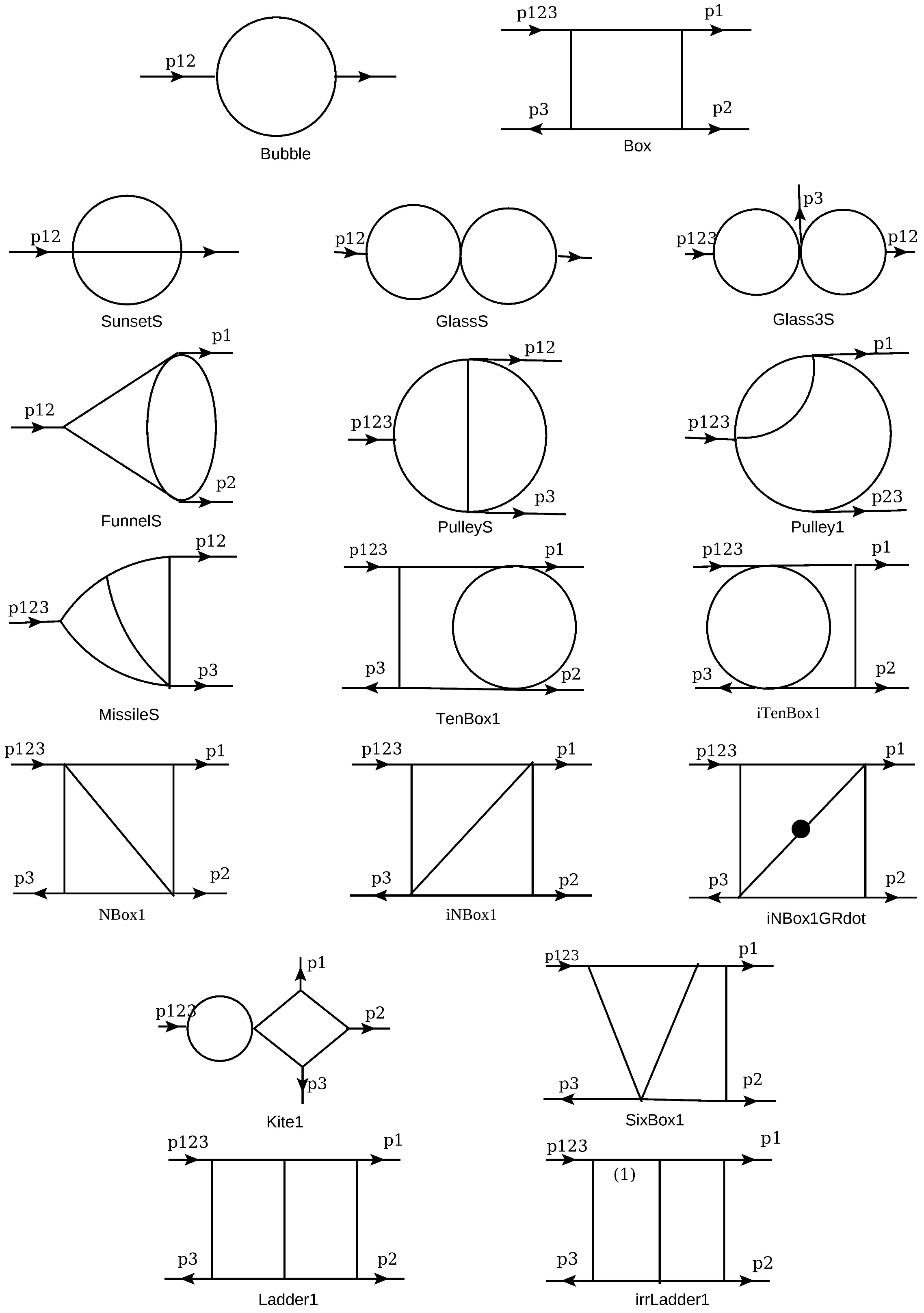}
\caption{Planar topologies of master integrals}
\label{fig:planar1}
\end{figure}
\begin{figure}
\centering 
\includegraphics[width=0.85\textwidth]{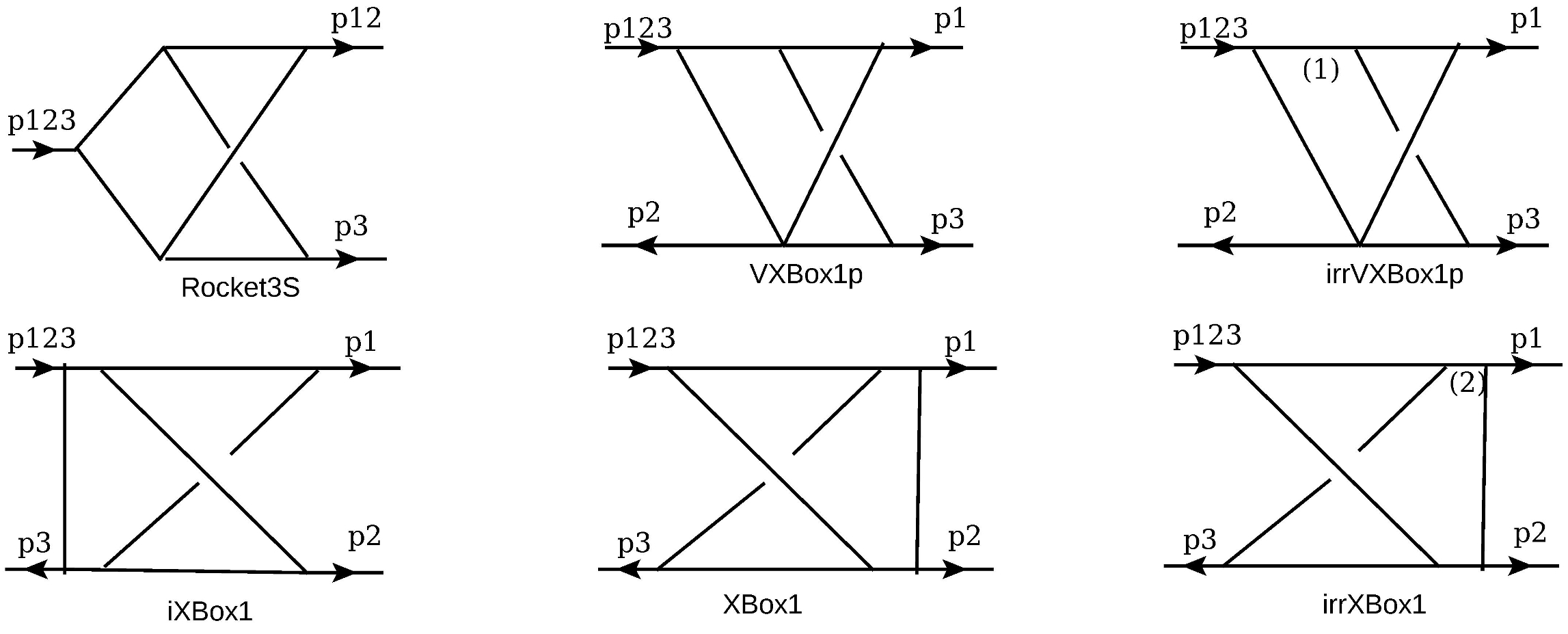}
\caption{Non-Planar topologies of master integrals}
\label{fig:planar2}
\end{figure}

Using these IBP and LI identities, we reduce all the integrals that appear in our computation to a minimal set of master integrals.  
For one loop, we get two topologically different master integrals namely box and bubble, see Fig.(\ref{fig:planar1}) and we find that the master integrals with three propagators are absent. For two loops, 
we encounter 16 planar and 5 non-planar topologies of master integrals. These master integrals can be related to those that were computed by
Gehrmann and Remiddi in their seminal papers \cite{gr1, gr2}.  In particular, our set of master integrals does not contain
integrals with irreducible numerator, instead we have higher power of propagators.  We use IBP and LI identities to express our set of master integrals
to those of \cite{gr1, gr2}. All the master integrals are drawn in Fig.(\ref{fig:planar1}) and Fig.(\ref{fig:planar2}) up to different permutations of 
the external momenta $p_1, p_2$ and $p_3$.
We also observe that some topologies like iXBox1 given in Fig.(\ref{fig:planar2}) are absent in our final result and 
find some new topologies namely Glass3S and Kite1 given in Fig.(\ref{fig:planar1}) which are absent in the \cite{gr1, gr2} and those are simply a product of two one loop topologies.

Substituting the master integrals computed by Gehrmann and Remiddi \cite{gr1, gr2} in terms of HPLs and 2dHPLs, we obtain the
unrenormalized matrix elements $\langle \unM^{(0)} |\unM^{(1)} \rangle$ and $\langle \unM^{(0)} |\unM^{(2)} \rangle$.  We use shuffle algebra to express product of
lower weight HPLs as a sum of higher weight HPLs.  In the next section we present our results.

\section{Results}

The UV renormalized matrix elements $\langle {\cal M}^{(0)} |{\cal M}^{(1)} \rangle$ and $\langle {\cal M}^{(0)} |{\cal M}^{(2)} \rangle$ 
are computed using the unrenormalized counter parts through
\begin{eqnarray}\label{resrenorm}
\langle \rnM^{(0)} |\rnM^{(1)} \rangle = \left(  \frac{1}{\mu^{\eps}_R} \right)^2 && \Big[ \langle \unM^{(0)} |\unM^{(1)} \rangle  
+ \mu^{\eps}_R  \frac{r_1}{2} \langle \unM^{(0)} |\unM^{(0)} \rangle \Big]\, ,
\nonumber\\[1ex]
\langle \rnM^{(0)} |\rnM^{(2)} \rangle = \left(  \frac{1}{\mu^{\eps}_R} \right)^3 && \Big[ \langle \unM^{(0)} |\unM^{(2)} \rangle  
+ \mu^{\eps}_R  \frac{3r_1}{2} \langle \unM^{(0)} |\unM^{(1)} \rangle 
\nonumber\\[1ex]
&&+ \hspace{0.1cm} \mu^{2\eps}_R \left( \frac{r_2}{2} - \frac{r_1^2}{8}  \right) \langle \unM^{(0)} |\unM^{(0)} \rangle \Big].
\end{eqnarray}
Using eqn.(\ref{catani1}), we can also express the renormalized matrix elements in terms of ${\bf I}^{(i)}_g(\epsilon)$, given by 
\begin{eqnarray}\label{rescat}
\langle \rnM^{(0)} |\rnM^{(1)}  \rangle &=& 2 \hspace{0.1cm} {\bf{I}}_{g}^{(1)} (\eps) \hspace{0.1cm}  \langle \rnM^{(0)}|\rnM^{(0)} \rangle
+  \langle \rnM^{(0)}|\rnM^{(1)fin}  \rangle \, ,
\nonumber\\[2ex] 
\langle \rnM^{(0)} |\rnM^{(2)} \rangle &=& 2 \hspace{0.1cm} {\bf{I}}_{g}^{(1)} (\eps) \hspace{0.1cm} \langle \rnM^{(0)} |\rnM^{(1)} \rangle
+ 4 \hspace{0.1cm} {\bf{I}}^{(2)}_{g} (\eps) \hspace{0.1cm}  \langle \rnM^{(0)} |\rnM^{(0)} \rangle
+ \langle \rnM^{(0)} |\rnM^{(2)fin} \rangle \, .
\end{eqnarray}
Expanding the right hand side of equations (\ref{resrenorm} \& \ref{rescat}) in powers of $\epsilon$ and comparing their coefficients of ${\cal O}(\epsilon^0)$, 
we obtain $ \langle \rnM^{(0)}|\rnM^{(1)fin}  \rangle$ and $ \langle \rnM^{(0)}|\rnM^{(2)fin}  \rangle$.   

We find that all the poles in $\epsilon$ resulting from the soft and collinear partons in eqn.(\ref{resrenorm}) are in agreement with
those of eqn.(\ref{rescat}).  
This serves as an important check on our result.  In addition, it establishes the universal structure of infrared
poles in QCD amplitudes involving tensorial operator insertion.  We also observe that the contributions resulting from the gauge fixing term
in eqn.(\ref{intlag}) cancel exactly with those of ghosts confirming the gauge independence of our result.  As we anticipated, the eqn.(\ref{resrenorm})
does not require any over all operator renormalization constant due to the conservation of energy momentum tensor 
and it can be made UV finite through strong coupling constant renormalization (eqn.(\ref{renas})) alone.    Below we present our final results

\begin{align}
&\langle \rnM^{(0)} |\rnM^{(0)} \rangle = {\cal F}_{h} \: {\cal A}^{(0)} \, , 
\nonumber\\[1ex]
&\langle \rnM^{(0)} |\rnM^{(1)fin} \rangle = {\cal F}_{h} \Bigg\{ - \frac{ \beta_0}{2} ~ {\cal A}^{(0)} ~ \ln \left( -\frac{Q^2}{\mu^2} \right) 
 + \left( {\cal A}^{(1)}_1 \zeta_2 + {\cal A}^{(1)}_2 \right)  \Bigg\}  \, ,
\nonumber\\[1ex]
&\langle \rnM^{(0)} |\rnM^{(2)fin} \rangle =  {\cal F}_{h}  \Bigg\{ \frac{3 \beta_0^2}{8} ~ {\cal A}^{(0)} ~ \ln^2 \left( -\frac{Q^2}{\mu^2} \right) 
\nonumber\\[1ex]
&{\white \langle \rnM^{(0)} |\rnM^{(2)fin} {\cal F}_{h} \rangle } + \left( - 3~ C_A^2~ \zeta_3~ {\cal A}^{(0)} + {\cal A}^{(2)}_1 \zeta_2 + {\cal A}^{(2)}_2 \right) \ln \left( -\frac{Q^2}{\mu^2} \right) 
\nonumber\\[1ex]
&{\white \langle \rnM^{(0)} |\rnM^{(2)fin} {\cal F}_{h} \rangle } +  \left(   {\cal A}^{(2)}_3 \zeta_2^2 + {\cal A}^{(2)}_4 \zeta_3  + {\cal A}^{(2)}_5 \zeta_2 + {\cal A}^{(2)}_6 \right)  \Bigg\}  
\end{align}

where,

\begin{eqnarray}
 {\cal F}_{h}  &=& 16 \pi^2 \kappa^2  N \Big(  N^2 - 1 \Big) \, ,
\nonumber\\
 {\cal A}^{(0)} &=& \frac{4} {s t u} \left(s^4+2 s^3 (t+u)+3 s^2 \left(t^2+u^2\right)+2 s \
\left(t^3+u^3\right)+\left(t^2+t u+u^2\right)^2\right) \, ,
\nonumber\\
 {\cal A}^{(1)}_{\alpha} &=& {\cal A}_{\alpha ; \ca}^{(1)} ~ C_A + {\cal A}_{\alpha ; n_f}^{(1)} ~ n_f  \, ,
\nonumber\\[1ex]
 {\cal A}^{(2)}_{\alpha} &=& {\cal A}^{(2)}_{\alpha ; \ca^2 } ~ C_A^2 + {\cal A}^{(2)}_{\alpha ; \ca n_f} ~ C_A n_f + {\cal A}^{(2)}_{\alpha ; \cf n_f} ~ C_F n_f + {\cal A}^{(2)}_{\alpha ; n_f^2} ~ n_f^2  \, .
\end{eqnarray}
The coefficients ${\cal A}_{\alpha ;{\cal C}}^{(i)}$ are given in the appendix except ${\cal A}_{6 ;C_A n_f}^{(2)}$, ${\cal A}_{6 ;C_F n_f}^{(2)}$ \& ${\cal A}_{6 ;n_f^2}^{(2)}$ which can be found
in the files A6Canf, A6Cfnf and A6nf2 respectively attached with this arXiv submission.

\section{Conclusion}

The computation of one and two loop QCD results for the process $h\rightarrow g+g+g$ is 
presented.  We use dimensional regularization to regulate both UV and IR divergences. 
Our result is very general in the sense that it can be used for any scattering process
involving production of a massive spin-2 particle that has a universal coupling with the SM fields.  
We can use it to study the production of a jet with missing energy due to KK
graviton escaping the detector or a process with resonant massive spin-2 particle in association with a jet.
Since the spin-2 field couples with the SM ones through rank-2 tensor, 
we not only encounter large number of Feynman diagrams but also the formidable challenge of reducing
one and two loop Feynman integrals with high powers of loop momenta to scalar ones.  The IBP and LI
identities reduce all such integrals to only few master integrals that were already
computed by Gehrmann and Remiddi.  This computation is the first of the kind involving
four point function at two loop level in QCD with tensorial insertion and one massive
external state.   We find that no overall UV renormalization is required
due to the conservation of energy momentum tensor.  
We also find that our results exhibit the right IR structure confirming
the factorization property of QCD amplitude even with tensorial insertion.

\section*{Acknowledgement}

MM, TA and NR thank for the hospitality provided by the Institute of Mathematical Sciences (IMSc)
where the work was carried out.  We thank the staff of IMSc computer center
for their help.  We sincerely thank T. Gehrmann for providing
us the master integrals required for our computation.  We thank R. N. Lee for his help 
with LiteRed, and A.V. Smirnov and V.A. Smirnov for their suggestions with FIRE.   
Finally, we would like to thank K. Hasegawa, M. K. Mandal and L. Tancredi for discussions and their valuable suggestions. 
NR, TA and MM also like to thank their parents and siblings for their wonderful love and continuous support.

\appendix
\section{Harmonic polylogarithms}
\setcounter{equation}{0}

In this section, we briefly describe the definition and properties of HPL and 2dHPL. HPL is represented by $H(\vec{m}_w;y)$ 
with a $w$-dimensional vector $\vec{m}_w$ of parameters and its argument $y$. $w$ is called the weight of the HPL. The elements of $\vec{m}_w$ belong to $\{ 1, 0, -1 \}$ through which 
the following rational functions are represented  
\begin{equation}
 f(1;y) \equiv \frac{1}{1-y}, \qquad f(0;y) \equiv \frac{1}{y},  \qquad f(-1;y) \equiv \frac{1}{1+y} \, .
\end{equation}
The weight 1 $(w = 1)$ HPLs are defined by
\begin{equation}
 H(1, y) \equiv - \ln (1 - y), \qquad  H(0, y) \equiv \ln y, \qquad  H(-1, y) \equiv \ln (1 + y) \, .
\end{equation}
For $w > 1$, $H(m, \vec{m}_{w};y)$ is defined by 
\begin{equation}\label{1dhpl}
 H(m, \vec{m}_w;y) \equiv \int_0^y dx ~ f(m, x) ~ H(\vec{m}_w;x),  \qquad \qquad  m \in 0, \pm 1  \, .
\end{equation}
The 2dHPLs are defined in the same way as eqn.(\ref{1dhpl}) with the new elements $\{ 2, 3 \}$ in $\vec{m}_w$ representing a new 
class of rational functions
\begin{equation}
 f(2;y) \equiv f(1-z;y) \equiv \frac{1}{1-y-z}, \qquad f(3;y) \equiv f(z;y) \equiv \frac{1}{y+z} 
\end{equation}
and correspondingly with the weight 1 $(w = 1)$ 2dHPLs
\begin{equation}
 H(2, y) \equiv - \ln \Big(1 - \frac{y}{1-z} \Big), \qquad  H(3, y) \equiv \ln \Big( \frac{y+z}{z} \Big) \, .
\end{equation}

\subsection*{Properties}
\underline{Shuffle algebra} : A product of two HPL with weights $w_1$ and $w_2$ of the same argument $y$ is a combination of HPLs with weight
$(w_1 + w_2)$ and argument $y$, such that all possible permutations of the elements of $\vec{m}_{w_1}$ and $\vec{m}_{w_2}$ are considered preserving the 
relative orders of the elements of $\vec{m}_{w_1}$ and $\vec{m}_{w_2}$,
\begin{equation}
 H(\vec{m}_{w_1};y) H(\vec{m}_{w_2};y) = \sum_{\text{\tiny $\vec{m}_{w} = \vec{m}_{w_1} \uplus \vec{m}_{w_2}$}}  H(\vec{m}_{w};y).
\end{equation}
\underline{Integration-by-parts identities} : The ordering of the elements of $\vec{m}_{w}$ in an HPL with weight $w$ and argument $y$ can be reversed using 
integration-by-parts and in the process, some products of two HPLs are generated in the following way
\begin{eqnarray}
 H(\vec{m}_{w};y) \equiv H(m_1, m_2, ... , m_w; y ) &=& H(m_1, y) H(m_2, ... , m_w; y )
\nonumber\\
 &-& H(m_2, m_1, y) H(m_3, ... , m_w; y )
\nonumber\\
 &+& ... + (-1)^{w+1} H ( m_w, ... , m_2, m_1 ; y ) \, .
\end{eqnarray}

\section{One-loop coefficients}
\setcounter{equation}{0}

{\tiny
\begin{dgroup*}
\begin{dmath*}
 {\cal A}^{(1)}_{1 ; C_A} =  -\frac{4}{s t u} \left(s^4+2 s^3 (t+u)+3 s^2 \left(t^2+u^2\right)+2 s \
\left(t^3+u^3\right)+t^4+u^4\right)
\end{dmath*}

\begin{dmath*}
 {\cal A}^{(1)}_{1 ; n_f} =  -\frac{2}{s} \left(t^2+u^2\right)
\end{dmath*}

\begin{dmath*}
 {\cal A}^{(1)}_{2 ; C_A} =   \Bigg\{-6 (t+u)^4 \Big(11 s^8+22 (3 t+u) s^7+\Big(187 t^2+64 u t+33 \
u^2\Big) s^6+\Big(330 t^3+60 u t^2+78 u^2 t+22 u^3\Big) s^5+\Big(396 \
t^4+14 u t^3+51 u^2 t^2+36 u^3 t+11 u^4\Big) s^4+2 t \Big(165 t^4+7 u \
t^3+6 u^2 t^2+7 u^3 t+10 u^4\Big) s^3+t^2 \Big(187 t^4+60 u t^3+51 u^2 \
t^2+14 u^3 t+24 u^4\Big) s^2+2 t^3 \Big(33 t^4+32 u t^3+39 u^2 t^2+18 u^3 \
t+10 u^4\Big) s+11 t^4 \Big(t^2+u t+u^2\Big)^2\Big) H(0,y) (s+u)^4-36 \
(s+t)^4 (t+u)^4 \Big(s^4+2 (t+u) s^3+3 \Big(t^2+u^2\Big) s^2+2 \
\Big(t^3+u^3\Big) s+t^4+u^4\Big) H(0,y) H(0,z) (s+u)^4+6 (s+t)^4 \
\Big(\Big(11 t^4+20 u t^3+24 u^2 t^2+20 u^3 t+11 u^4\Big) s^4+2 \Big(11 \
t^5+18 u t^4+7 u^2 t^3+7 u^3 t^2+18 u^4 t+11 u^5\Big) s^3+3 (t+u)^2 \
\Big(11 t^4+4 u t^3-2 u^2 t^2+4 u^3 t+11 u^4\Big) s^2+2 (t+u)^3 \Big(11 \
t^4-u t^3-u^3 t+11 u^4\Big) s+11 (t+u)^4 \Big(t^2+u t+u^2\Big)^2\Big) \
H(1,z) (s+u)^4+36 (s+t)^4 (t+u)^4 \Big(s^4+2 t s^3+3 t^2 s^2+2 t^3 \
s+\Big(t^2+u t+u^2\Big)^2\Big) H(0,y) H(1,z) (s+u)^4+6 (s+t)^4 \
\Big(\Big(11 t^4+20 u t^3+24 u^2 t^2+20 u^3 t+11 u^4\Big) s^4+2 \Big(11 \
t^5+18 u t^4+7 u^2 t^3+7 u^3 t^2+18 u^4 t+11 u^5\Big) s^3+3 (t+u)^2 \
\Big(11 t^4+4 u t^3-2 u^2 t^2+4 u^3 t+11 u^4\Big) s^2+2 (t+u)^3 \Big(11 \
t^4-u t^3-u^3 t+11 u^4\Big) s+11 (t+u)^4 \Big(t^2+u t+u^2\Big)^2\Big) \
H(2,y) (s+u)^4+36 (s+t)^4 (t+u)^4 \Big(s^4+2 u s^3+3 u^2 s^2+2 u^3 \
s+\Big(t^2+u t+u^2\Big)^2\Big) H(0,z) H(2,y) (s+u)^4-36 (s+t)^4 (t+u)^4 \
\Big(2 s^4+2 (t+u) s^3+3 \Big(t^2+u^2\Big) s^2+2 \Big(t^3+u^3\Big) \
s+2 \Big(t^2+u t+u^2\Big)^2\Big) H(1,z) H(3,y) (s+u)^4-36 (s+t)^4 \
(t+u)^4 \Big(s^4+2 t s^3+3 t^2 s^2+2 t^3 s+\Big(t^2+u \
t+u^2\Big)^2\Big) H(0,1,z) (s+u)^4+36 (s+t)^4 (t+u)^4 \Big(s^4+2 t \
s^3+3 t^2 s^2+2 t^3 s+\Big(t^2+u t+u^2\Big)^2\Big) H(0,2,y) (s+u)^4-36 \
(s+t)^4 (t+u)^4 \Big(2 s^4+2 (2 t+u) s^3+3 \Big(2 t^2+u^2\Big) s^2+2 \
\Big(2 t^3+u^3\Big) s+2 t^4+2 u^4+2 t u^3+3 t^2 u^2+2 t^3 u\Big) \
H(1,0,y) (s+u)^4-36 (s+t)^4 (t+u)^4 \Big(s^4+2 (t+u) s^3+3 \
\Big(t^2+u^2\Big) s^2+2 \Big(t^3+u^3\Big) s+t^4+u^4\Big) H(1,0,z) \
(s+u)^4+36 (s+t)^4 (t+u)^4 \Big(s^4+2 t s^3+3 t^2 s^2+2 t^3 s+\Big(t^2+u \
t+u^2\Big)^2\Big) H(2,0,y) (s+u)^4-36 (s+t)^4 (t+u)^4 \Big(2 s^4+2 \
(t+u) s^3+3 \Big(t^2+u^2\Big) s^2+2 \Big(t^3+u^3\Big) s+2 \Big(t^2+u \
t+u^2\Big)^2\Big) H(3,2,y) (s+u)^4
\end{dmath*}
\begin{dmath*}
{\white=}
-2 (s+t) (t+u) \Big(203 (t+u)^3 \
s^{10}+2 \Big(506 t^4+1988 u t^3+2973 u^2 t^2+1988 u^3 t+506 u^4\Big) \
s^9+3 \Big(807 t^5+3709 u t^4+7123 u^2 t^3+7123 u^3 t^2+3709 u^4 t+807 \
u^5\Big) s^8+3 \Big(1208 t^6+6186 u t^5+13887 u^2 t^4+17836 u^3 t^3+13887 \
u^4 t^2+6186 u^5 t+1208 u^6\Big) s^7+\Big(3624 t^7+21596 u t^6+54318 u^2 \
t^5+80567 u^3 t^4+80567 u^4 t^3+54318 u^5 t^2+21596 u^6 t+3624 u^7\Big) \
s^6+3 \Big(807 t^8+6186 u t^7+18106 u^2 t^6+29776 u^3 t^5+34116 u^4 \
t^4+29776 u^5 t^3+18106 u^6 t^2+6186 u^7 t+807 u^8\Big) s^5+\Big(1012 \
t^9+11127 u t^8+41661 u^2 t^7+80567 u^3 t^6+102348 u^4 t^5+102348 u^5 \
t^4+80567 u^6 t^3+41661 u^7 t^2+11127 u^8 t+1012 u^9\Big) s^4+\Big(203 \
t^{10}+3976 u t^9+21369 u^2 t^8+53508 u^3 t^7+80567 u^4 t^6+89328 u^5 \
t^5+80567 u^6 t^4+53508 u^7 t^3+21369 u^8 t^2+3976 u^9 t+203 u^{10}\Big) \
s^3+3 t u \Big(203 t^9+1982 u t^8+7123 u^2 t^7+13887 u^3 t^6+18106 u^4 \
t^5+18106 u^5 t^4+13887 u^6 t^3+7123 u^7 t^2+1982 u^8 t+203 u^9\Big) \
s^2+t^2 u^2 (t+u)^2 \Big(609 t^6+2758 u t^5+5002 u^2 t^4+5796 u^3 t^3+5002 \
u^4 t^2+2758 u^5 t+609 u^6\Big) s+t^3 u^3 (t+u)^3 \Big(203 t^4+403 u \
t^3+603 u^2 t^2+403 u^3 t+203 u^4\Big)\Big) (s+u)-6 (s+t)^4 (t+u)^4 \
\Big(11 s^8+22 (t+3 u) s^7+\Big(33 t^2+64 u t+187 u^2\Big) s^6+\Big(22 \
t^3+78 u t^2+60 u^2 t+330 u^3\Big) s^5+\Big(11 t^4+36 u t^3+51 u^2 t^2+14 \
u^3 t+396 u^4\Big) s^4+2 u \Big(10 t^4+7 u t^3+6 u^2 t^2+7 u^3 t+165 \
u^4\Big) s^3+u^2 \Big(24 t^4+14 u t^3+51 u^2 t^2+60 u^3 t+187 u^4\Big) \
s^2+2 u^3 \Big(10 t^4+18 u t^3+39 u^2 t^2+32 u^3 t+33 u^4\Big) s+11 u^4 \
\Big(t^2+u t+u^2\Big)^2\Big) H(0,z) \Bigg\} \Big/ \Big( {9 s t (s+t)^4 u (s+u)^4 (t+u)^4} \Big)
\end{dmath*}

\begin{dmath*}
 {\cal A}^{(1)}_{2 ; n_f} =   \Bigg\{-18 u (s+u)^4 (t+u)^4 \Big(s^2-t s+t^2+u^2\Big) H(1,0,y) (s+t)^5+6 \
(t+u)^4 \Big(2 s^8+4 (t+3 u) s^7+2 \Big(3 t^2+5 u t+17 u^2\Big) \
s^6+\Big(4 t^3+15 u t^2+6 u^2 t+60 u^3\Big) s^5+2 \Big(t^4+12 u^2 t^2-2 \
u^3 t+36 u^4\Big) s^4+2 u \Big(t^4-2 u t^3+15 u^2 t^2-2 u^3 t+30 \
u^4\Big) s^3+2 u^2 \Big(3 t^4-2 u t^3+12 u^2 t^2+3 u^3 t+17 u^4\Big) \
s^2+\Big(12 u^7+10 t u^6+15 t^2 u^5+2 t^4 u^3\Big) s+2 u^4 \Big(t^2+u \
t+u^2\Big)^2\Big) H(0,z) (s+t)^4-18 t u (s+u)^4 (t+u)^4 \
\Big(t^2+u^2\Big) H(0,y) H(0,z) (s+t)^4-6 (s+u)^4 \Big(2 \Big(t^4+u \
t^3+3 u^2 t^2+u^3 t+u^4\Big) s^4+4 \Big(t^5-u^2 t^3-u^3 t^2+u^5\Big) \
s^3+3 (t+u)^2 \Big(2 t^4+u t^3+4 u^2 t^2+u^3 t+2 u^4\Big) s^2+2 (t+u)^3 \
\Big(2 t^4-u t^3-u^3 t+2 u^4\Big) s+2 (t+u)^4 \Big(t^2+u \
t+u^2\Big)^2\Big) H(1,z) (s+t)^4+18 s u (s+u)^4 (t+u)^4 \
\Big(s^2+u^2\Big) H(0,y) H(1,z) (s+t)^4-6 (s+u)^4 \Big(2 \Big(t^4+u \
t^3+3 u^2 t^2+u^3 t+u^4\Big) s^4+4 \Big(t^5-u^2 t^3-u^3 t^2+u^5\Big) \
s^3+3 (t+u)^2 \Big(2 t^4+u t^3+4 u^2 t^2+u^3 t+2 u^4\Big) s^2+2 (t+u)^3 \
\Big(2 t^4-u t^3-u^3 t+2 u^4\Big) s+2 (t+u)^4 \Big(t^2+u \
t+u^2\Big)^2\Big) H(2,y) (s+t)^4+18 s t \Big(s^2+t^2\Big) (s+u)^4 \
(t+u)^4 H(0,z) H(2,y) (s+t)^4-18 s (s+u)^4 (t+u)^5 \Big(s^2+t^2+u^2-t \
u\Big) H(1,z) H(3,y) (s+t)^4-18 s u (s+u)^4 (t+u)^4 \Big(s^2+u^2\Big) \
H(0,1,z) (s+t)^4+18 s u (s+u)^4 (t+u)^4 \Big(s^2+u^2\Big) H(0,2,y) \
(s+t)^4-18 t u (s+u)^4 (t+u)^4 \Big(t^2+u^2\Big) H(1,0,z) (s+t)^4+18 s u \
(s+u)^4 (t+u)^4 \Big(s^2+u^2\Big) H(2,0,y) (s+t)^4
\end{dmath*}
\begin{dmath*}
{\white=}
-18 s (s+u)^4 (t+u)^5 \
\Big(s^2+t^2+u^2-t u\Big) H(3,2,y) (s+t)^4+2 (s+u) (t+u) \Big(35 (t+u)^3 \
s^{10}+2 \Big(86 t^4+335 u t^3+507 u^2 t^2+335 u^3 t+86 u^4\Big) s^9+3 \
\Big(135 t^5+607 u t^4+1177 u^2 t^3+1177 u^3 t^2+607 u^4 t+135 u^5\Big) \
s^8+3 \Big(200 t^6+990 u t^5+2211 u^2 t^4+2860 u^3 t^3+2211 u^4 t^2+990 u^5 \
t+200 u^6\Big) s^7+\Big(600 t^7+3428 u t^6+8436 u^2 t^5+12593 u^3 \
t^4+12593 u^4 t^3+8436 u^5 t^2+3428 u^6 t+600 u^7\Big) s^6+3 \Big(135 \
t^8+990 u t^7+2812 u^2 t^6+4612 u^3 t^5+5328 u^4 t^4+4612 u^5 t^3+2812 u^6 \
t^2+990 u^7 t+135 u^8\Big) s^5
\end{dmath*}
\intertext{}
\begin{dmath*}
{\white=}
+\Big(172 t^9+1821 u t^8+6633 u^2 t^7+12593 \
u^3 t^6+15984 u^4 t^5+15984 u^5 t^4+12593 u^6 t^3+6633 u^7 t^2+1821 u^8 \
t+172 u^9\Big) s^4+\Big(35 t^{10}+670 u t^9+3531 u^2 t^8+8580 u^3 \
t^7+12593 u^4 t^6+13836 u^5 t^5+12593 u^6 t^4+8580 u^7 t^3+3531 u^8 t^2+670 \
u^9 t+35 u^{10}\Big) s^3+3 t u \Big(35 t^9+338 u t^8+1177 u^2 t^7+2211 \
u^3 t^6+2812 u^4 t^5+2812 u^5 t^4+2211 u^6 t^3+1177 u^7 t^2+338 u^8 t+35 \
u^9\Big) s^2+t^2 u^2 (t+u)^2 \Big(105 t^6+460 u t^5+796 u^2 t^4+918 u^3 \
t^3+796 u^4 t^2+460 u^5 t+105 u^6\Big) s+t^3 u^3 (t+u)^3 \Big(35 t^4+67 u \
t^3+99 u^2 t^2+67 u^3 t+35 u^4\Big)\Big) (s+t)+6 (s+u)^4 (t+u)^4 \Big(2 \
s^8+4 (3 t+u) s^7+2 \Big(17 t^2+5 u t+3 u^2\Big) s^6+\Big(60 t^3+6 u \
t^2+15 u^2 t+4 u^3\Big) s^5+2 \Big(36 t^4-2 u t^3+12 u^2 t^2+u^4\Big) \
s^4+2 t \Big(30 t^4-2 u t^3+15 u^2 t^2-2 u^3 t+u^4\Big) s^3+2 t^2 \
\Big(17 t^4+3 u t^3+12 u^2 t^2-2 u^3 t+3 u^4\Big) s^2+\Big(12 t^7+10 u \
t^6+15 u^2 t^5+2 u^4 t^3\Big) s+2 t^4 \Big(t^2+u t+u^2\Big)^2\Big) \
H(0,y)  \Bigg\}   \Big/ \Big( {9 s t (s+t)^4 u (s+u)^4 (t+u)^4} \Big)
\end{dmath*}
\intertext{}
\end{dgroup*}
 }

\section{Two-loop coefficients}
\setcounter{equation}{0}

{\tiny
\begin{dgroup*}
\begin{dmath*}
 {\cal A}^{(2)}_{1 ; C_A^2} = \frac{11}{s t u} \Big(s^4+2 s^3 (t+u)+3 s^2 \Big(t^2+u^2\Big)+2 s \
\Big(t^3+u^3\Big)+t^4-2 t^3 u-3 t^2 u^2-2 t u^3+u^4\Big) 
\end{dmath*}

\begin{dmath*}
 {\cal A}^{(2)}_{1 ; C_A n_f} = -\frac{1}{s t u} \Big( 2 s^4+4 s^3 (t+u)+6 s^2 \Big(t^2+u^2\Big)+4 s \
\Big(t^3+u^3\Big)+2 t^4-15 t^3 u-6 t^2 u^2-15 t u^3+2 u^4 \Big) 
\end{dmath*}

\begin{dmath*}
 {\cal A}^{(2)}_{1 ; C_F n_f} = 0
\end{dmath*}

\begin{dmath*}
 {\cal A}^{(2)}_{1 ; n_f^2} = -\frac{2}{s} \Big(t^2+u^2\Big)
\end{dmath*}

\intertext{}

\begin{dmath*}
 {\cal A}^{(2)}_{2 ; C_A^2} =   \Bigg\{ 33 (t+u)^4 \Big(11 s^8+22 (3 t+u) s^7+\Big(187 t^2+64 u t+33 \
u^2\Big) s^6+\Big(330 t^3+60 u t^2+78 u^2 t+22 u^3\Big) s^5+\Big(396 \
t^4+14 u t^3+51 u^2 t^2
\end{dmath*}
\begin{dmath*}
{\white=}
+36 u^3 t+11 u^4\Big) s^4+2 t \Big(165 t^4+7 u \
t^3+6 u^2 t^2+7 u^3 t+10 u^4\Big) s^3+t^2 \Big(187 t^4+60 u t^3+51 u^2 \
t^2+14 u^3 t+24 u^4\Big) s^2+2 t^3 \Big(33 t^4+32 u t^3+39 u^2 t^2+18 u^3 \
t+10 u^4\Big) s+11 t^4 \Big(t^2+u t+u^2\Big)^2\Big) H(0,y) \
(s+u)^4+198 (s+t)^4 (t+u)^4 \Big(s^4+2 (t+u) s^3+3 \Big(t^2+u^2\Big) \
s^2+2 \Big(t^3+u^3\Big) s+t^4+u^4\Big) H(0,y) H(0,z) (s+u)^4-33 (s+t)^4 \
\Big(\Big(11 t^4+20 u t^3+24 u^2 t^2+20 u^3 t+11 u^4\Big) s^4+2 \Big(11 \
t^5+18 u t^4+7 u^2 t^3+7 u^3 t^2+18 u^4 t+11 u^5\Big) s^3+3 (t+u)^2 \
\Big(11 t^4+4 u t^3-2 u^2 t^2+4 u^3 t+11 u^4\Big) s^2+2 (t+u)^3 \Big(11 \
t^4-u t^3-u^3 t+11 u^4\Big) s+11 (t+u)^4 \Big(t^2+u t+u^2\Big)^2\Big) \
H(1,z) (s+u)^4-198 (s+t)^4 (t+u)^4 \Big(s^4+2 t s^3+3 t^2 s^2+2 t^3 \
s+\Big(t^2+u t+u^2\Big)^2\Big) H(0,y) H(1,z) (s+u)^4-33 (s+t)^4 \
\Big(\Big(11 t^4+20 u t^3+24 u^2 t^2+20 u^3 t+11 u^4\Big) s^4+2 \Big(11 \
t^5+18 u t^4+7 u^2 t^3+7 u^3 t^2+18 u^4 t+11 u^5\Big) s^3+3 (t+u)^2 \
\Big(11 t^4+4 u t^3-2 u^2 t^2+4 u^3 t+11 u^4\Big) s^2+2 (t+u)^3 \Big(11 \
t^4-u t^3-u^3 t+11 u^4\Big) s+11 (t+u)^4 \Big(t^2+u t+u^2\Big)^2\Big) \
H(2,y) (s+u)^4-198 (s+t)^4 (t+u)^4 \Big(s^4+2 u s^3+3 u^2 s^2+2 u^3 \
s+\Big(t^2+u t+u^2\Big)^2\Big) H(0,z) H(2,y) (s+u)^4+198 (s+t)^4 \
(t+u)^4 \Big(2 s^4+2 (t+u) s^3+3 \Big(t^2+u^2\Big) s^2+2 \
\Big(t^3+u^3\Big) s+2 \Big(t^2+u t+u^2\Big)^2\Big) H(1,z) H(3,y) \
(s+u)^4+198 (s+t)^4 (t+u)^4 \Big(s^4+2 t s^3+3 t^2 s^2+2 t^3 s+\Big(t^2+u \
t+u^2\Big)^2\Big) H(0,1,z) (s+u)^4
\end{dmath*}
\begin{dmath*}
{\white=}
-198 (s+t)^4 (t+u)^4 \Big(s^4+2 t \
s^3+3 t^2 s^2+2 t^3 s+\Big(t^2+u t+u^2\Big)^2\Big) H(0,2,y) (s+u)^4+198 \
(s+t)^4 (t+u)^4 \Big(2 s^4+2 (2 t+u) s^3+3 \Big(2 t^2+u^2\Big) s^2+2 \
\Big(2 t^3+u^3\Big) s+2 t^4+2 u^4+2 t u^3+3 t^2 u^2+2 t^3 u\Big) \
H(1,0,y) (s+u)^4+198 (s+t)^4 (t+u)^4 \Big(s^4+2 (t+u) s^3+3 \
\Big(t^2+u^2\Big) s^2+2 \Big(t^3+u^3\Big) s+t^4+u^4\Big) H(1,0,z) \
(s+u)^4-198 (s+t)^4 (t+u)^4 \Big(s^4+2 t s^3+3 t^2 s^2+2 t^3 s+\Big(t^2+u \
t+u^2\Big)^2\Big) H(2,0,y) (s+u)^4+198 (s+t)^4 (t+u)^4 \Big(2 s^4+2 \
(t+u) s^3+3 \Big(t^2+u^2\Big) s^2+2 \Big(t^3+u^3\Big) s+2 \Big(t^2+u \
t+u^2\Big)^2\Big) H(3,2,y) (s+u)^4
\end{dmath*}
\begin{dmath*}
{\white=}
+(s+t) (t+u) \Big(1939 (t+u)^3 \
s^{10}+\Big(9662 t^4+37856 u t^3+56586 u^2 t^2+37856 u^3 t+9662 u^4\Big) \
s^9+3 \Big(7701 t^5+35213 u t^4+67475 u^2 t^3+67475 u^3 t^2+35213 u^4 \
t+7701 u^5\Big) s^8+3 \Big(11524 t^6+58638 u t^5+131001 u^2 t^4+167972 \
u^3 t^3+131001 u^4 t^2+58638 u^5 t+11524 u^6\Big) s^7+\Big(34572 \
t^7+204628 u t^6+511062 u^2 t^5+754525 u^3 t^4+754525 u^4 t^3+511062 u^5 \
t^2+204628 u^6 t+34572 u^7\Big) s^6+3 \Big(7701 t^8+58638 u t^7+170354 \
u^2 t^6+278144 u^3 t^5+317652 u^4 t^4+278144 u^5 t^3+170354 u^6 t^2+58638 \
u^7 t+7701 u^8\Big) s^5+\Big(9662 t^9+105639 u t^8+393003 u^2 t^7+754525 \
u^3 t^6+952956 u^4 t^5+952956 u^5 t^4+754525 u^6 t^3+393003 u^7 t^2+105639 \
u^8 t+9662 u^9\Big) s^4+\Big(1939 t^{10}+37856 u t^9+202425 u^2 \
t^8+503916 u^3 t^7+754525 u^4 t^6+834432 u^5 t^5+754525 u^6 t^4+503916 u^7 \
t^3+202425 u^8 t^2+37856 u^9 t+1939 u^{10}\Big) s^3+3 t u \Big(1939 \
t^9+18862 u t^8+67475 u^2 t^7+131001 u^3 t^6+170354 u^4 t^5+170354 u^5 \
t^4+131001 u^6 t^3+67475 u^7 t^2+18862 u^8 t+1939 u^9\Big) s^2+t^2 u^2 \
(t+u)^2 \Big(5817 t^6+26222 u t^5+47378 u^2 t^4+54936 u^3 t^3+47378 u^4 \
t^2+26222 u^5 t+5817 u^6\Big) s+t^3 u^3 (t+u)^3 \Big(1939 t^4+3845 u \
t^3+5751 u^2 t^2+3845 u^3 t+1939 u^4\Big)\Big) (s+u)+33 (s+t)^4 (t+u)^4 \
\Big(11 s^8+22 (t+3 u) s^7+\Big(33 t^2+64 u t+187 u^2\Big) s^6+\Big(22 \
t^3+78 u t^2+60 u^2 t+330 u^3\Big) s^5+\Big(11 t^4+36 u t^3+51 u^2 t^2+14 \
u^3 t+396 u^4\Big) s^4+2 u \Big(10 t^4+7 u t^3+6 u^2 t^2+7 u^3 t+165 \
u^4\Big) s^3+u^2 \Big(24 t^4+14 u t^3+51 u^2 t^2+60 u^3 t+187 u^4\Big) \
s^2+2 u^3 \Big(10 t^4+18 u t^3+39 u^2 t^2+32 u^3 t+33 u^4\Big) s+11 u^4 \
\Big(t^2+u t+u^2\Big)^2\Big) H(0,z)  \Bigg\} \Big/ \Big( {9 s t (s+t)^4 u (s+u)^4 (t+u)^4} \Big)
\end{dmath*}

\intertext{}
\begin{dmath*}
 {\cal A}^{(2)}_{2 ; C_A n_f} =   -\Bigg\{3 (t+u)^4 \Big(44 s^8+88 (3 t+u) s^7+2 \Big(374 t^2+119 u t+66 \
u^2\Big) s^6+\Big(1320 t^3+186 u t^2+321 u^2 t+88 u^3\Big) s^5+2 \
\Big(792 t^4-8 u t^3
\end{dmath*}
\begin{dmath*}
{\white=}
+183 u^2 t^2+36 u^3 t+22 u^4\Big) s^4+2 t \Big(660 \
t^4-8 u t^3+177 u^2 t^2-8 u^3 t+31 u^4\Big) s^3+2 t^2 \Big(374 t^4+93 u \
t^3+183 u^2 t^2-8 u^3 t+57 u^4\Big) s^2+t^3 \Big(264 t^4+238 u t^3+321 \
u^2 t^2+72 u^3 t+62 u^4\Big) s+44 t^4 \Big(t^2+u t+u^2\Big)^2\Big) \
H(0,y) (s+u)^4+9 (s+t)^4 (t+u)^4 \Big(4 s^4+8 (t+u) s^3+12 \
\Big(t^2+u^2\Big) s^2+8 \Big(t^3+u^3\Big) s+4 t^4+4 u^4-11 t u^3-11 \
t^3 u\Big) H(0,y) H(0,z) (s+u)^4-3 (s+t)^4 \Big(2 \Big(22 t^4+31 u \
t^3+57 u^2 t^2+31 u^3 t+22 u^4\Big) s^4+8 \Big(11 t^5+9 u t^4-2 u^2 t^3-2 \
u^3 t^2+9 u^4 t+11 u^5\Big) s^3+3 (t+u)^2 \Big(44 t^4+19 u t^3+40 u^2 \
t^2+19 u^3 t+44 u^4\Big) s^2+2 (t+u)^3 \Big(44 t^4-13 u t^3-13 u^3 t+44 \
u^4\Big) s+44 (t+u)^4 \Big(t^2+u t+u^2\Big)^2\Big) H(1,z) (s+u)^4-9 \
(s+t)^4 (t+u)^4 \Big(4 s^4+(8 t-11 u) s^3+12 t^2 s^2+\Big(8 t^3-11 \
u^3\Big) s+4 \Big(t^2+u t+u^2\Big)^2\Big) H(0,y) H(1,z) (s+u)^4-3 \
(s+t)^4 \Big(2 \Big(22 t^4+31 u t^3+57 u^2 t^2+31 u^3 t+22 u^4\Big) \
s^4+8 \Big(11 t^5+9 u t^4-2 u^2 t^3-2 u^3 t^2+9 u^4 t+11 u^5\Big) s^3+3 \
(t+u)^2 \Big(44 t^4+19 u t^3+40 u^2 t^2+19 u^3 t+44 u^4\Big) s^2+2 \
(t+u)^3 \Big(44 t^4-13 u t^3-13 u^3 t+44 u^4\Big) s+44 (t+u)^4 \
\Big(t^2+u t+u^2\Big)^2\Big) H(2,y) (s+u)^4-9 (s+t)^4 (t+u)^4 \Big(4 \
s^4+(8 u-11 t) s^3+12 u^2 s^2+\Big(8 u^3-11 t^3\Big) s+4 \Big(t^2+u \
t+u^2\Big)^2\Big) H(0,z) H(2,y) (s+u)^4+9 (s+t)^4 (t+u)^4 \Big(8 s^4-3 \
(t+u) s^3+12 \Big(t^2+u^2\Big) s^2-3 \Big(t^3+u^3\Big) s+8 \Big(t^2+u \
t+u^2\Big)^2\Big) H(1,z) H(3,y) (s+u)^4+9 (s+t)^4 (t+u)^4 \Big(4 s^4+(8 \
t-11 u) s^3+12 t^2 s^2+\Big(8 t^3-11 u^3\Big) s+4 \Big(t^2+u \
t+u^2\Big)^2\Big) H(0,1,z) (s+u)^4-9 (s+t)^4 (t+u)^4 \Big(4 s^4+(8 t-11 \
u) s^3+12 t^2 s^2+\Big(8 t^3-11 u^3\Big) s+4 \Big(t^2+u \
t+u^2\Big)^2\Big) H(0,2,y) (s+u)^4+9 (s+t)^4 (t+u)^4 \Big(8 s^4+(16 t-3 \
u) s^3+12 \Big(2 t^2+u^2\Big) s^2+\Big(16 t^3-3 u^3\Big) s+8 t^4+8 \
u^4-3 t u^3+12 t^2 u^2-3 t^3 u\Big) H(1,0,y) (s+u)^4+9 (s+t)^4 (t+u)^4 \
\Big(4 s^4+8 (t+u) s^3+12 \Big(t^2+u^2\Big) s^2+8 \Big(t^3+u^3\Big) \
s+4 t^4+4 u^4-11 t u^3-11 t^3 u\Big) H(1,0,z) (s+u)^4-9 (s+t)^4 (t+u)^4 \
\Big(4 s^4+(8 t-11 u) s^3+12 t^2 s^2+\Big(8 t^3-11 u^3\Big) s+4 \
\Big(t^2+u t+u^2\Big)^2\Big) H(2,0,y) (s+u)^4+9 (s+t)^4 (t+u)^4 \Big(8 \
s^4-3 (t+u) s^3+12 \Big(t^2+u^2\Big) s^2-3 \Big(t^3+u^3\Big) s+8 \
\Big(t^2+u t+u^2\Big)^2\Big) H(3,2,y) (s+u)^4+(s+t) (t+u) \Big(499 \
(t+u)^3 s^{10}+2 \Big(1228 t^4+4741 u t^3
\end{dmath*}
\begin{dmath*}
{\white=}
+7143 u^2 t^2+4741 u^3 t+1228 \
u^4\Big) s^9+3 \Big(1931 t^5+8547 u t^4+16389 u^2 t^3+16389 u^3 t^2+8547 \
u^4 t+1931 u^5\Big) s^8+3 \Big(2864 t^6+13918 u t^5+30487 u^2 t^4+39100 \
u^3 t^3+30487 u^4 t^2+13918 u^5 t+2864 u^6\Big) s^7+\Big(8592 t^7+48196 u \
t^6+115584 u^2 t^5+168841 u^3 t^4+168841 u^4 t^3+115584 u^5 t^2+48196 u^6 \
t+8592 u^7\Big) s^6+3 \Big(1931 t^8+13918 u t^7+38528 u^2 t^6+61228 u^3 \
t^5+69608 u^4 t^4+61228 u^5 t^3+38528 u^6 t^2+13918 u^7 t+1931 u^8\Big) \
s^5+\Big(2456 t^9+25641 u t^8+91461 u^2 t^7+168841 u^3 t^6+208824 u^4 \
t^5+208824 u^5 t^4+168841 u^6 t^3+91461 u^7 t^2+25641 u^8 t+2456 u^9\Big) \
s^4+\Big(499 t^{10}+9482 u t^9+49167 u^2 t^8+117300 u^3 t^7+168841 u^4 \
t^6+183684 u^5 t^5+168841 u^6 t^4+117300 u^7 t^3+49167 u^8 t^2+9482 u^9 \
t+499 u^{10}\Big) s^3+3 t u \Big(499 t^9+4762 u t^8+16389 u^2 t^7+30487 \
u^3 t^6+38528 u^4 t^5+38528 u^5 t^4+30487 u^6 t^3+16389 u^7 t^2+4762 u^8 \
t+499 u^9\Big) s^2+t^2 u^2 (t+u)^2 \Big(1497 t^6+6488 u t^5+11168 u^2 \
t^4+12930 u^3 t^3+11168 u^4 t^2+6488 u^5 t+1497 u^6\Big) s+t^3 u^3 (t+u)^3 \
\Big(499 t^4+959 u t^3+1419 u^2 t^2+959 u^3 t+499 u^4\Big)\Big) (s+u)+3 \
(s+t)^4 (t+u)^4 \Big(44 s^8+88 (t+3 u) s^7+2 \Big(66 t^2+119 u t+374 \
u^2\Big) s^6+\Big(88 t^3+321 u t^2+186 u^2 t+1320 u^3\Big) s^5+2 \
\Big(22 t^4+36 u t^3+183 u^2 t^2-8 u^3 t+792 u^4\Big) s^4+2 u \Big(31 \
t^4-8 u t^3+177 u^2 t^2-8 u^3 t+660 u^4\Big) s^3+2 u^2 \Big(57 t^4-8 u \
t^3+183 u^2 t^2+93 u^3 t+374 u^4\Big) s^2+u^3 \Big(62 t^4+72 u t^3+321 \
u^2 t^2+238 u^3 t+264 u^4\Big) s+44 u^4 \Big(t^2+u t+u^2\Big)^2\Big) \
H(0,z)\Bigg\}  \Big/ \Big( {9 s t (s+t)^4 u (s+u)^4 (t+u)^4} \Big)
\end{dmath*}

\intertext{}
\begin{dmath*}
 {\cal A}^{(2)}_{2 ; C_F n_f} =   -\Bigg\{8 \Big(s^4+2 s^3 (t+u)+3 s^2 \Big(t^2+u^2\Big)+2 s \
\Big(t^3+u^3\Big)+\Big(t^2+t u+u^2\Big)^2\Big) \Bigg\} \Big/ \Big( {s t u} \Big)
\end{dmath*}

\intertext{}
\begin{dmath*}
 {\cal A}^{(2)}_{2 ; n_f^2} =   \Bigg\{-18 u (s+u)^4 (t+u)^4 \Big(s^2-t s+t^2+u^2\Big) H(1,0,y) (s+t)^5+6 \
(t+u)^4 \Big(2 s^8+4 (t+3 u) s^7+2 \Big(3 t^2+5 u t+17 u^2\Big) \
s^6
\end{dmath*}
\begin{dmath*}
{\white=}
+\Big(4 t^3+15 u t^2+6 u^2 t+60 u^3\Big) s^5+2 \Big(t^4+12 u^2 t^2-2 \
u^3 t+36 u^4\Big) s^4+2 u \Big(t^4-2 u t^3+15 u^2 t^2-2 u^3 t+30 \
u^4\Big) s^3+2 u^2 \Big(3 t^4-2 u t^3+12 u^2 t^2+3 u^3 t+17 u^4\Big) \
s^2+\Big(12 u^7+10 t u^6+15 t^2 u^5+2 t^4 u^3\Big) s+2 u^4 \Big(t^2+u \
t+u^2\Big)^2\Big) H(0,z) (s+t)^4-18 t u (s+u)^4 (t+u)^4 \
\Big(t^2+u^2\Big) H(0,y) H(0,z) (s+t)^4-6 (s+u)^4 \Big(2 \Big(t^4+u \
t^3+3 u^2 t^2+u^3 t+u^4\Big) s^4+4 \Big(t^5-u^2 t^3-u^3 t^2+u^5\Big) \
s^3+3 (t+u)^2 \Big(2 t^4+u t^3+4 u^2 t^2+u^3 t+2 u^4\Big) s^2+2 (t+u)^3 \
\Big(2 t^4-u t^3-u^3 t+2 u^4\Big) s+2 (t+u)^4 \Big(t^2+u \
t+u^2\Big)^2\Big) H(1,z) (s+t)^4+18 s u (s+u)^4 (t+u)^4 \
\Big(s^2+u^2\Big) H(0,y) H(1,z) (s+t)^4-6 (s+u)^4 \Big(2 \Big(t^4+u \
t^3+3 u^2 t^2+u^3 t+u^4\Big) s^4+4 \Big(t^5-u^2 t^3-u^3 t^2+u^5\Big) \
s^3+3 (t+u)^2 \Big(2 t^4+u t^3+4 u^2 t^2+u^3 t+2 u^4\Big) s^2+2 (t+u)^3 \
\Big(2 t^4-u t^3-u^3 t+2 u^4\Big) s+2 (t+u)^4 \Big(t^2+u \
t+u^2\Big)^2\Big) H(2,y) (s+t)^4+18 s t \Big(s^2+t^2\Big) (s+u)^4 \
(t+u)^4 H(0,z) H(2,y) (s+t)^4-18 s (s+u)^4 (t+u)^5 \Big(s^2+t^2+u^2-t \
u\Big) H(1,z) H(3,y) (s+t)^4-18 s u (s+u)^4 (t+u)^4 \Big(s^2+u^2\Big) \
H(0,1,z) (s+t)^4+18 s u (s+u)^4 (t+u)^4 \Big(s^2+u^2\Big) H(0,2,y) \
(s+t)^4-18 t u (s+u)^4 (t+u)^4 \Big(t^2+u^2\Big) H(1,0,z) (s+t)^4+18 s u \
(s+u)^4 (t+u)^4 \Big(s^2+u^2\Big) H(2,0,y) (s+t)^4-18 s (s+u)^4 (t+u)^5 \
\Big(s^2+t^2+u^2-t u\Big) H(3,2,y) (s+t)^4+6 (s+u) (t+u) \Big(5 (t+u)^3 \
s^{10}+6 \Big(4 t^4+15 u t^3+23 u^2 t^2+15 u^3 t+4 u^4\Big) s^9+\Big(55 \
t^5+227 u t^4+437 u^2 t^3+437 u^3 t^2+227 u^4 t+55 u^5\Big) s^8
\end{dmath*}

\begin{dmath*}
{\white =}
+\Big(80 \
t^6+350 u t^5+731 u^2 t^4+940 u^3 t^3+731 u^4 t^2+350 u^5 t+80 u^6\Big) \
s^7+\Big(80 t^7+396 u t^6+852 u^2 t^5+1211 u^3 t^4+1211 u^4 t^3+852 u^5 \
t^2+396 u^6 t+80 u^7\Big) s^6+\Big(55 t^8+350 u t^7+852 u^2 t^6+1252 u^3 \
t^5+1408 u^4 t^4+1252 u^5 t^3+852 u^6 t^2+350 u^7 t+55 u^8\Big) \
s^5+\Big(24 t^9+227 u t^8+731 u^2 t^7+1211 u^3 t^6+1408 u^4 t^5+1408 u^5 \
t^4+1211 u^6 t^3+731 u^7 t^2+227 u^8 t+24 u^9\Big) s^4+\Big(5 t^{10}+90 u \
t^9+437 u^2 t^8+940 u^3 t^7+1211 u^4 t^6+1252 u^5 t^5+1211 u^6 t^4+940 u^7 \
t^3+437 u^8 t^2+90 u^9 t+5 u^{10}\Big) s^3+t u \Big(15 t^9+138 u t^8+437 \
u^2 t^7+731 u^3 t^6+852 u^4 t^5+852 u^5 t^4+731 u^6 t^3+437 u^7 t^2+138 u^8 \
t+15 u^9\Big) s^2+t^2 u^2 (t+u)^2 \Big(15 t^6+60 u t^5+92 u^2 t^4+106 u^3 \
t^3+92 u^4 t^2+60 u^5 t+15 u^6\Big) s+t^3 u^3 (t+u)^3 \Big(5 t^4+9 u \
t^3+13 u^2 t^2+9 u^3 t+5 u^4\Big)\Big) (s+t)+6 (s+u)^4 (t+u)^4 \Big(2 \
s^8+4 (3 t+u) s^7+2 \Big(17 t^2+5 u t+3 u^2\Big) s^6+\Big(60 t^3+6 u \
t^2+15 u^2 t+4 u^3\Big) s^5+2 \Big(36 t^4-2 u t^3+12 u^2 t^2+u^4\Big) \
s^4+2 t \Big(30 t^4-2 u t^3+15 u^2 t^2-2 u^3 t+u^4\Big) s^3+2 t^2 \
\Big(17 t^4+3 u t^3+12 u^2 t^2-2 u^3 t+3 u^4\Big) s^2+\Big(12 t^7+10 u \
t^6+15 u^2 t^5+2 u^4 t^3\Big) s+2 t^4 \Big(t^2+u t+u^2\Big)^2\Big) \
H(0,y) \Bigg\} \Big/ \Big(  {9 s t (s+t)^4 u (s+u)^4 (t+u)^4} \Big)
\end{dmath*}

\intertext{}
\begin{dmath*}
 {\cal A}^{(2)}_{3 ; C_A^2} =   \Big\{ 49 s^4+98 s^3 \
(t+u)+147 s^2 \Big(t^2+u^2\Big)+98 s \Big(t^3+u^3\Big)+49 t^4-450 t^3 \
u+165 t^2 u^2-450 t u^3+49 u^4\Big\} \Big/ \Big( 5 s t u \Big)
\end{dmath*}

\begin{dmath*}
 {\cal A}^{(2)}_{3 ; C_A n_f} =  \left\{ 353 t^2-588 t u+353 u^2\right\} \Big/ \Big( 5 s\Big)
\end{dmath*}

\begin{dmath*}
 {\cal A}^{(2)}_{3 ; C_F n_f} =   - 18 \left\{ t^2+u^2\right\} \Big/ \Big( s\Big)
\end{dmath*}

\begin{dmath*}
 {\cal A}^{(2)}_{3 ; n_f^2} =  0
\end{dmath*}

\intertext{}
\begin{dmath*}
 {\cal A}^{(2)}_{4 ; C_A^2} =  \Bigg\{  9 s^2 t^2 u^2 \Big(231 s^{16}+1896 (t+u) s^{15}+6 \Big(1205 t^2+2582 u \
t+1205 u^2\Big) s^{14}+2 \Big(8485 t^3+28497 u t^2+28497 u^2 t+8485 \
u^3\Big) s^{13}
\end{dmath*}
\begin{dmath*}
{\white=}
+\Big(27247 t^4+132002 u t^3+187458 u^2 t^2+132002 u^3 \
t+27247 u^4\Big) s^{12}+12 \Big(2609 t^5+18137 u t^4+32471 u^2 t^3+32471 \
u^3 t^2+18137 u^4 t+2609 u^5\Big) s^{11}+\Big(26071 t^6+265278 u \
t^5+584223 u^2 t^4+758272 u^3 t^3+584223 u^4 t^2+265278 u^5 t+26071 \
u^6\Big) s^{10}+2 \Big(7765 t^7+118994 u t^6+329001 u^2 t^5+544812 u^3 \
t^4+544812 u^4 t^3+329001 u^5 t^2+118994 u^6 t+7765 u^7\Big) s^9+3 \
\Big(2110 t^8+50618 u t^7+186907 u^2 t^6+405032 u^3 t^5+468376 u^4 \
t^4+405032 u^5 t^3+186907 u^6 t^2+50618 u^7 t+2110 u^8\Big) s^8+4 \
\Big(398 t^9+16089 u t^8+86178 u^2 t^7+272710 u^3 t^6+333261 u^4 t^5+333261 \
u^5 t^4+272710 u^6 t^3+86178 u^7 t^2+16089 u^8 t+398 u^9\Big) \
s^7+\Big(187 t^{10}+16046 u t^9+137715 u^2 t^8+724384 u^3 t^7+1065452 u^4 \
t^6+947616 u^5 t^5+1065452 u^6 t^4+724384 u^7 t^3+137715 u^8 t^2+16046 u^9 \
t+187 u^{10}\Big) s^6+6 t u \Big(299 t^9+5160 u t^8+51597 u^2 t^7+106788 \
u^3 t^6+101934 u^4 t^5+101934 u^5 t^4+106788 u^6 t^3+51597 u^7 t^2+5160 u^8 \
t+299 u^9\Big) s^5+t^2 u^2 \Big(2973 t^8+74726 u t^7+236013 u^2 \
t^6+329784 u^3 t^5+314484 u^4 t^4+329784 u^5 t^3+236013 u^6 t^2+74726 u^7 \
t+2973 u^8\Big) s^4+12 t^3 u^3 \Big(651 t^7+3762 u t^6+9502 u^2 t^5+11138 \
u^3 t^4+11138 u^4 t^3+9502 u^5 t^2+3762 u^6 t+651 u^7\Big) s^3+3 t^4 u^4 \
\Big(1027 t^6+7446 u t^5+12323 u^2 t^4+11392 u^3 t^3+12323 u^4 t^2+7446 u^5 \
t+1027 u^6\Big) s^2+6 t^5 u^5 \Big(315 t^5+904 u t^4+837 u^2 t^3+837 u^3 \
t^2+904 u^4 t+315 u^5\Big) s+t^6 u^6 \Big(215 t^4+250 u t^3+357 u^2 \
t^2+250 u^3 t+215 u^4\Big)\Big) (t+u)^6
\end{dmath*}
\begin{dmath*}
{\white=}
-108 s^2 t^2 (s+t)^6 u^2 (s+u)^6 \
\Big(s^4+2 (t+u) s^3+3 \Big(t^2+u^2\Big) s^2+2 \Big(t^3+u^3\Big) \
s+t^4+u^4-12 t u^3+18 t^2 u^2-24 t^3 u\Big) H(0,y) (t+u)^6-108 s^2 t^2 \
(s+t)^6 u^2 (s+u)^6 \Big(s^4+2 (t+u) s^3+3 \Big(t^2+u^2\Big) s^2+2 \
\Big(t^3+u^3\Big) s+t^4+u^4-24 t u^3+18 t^2 u^2-12 t^3 u\Big) H(0,z) \
(t+u)^6+216 s^2 t^2 (s+t)^6 u^3 (s+u)^6 \Big(2 s^3-6 u s^2+10 u^2 s+3 t \
\Big(2 t^2-3 u t+4 u^2\Big)\Big) H(1,y) (t+u)^6+108 s^2 t^2 (s+t)^6 u^2 \
(s+u)^6 \Big(s^4+2 (t-u) s^3+3 \Big(t^2+5 u^2\Big) s^2+2 \Big(t^3-9 \
u^3\Big) s+t^4+u^4+18 t u^3-9 t^2 u^2+30 t^3 u\Big) H(1,z) (t+u)^6+108 \
s^2 t^2 (s+t)^6 u^2 (s+u)^6 \Big(s^4-2 (t+u) s^3+15 \Big(t^2+u^2\Big) \
s^2-18 \Big(t^3+u^3\Big) s+t^4+u^4+6 t u^3+9 t^2 u^2+6 t^3 u\Big) \
H(2,y) (t+u)^6    \Bigg\} \Big/ \Big(  {27 s^3 t^3 u^3 (s+t)^6 (s+u)^6 (t+u)^6} \Big)
\end{dmath*}

\intertext{}
\begin{dmath*}
 {\cal A}^{(2)}_{4 ; C_A n_f} =  \Bigg\{  9 s^2 t^2 u^2 \Big(54 s^{16}+420 (t+u) s^{15}+24 \Big(64 t^2+93 u t+64 \
u^2\Big) s^{14}+\Big(3520 t^3+6099 u t^2+6099 u^2 t+3520 u^3\Big) \
s^{13}
\end{dmath*}
\begin{dmath*}
{\white=}
+2 \Big(2821 t^4+4943 u t^3+9330 u^2 t^2+4943 u^3 t+2821 u^4\Big) \
s^{12}+6 \Big(1106 t^5+1697 u t^4+6163 u^2 t^3+6163 u^3 t^2+1697 u^4 t+1106 \
u^5\Big) s^{11}+\Big(5810 t^6+7365 u t^5+65076 u^2 t^4+38726 u^3 \
t^3+65076 u^4 t^2+7365 u^5 t+5810 u^6\Big) s^{10}+\Big(3736 t^7+3503 u \
t^6+105864 u^2 t^5+17977 u^3 t^4+17977 u^4 t^3+105864 u^5 t^2+3503 u^6 \
t+3736 u^7\Big) s^9+3 \Big(560 t^8+208 u t^7+38974 u^2 t^6+8059 u^3 \
t^5-15308 u^4 t^4+8059 u^5 t^3+38974 u^6 t^2+208 u^7 t+560 u^8\Big) \
s^8+\Big(472 t^9+60 u t^8+81318 u^2 t^7-8215 u^3 t^6+18453 u^4 t^5+18453 \
u^5 t^4-8215 u^6 t^3+81318 u^7 t^2+60 u^8 t+472 u^9\Big) s^7+\Big(62 \
t^{10}+229 u t^9+37632 u^2 t^8-50045 u^3 t^7+25292 u^4 t^6+177060 u^5 \
t^5+25292 u^6 t^4-50045 u^7 t^3+37632 u^8 t^2+229 u^9 t+62 u^{10}\Big) \
s^6+3 t u \Big(24 t^9+3679 u t^8-12820 u^2 t^7+195 u^3 t^6+50191 u^4 \
t^5+50191 u^5 t^4+195 u^6 t^3-12820 u^7 t^2+3679 u^8 t+24 u^9\Big) s^5+2 \
t^2 u^2 \Big(735 t^8-6322 u t^7+3033 u^2 t^6+30537 u^3 t^5+41982 u^4 \
t^4+30537 u^5 t^3+3033 u^6 t^2-6322 u^7 t+735 u^8\Big) s^4+t^3 u^3 \
\Big(-1700 t^7+5703 u t^6+17862 u^2 t^5+21167 u^3 t^4+21167 u^4 t^3+17862 \
u^5 t^2+5703 u^6 t-1700 u^7\Big) s^3+12 t^4 u^4 \Big(109 t^6+190 u \
t^5+297 u^2 t^4+423 u^3 t^3+297 u^4 t^2+190 u^5 t+109 u^6\Big) s^2-6 t^5 \
u^5 \Big(12 t^5-24 u t^4-209 u^2 t^3-209 u^3 t^2-24 u^4 t+12 u^5\Big) \
s
\end{dmath*}

\begin{dmath*}
{\white =}
+t^6 u^6 \Big(20 t^4+131 u t^3+174 u^2 t^2+131 u^3 t+20 u^4\Big)\Big) \
(t+u)^6-162 s^2 t^3 (s+t)^6 u^3 (s+u)^6 \Big(8 t^2-21 u t+17 u^2\Big) \
H(0,y) (t+u)^6-162 s^2 t^3 (s+t)^6 u^3 (s+u)^6 \Big(17 t^2-21 u t+8 \
u^2\Big) H(0,z) (t+u)^6-54 s^2 t^2 (s+t)^6 u^3 (s+u)^6 \Big(41 s^3-42 u \
s^2+5 u^2 s+t \Big(50 t^2-63 u t+23 u^2\Big)\Big) H(1,y) (t+u)^6+54 s^2 \
t^2 (s+t)^6 u^3 (s+u)^6 \Big(41 s^3-42 u s^2+5 u^2 s+t \Big(-25 t^2+63 u \
t-52 u^2\Big)\Big) H(1,z) (t+u)^6+54 s^2 t^2 (s+t)^6 u^2 (s+u)^6 \
\Big(41 (t+u) s^3-42 \Big(t^2+u^2\Big) s^2+5 \Big(t^3+u^3\Big) s-2 t \
u \Big(t^2+u^2\Big)\Big) H(2,y) (t+u)^6   \Bigg\} \Big/ \Big(  {27 s^3 t^3 u^3 (s+t)^6 (s+u)^6 (t+u)^6} \Big)
\end{dmath*}

\intertext{}
\begin{dmath*}
 {\cal A}^{(2)}_{4 ; C_F n_f} =  \Bigg\{ 432 s^3 t^2 (s+t)^6 u^2 (s+u)^6 \Big(-3 s^2+t^2+u^2-t u\Big) H(2,y) \
(t+u)^7-18 s^2 t^2 u^2 \Big(48 s^{16}+384 (t+u) s^{15}+12 \Big(120 t^2+227 \
u t
\end{dmath*}
\begin{dmath*}
{\white=}
+120 u^2\Big) s^{14}+\Big(3360 t^3+9087 u t^2+9087 u^2 t+3360 \
u^3\Big) s^{13}+3 \Big(1808 t^4+6521 u t^3+8604 u^2 t^2+6521 u^3 t+1808 \
u^4\Big) s^{12}+\Big(6336 t^5+30735 u t^4+47598 u^2 t^3+47598 u^3 \
t^2+30735 u^4 t+6336 u^5\Big) s^{11}+\Big(5424 t^6+36543 u t^5+67614 u^2 \
t^4+73464 u^3 t^3+67614 u^4 t^2+36543 u^5 t+5424 u^6\Big) \
s^{10}+\Big(3360 t^7+32505 u t^6+77922 u^2 t^5+91295 u^3 t^4+91295 u^4 \
t^3+77922 u^5 t^2+32505 u^6 t+3360 u^7\Big) s^9+3 \Big(480 t^8+6939 u \
t^7+23404 u^2 t^6+34497 u^3 t^5+31012 u^4 t^4+34497 u^5 t^3+23404 u^6 \
t^2+6939 u^7 t+480 u^8\Big) s^8+3 \Big(128 t^9+2999 u t^8+15226 u^2 \
t^7+33508 u^3 t^6+30137 u^4 t^5+30137 u^5 t^4+33508 u^6 t^3+15226 u^7 \
t^2+2999 u^8 t+128 u^9\Big) s^7+\Big(48 t^{10}+2337 u t^9+19494 u^2 \
t^8+70630 u^3 t^7+86678 u^4 t^6+65250 u^5 t^5+86678 u^6 t^4+70630 u^7 \
t^3+19494 u^8 t^2+2337 u^9 t+48 u^{10}\Big) s^6+3 t u \Big(92 t^9+1601 u \
t^8+10417 u^2 t^7+20343 u^3 t^6+17708 u^4 t^5+17708 u^5 t^4+20343 u^6 \
t^3+10417 u^7 t^2+1601 u^8 t+92 u^9\Big) s^5+3 t^2 u^2 \Big(172 t^8+2556 \
u t^7+8524 u^2 t^6+11797 u^3 t^5+10692 u^4 t^4+11797 u^5 t^3+8524 u^6 \
t^2+2556 u^7 t+172 u^8\Big) s^4+t^3 u^3 \Big(788 t^7+5538 u t^6+14103 u^2 \
t^5+16252 u^3 t^4+16252 u^4 t^3+14103 u^5 t^2+5538 u^6 t+788 u^7\Big) \
s^3+3 t^4 u^4 \Big(148 t^6+959 u t^5+1726 u^2 t^4+1748 u^3 t^3+1726 u^4 \
t^2+959 u^5 t+148 u^6\Big) s^2+3 t^5 u^5 \Big(72 t^5+265 u t^4+347 u^2 \
t^3+347 u^3 t^2+265 u^4 t+72 u^5\Big) s+2 t^6 u^6 \Big(16 t^4+39 u t^3+54 \
u^2 t^2+39 u^3 t+16 u^4\Big)\Big) (t+u)^6+108 s^2 t^3 (s+t)^6 u^3 \
(s+u)^6 \Big(t^2+13 u^2\Big) H(0,y) (t+u)^6+108 s^2 t^3 (s+t)^6 u^3 \
(s+u)^6 \Big(13 t^2+u^2\Big) H(0,z) (t+u)^6+108 s^2 t^2 (s+t)^6 u^3 \
(s+u)^6 \Big(12 s^3-4 u^2 s+13 t^3+t u^2\Big) H(1,y) (t+u)^6-108 s^2 t^2 \
(s+t)^6 u^3 (s+u)^6 \Big(12 s^3-4 u^2 s-t \Big(t^2+13 u^2\Big)\Big) \
H(1,z) (t+u)^6 \Bigg\} \Big/ \Big(  {27 s^3 t^3 u^3 (s+t)^6 (s+u)^6 (t+u)^6} \Big)
\end{dmath*}

\intertext{}
\begin{dmath*}
 {\cal A}^{(2)}_{4 ; n_f^2} = 0
\end{dmath*}

\intertext{}
\begin{dmath*}
 {\cal A}^{(2)}_{5 ; C_A^2} =  \Bigg\{  9 s^2 t^2 (s+t)^2 u^2 \Big(55 s^{14}+(330 t+416 u) s^{13}+\Big(935 \
t^2+1844 u t+1500 u^2\Big) s^{12}+30 \Big(55 t^3+135 u t^2+167 u^2 t+115 \
u^3\Big) s^{11}
\end{dmath*}
\begin{dmath*}
{\white=}
+\Big(1980 t^4+5670 u t^3+8826 u^2 t^2+8462 u^3 t+5643 \
u^4\Big) s^{10}+2 \Big(825 t^5+2487 u t^4+6495 u^2 t^3+3799 u^3 t^2+5061 \
u^4 t+3426 u^5\Big) s^9+\Big(935 t^6+2568 u t^5+15141 u^2 t^4+4070 u^3 \
t^3-1902 u^4 t^2+9678 u^5 t+6231 u^6\Big) s^8+2 \Big(165 t^7+443 u \
t^6+6261 u^2 t^5-139 u^3 t^4-8575 u^4 t^3-4065 u^5 t^2+3881 u^6 t+2085 \
u^7\Big) s^7+\Big(55 t^8+334 u t^7+7728 u^2 t^6-6262 u^3 t^5-17396 u^4 \
t^4-22194 u^5 t^3-7172 u^6 t^2+5130 u^7 t+1950 u^8\Big) s^6+2 u \Big(45 \
t^8+1695 u t^7-3483 u^2 t^6-4305 u^3 t^5-4851 u^4 t^4-8249 u^5 t^3-1455 u^6 \
t^2+1362 u^7 t+284 u^8\Big) s^5+u^2 \Big(717 t^8-2782 u t^7-1182 u^2 \
t^6+5322 u^3 t^5-4938 u^4 t^4-7502 u^5 t^3+711 u^6 t^2+998 u^7 t+77 \
u^8\Big) s^4-2 t u^3 \Big(158 t^7-735 u t^6-2541 u^2 t^5-2255 u^3 t^4+577 \
u^4 t^3+150 u^5 t^2-512 u^6 t-82 u^7\Big) s^3+t^2 u^4 \Big(717 t^6+1098 u \
t^5+2632 u^2 t^4+2622 u^3 t^3+1308 u^4 t^2+852 u^5 t+234 u^6\Big) s^2+2 \
t^3 u^5 \Big(45 t^5+157 u t^4+619 u^2 t^3+516 u^3 t^2+340 u^4 t+102 \
u^5\Big) s+t^4 u^6 \Big(55 t^4+134 u t^3+207 u^2 t^2+134 u^3 t+105 \
u^4\Big)\Big) H(0,y) (t+u)^6
\end{dmath*}
\begin{dmath*}
{\white=}
+9 s^2 t^2 u^2 (s+u)^2 \Big(55 s^{14}+(416 \
t+330 u) s^{13}+\Big(1500 t^2+1844 u t+935 u^2\Big) s^{12}+30 \Big(115 \
t^3+167 u t^2+135 u^2 t+55 u^3\Big) s^{11}+\Big(5643 t^4+8462 u t^3+8826 \
u^2 t^2+5670 u^3 t+1980 u^4\Big) s^{10}+2 \Big(3426 t^5+5061 u t^4+3799 \
u^2 t^3+6495 u^3 t^2+2487 u^4 t+825 u^5\Big) s^9+\Big(6231 t^6+9678 u \
t^5-1902 u^2 t^4+4070 u^3 t^3+15141 u^4 t^2+2568 u^5 t+935 u^6\Big) s^8+2 \
\Big(2085 t^7+3881 u t^6-4065 u^2 t^5-8575 u^3 t^4-139 u^4 t^3+6261 u^5 \
t^2+443 u^6 t+165 u^7\Big) s^7+\Big(1950 t^8+5130 u t^7-7172 u^2 \
t^6-22194 u^3 t^5-17396 u^4 t^4-6262 u^5 t^3+7728 u^6 t^2+334 u^7 t+55 \
u^8\Big) s^6+2 t \Big(284 t^8+1362 u t^7-1455 u^2 t^6-8249 u^3 t^5-4851 \
u^4 t^4-4305 u^5 t^3-3483 u^6 t^2+1695 u^7 t+45 u^8\Big) s^5+t^2 \Big(77 \
t^8+998 u t^7+711 u^2 t^6-7502 u^3 t^5-4938 u^4 t^4+5322 u^5 t^3-1182 u^6 \
t^2-2782 u^7 t+717 u^8\Big) s^4+2 t^3 u \Big(82 t^7+512 u t^6-150 u^2 \
t^5-577 u^3 t^4+2255 u^4 t^3+2541 u^5 t^2+735 u^6 t-158 u^7\Big) s^3+t^4 \
u^2 \Big(234 t^6+852 u t^5+1308 u^2 t^4+2622 u^3 t^3+2632 u^4 t^2+1098 u^5 \
t+717 u^6\Big) s^2+2 t^5 u^3 \Big(102 t^5+340 u t^4+516 u^2 t^3+619 u^3 \
t^2+157 u^4 t+45 u^5\Big) s+t^6 u^4 \Big(105 t^4+134 u t^3+207 u^2 \
t^2+134 u^3 t+55 u^4\Big)\Big) H(0,z) (t+u)^6+108 s^2 t^2 (s+t)^6 u^2 \
(s+u)^6 \Big(3 s^4+6 (t+u) s^3+9 \Big(t^2+u^2\Big) s^2+6 \
\Big(t^3+u^3\Big) s+3 t^4+3 u^4-4 t u^3-6 t^2 u^2-4 t^3 u\Big) H(0,y) \
H(0,z) (t+u)^6-18 s^2 t^2 (s+t)^2 u^2 (s+u)^6 \Big(3 s^8+2 (51 t+u) \
s^7+\Big(486 t^2+200 u t+30 u^2\Big) s^6+6 \Big(179 t^3+130 u t^2+96 u^2 \
t+7 u^3\Big) s^5+2 \Big(687 t^4+677 u t^3+945 u^2 t^2+265 u^3 t+29 \
u^4\Big) s^4+2 t \Big(537 t^4+677 u t^3+1344 u^2 t^2+658 u^3 t+170 \
u^4\Big) s^3+2 t^2 \Big(243 t^4+390 u t^3+945 u^2 t^2+658 u^3 t+270 \
u^4\Big) s^2+2 t^3 \Big(51 t^4+100 u t^3+288 u^2 t^2+265 u^3 t+170 \
u^4\Big) s+t^4 \Big(3 t^4+2 u t^3+30 u^2 t^2+42 u^3 t+58 \
u^4\Big)\Big) H(1,y) (t+u)^6
\end{dmath*}

\begin{dmath*}
{\white =}
-216 s^2 t^2 (s+t)^6 u^3 (s+u)^6 \Big(2 \
s^3-3 u s^2+4 u^2 s+t \Big(2 t^2-3 u t+4 u^2\Big)\Big) H(0,z) H(1,y) \
(t+u)^6-108 s^2 t^2 (s+t)^6 u^2 (s+u)^6 \Big(3 s^4+(6 t-8 u) s^3+\Big(9 \
t^2+6 u^2\Big) s^2+\Big(6 t^3-4 u^3\Big) s+3 t^4+3 u^4+12 t u^3+6 t^2 \
u^2+16 t^3 u\Big) H(0,y) H(1,z) (t+u)^6-1296 s^2 t^2 (s+t)^6 u^3 (s+u)^6 \
\Big(s^3-u s^2+u^2 s+t \Big(t^2-u t+u^2\Big)\Big) H(1,y) H(1,z) \
(t+u)^6-108 s^2 t^2 (s+t)^6 u^2 (s+u)^6 \Big(3 s^4+(2 u-8 t) s^3+3 \Big(2 \
t^2+5 u^2\Big) s^2-2 \Big(2 t^3+u^3\Big) s+3 t^4+3 u^4+8 t u^3+12 t^2 \
u^2+8 t^3 u\Big) H(0,z) H(2,y) (t+u)^6-108 s^2 t^2 (s+t)^6 u^2 (s+u)^6 \
\Big(8 s^4+(4 t-2 u) s^3+9 \Big(2 t^2+3 u^2\Big) s^2-2 u^3 s+8 t^4+8 \
u^4+18 t u^3+27 t^2 u^2+18 t^3 u\Big) H(1,z) H(2,y) (t+u)^6-108 s^2 t^2 \
(s+t)^6 u^2 (s+u)^6 \Big(2 s^4+10 (t+u) s^3-3 \Big(t^2+u^2\Big) s^2+6 \
\Big(t^3+u^3\Big) s+2 \Big(t^2+u t+u^2\Big)^2\Big) H(1,z) H(3,y) \
(t+u)^6-108 s^2 t^3 (s+t)^6 u^3 (s+u)^6 \Big(6 t^2-9 u t+2 u^2\Big) \
H(0,0,y) (t+u)^6-108 s^2 t^3 (s+t)^6 u^3 (s+u)^6 \Big(2 t^2-9 u t+6 \
u^2\Big) H(0,0,z) (t+u)^6+432 s^2 t^2 (s+t)^6 u^2 (s+u)^6 \Big(2 s^4+2 (2 \
t+u) s^3+3 \Big(2 t^2+u^2\Big) s^2+2 \Big(2 t^3+u^3\Big) s+2 t^4+2 \
u^4+2 t u^3+3 t^2 u^2+2 t^3 u\Big) H(0,1,y) (t+u)^6+108 s^2 t^2 (s+t)^6 \
u^2 (s+u)^6 \Big(3 s^4+(6 t-2 u) s^3+3 \Big(3 t^2+5 u^2\Big) s^2+2 \
\Big(3 t^3+u^3\Big) s+3 t^4+3 u^4-14 t u^3+9 t^2 u^2-14 t^3 u\Big) \
H(0,1,z) (t+u)^6-108 s^2 t^2 (s+t)^6 u^2 (s+u)^6 \Big(3 s^4+8 (t-u) s^3+6 \
\Big(2 t^2+u^2\Big) s^2+\Big(8 t^3-4 u^3\Big) s+3 \Big(t^4+6 u \
t^3-u^2 t^2+6 u^3 t+u^4\Big)\Big) H(0,2,y) (t+u)^6+108 s^2 t^2 (s+t)^6 \
u^2 (s+u)^6 \Big(6 s^4
\end{dmath*}
\begin{dmath*}
{\white=}
+2 (6 t-u) s^3+3 \Big(6 t^2+5 u^2\Big) s^2+2 \
\Big(6 t^3+u^3\Big) s+6 t^4+6 u^4+6 t u^3+9 t^2 u^2+6 t^3 u\Big) \
H(1,0,y) (t+u)^6+108 s^2 t^2 (s+t)^6 u^2 (s+u)^6 \Big(3 s^4+2 (3 t+5 u) \
s^3+3 \Big(3 t^2+u^2\Big) s^2+2 \Big(3 t^3+7 u^3\Big) s+3 t^4+3 u^4-2 \
t u^3-3 t^2 u^2-2 t^3 u\Big) H(1,0,z) (t+u)^6+216 s^2 t^2 (s+t)^6 u^2 \
(s+u)^6 \Big(4 s^4+8 (t+u) s^3+3 \Big(4 t^2+u^2\Big) s^2+\Big(8 t^3+6 \
u^3\Big) s+4 t^4+4 u^4+6 t u^3+3 t^2 u^2+8 t^3 u\Big) H(1,1,y) \
(t+u)^6+108 s^2 t^2 (s+t)^6 u^3 (s+u)^6 \Big(18 s^3-3 u s^2+18 u^2 s-t \
\Big(18 t^2+9 u t+14 u^2\Big)\Big) H(1,1,z) (t+u)^6-1296 s^2 t^2 \
(s+t)^6 u^3 (s+u)^6 \Big(s^3-u s^2+u^2 s+t \Big(t^2-u t+u^2\Big)\Big) \
H(1,2,y) (t+u)^6-108 s^2 t^2 (s+t)^6 u^2 (s+u)^6 \Big(3 s^4+2 (t-4 u) s^3+3 \
\Big(5 t^2+2 u^2\Big) s^2-2 \Big(t^3+2 u^3\Big) s+3 t^4+3 u^4+8 t \
u^3+12 t^2 u^2+8 t^3 u\Big) H(2,0,y) (t+u)^6-108 s^2 t^2 (s+t)^6 u^3 \
(s+u)^6 \Big(6 s^3-9 u s^2+2 u^2 s+t \Big(2 t^2+3 u t+2 u^2\Big)\Big) \
H(2,1,y) (t+u)^6-108 s^2 t^2 (s+t)^6 u^2 (s+u)^6 \Big(8 s^4-2 (t+u) s^3+27 \
\Big(t^2+u^2\Big) s^2-2 \Big(t^3+u^3\Big) s+8 \Big(t^2+u \
t+u^2\Big)^2\Big) H(2,2,y) (t+u)^6-108 s^2 t^2 (s+t)^6 u^2 (s+u)^6 \
\Big(2 s^4+10 (t+u) s^3-3 \Big(t^2+u^2\Big) s^2+6 \Big(t^3+u^3\Big) \
s+2 \Big(t^2+u t+u^2\Big)^2\Big) H(3,2,y) (t+u)^6
\end{dmath*}

\begin{dmath*}
{\white =}
+9 t^2 (s+t) u^2 (s+u) \
\Big(366 (t+u)^6 s^{16}+12 \Big(213 t^7+1475 u t^6+4463 u^2 t^5+7365 u^3 \
t^4+7365 u^4 t^3+4463 u^5 t^2+1475 u^6 t+213 u^7\Big) s^{15}+2 (t+u)^2 \
\Big(4186 t^6+23989 u t^5+61240 u^2 t^4+77078 u^3 t^3+61240 u^4 t^2+23989 \
u^5 t+4186 u^6\Big) s^{14}+2 \Big(8526 t^9+72111 u t^8+283103 u^2 \
t^7+646677 u^3 t^6+948359 u^4 t^5+948359 u^5 t^4+646677 u^6 t^3+283103 u^7 \
t^2+72111 u^8 t+8526 u^9\Big) s^{13}+(t+u)^2 \Big(23886 t^8+172264 u \
t^7+582649 u^2 t^6+1131132 u^3 t^5+1344162 u^4 t^4+1131132 u^5 t^3+582649 \
u^6 t^2+172264 u^7 t+23886 u^8\Big) s^{12}+\Big(23866 t^{11}+243086 u \
t^{10}+1157425 u^2 t^9+3382695 u^3 t^8+6575795 u^4 t^7+8973837 u^5 \
t^6+8973837 u^6 t^5+6575795 u^7 t^4+3382695 u^8 t^3+1157425 u^9 t^2+243086 \
u^{10} t+23866 u^{11}\Big) s^{11}+(t+u)^2 \Big(17016 t^{10}+164440 u \
t^9+724400 u^2 t^8+1917401 u^3 t^7+3236864 u^4 t^6+3710558 u^5 t^5+3236864 \
u^6 t^4+1917401 u^7 t^3+724400 u^8 t^2+164440 u^9 t+17016 u^{10}\Big) \
s^{10}+2 \Big(4176 t^{13}+58901 u t^{12}+382615 u^2 t^{11}+1458232 u^3 \
t^{10}+3634444 u^4 t^9+6299163 u^5 t^8+8074581 u^6 t^7+8074581 u^7 \
t^6+6299163 u^8 t^5+3634444 u^9 t^4+1458232 u^{10} t^3+382615 u^{11} \
t^2+58901 u^{12} t+4176 u^{13}\Big) s^9+(t+u)^2 \Big(2552 t^{12}+42616 u \
t^{11}+321337 u^2 t^{10}+1192960 u^3 t^9+2636700 u^4 t^8+3851544 u^5 \
t^7+4187542 u^6 t^6+3851544 u^7 t^5+2636700 u^8 t^4+1192960 u^9 t^3+321337 \
u^{10} t^2+42616 u^{11} t+2552 u^{12}\Big) s^8+\Big(366 t^{15}+11590 u \
t^{14}+150117 u^2 t^{13}+868535 u^3 t^{12}+2906129 u^4 t^{11}+6451075 u^5 \
t^{10}+10280514 u^6 t^9+12609962 u^7 t^8+12609962 u^8 t^7+10280514 u^9 \
t^6
\end{dmath*}
\begin{dmath*}
{\white=}
+6451075 u^{10} t^5+2906129 u^{11} t^4+868535 u^{12} t^3+150117 u^{13} \
t^2+11590 u^{14} t+366 u^{15}\Big) s^7+t u (t+u)^2 \Big(1254 t^{12}+30458 \
u t^{11}+192025 u^2 t^{10}+634366 u^3 t^9+1369170 u^4 t^8+2072140 u^5 \
t^7+2324670 u^6 t^6+2072140 u^7 t^5+1369170 u^8 t^4+634366 u^9 t^3+192025 \
u^{10} t^2+30458 u^{11} t+1254 u^{12}\Big) s^6+2 t^2 u^2 \Big(1605 \
t^{13}+19222 u t^{12}+110551 u^2 t^{11}+398318 u^3 t^{10}+989437 u^4 \
t^9+1764628 u^5 t^8+2329703 u^6 t^7+2329703 u^7 t^6+1764628 u^8 t^5+989437 \
u^9 t^4+398318 u^{10} t^3+110551 u^{11} t^2+19222 u^{12} t+1605 \
u^{13}\Big) s^5+t^3 u^3 (t+u)^2 \Big(1678 t^{10}+19862 u t^9+93582 u^2 \
t^8+253073 u^3 t^7+439352 u^4 t^6+520162 u^5 t^5+439352 u^6 t^4+253073 u^7 \
t^3+93582 u^8 t^2+19862 u^9 t+1678 u^{10}\Big) s^4+t^4 u^4 (t+u)^3 \
\Big(1260 t^8+9096 u t^7+30583 u^2 t^6+61244 u^3 t^5+75426 u^4 t^4+61244 \
u^5 t^3+30583 u^6 t^2+9096 u^7 t+1260 u^8\Big) s^3+t^5 u^5 (t+u)^4 \
\Big(324 t^6+2168 u t^5+6197 u^2 t^4+8622 u^3 t^3+6197 u^4 t^2+2168 u^5 \
t+324 u^6\Big) s^2+10 t^6 u^6 (t+u)^7 \Big(14 t^2+27 u t+14 u^2\Big) \
s+40 t^7 u^7 (t+u)^6 \Big(t^2+u t+u^2\Big)\Big)+9 s^2 t^2 u^2 \Big(3 \
\Big(11 t^6+130 u t^5+141 u^2 t^4+604 u^3 t^3+141 u^4 t^2+130 u^5 t+11 \
u^6\Big) s^{16}+8 \Big(45 t^7+521 u t^6+1056 u^2 t^5+2598 u^3 t^4+2488 \
u^4 t^3+858 u^5 t^2+411 u^6 t+23 u^7\Big) s^{15}+6 \Big(285 t^8+3120 u \
t^7+8470 u^2 t^6+17852 u^3 t^5+25104 u^4 t^4
\end{dmath*}
\begin{dmath*}
{\white=}
+15024 u^5 t^3+5642 u^6 t^2+1908 \
u^7 t+83 u^8\Big) s^{14}+2 \Big(2375 t^9+24561 u t^8+80646 u^2 t^7+164370 \
u^3 t^6+268500 u^4 t^5+242676 u^5 t^4+112746 u^6 t^3+43806 u^7 t^2+11685 u^8 \
t+539 u^9\Big) s^{13}+\Big(8669 t^{10}+85044 u t^9+320811 u^2 t^8+677420 \
u^3 t^7+1161930 u^4 t^6+1394268 u^5 t^5+895510 u^6 t^4+373044 u^7 t^3+149697 \
u^8 t^2+34384 u^9 t+2343 u^{10}\Big) s^{12}+12 \Big(915 t^{11}+8573 u \
t^{10}+35593 u^2 t^9+81037 u^3 t^8+140831 u^4 t^7+205671 u^5 t^6+184091 u^6 \
t^5+94279 u^7 t^4+41772 u^8 t^3+16990 u^9 t^2+3718 u^{10} t+370 \
u^{11}\Big) s^{11}+\Big(9845 t^{12}+88872 u t^{11}+394182 u^2 \
t^{10}+982076 u^3 t^9+1680477 u^4 t^8+2723040 u^5 t^7+3377880 u^6 \
t^6+2469456 u^7 t^5+1323771 u^8 t^4+704312 u^9 t^3+264210 u^{10} t^2+55188 \
u^{11} t+6179 u^{12}\Big) s^{10}+2 \Big(3095 t^{13}+27024 u t^{12}+127632 \
u^2 t^{11}+352506 u^3 t^{10}+563960 u^4 t^9+824355 u^5 t^8+1409056 u^6 \
t^7+1665520 u^7 t^6+1351875 u^8 t^5+966790 u^9 t^4+514400 u^{10} t^3+159702 \
u^{11} t^2+28798 u^{12} t+2919 u^{13}\Big) s^9+3 \Big(870 t^{14}+7314 u \
t^{13}+37677 u^2 t^{12}
\end{dmath*}

\intertext{}
\begin{dmath*}
{\white =}
+112780 u^3 t^{11}+143003 u^4 t^{10}+53282 u^5 \
t^9+249844 u^6 t^8+845260 u^7 t^7+1239890 u^8 t^6+1252614 u^9 t^5+905087 \
u^{10} t^4+401760 u^{11} t^3+101733 u^{12} t^2+14670 u^{13} t+1176 \
u^{14}\Big) s^8+4 \Big(166 t^{15}+1359 u t^{14}+8343 u^2 t^{13}+18170 u^3 \
t^{12}-29355 u^4 t^{11}-193428 u^5 t^{10}-186082 u^6 t^9+373479 u^7 \
t^8+1074360 u^8 t^7+1349559 u^9 t^6+1156713 u^{10} t^5+673434 u^{11} \
t^4+244731 u^{12} t^3+51600 u^{13} t^2+5604 u^{14} t+307 u^{15}\Big) \
s^7+\Big(77 t^{16}+712 u t^{15}+7926 u^2 t^{14}-22420 u^3 t^{13}-310304 u^4 \
t^{12}-964008 u^5 t^{11}-1004190 u^6 t^{10}+1290836 u^7 t^9+5018184 u^8 \
t^8+6920472 u^9 t^7+6102774 u^{10} t^6+3900588 u^{11} t^5+1764592 u^{12} \
t^4+528392 u^{13} t^3+92178 u^{14} t^2+6772 u^{15} t+187 u^{16}\Big) s^6+6 \
t u \Big(5 t^{15}+338 u t^{14}-3532 u^2 t^{13}-33260 u^3 t^{12}-97374 u^4 \
t^{11}-95800 u^5 t^{10}+181294 u^6 t^9+739773 u^7 t^8+1163599 u^8 \
t^7+1123652 u^9 t^6+758126 u^{10} t^5+368632 u^{11} t^4+127306 u^{12} \
t^3+29970 u^{13} t^2+4032 u^{14} t+151 u^{15}\Big) s^5+t^2 u^2 \Big(327 \
t^{14}-6804 u t^{13}-59454 u^2 t^{12}-167472 u^3 t^{11}-99128 u^4 \
t^{10}+699492 u^5 t^9+2552097 u^6 t^8+4572060 u^7 t^7+5195055 u^8 \
t^6+4069272 u^9 t^5+2242188 u^{10} t^4+847572 u^{11} t^3+212250 u^{12} \
t^2+33688 u^{13} t+2793 u^{14}\Big) s^4+4 t^3 u^3 \Big(-233 t^{13}-1616 u \
t^{12}-3051 u^2 t^{11}+10663 u^3 t^{10}
\end{dmath*}
\begin{dmath*}
{\white=}
+80192 u^4 t^9+249270 u^5 t^8+483611 \
u^6 t^7+633999 u^7 t^6+583800 u^8 t^5+382847 u^9 t^4+174178 u^{10} t^3+51444 \
u^{11} t^2+8727 u^{12} t+617 u^{13}\Big) s^3+3 t^4 u^4 (t+u)^2 \Big(109 \
t^{10}+662 u t^9+4205 u^2 t^8+17972 u^3 t^7+46152 u^4 t^6+74668 u^5 \
t^5+76440 u^6 t^4+53188 u^7 t^3+24579 u^8 t^2+6934 u^9 t+931 u^{10}\Big) \
s^2+2 t^5 u^5 (t+u)^3 \Big(15 t^8+481 u t^7+3156 u^2 t^6+8739 u^3 t^5+13780 \
u^4 t^4+12711 u^5 t^3+7500 u^6 t^2+2637 u^7 t+453 u^8\Big) s+t^6 u^6 \
(t+u)^6 \Big(77 t^4+270 u t^3+477 u^2 t^2+350 u^3 t+187 u^4\Big)\Big) \
H(1,z)+9 s^2 t^2 u^2 \Big(3 \Big(13 t^6+142 u t^5+171 u^2 t^4+644 u^3 \
t^3+171 u^4 t^2+142 u^5 t+13 u^6\Big) s^{16}+8 \Big(50 t^7+578 u t^6+1293 \
u^2 t^5+3103 u^3 t^4+3103 u^4 t^3+1293 u^5 t^2+578 u^6 t+50 u^7\Big) \
s^{15}+12 \Big(157 t^8+1789 u t^7+5420 u^2 t^6+12039 u^3 t^5+17342 u^4 \
t^4+12039 u^5 t^3+5420 u^6 t^2+1789 u^7 t+157 u^8\Big) s^{14}+\Big(5374 \
t^9+59826 u t^8+223092 u^2 t^7+513996 u^3 t^6+871536 u^4 t^5+871536 u^5 \
t^4+513996 u^6 t^3+223092 u^7 t^2+59826 u^8 t+5374 u^9\Big) \
s^{13}+\Big(10359 t^{10}+114484 u t^9+503079 u^2 t^8+1279432 u^3 \
t^7+2390090 u^4 t^6+3044424 u^5 t^5+2390090 u^6 t^4+1279432 u^7 t^3+503079 \
u^8 t^2+114484 u^9 t+10359 u^{10}\Big) s^{12}+12 \Big(1186 t^{11}+13429 u \
t^{10}+67778 u^2 t^9+197005 u^3 t^8+403359 u^4 t^7+605123 u^5 t^6+605123 u^6 \
t^5+403359 u^7 t^4+197005 u^8 t^3+67778 u^9 t^2+13429 u^{10} t+1186 \
u^{11}\Big) s^{11}+\Big(14195 t^{12}+171720 u t^{11}+991722 u^2 \
t^{10}+3343932 u^3 t^9+7591317 u^4 t^8+12777132 u^5 t^7+15424700 u^6 \
t^6+12777132 u^7 t^5+7591317 u^8 t^4+3343932 u^9 t^3+991722 u^{10} \
t^2+171720 u^{11} t+14195 u^{12}\Big) s^{10}+2 \Big(5067 t^{13}+68848 u \
t^{12}+464430 u^2 t^{11}+1834210 u^3 t^{10}+4687302 u^4 t^9+8645967 u^5 \
t^8+11943168 u^6 t^7+11943168 u^7 t^6+8645967 u^8 t^5+4687302 u^9 \
t^4+1834210 u^{10} t^3+464430 u^{11} t^2+68848 u^{12} t+5067 u^{13}\Big) \
s^9+3 \Big(1638 t^{14}+26822 u t^{13}+219527 u^2 t^{12}+1022692 u^3 \
t^{11}+2994269 u^4 t^{10}+6115342 u^5 t^9+9403374 u^6 t^8
\end{dmath*}
\begin{dmath*}
{\white=}
+10916304 u^7 \
t^7+9403374 u^8 t^6+6115342 u^9 t^5+2994269 u^{10} t^4+1022692 u^{11} \
t^3+219527 u^{12} t^2+26822 u^{13} t+1638 u^{14}\Big) s^8+4 \Big(361 \
t^{15}+8109 u t^{14}+85470 u^2 t^{13}+471328 u^3 t^{12}+1600674 u^4 \
t^{11}+3733632 u^5 t^{10}+6488383 u^6 t^9+8627643 u^7 t^8+8627643 u^8 \
t^7+6488383 u^9 t^6+3733632 u^{10} t^5+1600674 u^{11} t^4+471328 u^{12} \
t^3+85470 u^{13} t^2+8109 u^{14} t+361 u^{15}\Big) s^7+\Big(193 \
t^{16}+8108 u t^{15}+123366 u^2 t^{14}+816896 u^3 t^{13}+3259172 u^4 \
t^{12}+8952972 u^5 t^{11}+18149594 u^6 t^{10}+27988696 u^7 t^9+32478774 u^8 \
t^8+27988696 u^9 t^7+18149594 u^{10} t^6+8952972 u^{11} t^5+3259172 u^{12} \
t^4+816896 u^{13} t^3+123366 u^{14} t^2+8108 u^{15} t+193 u^{16}\Big) \
s^6+6 t u \Big(157 t^{15}+4612 u t^{14}+39024 u^2 t^{13}+191670 u^3 \
t^{12}+643658 u^4 t^{11}+1557030 u^5 t^{10}+2799334 u^6 t^9+3770803 u^7 \
t^8+3770803 u^8 t^7+2799334 u^9 t^6+1557030 u^{10} t^5+643658 u^{11} \
t^4+191670 u^{12} t^3+39024 u^{13} t^2+4612 u^{14} t+157 u^{15}\Big) \
s^5+t^2 u^2 \Big(2883 t^{14}+38608 u t^{13}+269730 u^2 t^{12}+1182108 u^3 \
t^{11}+3470348 u^4 t^{10}+7219608 u^5 t^9+11105895 u^6 t^8+12818744 u^7 \
t^7+11105895 u^8 t^6+7219608 u^9 t^5+3470348 u^{10} t^4+1182108 u^{11} \
t^3+269730 u^{12} t^2+38608 u^{13} t+2883 u^{14}\Big) s^4+4 t^3 u^3 \
\Big(647 t^{13}+9737 u t^{12}+60783 u^2 t^{11}+220492 u^3 t^{10}+533350 u^4 \
t^9+931704 u^5 t^8+1224463 u^6 t^7+1224463 u^7 t^6+931704 u^8 t^5+533350 u^9 \
t^4+220492 u^{10} t^3+60783 u^{11} t^2+9737 u^{12} t+647 u^{13}\Big) s^3+3 \
t^4 u^4 (t+u)^2 \Big(961 t^{10}+7506 u t^9+28145 u^2 t^8+66084 u^3 \
t^7+107838 u^4 t^6+127716 u^5 t^5+107838 u^6 t^4+66084 u^7 t^3+28145 u^8 \
t^2+7506 u^9 t+961 u^{10}\Big) s^2+6 t^5 u^5 (t+u)^3 \Big(157 t^8+937 u \
t^7+2766 u^2 t^6+5043 u^3 t^5+6226 u^4 t^4+5043 u^5 t^3+2766 u^6 t^2+937 u^7 \
t+157 u^8\Big) s+t^6 u^6 (t+u)^6 \Big(193 t^4+354 u t^3+537 u^2 t^2+354 \
u^3 t+193 u^4\Big)\Big) H(2,y)   \Bigg\}  \Big/ \Big(  {27 s^3 t^3 u^3 (s+t)^6 (s+u)^6 (t+u)^6} \Big)
\end{dmath*}

\intertext{}
\begin{dmath*}
 {\cal A}^{(2)}_{5 ; C_A n_f} =  \Bigg\{  216 s^2 t^2 u^3 (s+u)^6 (t+u)^6 \Big(s^2-t s+t^2+u^2\Big) H(0,1,y) \
(s+t)^7+162 s^2 t^3 u^3 (s+u)^6 (t+u)^6 \Big(t^2+u^2\Big) H(0,y) H(0,z) \
(s+t)^6
\end{dmath*}
\begin{dmath*}
{\white=}
+54 s^2 t^2 u^3 (s+u)^6 (t+u)^6 \Big(16 s^3-21 u s^2+7 u^2 s+t \
\Big(16 t^2-21 u t+7 u^2\Big)\Big) H(0,z) H(1,y) (s+t)^6-54 s^2 t^2 u^3 \
(s+u)^6 (t+u)^6 \Big(10 s^3-21 u s^2+19 u^2 s+t \Big(-8 t^2+21 u t-17 \
u^2\Big)\Big) H(0,y) H(1,z) (s+t)^6+1134 s^2 t^2 u^3 (s+u)^6 (t+u)^6 \
\Big(s^3-2 u s^2+u^2 s+t (t-u)^2\Big) H(1,y) H(1,z) (s+t)^6+54 s^2 t^2 \
u^2 (s+u)^6 (t+u)^6 \Big(-2 (5 t+8 u) s^3+21 \Big(t^2+u^2\Big) \
s^2-\Big(19 t^3+7 u^3\Big) s+t u \Big(t^2+u^2\Big)\Big) H(0,z) \
H(2,y) (s+t)^6+54 s^2 t^2 u^2 (s+u)^6 (t+u)^6 \Big(-4 (5 t+6 u) s^3+21 \
\Big(t^2+2 u^2\Big) s^2-\Big(11 t^3+24 u^3\Big) s+t u \
\Big(t^2+u^2\Big)\Big) H(1,z) H(2,y) (s+t)^6+162 s^3 t^2 u^2 (s+u)^6 \
(t+u)^6 \Big(2 (t+u) s^2-7 \Big(t^2+u^2\Big) s+5 \
\Big(t^3+u^3\Big)\Big) H(1,z) H(3,y) (s+t)^6+162 s^2 t^3 u^3 (s+u)^6 \
(t+u)^6 \Big(2 t^2-7 u t+5 u^2\Big) H(0,0,y) (s+t)^6
\end{dmath*}

\intertext{}
\begin{dmath*}
{\white =}
+162 s^2 t^3 u^3 \
(s+u)^6 (t+u)^6 \Big(5 t^2-7 u t+2 u^2\Big) H(0,0,z) (s+t)^6+54 s^2 t^2 \
u^3 (s+u)^6 (t+u)^6 \Big(7 s^3-21 u s^2+16 u^2 s+t \Big(25 t^2-42 u t+25 \
u^2\Big)\Big) H(0,1,z) (s+t)^6+54 s^2 t^2 u^2 (s+u)^6 (t+u)^6 \
\Big((t-10 u) s^3+21 u^2 s^2+\Big(t^3-19 u^3\Big) s+21 t (t-u)^2 \
u\Big) H(0,2,y) (s+t)^6+54 s^2 t^2 u^3 (s+u)^6 (t+u)^6 \Big(10 s^3-21 u \
s^2+19 u^2 s+t^3+t u^2\Big) H(1,0,y) (s+t)^6-54 s^2 t^2 u^3 (s+u)^6 \
(t+u)^6 \Big(16 s^3-21 u s^2+7 u^2 s-2 t \Big(t^2+u^2\Big)\Big) \
H(1,0,z) (s+t)^6-54 s^2 t^2 u^3 (s+u)^6 (t+u)^6 \Big(s^3-21 u s^2+10 u^2 \
s+t \Big(t^2-21 u t+10 u^2\Big)\Big) H(1,1,y) (s+t)^6-162 s^2 t^2 u^3 \
(s+u)^6 (t+u)^6 \Big(8 s^3-14 u s^2+8 u^2 s+t \Big(-4 t^2+7 u t-7 \
u^2\Big)\Big) H(1,1,z) (s+t)^6+1134 s^2 t^2 u^3 (s+u)^6 (t+u)^6 \
\Big(s^3-2 u s^2+u^2 s+t (t-u)^2\Big) H(1,2,y) (s+t)^6+54 s^2 t^2 u^2 \
(s+u)^6 (t+u)^6 \Big(-2 (8 t+5 u) s^3+21 \Big(t^2+u^2\Big) s^2-\Big(7 \
t^3+19 u^3\Big) s+t u \Big(t^2+u^2\Big)\Big) H(2,0,y) (s+t)^6+54 s^2 \
t^2 u^3 (s+u)^6 (t+u)^6 \Big(4 s^3-21 u s^2+13 u^2 s+t^3+t u^2\Big) \
H(2,1,y) (s+t)^6-324 s^3 t^2 u^2 (s+u)^6 (t+u)^6 \Big(4 (t+u) s^2-7 \
\Big(t^2+u^2\Big) s+4 \Big(t^3+u^3\Big)\Big) H(2,2,y) (s+t)^6+162 \
s^3 t^2 u^2 (s+u)^6 (t+u)^6 \Big(2 (t+u) s^2-7 \Big(t^2+u^2\Big) s+5 \
\Big(t^3+u^3\Big)\Big) H(3,2,y) (s+t)^6-9 s^2 t^2 u^2 (t+u)^6 \Big(10 \
s^{14}+(60 t+74 u) s^{13}+2 \Big(85 t^2+94 u t+129 u^2\Big) s^{12}+3 \
\Big(100 t^3+74 u t^2+123 u^2 t+190 u^3\Big) s^{11}+\Big(360 t^4+199 u \
t^3+954 u^2 t^2+467 u^3 t+900 u^4\Big) s^{10}+\Big(300 t^5-194 u t^4+4854 \
u^2 t^3+635 u^3 t^2+348 u^4 t+1074 u^5\Big) s^9+2 \Big(85 t^6-429 u \
t^5+4734 u^2 t^4+2414 u^3 t^3-498 u^4 t^2+135 u^5 t+492 u^6\Big) \
s^8+\Big(60 t^7-874 u t^6+8919 u^2 t^5+5677 u^3 t^4+1791 u^4 t^3-159 u^5 \
t^2+165 u^6 t+678 u^7\Big) s^7+\Big(10 t^8-337 u t^7+4848 u^2 t^6-2634 \
u^3 t^5+4506 u^4 t^4+5835 u^5 t^3+288 u^6 t^2+39 u^7 t+330 u^8\Big) s^6+u \
\Big(-36 t^8+1746 u t^7-7930 u^2 t^6+1614 u^3 t^5+14679 u^4 t^4+3583 u^5 \
t^3-339 u^6 t^2+150 u^7 t+100 u^8\Big) s^5+u^2 \Big(360 t^8
\end{dmath*}
\begin{dmath*}
{\white=}
-4417 u \
t^7-1014 u^2 t^6+13797 u^3 t^5+8278 u^4 t^4-689 u^5 t^3+306 u^6 t^2+172 u^7 \
t+14 u^8\Big) s^4+t u^3 \Big(-708 t^7-45 u t^6+4974 u^2 t^5+6885 u^3 \
t^4+1171 u^4 t^3-108 u^5 t^2+489 u^6 t+44 u^7\Big) s^3+3 t^2 u^4 \Big(120 \
t^6+114 u t^5+580 u^2 t^4+562 u^3 t^3+152 u^4 t^2+117 u^5 t+42 u^6\Big) \
s^2+t^3 u^5 \Big(-36 t^5-53 u t^4+526 u^2 t^3+402 u^3 t^2+271 u^4 t+104 \
u^5\Big) s+t^4 u^6 \Big(10 t^4+40 u t^3+72 u^2 t^2+49 u^3 t+56 \
u^4\Big)\Big) H(0,y) (s+t)^2+9 s^2 t^2 u^2 (s+u)^6 (t+u)^6 \Big(6 \
s^8+(78 t+22 u) s^7+\Big(324 t^2+202 u t+42 u^2\Big) s^6+3 \Big(230 \
t^3+202 u t^2+186 u^2 t+7 u^3\Big) s^5+\Big(876 t^4+962 u t^3+1710 u^2 \
t^2+103 u^3 t+58 u^4\Big) s^4+2 t \Big(345 t^4+481 u t^3+1194 u^2 t^2+38 \
u^3 t+122 u^4\Big) s^3+2 t^2 \Big(162 t^4+303 u t^3+855 u^2 t^2+38 u^3 \
t+162 u^4\Big) s^2+t^3 \Big(78 t^4+202 u t^3+558 u^2 t^2+103 u^3 t+244 \
u^4\Big) s+t^4 \Big(6 t^4+22 u t^3+42 u^2 t^2+21 u^3 t+58 \
u^4\Big)\Big) H(1,y) (s+t)^2
\end{dmath*}

\begin{dmath*}
{\white =}
-\frac{3}{2} t^2 u^2 (s+u) \Big(462 (t+u)^6 \
s^{16}+6 \Big(533 t^7+3647 u t^6+11355 u^2 t^5+18041 u^3 t^4+18041 u^4 \
t^3+11355 u^5 t^2+3647 u^6 t+533 u^7\Big) s^{15}+6 (t+u)^2 \Big(1725 \
t^6+9572 u t^5+26881 u^2 t^4+27580 u^3 t^3+26881 u^4 t^2+9572 u^5 t+1725 \
u^6\Big) s^{14}+\Big(20814 t^9+169590 u t^8+698271 u^2 t^7+1572623 u^3 \
t^6+2180718 u^4 t^5+2180718 u^5 t^4+1572623 u^6 t^3+698271 u^7 t^2+169590 \
u^8 t+20814 u^9\Big) s^{13}+(t+u)^2 \Big(28872 t^8+187896 u t^7+702435 \
u^2 t^6+1379212 u^3 t^5+1402562 u^4 t^4+1379212 u^5 t^3+702435 u^6 \
t^2+187896 u^7 t+28872 u^8\Big) s^{12}+\Big(28752 t^{11}+247032 u \
t^{10}+1200159 u^2 t^9+3905948 u^3 t^8+7929314 u^4 t^7+10661099 u^5 \
t^6+10661099 u^6 t^5+7929314 u^7 t^4+3905948 u^8 t^3+1200159 u^9 t^2+247032 \
u^{10} t+28752 u^{11}\Big) s^{11}+(t+u)^2 \Big(20598 t^{10}+137298 u \
t^9+671079 u^2 t^8+2337710 u^3 t^7+4295427 u^4 t^6+4577312 u^5 t^5+4295427 \
u^6 t^4+2337710 u^7 t^3+671079 u^8 t^2+137298 u^9 t+20598 u^{10}\Big) \
s^{10}+\Big(10230 t^{13}+94398 u t^{12}+670143 u^2 t^{11}+3176397 u^3 \
t^{10}+9253014 u^4 t^9+16795202 u^5 t^8+21138216 u^6 t^7+21138216 u^7 \
t^6+16795202 u^8 t^5+9253014 u^9 t^4+3176397 u^{10} t^3+670143 u^{11} \
t^2+94398 u^{12} t+10230 u^{13}\Big) s^9+(t+u)^2 \Big(3174 t^{12}+28308 u \
t^{11}+353829 u^2 t^{10}+1598364 u^3 t^9+3798910 u^4 t^8+5149552 u^5 \
t^7+4884942 u^6 t^6+5149552 u^7 t^5+3798910 u^8 t^4+1598364 u^9 t^3+353829 \
u^{10} t^2+28308 u^{11} t+3174 u^{12}\Big) s^8+\Big(462 t^{15}+7482 u \
t^{14}+190719 u^2 t^{13}+1263290 u^3 t^{12}+4354796 u^4 t^{11}+9652847 u^5 \
t^{10}+14817008 u^6 t^9+17262724 u^7 t^8+17262724 u^8 t^7+14817008 u^9 \
t^6+9652847 u^{10} t^5+4354796 u^{11} t^4+1263290 u^{12} t^3+190719 u^{13} \
t^2+7482 u^{14} t+462 u^{15}\Big) s^7+t u (t+u)^2 \Big(690 t^{12}+52407 u \
t^{11}
\end{dmath*}
\begin{dmath*}
{\white=}
+300320 u^2 t^{10}+958425 u^3 t^9+2341562 u^4 t^8+3965828 u^5 \
t^7+4535936 u^6 t^6+3965828 u^7 t^5+2341562 u^8 t^4+958425 u^9 t^3+300320 \
u^{10} t^2+52407 u^{11} t+690 u^{12}\Big) s^6+t^2 u^2 \Big(6786 \
t^{13}+57592 u t^{12}+322024 u^2 t^{11}+1440278 u^3 t^{10}+4550721 u^4 \
t^9+9574691 u^5 t^8+13690372 u^6 t^7+13690372 u^7 t^6+9574691 u^8 \
t^5+4550721 u^9 t^4+1440278 u^{10} t^3+322024 u^{11} t^2+57592 u^{12} t+6786 \
u^{13}\Big) s^5+t^3 u^3 (t+u)^2 \Big(74 t^{10}+34768 u t^9+270574 u^2 \
t^8+947815 u^3 t^7+1872940 u^4 t^6+2304682 u^5 t^5+1872940 u^6 t^4+947815 \
u^7 t^3+270574 u^8 t^2+34768 u^9 t+74 u^{10}\Big) s^4+t^4 u^4 (t+u)^3 \
\Big(4432 t^8+44590 u t^7+177657 u^2 t^6+382429 u^3 t^5+481348 u^4 \
t^4+382429 u^5 t^3+177657 u^6 t^2+44590 u^7 t+4432 u^8\Big) s^3+t^5 u^5 \
(t+u)^4 \Big(2948 t^6+20395 u t^5+53828 u^2 t^4+72306 u^3 t^3+53828 u^4 \
t^2+20395 u^5 t+2948 u^6\Big) s^2+720 t^6 u^6 (t+u)^7 \Big(2 t^2+3 u t+2 \
u^2\Big) s+288 t^7 u^7 (t+u)^6 \Big(t^2+u t+u^2\Big)\Big) (s+t)-9 s^2 \
t^2 u^2 (s+u)^2 (t+u)^6 \Big(10 s^{14}+(74 t+60 u) s^{13}+2 \Big(129 \
t^2+94 u t+85 u^2\Big) s^{12}+3 \Big(190 t^3+123 u t^2+74 u^2 t+100 \
u^3\Big) s^{11}+\Big(900 t^4+467 u t^3+954 u^2 t^2+199 u^3 t+360 \
u^4\Big) s^{10}+\Big(1074 t^5+348 u t^4+635 u^2 t^3+4854 u^3 t^2-194 u^4 \
t+300 u^5\Big) s^9+2 \Big(492 t^6+135 u t^5-498 u^2 t^4+2414 u^3 t^3+4734 \
u^4 t^2-429 u^5 t+85 u^6\Big) s^8+\Big(678 t^7+165 u t^6-159 u^2 t^5+1791 \
u^3 t^4+5677 u^4 t^3+8919 u^5 t^2-874 u^6 t+60 u^7\Big) s^7+\Big(330 \
t^8+39 u t^7+288 u^2 t^6+5835 u^3 t^5+4506 u^4 t^4-2634 u^5 t^3+4848 u^6 \
t^2-337 u^7 t+10 u^8\Big) s^6+t \Big(100 t^8+150 u t^7-339 u^2 t^6+3583 \
u^3 t^5+14679 u^4 t^4+1614 u^5 t^3-7930 u^6 t^2+1746 u^7 t-36 u^8\Big) \
s^5+t^2 \Big(14 t^8+172 u t^7+306 u^2 t^6-689 u^3 t^5+8278 u^4 t^4+13797 \
u^5 t^3-1014 u^6 t^2-4417 u^7 t+360 u^8\Big) s^4+t^3 u \Big(44 t^7+489 u \
t^6-108 u^2 t^5+1171 u^3 t^4+6885 u^4 t^3+4974 u^5 t^2-45 u^6 t-708 \
u^7\Big) s^3+3 t^4 u^2 \Big(42 t^6+117 u t^5+152 u^2 t^4+562 u^3 t^3+580 \
u^4 t^2+114 u^5 t+120 u^6\Big) s^2
\end{dmath*}

\begin{dmath*}
{\white =}
+t^5 u^3 \Big(104 t^5+271 u t^4+402 u^2 \
t^3+526 u^3 t^2-53 u^4 t-36 u^5\Big) s+t^6 u^4 \Big(56 t^4+49 u t^3+72 \
u^2 t^2+40 u^3 t+10 u^4\Big)\Big) H(0,z)-9 s^2 t^2 u^2 \Big(2 \Big(3 \
t^6+60 u t^5-78 u^2 t^4+490 u^3 t^3-78 u^4 t^2+60 u^5 t+3 u^6\Big) \
s^{16}+\Big(72 t^7+1289 u t^6+414 u^2 t^5+8005 u^3 t^4+7845 u^4 t^3+126 u^5 \
t^2+1129 u^6 t+40 u^7\Big) s^{15}+6 \Big(60 t^8+900 u t^7+1076 u^2 \
t^6+4412 u^3 t^5+9687 u^4 t^4+3894 u^5 t^3+558 u^6 t^2+678 u^7 t+23 \
u^8\Big) s^{14}+\Big(1024 t^9+12954 u t^8+24576 u^2 t^7+46761 u^3 \
t^6+163314 u^4 t^5+154677 u^5 t^4+29274 u^6 t^3+11781 u^7 t^2+8316 u^8 t+331 \
u^9\Big) s^{13}+\Big(1874 t^{10}+20222 u t^9+53220 u^2 t^8+61201 u^3 \
t^7+246846 u^4 t^6+451965 u^5 t^5+210152 u^6 t^4+17091 u^7 t^3+26370 u^8 \
t^2+11421 u^9 t+638 u^{10}\Big) s^{12}+3 \Big(780 t^{11}+7249 u \
t^{10}+25014 u^2 t^9+30013 u^3 t^8+86909 u^4 t^7+276042 u^5 t^6+269826 u^6 \
t^5+71475 u^7 t^4+13798 u^8 t^3+15298 u^9 t^2+4109 u^{10} t+359 \
u^{11}\Big) s^{11}+\Big(2042 t^{12}+16125 u t^{11}+76122 u^2 \
t^{10}+155760 u^3 t^9+296460 u^4 t^8+1068120 u^5 t^7+1786074 u^6 t^6+1133742 \
u^7 t^5+352110 u^8 t^4+166875 u^9 t^3+68808 u^{10} t^2+12114 u^{11} t+1520 \
u^{12}\Big) s^{10}+\Big(1240 t^{13}+7152 u t^{12}+58566 u^2 t^{11}+232980 \
u^3 t^{10}+464512 u^4 t^9+1090308 u^5 t^8+2341921 u^6 t^7+2520886 u^7 \
t^6+1455108 u^8 t^5+740077 u^9 t^4+345463 u^{10} t^3+84624 u^{11} t^2+10790 \
u^{12} t+1573 u^{13}\Big) s^9+3 \Big(168 t^{14}+208 u t^{13}+10592 u^2 \
t^{12}+73981 u^3 t^{11}+192464 u^4 t^{10}+350678 u^5 t^9+731518 u^6 \
t^8
\end{dmath*}
\begin{dmath*}
{\white=}
+1095507 u^7 t^7+938868 u^8 t^6+596751 u^9 t^5+345880 u^{10} t^4+133260 \
u^{11} t^3+25570 u^{12} t^2+2627 u^{13} t+352 u^{14}\Big) s^8+\Big(124 \
t^{15}-1050 u t^{14}+11760 u^2 t^{13}+109681 u^3 t^{12}+348969 u^4 \
t^{11}+744780 u^5 t^{10}+1812415 u^6 t^9+3571941 u^7 t^8+4045692 u^8 \
t^7+2838503 u^9 t^6+1635744 u^{10} t^5+818670 u^{11} t^4+274406 u^{12} \
t^3+50556 u^{13} t^2+4374 u^{14} t+403 u^{15}\Big) s^7+\Big(14 t^{16}-459 \
u t^{15}+4194 u^2 t^{14}+16667 u^3 t^{13}+21936 u^4 t^{12}+169446 u^5 \
t^{11}+1113116 u^6 t^{10}+3298442 u^7 t^9+4969926 u^8 t^8+4145565 u^9 \
t^7+2172862 u^{10} t^6+899067 u^{11} t^5+350974 u^{12} t^4+118004 u^{13} \
t^3+24360 u^{14} t^2+1596 u^{15} t+66 u^{16}\Big) s^6+3 t u \Big(-20 \
t^{15}+510 u t^{14}-2846 u^2 t^{13}-25726 u^3 t^{12}-36057 u^4 t^{11}+137862 \
u^5 t^{10}+687022 u^6 t^9+1363652 u^7 t^8+1488125 u^8 t^7+962136 u^9 \
t^6+387285 u^{10} t^5+104624 u^{11} t^4+27898 u^{12} t^3+10766 u^{13} \
t^2+2505 u^{14} t+88 u^{15}\Big) s^5+t^2 u^2 \Big(276 t^{14}-4915 u \
t^{13}-35832 u^2 t^{12}-60573 u^3 t^{11}+131822 u^4 t^{10}+839823 u^5 \
t^9+1989780 u^6 t^8+2753037 u^7 t^7+2394378 u^8 t^6+1355826 u^9 t^5+497554 \
u^{10} t^4+108936 u^{11} t^3+16362 u^{12} t^2+4782 u^{13} t+1056 \
u^{14}\Big) s^4+t^3 u^3 \Big(-820 t^{13}-4665 u t^{12}-3330 u^2 \
t^{11}+50014 u^3 t^{10}+234315 u^4 t^9+583398 u^5 t^8+954377 u^6 t^7+1055745 \
u^7 t^6+809880 u^8 t^5+443150 u^9 t^4+168838 u^{10} t^3+40614 u^{11} \
t^2+5192 u^{12} t+196 u^{13}\Big) s^3+3 t^4 u^4 (t+u)^2 \Big(92 \
t^{10}+476 u t^9+2090 u^2 t^8+7269 u^3 t^7+17036 u^4 t^6+25733 u^5 t^5+22780 \
u^6 t^4+14183 u^7 t^3+6446 u^8 t^2+2047 u^9 t+352 u^{10}\Big) s^2-t^5 u^5 \
(t+u)^3 \Big(60 t^8+29 u t^7-1449 u^2 t^6-4932 u^3 t^5-8177 u^4 t^4-6609 \
u^5 t^3-3504 u^6 t^2-1214 u^7 t-264 u^8\Big) s+t^6 u^6 (t+u)^6 \Big(14 \
t^4+90 u t^3+126 u^2 t^2+89 u^3 t+66 u^4\Big)\Big) H(1,z)
\end{dmath*}

\begin{dmath*}
{\white =}
-9 s^2 t^2 u^2 \
\Big(2 \Big(6 t^6+78 u t^5-33 u^2 t^4+550 u^3 t^3-33 u^4 t^2+78 u^5 t+6 \
u^6\Big) s^{16}+\Big(130 t^7+1727 u t^6+1824 u^2 t^5+10515 u^3 t^4+10515 \
u^4 t^3+1824 u^5 t^2+1727 u^6 t+130 u^7\Big) s^{15}+6 \Big(104 t^8+1295 u \
t^7+2603 u^2 t^6+7743 u^3 t^5+14182 u^4 t^4+7743 u^5 t^3+2603 u^6 t^2+1295 \
u^7 t+104 u^8\Big) s^{14}+\Big(1747 t^9+20760 u t^8+60177 u^2 t^7+138345 \
u^3 t^6+311235 u^4 t^5+311235 u^5 t^4+138345 u^6 t^3+60177 u^7 t^2+20760 u^8 \
t+1747 u^9\Big) s^{13}+\Big(3218 t^{10}+37773 u t^9+146652 u^2 t^8+339358 \
u^3 t^7+769808 u^4 t^6+1107894 u^5 t^5+769808 u^6 t^4+339358 u^7 t^3+146652 \
u^8 t^2+37773 u^9 t+3218 u^{10}\Big) s^{12}+3 \Big(1403 t^{11}+16697 u \
t^{10}+82856 u^2 t^9+227499 u^3 t^8+514057 u^4 t^7+898468 u^5 t^6+898468 u^6 \
t^5+514057 u^7 t^4+227499 u^8 t^3+82856 u^9 t^2+16697 u^{10} t+1403 \
u^{11}\Big) s^{11}+2 \Big(2050 t^{12}+24939 u t^{11}+154530 u^2 \
t^{10}+531155 u^3 t^9+1271385 u^4 t^8+2420490 u^5 t^7+3112438 u^6 \
t^6+2420490 u^7 t^5+1271385 u^8 t^4+531155 u^9 t^3+154530 u^{10} t^2+24939 \
u^{11} t+2050 u^{12}\Big) s^{10}+\Big(2989 t^{13}+37142 u t^{12}+287298 \
u^2 t^{11}+1240898 u^3 t^{10}+3321201 u^4 t^9+6609462 u^5 t^8+9856850 u^6 \
t^7+9856850 u^7 t^6+6609462 u^8 t^5+3321201 u^9 t^4+1240898 u^{10} \
t^3+287298 u^{11} t^2+37142 u^{12} t+2989 u^{13}\Big) s^9+3 \Big(514 \
t^{14}+6775 u t^{13}+65664 u^2 t^{12}+346961 u^3 t^{11}+1076100 u^4 \
t^{10}+2314869 u^5 t^9+3821756 u^6 t^8+4598458 u^7 t^7+3821756 u^8 \
t^6+2314869 u^9 t^5+1076100 u^{10} t^4+346961 u^{11} t^3+65664 u^{12} \
t^2+6775 u^{13} t+514 u^{14}\Big) s^8+\Big(493 t^{15}+8076 u t^{14}+98952 \
u^2 t^{13}+596673 u^3 t^{12}+2146416 u^4 t^{11}+5342982 u^5 t^{10}+10174467 \
u^6 t^9+14554545 u^7 t^8+14554545 u^8 t^7+10174467 u^9 t^6+5342982 u^{10} \
t^5+2146416 u^{11} t^4+596673 u^{12} t^3+98952 u^{13} t^2+8076 u^{14} t+493 \
u^{15}\Big) s^7+\Big(72 t^{16}+2194 u t^{15}+36630 u^2 t^{14}
\end{dmath*}
\begin{dmath*}
{\white=}
+227075 u^3 \
t^{13}+910630 u^4 t^{12}+2784993 u^5 t^{11}+6611664 u^6 t^{10}+11660494 u^7 \
t^9+14240640 u^8 t^8+11660494 u^9 t^7+6611664 u^{10} t^6+2784993 u^{11} \
t^5+910630 u^{12} t^4+227075 u^{13} t^3+36630 u^{14} t^2+2194 u^{15} t+72 \
u^{16}\Big) s^6+3 t u \Big(100 t^{15}+3071 u t^{14}+18464 u^2 \
t^{13}+80084 u^3 t^{12}+323267 u^4 t^{11}+1009711 u^5 t^{10}+2219756 u^6 \
t^9+3327843 u^7 t^8+3327843 u^8 t^7+2219756 u^9 t^6+1009711 u^{10} \
t^5+323267 u^{11} t^4+80084 u^{12} t^3+18464 u^{13} t^2+3071 u^{14} t+100 \
u^{15}\Big) s^5+t^2 u^2 \Big(1146 t^{14}+7452 u t^{13}+43332 u^2 \
t^{12}+256857 u^3 t^{11}+1020516 u^4 t^{10}+2637270 u^5 t^9+4640688 u^6 \
t^8+5609726 u^7 t^7+4640688 u^8 t^6+2637270 u^9 t^5+1020516 u^{10} \
t^4+256857 u^{11} t^3+43332 u^{12} t^2+7452 u^{13} t+1146 u^{14}\Big) \
s^4+t^3 u^3 \Big(316 t^{13}+7702 u t^{12}+60600 u^2 t^{11}+260422 u^3 \
t^{10}+721307 u^4 t^9+1402338 u^5 t^8+1962295 u^6 t^7+1962295 u^7 \
t^6+1402338 u^8 t^5+721307 u^9 t^4+260422 u^{10} t^3+60600 u^{11} t^2+7702 \
u^{12} t+316 u^{13}\Big) s^3+3 t^4 u^4 (t+u)^2 \Big(382 t^{10}+2457 u \
t^9+8650 u^2 t^8+21232 u^3 t^7+37622 u^4 t^6+46842 u^5 t^5+37622 u^6 \
t^4+21232 u^7 t^3+8650 u^8 t^2+2457 u^9 t+382 u^{10}\Big) s^2+t^5 u^5 \
(t+u)^3 \Big(300 t^8+1544 u t^7+4776 u^2 t^6+9573 u^3 t^5+12690 u^4 \
t^4+9573 u^5 t^3+4776 u^6 t^2+1544 u^7 t+300 u^8\Big) s+3 t^6 u^6 (t+u)^6 \
\Big(24 t^4+37 u t^3+56 u^2 t^2+37 u^3 t+24 u^4\Big)\Big) H(2,y)  \Bigg\} 
\Big/ \Big(  {27 s^3 t^3 u^3 (s+t)^6 (s+u)^6 (t+u)^6} \Big)
\end{dmath*}

\intertext{}
\begin{dmath*}
 {\cal A}^{(2)}_{5 ; C_F n_f} =  \Bigg\{ -432 s^2 t^2 \Big(s^2-t s+t^2\Big) u^3 (s+u)^6 (t+u)^6 H(0,z) H(1,y) \
(s+t)^7-324 s^2 t^2 u^3 (s+u)^6 (t+u)^6 \Big(s^2-t s
\end{dmath*}
\begin{dmath*}
{\white=}
+t^2+u^2\Big) H(1,y) \
H(1,z) (s+t)^7-108 s^2 t^2 u^3 (s+u)^6 (t+u)^6 \Big(s^2-t s+t^2-3 \
u^2\Big) H(1,1,y) (s+t)^7-324 s^2 t^2 u^3 (s+u)^6 (t+u)^6 \Big(s^2-t \
s+t^2+u^2\Big) H(1,2,y) (s+t)^7+108 s^2 t^3 u^3 (s+u)^6 (t+u)^6 \
\Big(t^2+u^2\Big) H(0,y) H(0,z) (s+t)^6+432 s^2 (s-t) t^2 u^5 (s+u)^6 \
(t+u)^6 H(0,y) H(1,z) (s+t)^6+432 s^3 t^2 u^2 (s+u)^6 (t+u)^6 \Big(t^3+s^2 \
u\Big) H(0,z) H(2,y) (s+t)^6
\end{dmath*}

\begin{dmath*}
{\white =}
+108 s^3 t^2 u^2 (s+u)^6 (t+u)^6 \Big(3 \
u^3+s^2 (4 t+3 u)\Big) H(1,z) H(2,y) (s+t)^6-432 s^3 t^2 u^2 (s+u)^6 \
(t+u)^7 \Big(t^2-u t+u^2\Big) H(1,z) H(3,y) (s+t)^6-432 s^2 t^3 u^5 \
(s+u)^6 (t+u)^6 H(0,0,y) (s+t)^6-432 s^2 t^5 u^3 (s+u)^6 (t+u)^6 H(0,0,z) \
(s+t)^6-108 s^2 t^2 u^3 (s+u)^6 (t+u)^6 \Big(3 t^3+3 u^2 t+4 s u^2\Big) \
H(0,1,z) (s+t)^6+108 s^2 t^2 u^3 (s+u)^6 (t+u)^6 \Big(-3 t^3-3 u^2 t+4 s \
u^2\Big) H(0,2,y) (s+t)^6+108 s^2 t^2 u^3 (s+u)^6 (t+u)^6 \Big(t^3+u^2 \
t-4 s u^2\Big) H(1,0,y) (s+t)^6+108 s^2 t^2 u^3 (s+u)^6 (t+u)^6 \Big(4 \
s^3+t^3+t u^2\Big) H(1,0,z) (s+t)^6+108 s^2 t^2 u^3 (s+u)^6 (t+u)^6 \
\Big(3 s^3+3 u^2 s-4 t u^2\Big) H(1,1,z) (s+t)^6+432 s^3 t^2 u^2 (s+u)^6 \
(t+u)^6 \Big(u^3+s^2 t\Big) H(2,0,y) (s+t)^6+108 s^3 t^2 u^3 (s+u)^6 \
(t+u)^6 \Big(s^2-3 u^2\Big) H(2,1,y) (s+t)^6+324 s^3 t^2 u^2 (s+u)^6 \
(t+u)^7 \Big(s^2+t^2+u^2-t u\Big) H(2,2,y) (s+t)^6-432 s^3 t^2 u^2 \
(s+u)^6 (t+u)^7 \Big(t^2-u t+u^2\Big) H(3,2,y) (s+t)^6-18 s^2 t^2 u^3 \
(s+u)^6 (t+u)^6 \Big(9 (t+2 u) s^4+9 \Big(3 t^2+6 u t+2 u^2\Big) \
s^3+\Big(27 t^3+72 u t^2+12 u^2 t+16 u^3\Big) s^2+t \Big(9 t^3+54 u \
t^2+12 u^2 t+20 u^3\Big) s+2 t^2 u \Big(9 t^2+9 u t+8 u^2\Big)\Big) \
H(1,y) (s+t)^4-18 s^2 t^3 u^3 (t+u)^6 \Big(18 s^{11}+51 (t+2 u) \
s^{10}+\Big(33 t^2+360 u t+203 u^2\Big) s^9+\Big(-27 t^3+498 u t^2+705 \
u^2 t+156 u^3\Big) s^8+\Big(-39 t^4+420 u t^3+776 u^2 t^2+546 u^3 t-63 \
u^4\Big) s^7-3 \Big(4 t^5-108 u t^4-91 u^2 t^3
\end{dmath*}
\begin{dmath*}
{\white=}
-106 u^3 t^2+21 u^4 t+98 \
u^5\Big) s^6-3 u \Big(-64 t^5-9 u t^4+98 u^2 t^3+142 u^3 t^2+236 u^4 \
t+121 u^5\Big) s^5-u \Big(-48 t^6-44 u t^5+174 u^2 t^4+711 u^3 t^3+1102 \
u^4 t^2+897 u^5 t+240 u^6\Big) s^4-u^2 \Big(-16 t^6-100 u t^5+333 u^2 \
t^4+948 u^3 t^3+1084 u^4 t^2+558 u^5 t+83 u^6\Big) s^3-3 u^3 \Big(-16 \
t^6+8 u t^5+124 u^2 t^4+199 u^3 t^3+182 u^4 t^2+60 u^5 t+4 u^6\Big) s^2-t \
u^5 \Big(52 t^4+135 u t^3+186 u^2 t^2+147 u^3 t+24 u^4\Big) s-2 t^2 u^6 \
\Big(4 t^3+9 u t^2+13 u^2 t+8 u^3\Big)\Big) H(0,y) (s+t)^3-12 t^3 u^3 \
(s+u) \Big(24 t u \Big(3 t^3+4 u t^2+4 u^2 t+3 u^3\Big) s^{15}-6 (t+u)^2 \
\Big(19 t^4-7 u t^3+86 u^2 t^2-7 u^3 t+19 u^4\Big) s^{14}-\Big(528 \
t^7+2241 u t^6+5905 u^2 t^5+10806 u^3 t^4+10806 u^4 t^3+5905 u^5 t^2+2241 \
u^6 t+528 u^7\Big) s^{13}-(t+u)^2 \Big(999 t^6+3708 u t^5+7757 u^2 \
t^4+13636 u^3 t^3+7757 u^4 t^2+3708 u^5 t+999 u^6\Big) s^{12}-\Big(945 \
t^9+6444 u t^8+18478 u^2 t^7+38524 u^3 t^6+60985 u^4 t^5+60985 u^5 t^4+38524 \
u^6 t^3+18478 u^7 t^2+6444 u^8 t+945 u^9\Big) s^{11}-(t+u)^2 \Big(408 \
t^8+1419 u t^7-737 u^2 t^6-1221 u^3 t^5+6934 u^4 t^4-1221 u^5 t^3-737 u^6 \
t^2+1419 u^7 t+408 u^8\Big) s^{10}+2 \Big(-9 t^{11}+1365 u t^{10}+9543 \
u^2 t^9+28308 u^3 t^8+43963 u^4 t^7+44166 u^5 t^6+44166 u^6 t^5+43963 u^7 \
t^4+28308 u^8 t^3+9543 u^9 t^2+1365 u^{10} t-9 u^{11}\Big) s^9+(t+u)^2 \
\Big(33 t^{10}+3726 u t^9+16908 u^2 t^8+31720 u^3 t^7+24859 u^4 t^6+8028 \
u^5 t^5+24859 u^6 t^4+31720 u^7 t^3+16908 u^8 t^2+3726 u^9 t+33 \
u^{10}\Big) s^8+\Big(3 t^{13}+2076 u t^{12}+13514 u^2 t^{11}+34430 u^3 \
t^{10}+33548 u^4 t^9-29287 u^5 t^8-114020 u^6 t^7-114020 u^7 t^6-29287 u^8 \
t^5+33548 u^9 t^4+34430 u^{10} t^3+13514 u^{11} t^2+2076 u^{12} t+3 \
u^{13}\Big) s^7+t u (t+u)^2 \Big(615 t^{10}
\end{dmath*}
\begin{dmath*}
{\white=}
+2129 u t^9-2862 u^2 t^8-27958 \
u^3 t^7-70915 u^4 t^6-96406 u^5 t^5-70915 u^6 t^4-27958 u^7 t^3-2862 u^8 \
t^2+2129 u^9 t+615 u^{10}\Big) s^6+t u \Big(87 t^{13}+151 u t^{12}-4613 \
u^2 t^{11}-33895 u^3 t^{10}-117246 u^4 t^9-252106 u^5 t^8-365426 u^6 \
t^7-365426 u^7 t^6-252106 u^8 t^5-117246 u^9 t^4-33895 u^{10} t^3-4613 \
u^{11} t^2+151 u^{12} t+87 u^{13}\Big) s^5-t^2 u^2 (t+u)^2 \Big(61 \
t^{10}+1421 u t^9+9539 u^2 t^8+31118 u^3 t^7+60284 u^4 t^6+74918 u^5 \
t^5+60284 u^6 t^4+31118 u^7 t^3+9539 u^8 t^2+1421 u^9 t+61 u^{10}\Big) \
s^4-t^3 u^3 (t+u)^3 \Big(119 t^8+1631 u t^7+6699 u^2 t^6+14006 u^3 \
t^5+17750 u^4 t^4+14006 u^5 t^3+6699 u^6 t^2+1631 u^7 t+119 u^8\Big) \
s^3-t^4 u^4 (t+u)^4 \Big(133 t^6+842 u t^5+1972 u^2 t^4+2532 u^3 t^3+1972 \
u^4 t^2+842 u^5 t+133 u^6\Big) s^2-6 t^5 u^5 (t+u)^7 \Big(8 t^2+9 u t+8 \
u^2\Big) s-6 t^6 u^6 (t+u)^6 \Big(t^2+u t+u^2\Big)\Big) (s+t)-18 s^2 \
t^3 u^3 (s+u)^3 (t+u)^6 \Big(18 s^{11}+51 (2 t+u) s^{10}+\Big(203 t^2+360 \
u t+33 u^2\Big) s^9+3 \Big(52 t^3+235 u t^2+166 u^2 t-9 u^3\Big) \
s^8+\Big(-63 t^4+546 u t^3+776 u^2 t^2+420 u^3 t-39 u^4\Big) s^7-3 \
\Big(98 t^5+21 u t^4-106 u^2 t^3-91 u^3 t^2-108 u^4 t+4 u^5\Big) s^6-3 t \
\Big(121 t^5+236 u t^4+142 u^2 t^3+98 u^3 t^2-9 u^4 t-64 u^5\Big) s^5-t \
\Big(240 t^6+897 u t^5+1102 u^2 t^4+711 u^3 t^3+174 u^4 t^2-44 u^5 t-48 \
u^6\Big) s^4-t^2 \Big(83 t^6+558 u t^5+1084 u^2 t^4+948 u^3 t^3+333 u^4 \
t^2-100 u^5 t-16 u^6\Big) s^3-3 t^3 \Big(4 t^6+60 u t^5+182 u^2 t^4+199 \
u^3 t^3+124 u^4 t^2+8 u^5 t-16 u^6\Big) s^2-t^5 u \Big(24 t^4+147 u \
t^3+186 u^2 t^2+135 u^3 t+52 u^4\Big) s-2 t^6 u^2 \Big(8 t^3+13 u t^2+9 \
u^2 t+4 u^3\Big)\Big) H(0,z)
\end{dmath*}

\begin{dmath*}
{\white =}
+18 s^2 t^2 u^3 \Big(16 t^2 u \Big(3 t^2+u \
t+3 u^2\Big) s^{16}-4 t \Big(t^5-96 u t^4-125 u^2 t^3-125 u^3 t^2-96 u^4 \
t+u^5\Big) s^{15}+3 \Big(-21 t^7+376 u t^6+969 u^2 t^5+1014 u^3 t^4+1053 \
u^4 t^3+460 u^5 t^2+15 u^6 t+6 u^7\Big) s^{14}+3 \Big(-91 t^8+434 u \
t^7+2445 u^2 t^6+3548 u^3 t^5+4121 u^4 t^4+3594 u^5 t^3+1259 u^6 t^2+200 u^7 \
t+42 u^8\Big) s^{13}+\Big(-507 t^9-228 u t^8+8958 u^2 t^7+21034 u^3 \
t^6+29016 u^4 t^5+37554 u^5 t^4+27842 u^6 t^3+10398 u^7 t^2+2643 u^8 t+394 \
u^9\Big) s^{12}+3 \Big(-139 t^{10}-658 u t^9+1402 u^2 t^8+7968 u^3 \
t^7+14642 u^4 t^6+25576 u^5 t^5+31288 u^6 t^4+20652 u^7 t^3+8197 u^8 \
t^2+2094 u^9 t+242 u^{10}\Big) s^{11}+\Big(-33 t^{11}-1944 u t^{10}-2626 \
u^2 t^9+15924 u^3 t^8+52455 u^4 t^7+115368 u^5 t^6+193671 u^6 t^5+189594 u^7 \
t^4+109994 u^8 t^3+41340 u^9 t^2+9291 u^{10} t+870 u^{11}\Big) \
s^{10}+\Big(229 t^{12}-1206 u t^{11}-7826 u^2 t^{10}+2016 u^3 t^9+60801 u^4 \
t^8+168742 u^5 t^7+312211 u^6 t^6+377892 u^7 t^5+285136 u^8 t^4+142036 u^9 \
t^3+47343 u^{10} t^2+9112 u^{11} t+698 u^{12}\Big) s^9+3 \Big(69 \
t^{13}-396 u t^{12}-3542 u^2 t^{11}-3876 u^3 t^{10}+19403 u^4 t^9+74332 u^5 \
t^8+145298 u^6 t^7+193458 u^7 t^6+172121 u^8 t^5+103198 u^9 t^4+43468 u^{10} \
t^3+12394 u^{11} t^2+2015 u^{12} t+122 u^{13}\Big) s^8+\Big(81 \
t^{14}-1338 u t^{13}-9602 u^2 t^{12}-15144 u^3 t^{11}+41583 u^4 \
t^{10}+217006 u^5 t^9+472374 u^6 t^8+692016 u^7 t^7+711667 u^8 t^6+501642 \
u^9 t^5+246150 u^{10} t^4+85492 u^{11} t^3+19953 u^{12} t^2+2646 u^{13} \
t+114 u^{14}\Big) s^7
\end{dmath*}

\begin{dmath*}
{\white =}
+\Big(12 t^{15}-936 u t^{14}-5785 u^2 t^{13}-9734 \
u^3 t^{12}+23766 u^4 t^{11}+151768 u^5 t^{10}+374411 u^6 t^9+611526 u^7 \
t^8+725603 u^8 t^7+609872 u^9 t^6+355422 u^{10} t^5+143724 u^{11} t^4+39635 \
u^{12} t^3+7044 u^{13} t^2+696 u^{14} t+16 u^{15}\Big) s^6+3 t u \
\Big(-112 t^{14}-711 u t^{13}-1428 u^2 t^{12}+2680 u^3 t^{11}+24408 u^4 \
t^{10}+73333 u^5 t^9+137580 u^6 t^8+184601 u^7 t^7+179614 u^8 t^6+124026 u^9 \
t^5+59992 u^{10} t^4+19831 u^{11} t^3+4174 u^{12} t^2+496 u^{13} t+28 \
u^{14}\Big) s^5+t^2 u \Big(-48 t^{14}-380 u t^{13}-1422 u^2 t^{12}-453 \
u^3 t^{11}+19778 u^4 t^{10}+90336 u^5 t^9+213354 u^6 t^8+330336 u^7 \
t^7+360906 u^8 t^6+281670 u^9 t^5+156716 u^{10} t^4+61605 u^{11} t^3+16236 \
u^{12} t^2+2470 u^{13} t+144 u^{14}\Big) s^4+t^3 u^2 \Big(-16 t^{13}-340 \
u t^{12}-1209 u^2 t^{11}+1822 u^3 t^{10}+23522 u^4 t^9+76164 u^5 t^8+142722 \
u^6 t^7+179964 u^7 t^6+159444 u^8 t^5+99732 u^9 t^4+43795 u^{10} t^3+13194 \
u^{11} t^2+2510 u^{12} t+232 u^{13}\Big) s^3-3 t^4 u^3 (t+u)^2 \Big(16 \
t^{10}+56 u t^9-124 u^2 t^8-1147 u^3 t^7-3208 u^4 t^6-5170 u^5 t^5-5528 u^6 \
t^4-3835 u^7 t^3-1676 u^8 t^2-424 u^9 t-48 u^{10}\Big) s^2+t^5 u^5 (t+u)^3 \
\Big(52 t^7+321 u t^6+921 u^2 t^5+1569 u^3 t^4+1779 u^4 t^3+1230 u^5 \
t^2+484 u^6 t+84 u^7\Big) s+2 t^6 u^6 (t+u)^7 \Big(4 t^2+5 u t+8 \
u^2\Big)\Big) H(1,z)+18 s^2 t^2 u^2 \Big(16 t^2 u^2 \Big(3 t^2+u t+3 \
u^2\Big) s^{16}-4 t u \Big(t^5-96 u t^4-125 u^2 t^3-125 u^3 t^2-96 u^4 \
t+u^5\Big) s^{15}+6 \Big(3 t^8+9 u t^7+242 u^2 t^6+567 u^3 t^5+582 u^4 \
t^4+567 u^5 t^3+242 u^6 t^2+9 u^7 t+3 u^8\Big) s^{14}+3 \Big(42 t^9+221 u \
t^8+1445 u^2 t^7+4311 u^3 t^6+5693 u^4 t^5+5693 u^5 t^4+4311 u^6 t^3+1445 \
u^7 t^2+221 u^8 t+42 u^9\Big) s^{13}+2 \Big(197 t^{10}+1416 u t^9+6153 \
u^2 t^8+18148 u^3 t^7+29594 u^4 t^6+32160 u^5 t^5+29594 u^6 t^4+18148 u^7 \
t^3+6153 u^8 t^2+1416 u^9 t+197 u^{10}\Big) s^{12}+3 \Big(242 t^{11}+2199 \
u t^{10}+9451 u^2 t^9+27134 u^3 t^8
\end{dmath*}
\begin{dmath*}
{\white=}
+50612 u^4 t^7+62634 u^5 t^6+62634 u^6 \
t^5+50612 u^7 t^4+27134 u^8 t^3+9451 u^9 t^2+2199 u^{10} t+242 u^{11}\Big) \
s^{11}+2 \Big(435 t^{12}+4803 u t^{11}+23010 u^2 t^{10}+69466 u^3 \
t^9+145491 u^4 t^8+210555 u^5 t^7+230976 u^6 t^6+210555 u^7 t^5+145491 u^8 \
t^4+69466 u^9 t^3+23010 u^{10} t^2+4803 u^{11} t+435 u^{12}\Big) \
s^{10}+\Big(698 t^{13}+9301 u t^{12}+51105 u^2 t^{11}+170974 u^3 \
t^{10}+406704 u^4 t^9+698997 u^5 t^8+884941 u^6 t^7+884941 u^7 t^6+698997 \
u^8 t^5+406704 u^9 t^4+170974 u^{10} t^3+51105 u^{11} t^2+9301 u^{12} t+698 \
u^{13}\Big) s^9+6 \Big(61 t^{14}+1018 u t^{13}+6515 u^2 t^{12}+24975 u^3 \
t^{11}+68497 u^4 t^{10}+139578 u^5 t^9+209143 u^6 t^8+237090 u^7 t^7+209143 \
u^8 t^6+139578 u^9 t^5+68497 u^{10} t^4+24975 u^{11} t^3+6515 u^{12} \
t^2+1018 u^{13} t+61 u^{14}\Big) s^8+\Big(114 t^{15}+2655 u t^{14}+20511 \
u^2 t^{13}+93946 u^3 t^{12}+304122 u^4 t^{11}+729081 u^5 t^{10}+1284397 u^6 \
t^9+1678662 u^7 t^8+1678662 u^8 t^7+1284397 u^9 t^6+729081 u^{10} t^5+304122 \
u^{11} t^4+93946 u^{12} t^3+20511 u^{13} t^2+2655 u^{14} t+114 u^{15}\Big) \
s^7+2 \Big(8 t^{16}+348 u t^{15}+3558 u^2 t^{14}+20893 u^3 t^{13}+82679 u^4 \
t^{12}+233298 u^5 t^{11}+478228 u^6 t^{10}+720901 u^7 t^9+821694 u^8 \
t^8+720901 u^9 t^7+478228 u^{10} t^6+233298 u^{11} t^5+82679 u^{12} \
t^4+20893 u^{13} t^3+3558 u^{14} t^2+348 u^{15} t+8 u^{16}\Big) s^6+3 t u \
\Big(28 t^{15}+496 u t^{14}+4255 u^2 t^{13}+21403 u^3 t^{12}
\end{dmath*}
\begin{dmath*}
{\white=}
+71760 u^4 \
t^{11}+172018 u^5 t^{10}+302499 u^6 t^9+397333 u^7 t^8+397333 u^8 t^7+302499 \
u^9 t^6+172018 u^{10} t^5+71760 u^{11} t^4+21403 u^{12} t^3+4255 u^{13} \
t^2+496 u^{14} t+28 u^{15}\Big) s^5+2 t^2 u^2 \Big(72 t^{14}+1235 u \
t^{13}+8343 u^2 t^{12}+34020 u^3 t^{11}+97435 u^4 t^{10}+204801 u^5 \
t^9+317982 u^6 t^8+367512 u^7 t^7+317982 u^8 t^6+204801 u^9 t^5+97435 u^{10} \
t^4+34020 u^{11} t^3+8343 u^{12} t^2+1235 u^{13} t+72 u^{14}\Big) s^4+t^3 \
u^3 \Big(232 t^{13}+2510 u t^{12}+13689 u^2 t^{11}+49393 u^3 t^{10}+127070 \
u^4 t^9+236640 u^5 t^8+321522 u^6 t^7+321522 u^7 t^6+236640 u^8 t^5+127070 \
u^9 t^4+49393 u^{10} t^3+13689 u^{11} t^2+2510 u^{12} t+232 u^{13}\Big) \
s^3+6 t^4 u^4 (t+u)^2 \Big(24 t^{10}+212 u t^9+892 u^2 t^8+2315 u^3 \
t^7+4004 u^4 t^6+4762 u^5 t^5+4004 u^6 t^4+2315 u^7 t^3+892 u^8 t^2+212 u^9 \
t+24 u^{10}\Big) s^2+t^5 u^5 (t+u)^3 \Big(84 t^8+484 u t^7+1347 u^2 \
t^6+2364 u^3 t^5+2802 u^4 t^4+2364 u^5 t^3+1347 u^6 t^2+484 u^7 t+84 \
u^8\Big) s+2 t^6 u^6 (t+u)^6 \Big(8 t^4+13 u t^3+18 u^2 t^2+13 u^3 t+8 \
u^4\Big)\Big) H(2,y)  \Bigg\}  \Big/ \Big(  {27 s^3 t^3 u^3 (s+t)^6 (s+u)^6 (t+u)^6} \Big)
\end{dmath*}

\intertext{}
\begin{dmath*}
 {\cal A}^{(2)}_{5 ; n_f^2} =    2 \Bigg\{ t^3 u H(2,y)+t^3 u H(1,z)-t u \
\Big(t^2+u^2\Big) H(0,y)-t u \Big(t^2+u^2\Big) H(0,z)+t u^3 H(2,y)+t \
u^3 H(1,z)+s^4+2 s^3 t+2 s^3 u+3 s^2 t^2+3 s^2 u^2+2 s t^3+2 s u^3+t^4+2 t^3 \
u+3 t^2 u^2+2 t u^3+u^4\Bigg\} \Big/ \Big(3 s t u \Big)
\end{dmath*}

\intertext{}
\begin{dmath*}
 {\cal A}^{(2)}_{6 ; C_A^2} =  \Bigg\{  3 s^2 t^2 (s+t)^6 u^2 (t+u)^2 \Big(\Big(363 t^4+492 u t^3+504 u^2 t^2+492 \
u^3 t+363 u^4\Big) s^4+2 \Big(363 t^5+479 u t^4-58 u^2 t^3-58 u^3 t^2+479 \
u^4 t
\end{dmath*}
\begin{dmath*}
{\white=}
+363 u^5\Big) s^3+9 (t+u)^2 \Big(121 t^4-10 u t^3-134 u^2 t^2-10 u^3 \
t+121 u^4\Big) s^2+6 (t+u)^3 \Big(121 t^4+35 u t^3+12 u^2 t^2+35 u^3 \
t+121 u^4\Big) s+363 (t+u)^4 \Big(t^2+u t+u^2\Big)^2\Big) H(1,z) \
H(2,y) (s+u)^6+3 s^2 t^2 (s+t)^2 u^2 (t+u)^6 \Big(363 s^8+726 (3 t+u) s^7+3 \
\Big(2057 t^2+796 u t+363 u^2\Big) s^6+6 \Big(1815 t^3+480 u t^2+348 u^2 \
t+121 u^3\Big) s^5+\Big(13068 t^4+1782 u t^3-297 u^2 t^2+958 u^3 t+363 \
u^4\Big) s^4+2 t \Big(5445 t^4+891 u t^3-1296 u^2 t^2-58 u^3 t+246 \
u^4\Big) s^3+t^2 \Big(6171 t^4+2880 u t^3-297 u^2 t^2-116 u^3 t+504 \
u^4\Big) s^2+2 t^3 \Big(1089 t^4+1194 u t^3+1044 u^2 t^2+479 u^3 t+246 \
u^4\Big) s+363 t^4 \Big(t^2+u t+u^2\Big)^2\Big) H(0,0,y) (s+u)^6+36 \
s^2 t^2 (s+t)^2 u^2 (t+u)^6 \Big(22 s^8+44 (3 t+u) s^7+2 \Big(187 t^2+70 u \
t+33 u^2\Big) s^6+\Big(660 t^3+147 u t^2+246 u^2 t+44 u^3\Big) \
s^5+\Big(792 t^4+41 u t^3+360 u^2 t^2+115 u^3 t+22 u^4\Big) s^4+t \
\Big(660 t^4+2 u t^3+246 u^2 t^2+95 u^3 t+64 u^4\Big) s^3+t^2 \Big(374 \
t^4+54 u t^3+84 u^2 t^2+9 u^3 t+84 u^4\Big) s^2+t^3 \Big(132 t^4+63 u \
t^3+36 u^2 t^2-11 u^3 t+54 u^4\Big) s+t^4 \Big(22 t^4+21 u t^3+18 u^2 \
t^2+4 u^3 t+18 u^4\Big)\Big) H(0,z) H(0,0,y) (s+u)^6-36 s^2 t^2 (s+t)^2 \
u^2 (t+u)^6 \Big(22 s^8+3 (44 t+7 u) s^7+\Big(374 t^2+63 u t+18 u^2\Big) \
s^6+\Big(660 t^3+54 u t^2+36 u^2 t+4 u^3\Big) s^5+\Big(792 t^4+2 u \
t^3+84 u^2 t^2-11 u^3 t+18 u^4\Big) s^4+t \Big(660 t^4+41 u t^3+246 u^2 \
t^2+9 u^3 t+54 u^4\Big) s^3+t^2 \Big(374 t^4+147 u t^3+360 u^2 t^2+95 u^3 \
t+84 u^4\Big) s^2+t^3 \Big(132 t^4+140 u t^3+246 u^2 t^2+115 u^3 t+64 \
u^4\Big) s+22 t^4 \Big(t^2+u t+u^2\Big)^2\Big) H(1,z) H(0,0,y) \
(s+u)^6
\end{dmath*}

\intertext{}
\begin{dmath*}
{\white =}
+108 s^2 t^2 (s+t)^6 u^2 (t+u)^6 \Big(2 s^4+4 (t+u) s^3+6 \
\Big(t^2+u^2\Big) s^2+4 \Big(t^3+u^3\Big) s+2 t^4+2 u^4-2 t u^3-3 t^2 \
u^2-2 t^3 u\Big) H(0,0,y) H(0,0,z) (s+u)^6+18 s^2 t^2 (s+t)^6 u^2 (t+u)^2 \
\Big(\Big(44 t^4+296 u t^3+486 u^2 t^2+316 u^3 t+58 u^4\Big) s^4+2 \
\Big(15 t^5+247 u t^4+634 u^2 t^3+646 u^3 t^2+271 u^4 t+27 u^5\Big) s^3+3 \
(t+u)^2 \Big(5 t^4+150 u t^3+228 u^2 t^2+146 u^3 t+15 u^4\Big) s^2-2 \
(t+u)^3 \Big(2 t^4-91 u t^3-90 u^2 t^2-97 u^3 t-4 u^4\Big) s+(t+u)^4 \
\Big(11 t^4+94 u t^3+105 u^2 t^2+72 u^3 t+6 u^4\Big)\Big) H(3,y) \
H(0,1,z) (s+u)^6-108 s^2 t^2 (s+t)^6 u^2 (t+u)^6 \Big(2 s^4+2 (2 t+u) s^3+3 \
\Big(2 t^2+u^2\Big) s^2+2 \Big(2 t^3+u^3\Big) s+2 \Big(t^4-2 u t^3+6 \
u^2 t^2+u^4\Big)\Big) H(0,0,y) H(0,1,z) (s+u)^6
\end{dmath*}

\begin{dmath*}
{\white =}
+36 s^2 t^2 (s+t)^6 u^2 \
(t+u)^2 \Big(2 \Big(11 t^4+32 u t^3+42 u^2 t^2+27 u^3 t+9 u^4\Big) \
s^4+\Big(44 t^5+115 u t^4+95 u^2 t^3+9 u^3 t^2-11 u^4 t+4 u^5\Big) s^3+6 \
(t+u)^2 \Big(11 t^4+19 u t^3+11 u^2 t^2+3 u^4\Big) s^2+(t+u)^3 \Big(44 \
t^4+8 u t^3-9 u^2 t^2+21 u^4\Big) s+22 (t+u)^4 \Big(t^2+u \
t+u^2\Big)^2\Big) H(1,z) H(0,2,y) (s+u)^6-108 s^2 t^2 (s+t)^6 u^2 \
(t+u)^6 \Big(2 s^4+2 (t+2 u) s^3+3 \Big(t^2+2 u^2\Big) s^2+2 \Big(t^3+2 \
u^3\Big) s+2 \Big(t^4+6 u^2 t^2-2 u^3 t+u^4\Big)\Big) H(0,0,z) \
H(0,2,y) (s+u)^6+108 s^2 t^2 (s+t)^6 u^2 (t+u)^6 \Big(-2 (t-2 u) s^3-3 \
\Big(t^2-2 u^2\Big) s^2-2 \Big(t^3-2 u^3\Big) s+12 t u \Big(t^2-u \
t+u^2\Big)\Big) H(0,1,z) H(0,2,y) (s+u)^6+216 s^2 t^2 (s+t)^6 u^2 \
(t+u)^6 \Big(s^4+2 u s^3+3 u^2 s^2+2 u^3 s+\Big(t^2+u \
t+u^2\Big)^2\Big) H(0,1,z) H(0,3,y) (s+u)^6+18 s^2 t^2 (s+t)^3 u^2 \
(t+u)^6 \Big(3 s^7+(6 u-9 t) s^6+\Big(-48 t^2+12 u t+15 u^2\Big) \
s^5-\Big(56 t^3+6 u t^2+63 u^2 t-12 u^3\Big) s^4-t^2 \Big(7 t^2+36 u \
t+222 u^2\Big) s^3+3 t \Big(11 t^4-14 u t^3-70 u^2 t^2-8 u^3 t-8 \
u^4\Big) s^2+t^2 \Big(28 t^4-24 u t^3-81 u^2 t^2-24 u^3 t-30 u^4\Big) \
s+t^3 \Big(8 t^4-6 u t^3-15 u^2 t^2-12 u^3 t
\end{dmath*}
\begin{dmath*}
{\white=}
-14 u^4\Big)\Big) H(0,z) \
H(1,0,y) (s+u)^6-18 s^2 t^2 (s+t)^3 u^2 (t+u)^6 \Big(8 s^7+(28 t-6 u) s^6+3 \
\Big(11 t^2-8 u t-5 u^2\Big) s^5-\Big(7 t^3+42 u t^2+81 u^2 t+12 \
u^3\Big) s^4-2 \Big(28 t^4+18 u t^3+105 u^2 t^2+12 u^3 t+7 u^4\Big) \
s^3-6 t \Big(8 t^4+u t^3+37 u^2 t^2+4 u^3 t+5 u^4\Big) s^2-3 \Big(3 \
t^6-4 u t^5+21 u^2 t^4+8 u^4 t^2\Big) s+3 t^4 \Big(t^3+2 u t^2+5 u^2 t+4 \
u^3\Big)\Big) H(1,z) H(1,0,y) (s+u)^6+108 s^2 t^2 (s+t)^6 u^2 (t+u)^6 \
\Big(2 s^4+2 (2 t+u) s^3+3 \Big(2 t^2+u^2\Big) s^2+2 \Big(2 \
t^3+u^3\Big) s+2 t^4+2 u^4-2 t u^3+9 t^2 u^2-6 t^3 u\Big) H(0,1,z) \
H(1,0,y) (s+u)^6-18 s^2 t^2 (s+t)^6 u^2 (t+u)^3 \Big(2 t \Big(7 t^2+15 u \
t+12 u^2\Big) s^4+12 \Big(t^4+2 u t^3+2 u^2 t^2-u^4\Big) s^3+3 (t+u)^2 \
\Big(5 t^3+17 u t^2+31 u^2 t-5 u^3\Big) s^2+6 (t+u)^3 \Big(t^3+u t^2+u^2 \
t-u^3\Big) s-(t+u)^3 \Big(8 t^4+4 u t^3-3 u^2 t^2-18 u^3 t+3 \
u^4\Big)\Big) H(3,y) H(1,0,z) (s+u)^6-108 s^2 t^2 (s+t)^6 u^2 (t+u)^6 \
\Big(2 s^4+4 (t-u) s^3+6 \Big(t^2+2 u^2\Big) s^2+4 t^3 s+2 t^4+2 u^4+2 t \
u^3+3 t^2 u^2+2 t^3 u\Big) H(0,0,y) H(1,0,z) (s+u)^6-216 s^3 t^2 (s+t)^6 \
u^2 (t+u)^6 \Big(2 (t-2 u) s^2+3 \Big(t^2+u^2\Big) s+2 \
\Big(t^3-u^3\Big)\Big) H(0,2,y) H(1,0,z) (s+u)^6+108 s^2 t^2 (s+t)^6 \
u^2 (t+u)^6 \Big(2 s^4+2 (t+3 u) s^3+3 \Big(t^2-u^2\Big) s^2+2 \
\Big(t^3+5 u^3\Big) s+2 \Big(t^2+u t+u^2\Big)^2\Big) H(0,3,y) \
H(1,0,z) (s+u)^6
\end{dmath*}

\begin{dmath*}
{\white =}
+108 s^2 t^2 (s+t)^6 u^2 (t+u)^6 \Big(2 s^4+(4 t-6 u) \
s^3+\Big(6 t^2+9 u^2\Big) s^2+\Big(4 t^3-2 u^3\Big) s+2 t^4+2 u^4+2 t \
u^3+3 t^2 u^2+2 t^3 u\Big) H(1,0,y) H(1,0,z) (s+u)^6+3 s^2 t^2 (s+t)^6 u^2 \
(t+u)^2 \Big(\Big(363 t^4+492 u t^3+504 u^2 t^2+492 u^3 t+363 u^4\Big) \
s^4+2 \Big(363 t^5+479 u t^4-58 u^2 t^3-58 u^3 t^2+479 u^4 t+363 u^5\Big) \
s^3+9 (t+u)^2 \Big(121 t^4-10 u t^3-134 u^2 t^2-10 u^3 t+121 u^4\Big) \
s^2+6 (t+u)^3 \Big(121 t^4+35 u t^3+12 u^2 t^2+35 u^3 t+121 u^4\Big) \
s+363 (t+u)^4 \Big(t^2+u t+u^2\Big)^2\Big) H(1,1,z) (s+u)^6+36 s^2 t^2 \
(s+t)^6 u^2 (t+u)^2 \Big(2 \Big(11 t^4+32 u t^3+42 u^2 t^2+27 u^3 t+9 \
u^4\Big) s^4+\Big(44 t^5+115 u t^4+95 u^2 t^3+9 u^3 t^2-11 u^4 t+4 \
u^5\Big) s^3+6 (t+u)^2 \Big(11 t^4+19 u t^3+11 u^2 t^2+3 u^4\Big) \
s^2+(t+u)^3 \Big(44 t^4+8 u t^3-9 u^2 t^2+21 u^4\Big) s+22 (t+u)^4 \
\Big(t^2+u t+u^2\Big)^2\Big) H(0,y) H(1,1,z) (s+u)^6-36 s^2 t^2 (s+t)^6 \
u^2 (t+u)^2 \Big(2 \Big(20 t^4+59 u t^3+84 u^2 t^2+59 u^3 t+20 u^4\Big) \
s^4+8 \Big(6 t^5+13 u t^4+13 u^2 t^3+13 u^3 t^2+13 u^4 t+6 u^5\Big) s^3+6 \
(t+u)^2 \Big(14 t^4+19 u t^3+22 u^2 t^2+19 u^3 t+14 u^4\Big) s^2+(t+u)^3 \
\Big(65 t^4+8 u t^3-18 u^2 t^2+8 u^3 t+65 u^4\Big) s+44 (t+u)^4 \
\Big(t^2+u t+u^2\Big)^2\Big) H(3,y) H(1,1,z) (s+u)^6+108 s^2 t^2 \
(s+t)^6 u^2 (t+u)^6 \Big(2 s^4+(4 t-2 u) s^3+\Big(6 t^2-3 u^2\Big) \
s^2+\Big(4 t^3-2 u^3\Big) s+2 \Big(t^2+u t+u^2\Big)^2\Big) H(0,0,y) \
H(1,1,z) (s+u)^6-108 s^2 t^2 (s+t)^6 u^2 (t+u)^6 \Big(4 s^4+(6 t-2 u) s^3+9 \
\Big(t^2+u^2\Big) s^2+6 \Big(t^3-u^3\Big) s+4 \Big(t^2+u \
t+u^2\Big)^2\Big) H(0,3,y) H(1,1,z) (s+u)^6+216 s^2 t^2 (s+t)^6 u^3 \
(t+u)^6 \Big(2 s^3-3 u s^2+4 u^2 s+6 t \Big(t^2-u t+u^2\Big)\Big) \
H(0,1,z) H(1,2,y) (s+u)^6-216 s^2 t^2 (s+t)^6 u^3 (t+u)^6 \Big(6 s^3-6 u \
s^2+6 u^2 s+t \Big(2 t^2-3 u t
\end{dmath*}
\begin{dmath*}
{\white=}
+4 u^2\Big)\Big) H(1,0,z) H(1,2,y) \
(s+u)^6+36 s^2 t^2 (s+t)^6 u^2 (t+u)^2 \Big(2 \Big(11 t^4+32 u t^3+42 u^2 \
t^2+27 u^3 t+9 u^4\Big) s^4+\Big(44 t^5+115 u t^4+95 u^2 t^3+9 u^3 t^2-11 \
u^4 t+4 u^5\Big) s^3+6 (t+u)^2 \Big(11 t^4+19 u t^3+11 u^2 t^2+3 \
u^4\Big) s^2+(t+u)^3 \Big(44 t^4+8 u t^3-9 u^2 t^2+21 u^4\Big) s+22 \
(t+u)^4 \Big(t^2+u t+u^2\Big)^2\Big) H(1,z) H(2,0,y) (s+u)^6-108 s^2 \
t^2 (s+t)^6 u^2 (t+u)^6 \Big(2 s^4-4 (t-u) s^3+6 \Big(2 t^2+u^2\Big) \
s^2+4 u^3 s+2 t^4+2 u^4+2 t u^3+3 t^2 u^2+2 t^3 u\Big) H(0,0,z) H(2,0,y) \
(s+u)^6-108 s^2 t^2 (s+t)^6 u^2 (t+u)^6 \Big(4 s^4+2 (5 t+2 u) s^3+3 \
\Big(t^2+2 u^2\Big) s^2+2 \Big(7 t^3+2 u^3\Big) s+4 t^4+4 u^4+4 t \
u^3+6 t^2 u^2+4 t^3 u\Big) H(0,1,z) H(2,0,y) (s+u)^6-108 s^2 t^2 (s+t)^6 \
u^2 (t+u)^6 \Big(4 s^4+6 t s^3+9 \Big(t^2+2 u^2\Big) s^2+\Big(6 t^3+4 \
u^3\Big) s+4 t^4+4 u^4+6 t u^3+9 t^2 u^2+6 t^3 u\Big) H(1,0,z) H(2,0,y) \
(s+u)^6+3 s^2 t^2 (s+t)^6 u^2 (t+u)^2 \Big(\Big(363 t^4+492 u t^3+504 u^2 \
t^2+492 u^3 t+363 u^4\Big) s^4+2 \Big(363 t^5+479 u t^4-58 u^2 t^3-58 u^3 \
t^2+479 u^4 t+363 u^5\Big) s^3+9 (t+u)^2 \Big(121 t^4-10 u t^3-134 u^2 \
t^2-10 u^3 t+121 u^4\Big) s^2+6 (t+u)^3 \Big(121 t^4+35 u t^3+12 u^2 \
t^2+35 u^3 t+121 u^4\Big) s+363 (t+u)^4 \Big(t^2+u t+u^2\Big)^2\Big) \
H(2,2,y) (s+u)^6+36 s^2 t^2 (s+t)^6 u^2 (t+u)^2 \Big(2 \Big(9 t^4+27 u \
t^3+42 u^2 t^2+32 u^3 t+11 u^4\Big) s^4+\Big(4 t^5-11 u t^4+9 u^2 t^3+95 \
u^3 t^2+115 u^4 t+44 u^5\Big) s^3+6 (t+u)^2 \Big(3 t^4+11 u^2 t^2+19 u^3 \
t+11 u^4\Big) s^2+(t+u)^3 \Big(21 t^4-9 u^2 t^2+8 u^3 t+44 u^4\Big) \
s+22 (t+u)^4 \Big(t^2+u t+u^2\Big)^2\Big) H(0,z) H(2,2,y) (s+u)^6+108 \
s^2 t^2 (s+t)^6 u^2 (t+u)^6 \Big(2 s^4-2 (t-2 u) s^3-3 \Big(t^2-2 \
u^2\Big) s^2-2 \Big(t^3-2 u^3\Big) s+2 \Big(t^2+u \
t+u^2\Big)^2\Big) H(0,0,z) H(2,2,y) (s+u)^6
\end{dmath*}
\begin{dmath*}
{\white=}
+432 s^2 t^2 (s+t)^6 u^2 \
(t+u)^6 \Big(s^4+3 \Big(t^2+u^2\Big) s^2-\Big(t^3+u^3\Big) \
s+\Big(t^2+u t+u^2\Big)^2\Big) H(0,1,z) H(2,2,y) (s+u)^6-108 s^2 t^2 \
(s+t)^6 u^2 (t+u)^6 \Big(4 s^4-2 (t+2 u) s^3+3 \Big(5 t^2+8 u^2\Big) \
s^2+2 \Big(t^3-2 u^3\Big) s+4 \Big(t^2+u t+u^2\Big)^2\Big) H(1,0,z) \
H(2,2,y) (s+u)^6+324 s^2 t^2 (s+t)^6 u^2 (t+u)^6 \Big(2 s^4+2 (t+u) s^3+3 \
\Big(t^2+u^2\Big) s^2+2 \Big(t^3+u^3\Big) s+2 \Big(t^2+u \
t+u^2\Big)^2\Big) H(0,1,z) H(2,3,y) (s+u)^6
\end{dmath*}

\intertext{}
\begin{dmath*}
{\white =}
-108 s^2 t^2 (s+t)^6 u^2 \
(t+u)^6 \Big(2 s^4+2 (t+u) s^3+3 \Big(t^2+u^2\Big) s^2+2 \
\Big(t^3+u^3\Big) s+2 \Big(t^2+u t+u^2\Big)^2\Big) H(1,0,z) H(2,3,y) \
(s+u)^6-18 s^2 t^2 (s+t)^6 u^2 (t+u)^3 \Big(2 u \Big(12 t^2+15 u t+7 \
u^2\Big) s^4-12 \Big(t^4-2 u^2 t^2-2 u^3 t-u^4\Big) s^3-3 (t+u)^2 \
\Big(5 t^3-31 u t^2-17 u^2 t-5 u^3\Big) s^2-6 (t+u)^3 \Big(t^3-u t^2-u^2 \
t-u^3\Big) s-(t+u)^3 \Big(3 t^4-18 u t^3-3 u^2 t^2+4 u^3 t+8 \
u^4\Big)\Big) H(1,z) H(3,0,y) (s+u)^6-108 s^2 t^2 (s+t)^6 u^2 (t+u)^6 \
\Big(2 s^4+2 (t+u) s^3+3 \Big(t^2+u^2\Big) s^2+2 \Big(t^3+u^3\Big) \
s+2 \Big(t^2+u t+u^2\Big)^2\Big) H(0,1,z) H(3,0,y) (s+u)^6-108 s^2 t^2 \
(s+t)^6 u^2 (t+u)^6 \Big(2 s^4+10 (t+u) s^3-3 \Big(t^2+u^2\Big) s^2+6 \
\Big(t^3+u^3\Big) s+2 \Big(t^2+u t+u^2\Big)^2\Big) H(1,0,z) H(3,0,y) \
(s+u)^6-216 s^2 t^2 (s+t)^6 u^2 (t+u)^6 \Big(2 s^4+2 t s^3+9 t^2 s^2+2 \
\Big(t^2+u t+u^2\Big)^2\Big) H(1,1,z) H(3,0,y) (s+u)^6
\end{dmath*}

\begin{dmath*}
{\white =}
-18 s^2 t^2 \
(s+t)^6 u^2 (t+u)^3 \Big(2 t \Big(7 t^2+15 u t+12 u^2\Big) s^4+12 \
\Big(t^4+2 u t^3+2 u^2 t^2-u^4\Big) s^3+3 (t+u)^2 \Big(5 t^3+17 u t^2+31 \
u^2 t-5 u^3\Big) s^2+6 (t+u)^3 \Big(t^3+u t^2+u^2 t-u^3\Big) s-(t+u)^3 \
\Big(8 t^4+4 u t^3-3 u^2 t^2-18 u^3 t+3 u^4\Big)\Big) H(0,z) H(3,2,y) \
(s+u)^6-36 s^2 t^2 (s+t)^6 u^2 (t+u)^2 \Big(2 \Big(20 t^4+59 u t^3+84 u^2 \
t^2+59 u^3 t+20 u^4\Big) s^4+8 \Big(6 t^5+13 u t^4+13 u^2 t^3+13 u^3 \
t^2+13 u^4 t+6 u^5\Big) s^3+6 (t+u)^2 \Big(14 t^4+19 u t^3+22 u^2 t^2+19 \
u^3 t+14 u^4\Big) s^2+(t+u)^3 \Big(65 t^4+8 u t^3-18 u^2 t^2+8 u^3 t+65 \
u^4\Big) s+44 (t+u)^4 \Big(t^2+u t+u^2\Big)^2\Big) H(1,z) H(3,2,y) \
(s+u)^6+216 s^2 t^2 (s+t)^6 u^2 (t+u)^6 \Big(2 s^4+2 t s^3+9 t^2 s^2+2 \
\Big(t^2+u t+u^2\Big)^2\Big) H(0,1,z) H(3,2,y) (s+u)^6-216 s^2 t^2 \
(s+t)^6 u^2 (t+u)^6 \Big(2 s^4+4 (t+2 u) s^3-3 \Big(t^2-u^2\Big) s^2+2 \
\Big(t^3+3 u^3\Big) s+2 \Big(t^2+u t+u^2\Big)^2\Big) H(1,0,z) \
H(3,2,y) (s+u)^6+18 s^2 t^2 (s+t)^6 u^2 (t+u)^2 \Big(\Big(58 t^4+340 u \
t^3+540 u^2 t^2+340 u^3 t+58 u^4\Big) s^4+2 \Big(21 t^5+265 u t^4+658 u^2 \
t^3+658 u^3 t^2+265 u^4 t+21 u^5\Big) s^3+6 (t+u)^2 \Big(5 t^4+86 u \
t^3+138 u^2 t^2+86 u^3 t+5 u^4\Big) s^2+2 (t+u)^3 \Big(t^4+97 u t^3+96 \
u^2 t^2+97 u^3 t+u^4\Big) s+3 (t+u)^4 \Big(t^4+30 u t^3+36 u^2 t^2+30 u^3 \
t+u^4\Big)\Big) H(1,z) H(3,3,y) (s+u)^6+216 s^2 t^2 (s+t)^6 u^2 (t+u)^6 \
\Big(4 s^4+(6 t+4 u) s^3+3 \Big(t^2+2 u^2\Big) s^2+4 \Big(2 \
t^3+u^3\Big) s+4 \Big(t^2+u t+u^2\Big)^2\Big) H(0,1,z) H(3,3,y) \
(s+u)^6-216 s^3 t^2 (s+t)^6 u^3 (t+u)^6 \Big(2 s^2-3 u s+4 u^2\Big) \
H(1,0,z) H(3,3,y) (s+u)^6+216 s^2 t^2 (s+t)^6 u^2 (t+u)^6 \Big(4 s^4+2 \
(t+u) s^3+9 \Big(t^2+u^2\Big) s^2+4 \Big(t^2+u t+u^2\Big)^2\Big) \
H(1,1,z) H(3,3,y) (s+u)^6
\end{dmath*}
\begin{dmath*}
{\white=}
+216 s^2 t^2 (s+t)^6 u^2 (t+u)^6 \Big(s^4+2 (t+u) \
s^3+3 \Big(t^2+u^2\Big) s^2+2 \Big(t^3+u^3\Big) s+t^4+u^4+6 t^2 u^2-2 \
t^3 u\Big) H(0,y) H(0,0,1,z) (s+u)^6+216 s^2 t^2 (s+t)^6 u^3 (t+u)^6 \
\Big(2 s^3-3 u s^2+4 u^2 s+6 t \Big(t^2-u t+u^2\Big)\Big) H(1,y) \
H(0,0,1,z) (s+u)^6+108 s^2 t^2 (s+t)^6 u^2 (t+u)^6 \Big(6 s^4+2 (4 t+u) \
s^3+15 u^2 s^2+2 \Big(6 t^3-u^3\Big) s+6 \Big(t^2+u \
t+u^2\Big)^2\Big) H(2,y) H(0,0,1,z) (s+u)^6+432 s^2 t^2 (s+t)^6 u^2 \
(t+u)^6 \Big(2 s^4+2 (t+u) s^3+3 \Big(t^2+u^2\Big) s^2+2 \
\Big(t^3+u^3\Big) s+2 \Big(t^2+u t+u^2\Big)^2\Big) H(3,y) H(0,0,1,z) \
(s+u)^6-36 s^2 t^2 (s+t)^2 u^2 (t+u)^6 \Big(22 s^8+3 (44 t+7 u) \
s^7+\Big(374 t^2+63 u t+18 u^2\Big) s^6+\Big(660 t^3+54 u t^2+36 u^2 t+4 \
u^3\Big) s^5+\Big(792 t^4+2 u t^3+84 u^2 t^2-11 u^3 t+18 u^4\Big) s^4+t \
\Big(660 t^4+41 u t^3+246 u^2 t^2+9 u^3 t+54 u^4\Big) s^3+t^2 \Big(374 \
t^4+147 u t^3+360 u^2 t^2+95 u^3 t+84 u^4\Big) s^2+t^3 \Big(132 t^4+140 u \
t^3+246 u^2 t^2+115 u^3 t+64 u^4\Big) s+22 t^4 \Big(t^2+u \
t+u^2\Big)^2\Big) H(0,0,2,y) (s+u)^6+216 s^2 t^2 (s+t)^6 u^2 (t+u)^6 \
\Big(s^4+2 t s^3+3 \Big(t^2+2 u^2\Big) s^2+2 \Big(t^3-u^3\Big) \
s+t^4+u^4-2 t u^3+6 t^2 u^2\Big) H(0,z) H(0,0,2,y) (s+u)^6+108 s^2 t^2 \
(s+t)^6 u^2 (t+u)^6 \Big(2 s^4+(4 t-2 u) s^3+\Big(6 t^2-3 u^2\Big) \
s^2+\Big(4 t^3-2 u^3\Big) s+2 \Big(t^2+u t+u^2\Big)^2\Big) H(1,z) \
H(0,0,2,y) (s+u)^6-108 s^2 t^2 (s+t)^6 u^2 (t+u)^6 \Big(4 s^4+2 (4 t+u) \
s^3+3 \Big(4 t^2+5 u^2\Big) s^2+\Big(8 t^3-2 u^3\Big) s+4 \Big(t^4-u \
t^3+6 u^2 t^2-u^3 t+u^4\Big)\Big) H(1,z) H(0,0,3,y) (s+u)^6+108 s^2 t^2 \
(s+t)^6 u^2 (t+u)^6 \Big(2 s^4+2 (2 t+u) s^3+3 \Big(2 t^2+u^2\Big) s^2+2 \
\Big(2 t^3+u^3\Big) s+2 t^4+2 u^4+2 t u^3+3 t^2 u^2+2 t^3 u\Big) H(0,z) \
H(0,1,0,y) (s+u)^6-108 s^2 t^2 (s+t)^6 u^2 (t+u)^6 \Big(2 s^4+2 (2 t+u) \
s^3+3 \Big(2 t^2+u^2\Big) s^2+2 \Big(2 t^3+u^3\Big) s+2 t^4+2 u^4+2 t \
u^3+3 t^2 u^2+2 t^3 u\Big) H(1,z) H(0,1,0,y) (s+u)^6+108 s^2 t^2 (s+t)^6 \
u^2 (t+u)^6 \Big(2 s^4
\end{dmath*}
\begin{dmath*}
{\white=}
+(4 t-6 u) s^3+\Big(6 t^2+9 u^2\Big) s^2+\Big(4 \
t^3-2 u^3\Big) s+2 \Big(t^4+u^4\Big)\Big) H(0,y) H(0,1,0,z) \
(s+u)^6-216 s^2 t^2 (s+t)^6 u^3 (t+u)^6 \Big(6 s^3-6 u s^2+6 u^2 s+t \
\Big(2 t^2-3 u t+4 u^2\Big)\Big) H(1,y) H(0,1,0,z) (s+u)^6-108 s^2 t^2 \
(s+t)^6 u^2 (t+u)^6 \Big(2 s^4-2 (2 t+5 u) s^3+3 \Big(4 t^2+5 u^2\Big) \
s^2-10 u^3 s+2 \Big(t^2+u t+u^2\Big)^2\Big) H(2,y) H(0,1,0,z) \
(s+u)^6+216 s^3 t^2 (s+t)^6 u^2 (t+u)^6 \Big(4 (t+u) s^2-3 \
\Big(t^2+u^2\Big) s+2 \Big(t^3+u^3\Big)\Big) H(3,y) H(0,1,0,z) \
(s+u)^6-36 s^2 t^2 (s+t)^6 u^2 (t+u)^2 \Big(2 \Big(11 t^4+32 u t^3+42 u^2 \
t^2+27 u^3 t+9 u^4\Big) s^4+\Big(44 t^5+115 u t^4+95 u^2 t^3+9 u^3 t^2-11 \
u^4 t+4 u^5\Big) s^3+6 (t+u)^2 \Big(11 t^4+19 u t^3+11 u^2 t^2+3 \
u^4\Big) s^2+(t+u)^3 \Big(44 t^4+8 u t^3-9 u^2 t^2+21 u^4\Big) s+22 \
(t+u)^4 \Big(t^2+u t+u^2\Big)^2\Big) H(0,1,1,z) (s+u)^6-108 s^2 t^2 \
(s+t)^6 u^2 (t+u)^6 \Big(2 s^4+(4 t-2 u) s^3+\Big(6 t^2-3 u^2\Big) \
s^2+\Big(4 t^3-2 u^3\Big) s+2 \Big(t^2+u t+u^2\Big)^2\Big) H(0,y) \
H(0,1,1,z) (s+u)^6+216 s^2 t^2 (s+t)^6 u^2 (t+u)^6 \Big(2 s^4+2 t s^3+9 t^2 \
s^2+2 \Big(t^2+u t+u^2\Big)^2\Big) H(3,y) H(0,1,1,z) (s+u)^6
\end{dmath*}

\begin{dmath*}
{\white =}
-36 s^2 t^2 \
(s+t)^2 u^2 (t+u)^6 \Big(22 s^8+3 (44 t+7 u) s^7+\Big(374 t^2+63 u t+18 \
u^2\Big) s^6+\Big(660 t^3+54 u t^2+36 u^2 t+4 u^3\Big) s^5+\Big(792 \
t^4+2 u t^3+84 u^2 t^2-11 u^3 t+18 u^4\Big) s^4+t \Big(660 t^4+41 u \
t^3+246 u^2 t^2+9 u^3 t+54 u^4\Big) s^3+t^2 \Big(374 t^4+147 u t^3+360 \
u^2 t^2+95 u^3 t+84 u^4\Big) s^2+t^3 \Big(132 t^4+140 u t^3+246 u^2 \
t^2+115 u^3 t+64 u^4\Big) s+22 t^4 \Big(t^2+u t+u^2\Big)^2\Big) \
H(0,2,0,y) (s+u)^6+108 s^2 t^2 (s+t)^6 u^2 (t+u)^6 \Big(2 s^4+12 u s^3+8 \
u^3 s+2 t^4+2 u^4+2 t u^3+3 t^2 u^2+2 t^3 u\Big) H(0,z) H(0,2,0,y) \
(s+u)^6+108 s^2 t^2 (s+t)^6 u^2 (t+u)^6 \Big(2 s^4+(4 t-2 u) s^3+\Big(6 \
t^2-3 u^2\Big) s^2+\Big(4 t^3-2 u^3\Big) s+2 \Big(t^2+u \
t+u^2\Big)^2\Big) H(1,z) H(0,2,0,y) (s+u)^6+36 s^2 t^2 (s+t)^6 u^2 \
(t+u)^2 \Big(2 \Big(11 t^4+32 u t^3+42 u^2 t^2+27 u^3 t+9 u^4\Big) \
s^4+\Big(44 t^5+115 u t^4+95 u^2 t^3+9 u^3 t^2-11 u^4 t+4 u^5\Big) s^3+6 \
(t+u)^2 \Big(11 t^4+19 u t^3+11 u^2 t^2+3 u^4\Big) s^2+(t+u)^3 \Big(44 \
t^4+8 u t^3-9 u^2 t^2+21 u^4\Big) s+22 (t+u)^4 \Big(t^2+u \
t+u^2\Big)^2\Big) H(0,2,2,y) (s+u)^6+108 s^2 t^2 (s+t)^6 u^2 (t+u)^6 \
\Big(2 s^4+2 t s^3+3 \Big(t^2+4 u^2\Big) s^2+2 \Big(t^3-2 u^3\Big) \
s+2 \Big(t^2+u t+u^2\Big)^2\Big) H(0,z) H(0,2,2,y) (s+u)^6
\end{dmath*}

\intertext{}
\begin{dmath*}
{\white =}
-216 s^2 t^2 \
(s+t)^6 u^2 (t+u)^6 \Big(s^4+2 (t-u) s^3+3 \Big(t^2+u^2\Big) s^2+2 \
\Big(t^3-2 u^3\Big) s+t^4+u^4-4 t u^3+9 t^2 u^2-4 t^3 u\Big) H(1,z) \
H(0,2,3,y) (s+u)^6-216 s^2 t^2 (s+t)^6 u^2 (t+u)^6 \Big(2 s^4+2 (t+u) s^3+3 \
\Big(t^2+u^2\Big) s^2+2 \Big(t^3+u^3\Big) s+2 \Big(t^2+u \
t+u^2\Big)^2\Big) H(1,z) H(0,3,0,y) (s+u)^6+108 s^2 t^2 (s+t)^6 u^2 \
(t+u)^6 \Big(2 s^4+2 (t+3 u) s^3+3 \Big(t^2-u^2\Big) s^2+2 \Big(t^3+5 \
u^3\Big) s+2 \Big(t^2+u t+u^2\Big)^2\Big) H(0,z) H(0,3,2,y) \
(s+u)^6-108 s^2 t^2 (s+t)^6 u^2 (t+u)^6 \Big(4 s^4+(6 t-2 u) s^3+9 \
\Big(t^2+u^2\Big) s^2+6 \Big(t^3-u^3\Big) s+4 \Big(t^2+u \
t+u^2\Big)^2\Big) H(1,z) H(0,3,2,y) (s+u)^6-108 s^3 t^2 (s+t)^6 u^2 \
(t+u)^6 \Big(2 (t+u) s^2+3 \Big(t^2-3 u^2\Big) s+2 \Big(t^3+3 \
u^3\Big)\Big) H(1,z) H(0,3,3,y) (s+u)^6+36 s^2 t^2 (s+t)^2 u^2 (t+u)^6 \
\Big(44 s^8+(264 t+65 u) s^7+\Big(748 t^2+203 u t+84 u^2\Big) s^6+3 \
\Big(440 t^3+67 u t^2+94 u^2 t+16 u^3\Big) s^5+\Big(1584 t^4+43 u \
t^3+444 u^2 t^2+104 u^3 t+40 u^4\Big) s^4+t \Big(1320 t^4+43 u t^3+492 \
u^2 t^2+104 u^3 t+118 u^4\Big) s^3+t^2 \Big(748 t^4+201 u t^3+444 u^2 \
t^2+104 u^3 t+168 u^4\Big) s^2+t^3 \Big(264 t^4+203 u t^3+282 u^2 t^2+104 \
u^3 t+118 u^4\Big) s+t^4 \Big(44 t^4+65 u t^3+84 u^2 t^2+48 u^3 t+40 \
u^4\Big)\Big) H(1,0,0,y) (s+u)^6
\end{dmath*}

\begin{dmath*}
{\white =}
+216 s^2 t^2 (s+t)^6 u^2 (t+u)^6 \Big(2 \
s^4+4 t s^3+\Big(6 t^2+9 u^2\Big) s^2+2 \Big(2 t^3+u^3\Big) s+2 \
\Big(t^4+u^4\Big)\Big) H(0,z) H(1,0,0,y) (s+u)^6-216 s^2 t^2 (s+t)^6 \
u^2 (t+u)^6 \Big(2 s^4+4 t s^3+6 t^2 s^2+4 t^3 s+2 t^4+2 u^4+2 t u^3+9 t^2 \
u^2\Big) H(1,z) H(1,0,0,y) (s+u)^6+108 s^2 t^2 (s+t)^6 u^2 (t+u)^6 \Big(2 \
s^4+4 (t+u) s^3+6 \Big(t^2+u^2\Big) s^2+4 \Big(t^3+u^3\Big) s+2 t^4+2 \
u^4-2 t u^3-3 t^2 u^2-2 t^3 u\Big) H(0,y) H(1,0,0,z) (s+u)^6-108 s^2 t^2 \
(s+t)^6 u^2 (t+u)^6 \Big(2 s^4-4 (t-u) s^3+6 \Big(2 t^2+u^2\Big) s^2+4 \
u^3 s+2 t^4+2 u^4+2 t u^3+3 t^2 u^2+2 t^3 u\Big) H(2,y) H(1,0,0,z) \
(s+u)^6-108 s^2 t^2 (s+t)^6 u^2 (t+u)^6 \Big(2 s^4+4 t s^3+6 t^2 s^2+4 t^3 \
s+2 t^4+2 u^4-2 t u^3+9 t^2 u^2-6 t^3 u\Big) H(0,y) H(1,0,1,z) (s+u)^6+216 \
s^2 t^2 (s+t)^6 u^3 (t+u)^6 \Big(2 s^3-3 u s^2+4 u^2 s+6 t \Big(t^2-u \
t+u^2\Big)\Big) H(1,y) H(1,0,1,z) (s+u)^6-216 s^2 t^2 (s+t)^6 u^2 \
(t+u)^6 \Big(2 (t+2 u) s^3-3 t^2 s^2+\Big(4 t^3+6 u^3\Big) s-t u \Big(2 \
t^2+3 u t+2 u^2\Big)\Big) H(2,y) H(1,0,1,z) (s+u)^6+216 s^2 t^2 (s+t)^6 \
u^2 (t+u)^6 \Big(2 s^4+2 t s^3+9 t^2 s^2+2 \Big(t^2+u \
t+u^2\Big)^2\Big) H(3,y) H(1,0,1,z) (s+u)^6
\end{dmath*}
\begin{dmath*}
{\white=}
-18 s^2 t^2 (s+t)^3 u^2 \
(t+u)^6 \Big(8 s^7+(28 t-6 u) s^6+3 \Big(11 t^2-8 u t-5 u^2\Big) \
s^5-\Big(7 t^3+42 u t^2+81 u^2 t+12 u^3\Big) s^4-2 \Big(28 t^4+18 u \
t^3+105 u^2 t^2+12 u^3 t+7 u^4\Big) s^3-6 t \Big(8 t^4+u t^3+37 u^2 t^2+4 \
u^3 t+5 u^4\Big) s^2-3 \Big(3 t^6-4 u t^5+21 u^2 t^4+8 u^4 t^2\Big) s+3 \
t^4 \Big(t^3+2 u t^2+5 u^2 t+4 u^3\Big)\Big) H(1,0,2,y) (s+u)^6+108 s^2 \
t^2 (s+t)^6 u^2 (t+u)^6 \Big(2 s^4+(4 t+6 u) s^3+\Big(6 t^2-3 u^2\Big) \
s^2+2 \Big(2 t^3+5 u^3\Big) s+2 t^4+2 u^4+10 t u^3-3 t^2 u^2+6 t^3 \
u\Big) H(0,z) H(1,0,2,y) (s+u)^6-216 s^2 t^2 (s+t)^6 u^3 (t+u)^6 \Big(2 \
s^3-3 u s^2+4 u^2 s+6 t \Big(t^2-u t+u^2\Big)\Big) H(1,z) H(1,0,3,y) \
(s+u)^6-18 s^2 t^2 (s+t)^2 u^2 (t+u)^6 \Big(3 s^8+2 (51 t+u) s^7+\Big(486 \
t^2+200 u t+30 u^2\Big) s^6+6 \Big(179 t^3+130 u t^2+96 u^2 t+7 \
u^3\Big) s^5+2 \Big(687 t^4+677 u t^3+945 u^2 t^2+265 u^3 t+29 u^4\Big) \
s^4+2 t \Big(537 t^4+677 u t^3+1344 u^2 t^2+658 u^3 t+170 u^4\Big) s^3+2 \
t^2 \Big(243 t^4+390 u t^3+945 u^2 t^2+658 u^3 t+270 u^4\Big) s^2+2 t^3 \
\Big(51 t^4+100 u t^3+288 u^2 t^2+265 u^3 t+170 u^4\Big) s+t^4 \Big(3 \
t^4+2 u t^3+30 u^2 t^2+42 u^3 t+58 u^4\Big)\Big) H(1,1,0,y) (s+u)^6+216 \
s^3 t^2 (s+t)^6 u^3 (t+u)^6 \Big(4 s^2-3 u s+2 u^2\Big) H(0,z) H(1,1,0,y) \
(s+u)^6
\end{dmath*}

\begin{dmath*}
{\white =}
-216 s^2 t^3 (s+t)^6 u^3 (t+u)^6 \Big(4 t^2-3 u t+2 u^2\Big) \
H(1,z) H(1,1,0,y) (s+u)^6-216 s^2 t^2 (s+t)^6 u^2 (t+u)^6 \Big(s^4+2 (t-u) \
s^3+3 \Big(t^2+2 u^2\Big) s^2+2 t^3 s+\Big(t^2+u t+u^2\Big)^2\Big) \
H(0,y) H(1,1,0,z) (s+u)^6-216 s^2 t^2 (s+t)^6 u^3 (t+u)^6 \Big(6 s^3-6 u \
s^2+6 u^2 s+t \Big(2 t^2-3 u t+4 u^2\Big)\Big) H(1,y) H(1,1,0,z) \
(s+u)^6-108 s^2 t^2 (s+t)^6 u^2 (t+u)^6 \Big(8 s^4+4 (2 t+u) s^3+12 \
\Big(t^2+3 u^2\Big) s^2+4 \Big(2 t^3+u^3\Big) s+8 t^4+8 u^4+14 t \
u^3+21 t^2 u^2+14 t^3 u\Big) H(2,y) H(1,1,0,z) (s+u)^6-216 s^2 t^2 (s+t)^6 \
u^2 (t+u)^6 \Big(2 s^4+4 (t+2 u) s^3-3 \Big(t^2-u^2\Big) s^2+2 \
\Big(t^3+3 u^3\Big) s+2 \Big(t^2+u t+u^2\Big)^2\Big) H(3,y) \
H(1,1,0,z) (s+u)^6-18 s^2 t^2 (s+t)^3 u^2 (t+u)^6 \Big(8 s^7+(28 t-6 u) \
s^6+3 \Big(11 t^2-8 u t-5 u^2\Big) s^5-\Big(7 t^3+42 u t^2+81 u^2 t+12 \
u^3\Big) s^4-2 \Big(28 t^4+18 u t^3+105 u^2 t^2+12 u^3 t+7 u^4\Big) \
s^3-6 t \Big(8 t^4+u t^3+37 u^2 t^2+4 u^3 t+5 u^4\Big) s^2-3 \Big(3 \
t^6-4 u t^5+21 u^2 t^4+8 u^4 t^2\Big) s+3 t^4 \Big(t^3+2 u t^2+5 u^2 t+4 \
u^3\Big)\Big) H(1,2,0,y) (s+u)^6+108 s^2 t^2 (s+t)^6 u^2 (t+u)^6 \Big(2 \
s^4+(4 t-6 u) s^3+\Big(6 t^2+9 u^2\Big) s^2+\Big(4 t^3-2 u^3\Big) s+2 \
t^4+2 u^4+2 t u^3+3 t^2 u^2+2 t^3 u\Big) H(0,z) H(1,2,0,y) (s+u)^6+216 s^2 \
t^2 (s+t)^6 u^3 (t+u)^6 \Big(2 s^3
\end{dmath*}
\begin{dmath*}
{\white=}
-3 u s^2+4 u^2 s+6 t \Big(t^2-u \
t+u^2\Big)\Big) H(1,z) H(1,2,3,y) (s+u)^6-36 s^2 t^2 (s+t)^2 u^2 (t+u)^6 \
\Big(22 s^8+3 (44 t+7 u) s^7+\Big(374 t^2+63 u t+18 u^2\Big) \
s^6+\Big(660 t^3+54 u t^2+36 u^2 t+4 u^3\Big) s^5+\Big(792 t^4+2 u \
t^3+84 u^2 t^2-11 u^3 t+18 u^4\Big) s^4+t \Big(660 t^4+41 u t^3+246 u^2 \
t^2+9 u^3 t+54 u^4\Big) s^3+t^2 \Big(374 t^4+147 u t^3+360 u^2 t^2+95 u^3 \
t+84 u^4\Big) s^2+t^3 \Big(132 t^4+140 u t^3+246 u^2 t^2+115 u^3 t+64 \
u^4\Big) s+22 t^4 \Big(t^2+u t+u^2\Big)^2\Big) H(2,0,0,y) (s+u)^6-108 \
s^2 t^2 (s+t)^6 u^2 (t+u)^6 \Big(2 s^4+4 (t-u) s^3+6 \Big(t^2+2 u^2\Big) \
s^2+4 t^3 s+2 t^4+2 u^4+2 t u^3+3 t^2 u^2+2 t^3 u\Big) H(0,z) H(2,0,0,y) \
(s+u)^6+108 s^2 t^2 (s+t)^6 u^2 (t+u)^6 \Big(2 s^4+(4 t-2 u) s^3+\Big(6 \
t^2-3 u^2\Big) s^2+\Big(4 t^3-2 u^3\Big) s+2 \Big(t^2+u \
t+u^2\Big)^2\Big) H(1,z) H(2,0,0,y) (s+u)^6+36 s^2 t^2 (s+t)^6 u^2 \
(t+u)^2 \Big(2 \Big(11 t^4+32 u t^3+42 u^2 t^2+27 u^3 t+9 u^4\Big) \
s^4+\Big(44 t^5+115 u t^4+95 u^2 t^3+9 u^3 t^2-11 u^4 t+4 u^5\Big) s^3+6 \
(t+u)^2 \Big(11 t^4+19 u t^3+11 u^2 t^2+3 u^4\Big) s^2+(t+u)^3 \Big(44 \
t^4+8 u t^3-9 u^2 t^2+21 u^4\Big) s+22 (t+u)^4 \Big(t^2+u \
t+u^2\Big)^2\Big) H(2,0,2,y) (s+u)^6-108 s^2 t^2 (s+t)^6 u^2 (t+u)^6 \
\Big(2 s^4+2 (t+4 u) s^3+3 t^2 s^2+2 \Big(t^3+6 u^3\Big) s+2 \
\Big(t^4+u^4\Big)\Big) H(0,z) H(2,0,2,y) (s+u)^6-216 s^2 t^2 (s+t)^6 \
u^2 (t+u)^6 \Big(s^4+(4 t-2 u) s^3+3 u^2 s^2+\Big(6 t^3-4 u^3\Big) \
s+\Big(t^2+u t+u^2\Big)^2\Big) H(1,z) H(2,0,3,y) (s+u)^6-108 s^2 t^2 \
(s+t)^6 u^2 (t+u)^6 \Big(2 s^4+2 (2 t+5 u) s^3+\Big(6 t^2-3 u^2\Big) \
s^2+\Big(4 t^3+6 u^3\Big) s+2 t^4+2 u^4+2 t u^3+3 t^2 u^2+2 t^3 u\Big) \
H(0,z) H(2,1,0,y) (s+u)^6-216 s^2 t^2 (s+t)^6 u^2 (t+u)^6 \Big(2 s^4+2 (2 \
t+u) s^3+3 \Big(2 t^2+u^2\Big) s^2+2 \Big(2 t^3+u^3\Big) s+2 t^4+2 \
u^4+2 t u^3+3 t^2 u^2+2 t^3 u\Big) H(1,z) H(2,1,0,y) (s+u)^6
\end{dmath*}

\intertext{}
\begin{dmath*}
{\white =}
+36 s^2 t^2 \
(s+t)^6 u^2 (t+u)^2 \Big(2 \Big(11 t^4+32 u t^3+42 u^2 t^2+27 u^3 t+9 \
u^4\Big) s^4+\Big(44 t^5+115 u t^4+95 u^2 t^3+9 u^3 t^2-11 u^4 t+4 \
u^5\Big) s^3+6 (t+u)^2 \Big(11 t^4+19 u t^3+11 u^2 t^2+3 u^4\Big) \
s^2+(t+u)^3 \Big(44 t^4+8 u t^3-9 u^2 t^2+21 u^4\Big) s+22 (t+u)^4 \
\Big(t^2+u t+u^2\Big)^2\Big) H(2,2,0,y) (s+u)^6-216 s^2 t^2 (s+t)^6 u^2 \
(t+u)^6 \Big(s^4-2 (t+u) s^3+6 \Big(t^2+u^2\Big) s^2+\Big(t^2+u \
t+u^2\Big)^2\Big) H(0,z) H(2,2,0,y) (s+u)^6+432 s^2 t^2 (s+t)^6 u^2 \
(t+u)^6 \Big(s^4+3 \Big(t^2+u^2\Big) s^2-\Big(t^3+u^3\Big) \
s+\Big(t^2+u t+u^2\Big)^2\Big) H(1,z) H(2,2,3,y) (s+u)^6-108 s^2 t^2 \
(s+t)^6 u^2 (t+u)^6 \Big(2 s^4+2 (t+u) s^3+3 \Big(t^2+u^2\Big) s^2+2 \
\Big(t^3+u^3\Big) s+2 \Big(t^2+u t+u^2\Big)^2\Big) H(1,z) H(2,3,0,y) \
(s+u)^6-108 s^2 t^2 (s+t)^6 u^2 (t+u)^6 \Big(2 s^4+2 (t+u) s^3+3 \
\Big(t^2+u^2\Big) s^2+2 \Big(t^3+u^3\Big) s+2 \Big(t^2+u \
t+u^2\Big)^2\Big) H(0,z) H(2,3,2,y) (s+u)^6+432 s^2 t^2 (s+t)^6 u^2 \
(t+u)^6 \Big(2 s^4+2 (t+u) s^3+3 \Big(t^2+u^2\Big) s^2+2 \
\Big(t^3+u^3\Big) s+2 \Big(t^2+u t+u^2\Big)^2\Big) H(1,z) H(2,3,3,y) \
(s+u)^6-18 s^2 t^2 (s+t)^6 u^2 (t+u)^3 \Big(2 u \Big(12 t^2+15 u t+7 \
u^2\Big) s^4-12 \Big(t^4-2 u^2 t^2-2 u^3 t-u^4\Big) s^3-3 (t+u)^2 \
\Big(5 t^3-31 u t^2-17 u^2 t-5 u^3\Big) s^2-6 (t+u)^3 \Big(t^3-u t^2-u^2 \
t-u^3\Big) s-(t+u)^3 \Big(3 t^4-18 u t^3-3 u^2 t^2+4 u^3 t+8 \
u^4\Big)\Big) H(3,0,2,y) (s+u)^6
\end{dmath*}

\begin{dmath*}
{\white =}
-108 s^2 t^2 (s+t)^6 u^2 (t+u)^6 \Big(2 \
s^4+2 (t-u) s^3+3 \Big(t^2+3 u^2\Big) s^2+2 \Big(t^3-3 u^3\Big) s+2 \
\Big(t^2+u t+u^2\Big)^2\Big) H(0,z) H(3,0,2,y) (s+u)^6-216 s^2 t^2 \
(s+t)^6 u^2 (t+u)^6 \Big(2 s^4+2 t s^3+9 t^2 s^2+2 \Big(t^2+u \
t+u^2\Big)^2\Big) H(1,z) H(3,0,2,y) (s+u)^6-216 s^3 t^2 (s+t)^6 u^3 \
(t+u)^6 \Big(2 s^2-3 u s+4 u^2\Big) H(1,z) H(3,0,3,y) (s+u)^6-18 s^2 t^2 \
(s+t)^6 u^2 (t+u)^3 \Big(2 u \Big(12 t^2+15 u t+7 u^2\Big) s^4-12 \
\Big(t^4-2 u^2 t^2-2 u^3 t-u^4\Big) s^3-3 (t+u)^2 \Big(5 t^3-31 u t^2-17 \
u^2 t-5 u^3\Big) s^2-6 (t+u)^3 \Big(t^3-u t^2-u^2 t-u^3\Big) s-(t+u)^3 \
\Big(3 t^4-18 u t^3-3 u^2 t^2+4 u^3 t+8 u^4\Big)\Big) H(3,2,0,y) \
(s+u)^6-108 s^2 t^2 (s+t)^6 u^2 (t+u)^6 \Big(2 s^4+10 (t+u) s^3-3 \
\Big(t^2+u^2\Big) s^2+6 \Big(t^3+u^3\Big) s+2 \Big(t^2+u \
t+u^2\Big)^2\Big) H(0,z) H(3,2,0,y) (s+u)^6-216 s^2 t^2 (s+t)^6 u^2 \
(t+u)^6 \Big(2 s^4+2 t s^3+9 t^2 s^2+2 \Big(t^2+u t+u^2\Big)^2\Big) \
H(1,z) H(3,2,0,y) (s+u)^6-36 s^2 t^2 (s+t)^6 u^2 (t+u)^2 \Big(2 \Big(20 \
t^4+59 u t^3+84 u^2 t^2+59 u^3 t+20 u^4\Big) s^4+8 \Big(6 t^5+13 u t^4+13 \
u^2 t^3+13 u^3 t^2+13 u^4 t+6 u^5\Big) s^3+6 (t+u)^2 \Big(14 t^4+19 u \
t^3+22 u^2 t^2+19 u^3 t+14 u^4\Big) s^2+(t+u)^3 \Big(65 t^4+8 u t^3-18 \
u^2 t^2+8 u^3 t+65 u^4\Big) s+44 (t+u)^4 \Big(t^2+u t+u^2\Big)^2\Big) \
H(3,2,2,y) (s+u)^6-216 s^2 t^2 (s+t)^6 u^2 (t+u)^6 \Big(2 s^4+2 u s^3+9 u^2 \
s^2+2 \Big(t^2+u t+u^2\Big)^2\Big) H(0,z) H(3,2,2,y) (s+u)^6
\end{dmath*}
\begin{dmath*}
{\white=}
+216 s^2 \
t^2 (s+t)^6 u^2 (t+u)^6 \Big(4 s^4+2 (t+u) s^3+9 \Big(t^2+u^2\Big) s^2+4 \
\Big(t^2+u t+u^2\Big)^2\Big) H(1,z) H(3,2,3,y) (s+u)^6-216 s^3 t^3 \
(s+t)^6 \Big(2 s^2-3 t s+4 t^2\Big) u^2 (t+u)^6 H(1,z) H(3,3,0,y) \
(s+u)^6+18 s^2 t^2 (s+t)^6 u^2 (t+u)^2 \Big(\Big(58 t^4+340 u t^3+540 u^2 \
t^2+340 u^3 t+58 u^4\Big) s^4+2 \Big(21 t^5+265 u t^4+658 u^2 t^3+658 u^3 \
t^2+265 u^4 t+21 u^5\Big) s^3+6 (t+u)^2 \Big(5 t^4+86 u t^3+138 u^2 \
t^2+86 u^3 t+5 u^4\Big) s^2+2 (t+u)^3 \Big(t^4+97 u t^3+96 u^2 t^2+97 u^3 \
t+u^4\Big) s+3 (t+u)^4 \Big(t^4+30 u t^3+36 u^2 t^2+30 u^3 \
t+u^4\Big)\Big) H(3,3,2,y) (s+u)^6-216 s^3 t^2 (s+t)^6 u^3 (t+u)^6 \
\Big(2 s^2-3 u s+4 u^2\Big) H(0,z) H(3,3,2,y) (s+u)^6+216 s^2 t^2 (s+t)^6 \
u^2 (t+u)^6 \Big(4 s^4+2 (t+u) s^3+9 \Big(t^2+u^2\Big) s^2+4 \Big(t^2+u \
t+u^2\Big)^2\Big) H(1,z) H(3,3,2,y) (s+u)^6+216 s^2 t^2 (s+t)^6 u^2 \
(t+u)^6 \Big(4 s^4+6 (t+u) s^3+3 \Big(t^2+u^2\Big) s^2+8 \
\Big(t^3+u^3\Big) s+4 \Big(t^2+u t+u^2\Big)^2\Big) H(1,z) H(3,3,3,y) \
(s+u)^6-324 s^2 t^2 (s+t)^6 u^3 (t+u)^6 \Big(2 s^3+3 u s^2+2 u^2 s-4 t \
\Big(t^2-u t+u^2\Big)\Big) H(0,0,0,1,z) (s+u)^6
\end{dmath*}

\begin{dmath*}
{\white =}
+432 s^2 t^2 (s+t)^6 u^2 \
(t+u)^6 \Big(s^4+(2 t-u) s^3+3 \Big(t^2+u^2\Big) s^2+2 t^3 s+t^4+u^4+3 \
t^2 u^2-t^3 u\Big) H(0,0,1,0,y) (s+u)^6+432 s^2 t^2 (s+t)^6 u^3 (t+u)^6 \
\Big(3 s^3+2 u^2 s-2 t^3-3 t u^2\Big) H(0,0,1,0,z) (s+u)^6+108 s^2 t^2 \
(s+t)^6 u^2 (t+u)^6 \Big(2 s^4+(4 t-2 u) s^3+\Big(6 t^2-3 u^2\Big) \
s^2+\Big(4 t^3-2 u^3\Big) s+2 \Big(t^2+u t+u^2\Big)^2\Big) \
H(0,0,1,1,z) (s+u)^6+108 s^2 t^2 (s+t)^6 u^2 (t+u)^6 \Big(2 s^4+(4 t-2 u) \
s^3+\Big(6 t^2-3 u^2\Big) s^2+\Big(4 t^3-2 u^3\Big) s+2 \Big(t^2+u \
t+u^2\Big)^2\Big) H(0,0,2,2,y) (s+u)^6-108 s^2 t^2 (s+t)^6 u^2 (t+u)^6 \
\Big(4 s^4+2 (4 t+u) s^3+3 \Big(4 t^2+5 u^2\Big) s^2+\Big(8 t^3-2 \
u^3\Big) s+4 \Big(t^4-u t^3+6 u^2 t^2-u^3 t+u^4\Big)\Big) \
H(0,0,3,2,y) (s+u)^6+108 s^2 t^2 (s+t)^6 u^2 (t+u)^6 \Big(2 s^4+4 (t-u) \
s^3+6 \Big(t^2-u^2\Big) s^2+4 \Big(t^3-u^3\Big) s+2 t^4+2 u^4+18 t \
u^3-3 t^2 u^2+18 t^3 u\Big) H(0,1,0,1,z) (s+u)^6-108 s^2 t^2 (s+t)^6 u^2 \
(t+u)^6 \Big(2 s^4+2 (2 t+u) s^3+3 \Big(2 t^2+u^2\Big) s^2+2 \Big(2 \
t^3+u^3\Big) s+2 t^4+2 u^4+2 t u^3+3 t^2 u^2+2 t^3 u\Big) H(0,1,0,2,y) \
(s+u)^6+432 s^2 t^2 (s+t)^6 u^2 (t+u)^6 \Big(2 s^4+2 (2 t+u) s^3+3 \Big(2 \
t^2+u^2\Big) s^2+2 \Big(2 t^3+u^3\Big) s+2 t^4+2 u^4+2 t u^3+3 t^2 \
u^2+2 t^3 u\Big) H(0,1,1,0,y) (s+u)^6+108 s^2 t^2 (s+t)^6 u^2 (t+u)^6 \
\Big(2 s^4+4 (t-u) s^3+6 \Big(t^2+2 u^2\Big) s^2+4 t^3 s+2 t^4+2 u^4-10 \
t u^3+3 t^2 u^2-6 t^3 u\Big) H(0,1,1,0,z) (s+u)^6-108 s^2 t^2 (s+t)^6 u^2 \
(t+u)^6 \Big(2 s^4+2 (2 t+u) s^3+3 \Big(2 t^2+u^2\Big) s^2+2 \Big(2 \
t^3+u^3\Big) s+2 t^4+2 u^4+2 t u^3+3 t^2 u^2+2 t^3 u\Big) H(0,1,2,0,y) \
(s+u)^6+108 s^2 t^2 (s+t)^6 u^2 (t+u)^6 \Big(2 s^4+(4 t-2 u) s^3+\Big(6 \
t^2-3 u^2\Big) s^2+\Big(4 t^3-2 u^3\Big) s+2 \Big(t^2+u \
t+u^2\Big)^2\Big) H(0,2,0,2,y) (s+u)^6-216 s^2 t^2 (s+t)^6 u^2 (t+u)^6 \
\Big(s^4+2 (t-2 u) s^3+3 \Big(t^2+u^2\Big) s^2+2 \Big(t^3-u^3\Big) \
s+t^4+u^4+4 t u^3+6 t^3 u\Big) H(0,2,1,0,y) (s+u)^6+108 s^2 t^2 (s+t)^6 \
u^2 (t+u)^6 \Big(2 s^4
\end{dmath*}
\begin{dmath*}
{\white=}
+(4 t-2 u) s^3+\Big(6 t^2-3 u^2\Big) s^2+\Big(4 \
t^3-2 u^3\Big) s+2 \Big(t^2+u t+u^2\Big)^2\Big) H(0,2,2,0,y) \
(s+u)^6-216 s^2 t^2 (s+t)^6 u^2 (t+u)^6 \Big(s^4+2 (t-u) s^3+3 \
\Big(t^2+u^2\Big) s^2+2 \Big(t^3-2 u^3\Big) s+t^4+u^4-4 t u^3+9 t^2 \
u^2-4 t^3 u\Big) H(0,2,3,2,y) (s+u)^6-216 s^2 t^2 (s+t)^6 u^2 (t+u)^6 \
\Big(2 s^4+2 (t+u) s^3+3 \Big(t^2+u^2\Big) s^2+2 \Big(t^3+u^3\Big) \
s+2 \Big(t^2+u t+u^2\Big)^2\Big) H(0,3,0,2,y) (s+u)^6-216 s^2 t^2 \
(s+t)^6 u^2 (t+u)^6 \Big(2 s^4+2 (t+u) s^3+3 \Big(t^2+u^2\Big) s^2+2 \
\Big(t^3+u^3\Big) s+2 \Big(t^2+u t+u^2\Big)^2\Big) H(0,3,2,0,y) \
(s+u)^6-108 s^2 t^2 (s+t)^6 u^2 (t+u)^6 \Big(4 s^4+(6 t-2 u) s^3+9 \
\Big(t^2+u^2\Big) s^2+6 \Big(t^3-u^3\Big) s+4 \Big(t^2+u \
t+u^2\Big)^2\Big) H(0,3,2,2,y) (s+u)^6-108 s^3 t^2 (s+t)^6 u^2 (t+u)^6 \
\Big(2 (t+u) s^2+3 \Big(t^2-3 u^2\Big) s+2 \Big(t^3+3 u^3\Big)\Big) \
H(0,3,3,2,y) (s+u)^6+108 s^2 t^2 (s+t)^6 u^2 (t+u)^6 \Big(2 s^4+(4 t-6 u) \
s^3+3 \Big(2 t^2+u^2\Big) s^2+2 \Big(2 t^3-5 u^3\Big) s+2 \Big(t^4-2 \
u t^3+6 u^2 t^2+u^4\Big)\Big) H(1,0,0,1,z) (s+u)^6-216 s^2 t^2 (s+t)^6 \
u^2 (t+u)^6 \Big(2 s^4+4 t s^3+6 t^2 s^2+4 t^3 s+2 t^4+2 u^4+2 t u^3+9 t^2 \
u^2\Big) H(1,0,0,2,y) (s+u)^6+216 s^2 t^2 (s+t)^6 u^2 (t+u)^6 \Big(4 \
s^4+8 t s^3+3 \Big(4 t^2+3 u^2\Big) s^2+2 \Big(4 t^3+u^3\Big) s+4 \
t^4+4 u^4+2 t u^3+9 t^2 u^2\Big) H(1,0,1,0,y) (s+u)^6
\end{dmath*}

\intertext{}
\begin{dmath*}
{\white =}
+108 s^2 t^2 (s+t)^6 \
u^2 (t+u)^6 \Big(2 s^4+2 (2 t+9 u) s^3+\Big(6 t^2-3 u^2\Big) s^2+2 \
\Big(2 t^3+9 u^3\Big) s+2 \Big(t^4-2 u t^3-3 u^2 t^2-2 u^3 \
t+u^4\Big)\Big) H(1,0,1,0,z) (s+u)^6-216 s^2 t^2 (s+t)^6 u^2 (t+u)^6 \
\Big(2 s^4+4 t s^3+6 t^2 s^2+4 t^3 s+2 t^4+2 u^4+2 t u^3+9 t^2 u^2\Big) \
H(1,0,2,0,y) (s+u)^6-216 s^2 t^2 (s+t)^6 u^3 (t+u)^6 \Big(2 s^3-3 u s^2+4 \
u^2 s+6 t \Big(t^2-u t+u^2\Big)\Big) H(1,0,3,2,y) (s+u)^6+216 s^2 t^2 \
(s+t)^6 u^2 (t+u)^6 \Big(4 s^4+8 t s^3+3 \Big(4 t^2+3 u^2\Big) s^2+2 \
\Big(4 t^3+u^3\Big) s+4 t^4+4 u^4+2 t u^3+9 t^2 u^2\Big) H(1,1,0,0,y) \
(s+u)^6+108 s^2 t^2 (s+t)^6 u^2 (t+u)^6 \Big(2 s^4+4 (t+u) s^3+6 \
\Big(t^2+u^2\Big) s^2+4 \Big(t^3+u^3\Big) s+2 t^4+2 u^4-2 t u^3-3 t^2 \
u^2-2 t^3 u\Big) H(1,1,0,0,z) (s+u)^6-432 s^2 t^2 (s+t)^6 u^3 (t+u)^6 \
\Big(2 s^3+3 u^2 s-3 t^3-2 t u^2\Big) H(1,1,0,1,z) (s+u)^6-216 s^2 t^3 \
(s+t)^6 u^3 (t+u)^6 \Big(4 t^2-3 u t+2 u^2\Big) H(1,1,0,2,y) (s+u)^6+216 \
s^2 t^2 (s+t)^6 u^2 (t+u)^6 \Big(4 s^4+8 (t+u) s^3+3 \Big(4 t^2+u^2\Big) \
s^2+\Big(8 t^3+6 u^3\Big) s+4 t^4+4 u^4+6 t u^3+3 t^2 u^2+8 t^3 u\Big) \
H(1,1,1,0,y) (s+u)^6+324 s^2 t^2 (s+t)^6 u^3 (t+u)^6 \Big(4 s^3-4 u s^2+4 \
u^2 s-t \Big(2 t^2+3 u t+2 u^2\Big)\Big) H(1,1,1,0,z) (s+u)^6-216 s^2 \
t^3 (s+t)^6 u^3 (t+u)^6 \Big(4 t^2-3 u t+2 u^2\Big) H(1,1,2,0,y) \
(s+u)^6-216 s^2 t^2 (s+t)^6 u^2 (t+u)^6 \Big(2 s^4+4 t s^3+6 t^2 s^2+4 t^3 \
s+2 t^4+2 u^4+2 t u^3+9 t^2 u^2\Big) H(1,2,0,0,y) (s+u)^6
\end{dmath*}

\begin{dmath*}
{\white =}
-216 s^3 t^2 \
(s+t)^6 u^3 (t+u)^6 \Big(4 s^2-3 u s+2 u^2\Big) H(1,2,1,0,y) (s+u)^6+216 \
s^2 t^2 (s+t)^6 u^3 (t+u)^6 \Big(2 s^3-3 u s^2+4 u^2 s+6 t \Big(t^2-u \
t+u^2\Big)\Big) H(1,2,3,2,y) (s+u)^6+108 s^2 t^2 (s+t)^6 u^2 (t+u)^6 \
\Big(2 s^4+(4 t-2 u) s^3+\Big(6 t^2-3 u^2\Big) s^2+\Big(4 t^3-2 \
u^3\Big) s+2 \Big(t^2+u t+u^2\Big)^2\Big) H(2,0,0,2,y) (s+u)^6-216 \
s^2 t^2 (s+t)^6 u^2 (t+u)^6 \Big(s^4-4 (t+u) s^3+3 \Big(3 t^2+u^2\Big) \
s^2-2 \Big(2 t^3+u^3\Big) s+\Big(t^2+u t+u^2\Big)^2\Big) \
H(2,0,1,0,y) (s+u)^6+108 s^2 t^2 (s+t)^6 u^2 (t+u)^6 \Big(2 s^4+(4 t-2 u) \
s^3+\Big(6 t^2-3 u^2\Big) s^2+\Big(4 t^3-2 u^3\Big) s+2 \Big(t^2+u \
t+u^2\Big)^2\Big) H(2,0,2,0,y) (s+u)^6-216 s^2 t^2 (s+t)^6 u^2 (t+u)^6 \
\Big(s^4+(4 t-2 u) s^3+3 u^2 s^2+\Big(6 t^3-4 u^3\Big) s+\Big(t^2+u \
t+u^2\Big)^2\Big) H(2,0,3,2,y) (s+u)^6-108 s^2 t^2 (s+t)^6 u^2 (t+u)^6 \
\Big(4 s^4+(8 t-6 u) s^3+3 \Big(4 t^2+3 u^2\Big) s^2+\Big(8 t^3-2 \
u^3\Big) s+4 t^4+4 u^4+6 t u^3+9 t^2 u^2+6 t^3 u\Big) H(2,1,0,0,y) \
(s+u)^6-216 s^2 t^2 (s+t)^6 u^2 (t+u)^6 \Big(2 s^4+2 (2 t+u) s^3+3 \Big(2 \
t^2+u^2\Big) s^2+2 \Big(2 t^3+u^3\Big) s+2 t^4+2 u^4+2 t u^3+3 t^2 \
u^2+2 t^3 u\Big) H(2,1,0,2,y) (s+u)^6-108 s^2 t^2 (s+t)^6 u^3 (t+u)^6 \
\Big(6 s^3-9 u s^2+2 u^2 s+t \Big(2 t^2+3 u t+2 u^2\Big)\Big) \
H(2,1,1,0,y) (s+u)^6-216 s^2 t^2 (s+t)^6 u^2 (t+u)^6 \Big(2 s^4+2 (2 t+u) \
s^3+3 \Big(2 t^2+u^2\Big) s^2+2 \Big(2 t^3+u^3\Big) s+2 t^4+2 u^4+2 t \
u^3+3 t^2 u^2+2 t^3 u\Big) H(2,1,2,0,y) (s+u)^6+108 s^2 t^2 (s+t)^6 u^2 \
(t+u)^6 \Big(2 s^4+(4 t-2 u) s^3+\Big(6 t^2-3 u^2\Big) s^2+\Big(4 t^3-2 \
u^3\Big) s+2 \Big(t^2+u t+u^2\Big)^2\Big) H(2,2,0,0,y) (s+u)^6
\end{dmath*}
\begin{dmath*}
{\white=}
-108 \
s^2 t^2 (s+t)^6 u^2 (t+u)^6 \Big(4 s^4-2 (2 t+u) s^3+3 \Big(8 t^2+5 \
u^2\Big) s^2+\Big(2 u^3-4 t^3\Big) s+4 \Big(t^2+u \
t+u^2\Big)^2\Big) H(2,2,1,0,y) (s+u)^6+432 s^2 t^2 (s+t)^6 u^2 (t+u)^6 \
\Big(s^4+3 \Big(t^2+u^2\Big) s^2-\Big(t^3+u^3\Big) s+\Big(t^2+u \
t+u^2\Big)^2\Big) H(2,2,3,2,y) (s+u)^6-108 s^2 t^2 (s+t)^6 u^2 (t+u)^6 \
\Big(2 s^4+2 (t+u) s^3+3 \Big(t^2+u^2\Big) s^2+2 \Big(t^3+u^3\Big) \
s+2 \Big(t^2+u t+u^2\Big)^2\Big) H(2,3,0,2,y) (s+u)^6-108 s^2 t^2 \
(s+t)^6 u^2 (t+u)^6 \Big(2 s^4+2 (t+u) s^3+3 \Big(t^2+u^2\Big) s^2+2 \
\Big(t^3+u^3\Big) s+2 \Big(t^2+u t+u^2\Big)^2\Big) H(2,3,2,0,y) \
(s+u)^6+432 s^2 t^2 (s+t)^6 u^2 (t+u)^6 \Big(2 s^4+2 (t+u) s^3+3 \
\Big(t^2+u^2\Big) s^2+2 \Big(t^3+u^3\Big) s+2 \Big(t^2+u \
t+u^2\Big)^2\Big) H(2,3,3,2,y) (s+u)^6+216 s^3 t^2 (s+t)^6 u^2 (t+u)^6 \
\Big((6 t+4 u) s^2-3 \Big(2 t^2+u^2\Big) s+2 \Big(3 \
t^3+u^3\Big)\Big) H(3,0,1,0,y) (s+u)^6-216 s^2 t^2 (s+t)^6 u^2 (t+u)^6 \
\Big(2 s^4+2 t s^3+9 t^2 s^2+2 \Big(t^2+u t+u^2\Big)^2\Big) \
H(3,0,2,2,y) (s+u)^6
\end{dmath*}

\begin{dmath*}
{\white =}
-216 s^3 t^2 (s+t)^6 u^3 (t+u)^6 \Big(2 s^2-3 u s+4 \
u^2\Big) H(3,0,3,2,y) (s+u)^6-216 s^2 t^2 (s+t)^6 u^2 (t+u)^6 \Big(2 \
s^4+2 t s^3+9 t^2 s^2+2 \Big(t^2+u t+u^2\Big)^2\Big) H(3,2,0,2,y) \
(s+u)^6-216 s^3 t^2 (s+t)^6 u^2 (t+u)^6 \Big((6 t+4 u) s^2-3 \Big(2 \
t^2+u^2\Big) s+2 \Big(3 t^3+u^3\Big)\Big) H(3,2,1,0,y) (s+u)^6-216 \
s^2 t^2 (s+t)^6 u^2 (t+u)^6 \Big(2 s^4+2 t s^3+9 t^2 s^2+2 \Big(t^2+u \
t+u^2\Big)^2\Big) H(3,2,2,0,y) (s+u)^6+216 s^2 t^2 (s+t)^6 u^2 (t+u)^6 \
\Big(4 s^4+2 (t+u) s^3+9 \Big(t^2+u^2\Big) s^2+4 \Big(t^2+u \
t+u^2\Big)^2\Big) H(3,2,3,2,y) (s+u)^6-216 s^3 t^3 (s+t)^6 \Big(2 s^2-3 \
t s+4 t^2\Big) u^2 (t+u)^6 H(3,3,0,2,y) (s+u)^6-216 s^3 t^3 (s+t)^6 \
\Big(2 s^2-3 t s+4 t^2\Big) u^2 (t+u)^6 H(3,3,2,0,y) (s+u)^6+216 s^2 t^2 \
(s+t)^6 u^2 (t+u)^6 \Big(4 s^4+2 (t+u) s^3+9 \Big(t^2+u^2\Big) s^2+4 \
\Big(t^2+u t+u^2\Big)^2\Big) H(3,3,2,2,y) (s+u)^6+216 s^2 t^2 (s+t)^6 \
u^2 (t+u)^6 \Big(4 s^4+6 (t+u) s^3+3 \Big(t^2+u^2\Big) s^2+8 \
\Big(t^3+u^3\Big) s+4 \Big(t^2+u t+u^2\Big)^2\Big) H(3,3,3,2,y) \
(s+u)^6-s^2 t^2 (s+t)^3 u^2 (t+u)^3 \Big(7948 (t+u)^3 s^{10}+\Big(39359 \
t^4+155744 u t^3+236082 u^2 t^2+155744 u^3 t+39359 u^4\Big) \
s^9+\Big(93651 t^5+434331 u t^4+850587 u^2 t^3+850587 u^3 t^2+434331 u^4 \
t+93651 u^5\Big) s^8+3 \Big(46598 t^6+240304 u t^5+550627 u^2 t^4+717154 \
u^3 t^3+550627 u^4 t^2+240304 u^5 t+46598 u^6\Big) s^7+\Big(139794 \
t^7+836962 u t^6+2146599 u^2 t^5+3271156 u^3 t^4+3271156 u^4 t^3+2146599 u^5 \
t^2+836962 u^6 t+139794 u^7\Big) s^6+3 \Big(31217 t^8+240304 u t^7+715533 \
u^2 t^6+1216094 u^3 t^5+1422608 u^4 t^4+1216094 u^5 t^3+715533 u^6 \
t^2+240304 u^7 t+31217 u^8\Big) s^5+\Big(39359 t^9+434331 u t^8+1651881 \
u^2 t^7+3271156 u^3 t^6+4267824 u^4 t^5+4267824 u^5 t^4
\end{dmath*}
\begin{dmath*}
{\white=}
+3271156 u^6 \
t^3+1651881 u^7 t^2+434331 u^8 t+39359 u^9\Big) s^4+\Big(7948 \
t^{10}+155744 u t^9+850587 u^2 t^8+2151462 u^3 t^7+3271156 u^4 t^6+3648282 \
u^5 t^5+3271156 u^6 t^4+2151462 u^7 t^3+850587 u^8 t^2+155744 u^9 t+7948 \
u^{10}\Big) s^3+3 t u \Big(7948 t^9+78694 u t^8+283529 u^2 t^7+550627 u^3 \
t^6+715533 u^4 t^5+715533 u^5 t^4+550627 u^6 t^3+283529 u^7 t^2+78694 u^8 \
t+7948 u^9\Big) s^2+t^2 u^2 (t+u)^2 \Big(23844 t^6+108056 u t^5+194375 \
u^2 t^4+224106 u^3 t^3+194375 u^4 t^2+108056 u^5 t+23844 u^6\Big) s+t^3 \
u^3 (t+u)^3 \Big(7948 t^4+15515 u t^3+23262 u^2 t^2+15515 u^3 t+7948 \
u^4\Big)\Big) (s+u)^3+18 s^2 t^2 u^2 \Big(\Big(11 t^6+162 u t^5+153 \
u^2 t^4+884 u^3 t^3+255 u^4 t^2+246 u^5 t+33 u^6\Big) s^{13}+\Big(62 \
t^7+1339 u t^6+2160 u^2 t^5+8525 u^3 t^4+7390 u^4 t^3+3807 u^5 t^2+1728 u^6 \
t+189 u^7\Big) s^{12}+6 \Big(26 t^8+774 u t^7+1639 u^2 t^6+5548 u^3 \
t^5+7678 u^4 t^4+4900 u^5 t^3+2621 u^6 t^2+826 u^7 t+84 u^8\Big) s^{11}+2 \
\Big(140 t^9+4728 u t^8+12480 u^2 t^7+37199 u^3 t^6+69840 u^4 t^5+60090 u^5 \
t^4+36224 u^6 t^3+17091 u^7 t^2+4188 u^8 t+404 u^9\Big) s^{10}+\Big(534 \
t^{10}+13578 u t^9+45012 u^2 t^8+119424 u^3 t^7+270061 u^4 t^6+311178 u^5 \
t^5+220301 u^6 t^4+125964 u^7 t^3+47997 u^8 t^2+9560 u^9 t+839 u^{10}\Big) \
s^9+3 \Big(340 t^{11}+5526 u t^{10}+22370 u^2 t^9+55138 u^3 t^8+129254 u^4 \
t^7+190253 u^5 t^6
\end{dmath*}

\intertext{}
\begin{dmath*}
{\white =}
+166896 u^6 t^5+109793 u^7 t^4+53466 u^8 t^3+15573 u^9 \
t^2+2466 u^{10} t+189 u^{11}\Big) s^8+2 \Big(736 t^{12}+9471 u \
t^{11}+44292 u^2 t^{10}+110988 u^3 t^9+235059 u^4 t^8+387831 u^5 t^7+413404 \
u^6 t^6+317445 u^7 t^5+187035 u^8 t^4+72214 u^9 t^3+15774 u^{10} t^2+1827 \
u^{11} t+116 u^{12}\Big) s^7+2 \Big(708 t^{13}+9054 u t^{12}+47823 u^2 \
t^{11}+134712 u^3 t^{10}+269098 u^4 t^9+435561 u^5 t^8+508839 u^6 t^7+430834 \
u^7 t^6+288621 u^8 t^5+141397 u^9 t^4+43186 u^{10} t^3+7236 u^{11} t^2+541 \
u^{12} t+22 u^{13}\Big) s^6+3 t \Big(285 t^{13}+4132 u t^{12}+24795 u^2 \
t^{11}+81510 u^3 t^{10}+175113 u^4 t^9+286188 u^5 t^8+351323 u^6 t^7+311558 \
u^7 t^6+211822 u^8 t^5+112744 u^9 t^4+42788 u^{10} t^3+10552 u^{11} t^2+1454 \
u^{12} t+56 u^{13}\Big) s^5+t^2 \Big(294 t^{13}+5499 u t^{12}+38730 u^2 \
t^{11}+148723 u^3 t^{10}+362736 u^4 t^9+642597 u^5 t^8+864762 u^6 t^7+853029 \
u^7 t^6+614934 u^8 t^5+328490 u^9 t^4+124788 u^{10} t^3+32430 u^{11} \
t^2+6072 u^{12} t+708 u^{13}\Big) s^4+2 t^3 \Big(22 t^{13}+707 u \
t^{12}+6399 u^2 t^{11}+29412 u^3 t^{10}+82842 u^4 t^9+162087 u^5 t^8+239183 \
u^6 t^7+267073 u^7 t^6+222465 u^8 t^5+137118 u^9 t^4+59592 u^{10} t^3+16485 \
u^{11} t^2+2553 u^{12} t+174 u^{13}\Big) s^3+6 t^4 u (t+u)^2 \Big(26 \
t^{10}+335 u t^9+1643 u^2 t^8+4383 u^3 t^7+7465 u^4 t^6+9430 u^5 t^5+8350 \
u^6 t^4+5555 u^7 t^3+2649 u^8 t^2+809 u^9 t+127 u^{10}\Big) s^2
\end{dmath*}

\begin{dmath*}
{\white =}
+t^5 u^2 \
(t+u)^3 \Big(162 t^8+1164 u t^7+3573 u^2 t^6+6255 u^3 t^5+7589 u^4 t^4+5865 \
u^5 t^3+3192 u^6 t^2+1112 u^7 t+216 u^8\Big) s+t^6 u^3 (t+u)^6 \Big(58 \
t^4+100 u t^3+153 u^2 t^2+100 u^3 t+58 u^4\Big)\Big) H(2,y) H(1,0,z) \
(s+u)^3+s t (s+t)^2 u^2 (t+u)^2 \Big(1531 t (t+u)^4 s^{13}+3 \Big(3029 \
t^6+15191 u t^5+28746 u^2 t^4+30440 u^3 t^3+14567 u^4 t^2+3351 u^5 t+120 \
u^6\Big) s^{12}+\Big(25433 t^7+151345 u t^6+347338 u^2 t^5+471508 u^3 \
t^4+364546 u^4 t^3+144937 u^5 t^2+32273 u^6 t+1800 u^7\Big) \
s^{11}+\Big(44445 t^8+300713 u t^7+814816 u^2 t^6+1300511 u^3 t^5+1362455 \
u^4 t^4+833113 u^5 t^3+303956 u^6 t^2+65055 u^7 t+3960 u^8\Big) \
s^{10}+\Big(53136 t^9+409309 u t^8+1281415 u^2 t^7+2291071 u^3 t^6+2894942 \
u^4 t^5+2471851 u^5 t^4+1320295 u^6 t^3+459769 u^7 t^2+87252 u^8 t+5040 \
u^9\Big) s^9+\Big(44445 t^{10}+409309 u t^9+1473770 u^2 t^8+2915618 u^3 \
t^7+4102638 u^4 t^6+4464012 u^5 t^5+3318386 u^6 t^4+1630106 u^7 t^3+508285 \
u^8 t^2+77799 u^9 t+3960 u^{10}\Big) s^8+\Big(25433 t^{11}+300713 u \
t^{10}+1281415 u^2 t^9+2915618 u^3 t^8+4495118 u^4 t^7+5664674 u^5 \
t^6+5440250 u^6 t^5+3552410 u^7 t^4+1539427 u^8 t^3+391013 u^9 t^2+44585 \
u^{10} t+1800 u^{11}\Big) s^7+\Big(9087 t^{12}+151345 u t^{11}+814816 u^2 \
t^{10}+2291071 u^3 t^9+4102638 u^4 t^8+5664674 u^5 t^7+6338504 u^6 \
t^6+5147724 u^7 t^5+2865079 u^8 t^4+1027309 u^9 t^3+196900 u^{10} t^2+15021 \
u^{11} t+360 u^{12}\Big) s^6+t \Big(1531 t^{12}+45573 u t^{11}+347338 u^2 \
t^{10}+1300511 u^3 t^9+2894942 u^4 t^8+4464012 u^5 t^7+5440250 u^6 \
t^6+5147724 u^7 t^5+3498698 u^8 t^4+1616591 u^9 t^3+448090 u^{10} t^2+58725 \
u^{11} t+2287 u^{12}\Big) s^5+t^2 u \Big(6124 t^{11}+86238 u \
t^{10}+471508 u^2 t^9+1362455 u^3 t^8+2471851 u^4 t^7+3318386 u^5 \
t^6+3552410 u^6 t^5+2865079 u^7 t^4+1616591 u^8 t^3+581902 u^9 t^2
\end{dmath*}
\begin{dmath*}
{\white=}
+110862 \
u^{10} t+7726 u^{11}\Big) s^4+t^3 u^2 \Big(9186 t^{10}+91320 u t^9+364546 \
u^2 t^8+833113 u^3 t^7+1320295 u^4 t^6+1630106 u^5 t^5+1539427 u^6 \
t^4+1027309 u^7 t^3+448090 u^8 t^2+110862 u^9 t+11166 u^{10}\Big) s^3+t^4 \
u^3 (t+u)^2 \Big(6124 t^7+31453 u t^6+75907 u^2 t^5+120689 u^3 t^4+142484 \
u^4 t^3+102628 u^5 t^2+43273 u^6 t+7726 u^7\Big) s^2+t^5 u^4 (t+u)^4 \
\Big(1531 t^4+3929 u t^3+7371 u^2 t^2+5873 u^3 t+2287 u^4\Big) s+360 t^6 \
u^6 (t+u)^4 \Big(t^2+u t+u^2\Big)\Big) H(0,y) (s+u)^2+s t^2 (s+t)^2 u \
(t+u)^2 \Big(1531 u (t+u)^4 s^{13}+3 \Big(120 t^6+3351 u t^5+14567 u^2 \
t^4+30440 u^3 t^3+28746 u^4 t^2+15191 u^5 t+3029 u^6\Big) \
s^{12}+\Big(1800 t^7+32273 u t^6+144937 u^2 t^5+364546 u^3 t^4+471508 u^4 \
t^3+347338 u^5 t^2+151345 u^6 t+25433 u^7\Big) s^{11}+\Big(3960 t^8+65055 \
u t^7+303956 u^2 t^6+833113 u^3 t^5+1362455 u^4 t^4+1300511 u^5 t^3+814816 \
u^6 t^2+300713 u^7 t+44445 u^8\Big) s^{10}+\Big(5040 t^9+87252 u \
t^8+459769 u^2 t^7+1320295 u^3 t^6+2471851 u^4 t^5+2894942 u^5 t^4+2291071 \
u^6 t^3+1281415 u^7 t^2+409309 u^8 t+53136 u^9\Big) s^9+\Big(3960 \
t^{10}+77799 u t^9+508285 u^2 t^8+1630106 u^3 t^7+3318386 u^4 t^6+4464012 \
u^5 t^5+4102638 u^6 t^4+2915618 u^7 t^3
\end{dmath*}
\begin{dmath*}
{\white=}
+1473770 u^8 t^2+409309 u^9 t+44445 \
u^{10}\Big) s^8+\Big(1800 t^{11}+44585 u t^{10}+391013 u^2 t^9+1539427 \
u^3 t^8+3552410 u^4 t^7+5440250 u^5 t^6+5664674 u^6 t^5+4495118 u^7 \
t^4+2915618 u^8 t^3+1281415 u^9 t^2+300713 u^{10} t+25433 u^{11}\Big) \
s^7+\Big(360 t^{12}+15021 u t^{11}+196900 u^2 t^{10}+1027309 u^3 \
t^9+2865079 u^4 t^8+5147724 u^5 t^7+6338504 u^6 t^6+5664674 u^7 t^5+4102638 \
u^8 t^4+2291071 u^9 t^3+814816 u^{10} t^2+151345 u^{11} t+9087 u^{12}\Big) \
s^6+u \Big(2287 t^{12}+58725 u t^{11}+448090 u^2 t^{10}+1616591 u^3 \
t^9+3498698 u^4 t^8+5147724 u^5 t^7+5440250 u^6 t^6+4464012 u^7 t^5+2894942 \
u^8 t^4+1300511 u^9 t^3+347338 u^{10} t^2+45573 u^{11} t+1531 u^{12}\Big) \
s^5+t u^2 \Big(7726 t^{11}+110862 u t^{10}+581902 u^2 t^9+1616591 u^3 \
t^8+2865079 u^4 t^7+3552410 u^5 t^6+3318386 u^6 t^5+2471851 u^7 t^4+1362455 \
u^8 t^3+471508 u^9 t^2+86238 u^{10} t+6124 u^{11}\Big) s^4+t^2 u^3 \
\Big(11166 t^{10}+110862 u t^9+448090 u^2 t^8+1027309 u^3 t^7+1539427 u^4 \
t^6+1630106 u^5 t^5+1320295 u^6 t^4+833113 u^7 t^3+364546 u^8 t^2+91320 u^9 \
t+9186 u^{10}\Big) s^3+t^3 u^4 (t+u)^2 \Big(7726 t^7+43273 u t^6+102628 \
u^2 t^5+142484 u^3 t^4+120689 u^4 t^3+75907 u^5 t^2+31453 u^6 t+6124 \
u^7\Big) s^2+t^4 u^5 (t+u)^4 \Big(2287 t^4+5873 u t^3+7371 u^2 t^2+3929 \
u^3 t+1531 u^4\Big) s+360 t^6 u^6 (t+u)^4 \Big(t^2+u t+u^2\Big)\Big) \
H(0,z) (s+u)^2
\end{dmath*}

\begin{dmath*}
{\white =}
+3 t^2 (s+t)^2 u^2 (t+u) \Big(259 (t+u)^5 s^{14}+36 \Big(43 \
t^6+242 u t^5+651 u^2 t^4+764 u^3 t^3+651 u^4 t^2+242 u^5 t+43 u^6\Big) \
s^{13}+\Big(4367 t^7+26163 u t^6+84905 u^2 t^5+124733 u^3 t^4+124733 u^4 \
t^3+84905 u^5 t^2+26163 u^6 t+4367 u^7\Big) s^{12}+4 \Big(1920 t^8+12056 \
u t^7+43065 u^2 t^6+78684 u^3 t^5+86470 u^4 t^4+78684 u^5 t^3+43065 u^6 \
t^2+12056 u^7 t+1920 u^8\Big) s^{11}+6 \Big(1534 t^9+10415 u t^8+38108 \
u^2 t^7+83108 u^3 t^6+101419 u^4 t^5+101419 u^5 t^4+83108 u^6 t^3+38108 u^7 \
t^2+10415 u^8 t+1534 u^9\Big) s^{10}+12 \Big(640 t^{10}+4992 u t^9+18355 \
u^2 t^8+44658 u^3 t^7+62385 u^4 t^6+60940 u^5 t^5+62385 u^6 t^4+44658 u^7 \
t^3+18355 u^8 t^2+4992 u^9 t+640 u^{10}\Big) s^9+\Big(4367 t^{11}+41875 u \
t^{10}+163877 u^2 t^9+417437 u^3 t^8+677572 u^4 t^7+645896 u^5 t^6+645896 \
u^6 t^5+677572 u^7 t^4+417437 u^8 t^3+163877 u^9 t^2+41875 u^{10} t+4367 \
u^{11}\Big) s^8+2 \Big(774 t^{12}+9924 u t^{11}+46320 u^2 t^{10}+124333 \
u^3 t^9+232260 u^4 t^8+249279 u^5 t^7+198284 u^6 t^6+249279 u^7 t^5+232260 \
u^8 t^4+124333 u^9 t^3
\end{dmath*}

\intertext{}
\begin{dmath*}
{\white =}
+46320 u^{10} t^2+9924 u^{11} t+774 u^{12}\Big) \
s^7+\Big(259 t^{13}+5481 u t^{12}+35278 u^2 t^{11}+110676 u^3 t^{10}+240564 \
u^4 t^9+349066 u^5 t^8+300068 u^6 t^7+300068 u^7 t^6+349066 u^8 t^5+240564 \
u^9 t^4+110676 u^{10} t^3+35278 u^{11} t^2+5481 u^{12} t+259 u^{13}\Big) \
s^6+2 t u \Big(308 t^{12}+3564 u t^{11}+15365 u^2 t^{10}+42591 u^3 \
t^9+94706 u^4 t^8+136033 u^5 t^7+141098 u^6 t^6+136033 u^7 t^5+94706 u^8 \
t^4+42591 u^9 t^3+15365 u^{10} t^2+3564 u^{11} t+308 u^{12}\Big) s^5+t^2 \
u^2 \Big(426 t^{11}+3380 u t^{10}+15835 u^2 t^9+65921 u^3 t^8+171209 u^4 \
t^7+259557 u^5 t^6+259557 u^6 t^5+171209 u^7 t^4+65921 u^8 t^3+15835 u^9 \
t^2+3380 u^{10} t+426 u^{11}\Big) s^4-2 t^3 u^3 (t+u)^2 \Big(34 t^8-527 u \
t^7-6013 u^2 t^6-19815 u^3 t^5-26374 u^4 t^4-19815 u^5 t^3-6013 u^6 t^2-527 \
u^7 t+34 u^8\Big) s^3+t^4 u^4 (t+u)^3 \Big(109 t^6+2242 u t^5+8867 u^2 \
t^4+13216 u^3 t^3+8867 u^4 t^2+2242 u^5 t+109 u^6\Big) s^2+30 t^5 u^5 \
(t+u)^6 \Big(10 t^2+23 u t+10 u^2\Big) s+120 t^6 u^6 (t+u)^5 \Big(t^2+u \
t+u^2\Big)\Big) H(0,y) H(0,z) (s+u)^2-s^2 t (s+t)^2 u (t+u)^2 \
\Big(\Big(360 t^6+2287 u t^5+7726 u^2 t^4+11166 u^3 t^3+7726 u^4 t^2+2287 \
u^5 t+360 u^6\Big) s^{12}+9 \Big(200 t^7+1669 u t^6+6525 u^2 t^5+12318 \
u^3 t^4+12318 u^4 t^3+6525 u^5 t^2+1669 u^6 t+200 u^7\Big) \
s^{11}+\Big(3960 t^8+44585 u t^7+196900 u^2 t^6+448090 u^3 t^5+581902 u^4 \
t^4+448090 u^5 t^3+196900 u^6 t^2+44585 u^7 t+3960 u^8\Big) \
s^{10}
\end{dmath*}

\begin{dmath*}
{\white =}
+\Big(5040 t^9+77799 u t^8+391013 u^2 t^7+1027309 u^3 t^6+1616591 u^4 \
t^5+1616591 u^5 t^4+1027309 u^6 t^3+391013 u^7 t^2+77799 u^8 t+5040 \
u^9\Big) s^9+\Big(3960 t^{10}+87252 u t^9+508285 u^2 t^8+1539427 u^3 \
t^7+2865079 u^4 t^6+3498698 u^5 t^5+2865079 u^6 t^4+1539427 u^7 t^3+508285 \
u^8 t^2+87252 u^9 t+3960 u^{10}\Big) s^8+\Big(1800 t^{11}+65055 u \
t^{10}+459769 u^2 t^9+1630106 u^3 t^8+3552410 u^4 t^7+5147724 u^5 \
t^6+5147724 u^6 t^5+3552410 u^7 t^4+1630106 u^8 t^3+459769 u^9 t^2+65055 \
u^{10} t+1800 u^{11}\Big) s^7+\Big(360 t^{12}+32273 u t^{11}+303956 u^2 \
t^{10}+1320295 u^3 t^9+3318386 u^4 t^8+5440250 u^5 t^7+6338504 u^6 \
t^6+5440250 u^7 t^5+3318386 u^8 t^4+1320295 u^9 t^3+303956 u^{10} t^2+32273 \
u^{11} t+360 u^{12}\Big) s^6+t u \Big(10053 t^{11}+144937 u t^{10}+833113 \
u^2 t^9+2471851 u^3 t^8+4464012 u^4 t^7+5664674 u^5 t^6+5664674 u^6 \
t^5+4464012 u^7 t^4+2471851 u^8 t^3+833113 u^9 t^2+144937 u^{10} t+10053 \
u^{11}\Big) s^5+t u \Big(1531 t^{12}+43701 u t^{11}+364546 u^2 \
t^{10}+1362455 u^3 t^9+2894942 u^4 t^8+4102638 u^5 t^7+4495118 u^6 \
t^6+4102638 u^7 t^5+2894942 u^8 t^4+1362455 u^9 t^3+364546 u^{10} t^2+43701 \
u^{11} t+1531 u^{12}\Big) s^4+t^2 u^2 \Big(6124 t^{11}+91320 u \
t^{10}+471508 u^2 t^9+1300511 u^3 t^8+2291071 u^4 t^7+2915618 u^5 \
t^6+2915618 u^6 t^5+2291071 u^7 t^4+1300511 u^8 t^3+471508 u^9 t^2+91320 \
u^{10} t+6124 u^{11}\Big) s^3+t^3 u^3 (t+u)^2 \Big(9186 t^8+67866 u \
t^7+202420 u^2 t^6+342110 u^3 t^5+394775 u^4 t^4+342110 u^5 t^3+202420 u^6 \
t^2+67866 u^7 t+9186 u^8\Big) s^2+t^4 u^4 (t+u)^3 \Big(6124 t^6+27201 u \
t^5+51370 u^2 t^4+58876 u^3 t^3+51370 u^4 t^2+27201 u^5 t+6124 u^6\Big) \
s+t^5 u^5 (t+u)^4 \Big(1531 t^4+2963 u t^3+4395 u^2 t^2+2963 u^3 t+1531 \
u^4\Big)\Big) H(1,z) (s+u)^2-s^2 t (s+t)^2 u (t+u)^2 \Big(\Big(360 \
t^6+2287 u t^5
\end{dmath*}
\begin{dmath*}
{\white=}
+7726 u^2 t^4+11166 u^3 t^3+7726 u^4 t^2+2287 u^5 t+360 \
u^6\Big) s^{12}+9 \Big(200 t^7+1669 u t^6+6525 u^2 t^5+12318 u^3 \
t^4+12318 u^4 t^3+6525 u^5 t^2+1669 u^6 t+200 u^7\Big) s^{11}+\Big(3960 \
t^8+44585 u t^7+196900 u^2 t^6+448090 u^3 t^5+581902 u^4 t^4+448090 u^5 \
t^3+196900 u^6 t^2+44585 u^7 t+3960 u^8\Big) s^{10}+\Big(5040 t^9+77799 u \
t^8+391013 u^2 t^7+1027309 u^3 t^6+1616591 u^4 t^5+1616591 u^5 t^4+1027309 \
u^6 t^3+391013 u^7 t^2+77799 u^8 t+5040 u^9\Big) s^9+\Big(3960 \
t^{10}+87252 u t^9+508285 u^2 t^8+1539427 u^3 t^7+2865079 u^4 t^6+3498698 \
u^5 t^5+2865079 u^6 t^4+1539427 u^7 t^3+508285 u^8 t^2+87252 u^9 t+3960 \
u^{10}\Big) s^8+\Big(1800 t^{11}+65055 u t^{10}+459769 u^2 t^9+1630106 \
u^3 t^8+3552410 u^4 t^7+5147724 u^5 t^6+5147724 u^6 t^5+3552410 u^7 \
t^4+1630106 u^8 t^3+459769 u^9 t^2+65055 u^{10} t+1800 u^{11}\Big) \
s^7+\Big(360 t^{12}+32273 u t^{11}+303956 u^2 t^{10}+1320295 u^3 \
t^9+3318386 u^4 t^8+5440250 u^5 t^7+6338504 u^6 t^6+5440250 u^7 t^5+3318386 \
u^8 t^4+1320295 u^9 t^3+303956 u^{10} t^2+32273 u^{11} t+360 u^{12}\Big) \
s^6+t u \Big(10053 t^{11}+144937 u t^{10}+833113 u^2 t^9+2471851 u^3 \
t^8+4464012 u^4 t^7+5664674 u^5 t^6+5664674 u^6 t^5+4464012 u^7 t^4+2471851 \
u^8 t^3+833113 u^9 t^2+144937 u^{10} t+10053 u^{11}\Big) s^5+t u \
\Big(1531 t^{12}+43701 u t^{11}+364546 u^2 t^{10}+1362455 u^3 t^9+2894942 \
u^4 t^8+4102638 u^5 t^7+4495118 u^6 t^6+4102638 u^7 t^5+2894942 u^8 \
t^4+1362455 u^9 t^3+364546 u^{10} t^2+43701 u^{11} t+1531 u^{12}\Big) \
s^4+t^2 u^2 \Big(6124 t^{11}+91320 u t^{10}+471508 u^2 t^9+1300511 u^3 \
t^8+2291071 u^4 t^7+2915618 u^5 t^6+2915618 u^6 t^5+2291071 u^7 t^4+1300511 \
u^8 t^3+471508 u^9 t^2+91320 u^{10} t+6124 u^{11}\Big) s^3+t^3 u^3 (t+u)^2 \
\Big(9186 t^8+67866 u t^7+202420 u^2 t^6+342110 u^3 t^5+394775 u^4 \
t^4+342110 u^5 t^3+202420 u^6 t^2+67866 u^7 t+9186 u^8\Big) s^2+t^4 u^4 \
(t+u)^3 \Big(6124 t^6+27201 u t^5+51370 u^2 t^4+58876 u^3 t^3+51370 u^4 \
t^2+27201 u^5 t+6124 u^6\Big) s+t^5 u^5 (t+u)^4 \Big(1531 t^4+2963 u \
t^3+4395 u^2 t^2+2963 u^3 t+1531 u^4\Big)\Big) H(2,y) (s+u)^2
\end{dmath*}

\begin{dmath*}
{\white =}
-3 s^2 t^2 \
(t+u)^2 \Big(\Big(120 t^6+300 u t^5+109 u^2 t^4-68 u^3 t^3+426 u^4 t^2+616 \
u^5 t+259 u^6\Big) s^{14}+2 \Big(420 t^7+1395 u t^6+1339 u^2 t^5+425 u^3 \
t^4+1903 u^4 t^3+3872 u^5 t^2+2870 u^6 t+774 u^7\Big) s^{13}+\Big(2640 \
t^8+11430 u t^7+18489 u^2 t^6+14984 u^3 t^5+19215 u^4 t^4+37858 u^5 \
t^3+40759 u^6 t^2+21396 u^7 t+4367 u^8\Big) s^{12}+2 \Big(2460 t^9+13545 \
u t^8+31286 u^2 t^7+39401 u^3 t^6+40878 u^4 t^5+57956 u^5 t^4+72977 u^6 \
t^3+56244 u^7 t^2+23121 u^8 t+3840 u^9\Big) s^{11}+2 \Big(3000 \
t^{10}+20475 u t^9+62005 u^2 t^8+104385 u^3 t^7+118565 u^4 t^6+137297 u^5 \
t^5+175620 u^6 t^4+170653 u^7 t^3+102876 u^8 t^2+33792 u^9 t+4602 \
u^{10}\Big) s^{10}+2 \Big(2460 t^{11}+20475 u t^{10}+77358 u^2 t^9+164395 \
u^3 t^8+215383 u^4 t^7+230739 u^5 t^6+294815 u^6 t^5+356593 u^7 t^4+290657 \
u^8 t^3+140082 u^9 t^2+35847 u^{10} t+3840 u^{11}\Big) s^9+\Big(2640 \
t^{12}+27090 u t^{11}+124010 u^2 t^{10}+328790 u^3 t^9+519114 u^4 t^8+554262 \
u^5 t^7+649134 u^6 t^6+963078 u^7 t^5+1095009 u^8 t^4+756156 u^9 t^3+291138 \
u^{10} t^2+55904 u^{11} t+4367 u^{12}\Big) s^8+2 \Big(420 t^{13}+5715 u \
t^{12}+31286 u^2 t^{11}+104385 u^3 t^{10}+215383 u^4 t^9+277131 u^5 \
t^8+300068 u^6 t^7+447563 u^7 t^6+661734 u^8 t^5+642258 u^9 t^4+363648 \
u^{10} t^3+110242 u^{11} t^2+15265 u^{12} t+774 u^{13}\Big) s^7+\Big(120 \
t^{14}+2790 u t^{13}+18489 u^2 t^{12}+78802 u^3 t^{11}+237130 u^4 \
t^{10}+461478 u^5 t^9+649134 u^6 t^8+895126 u^7 t^7+1291792 u^8 t^6+1479900 \
u^9 t^5+1107162 u^{10} t^4+486996 u^{11} t^3+111068 u^{12} t^2
\end{dmath*}

\intertext{}
\begin{dmath*}
{\white =}
+10260 u^{13} \
t+259 u^{14}\Big) s^6+2 t u \Big(150 t^{13}+1339 u t^{12}+7492 u^2 \
t^{11}+40878 u^3 t^{10}+137297 u^4 t^9+294815 u^5 t^8+481539 u^6 t^7+661734 \
u^7 t^6+739950 u^8 t^5+608514 u^9 t^4+330308 u^{10} t^3+104819 u^{11} \
t^2+16074 u^{12} t+777 u^{13}\Big) s^5+t^2 u^2 \Big(109 t^{12}+850 u \
t^{11}+19215 u^2 t^{10}+115912 u^3 t^9+351240 u^4 t^8+713186 u^5 t^7+1095009 \
u^6 t^6+1284516 u^7 t^5+1107162 u^8 t^4+660616 u^9 t^3+249466 u^{10} \
t^2+50940 u^{11} t+3885 u^{12}\Big) s^4+2 t^3 u^3 \Big(-34 t^{11}+1903 u \
t^{10}+18929 u^2 t^9+72977 u^3 t^8+170653 u^4 t^7+290657 u^5 t^6+378078 u^6 \
t^5+363648 u^7 t^4+243498 u^8 t^3+104819 u^9 t^2+25470 u^{10} t+2590 \
u^{11}\Big) s^3+t^4 u^4 (t+u)^2 \Big(426 t^8+6892 u t^7+26549 u^2 \
t^6+52498 u^3 t^5+74207 u^4 t^4+79252 u^5 t^3+58427 u^6 t^2+24378 u^7 t+3885 \
u^8\Big) s^2+2 t^5 u^5 (t+u)^3 \Big(308 t^6+1946 u t^5+3936 u^2 t^4+5167 \
u^3 t^3+4537 u^4 t^2+2799 u^5 t+777 u^6\Big) s+t^6 u^6 (t+u)^4 \Big(259 \
t^4+512 u t^3+765 u^2 t^2+512 u^3 t+259 u^4\Big)\Big) H(0,z) H(2,y) \
(s+u)^2+3 s^2 t^2 (s+t)^6 u^2 (t+u)^6 \Big(363 s^8+726 (t+3 u) s^7+3 \
\Big(363 t^2+796 u t+2057 u^2\Big) s^6+6 \Big(121 t^3+348 u t^2+480 u^2 \
t+1815 u^3\Big) s^5+\Big(363 t^4+958 u t^3-297 u^2 t^2+1782 u^3 t+13068 \
u^4\Big) s^4+2 u \Big(246 t^4-58 u t^3-1296 u^2 t^2+891 u^3 t+5445 \
u^4\Big) s^3+u^2 \Big(504 t^4-116 u t^3-297 u^2 t^2+2880 u^3 t+6171 \
u^4\Big) s^2
\end{dmath*}

\begin{dmath*}
{\white =}
+2 u^3 \Big(246 t^4+479 u t^3+1044 u^2 t^2+1194 u^3 t+1089 \
u^4\Big) s+363 u^4 \Big(t^2+u t+u^2\Big)^2\Big) H(0,0,z) (s+u)^2+36 \
s^2 t^2 (s+t)^6 u^2 (t+u)^6 \Big(22 s^8+44 (t+3 u) s^7+2 \Big(33 t^2+70 u \
t+187 u^2\Big) s^6+\Big(44 t^3+246 u t^2+147 u^2 t+660 u^3\Big) \
s^5+\Big(22 t^4+115 u t^3+360 u^2 t^2+41 u^3 t+792 u^4\Big) s^4+u \
\Big(64 t^4+95 u t^3+246 u^2 t^2+2 u^3 t+660 u^4\Big) s^3+u^2 \Big(84 \
t^4+9 u t^3+84 u^2 t^2+54 u^3 t+374 u^4\Big) s^2+u^3 \Big(54 t^4-11 u \
t^3+36 u^2 t^2+63 u^3 t+132 u^4\Big) s+u^4 \Big(18 t^4+4 u t^3+18 u^2 \
t^2+21 u^3 t+22 u^4\Big)\Big) H(0,y) H(0,0,z) (s+u)^2-36 s^2 t^2 (s+t)^6 \
u^2 (t+u)^6 \Big(22 s^8+3 (7 t+44 u) s^7+\Big(18 t^2+63 u t+374 u^2\Big) \
s^6+\Big(4 t^3+36 u t^2+54 u^2 t+660 u^3\Big) s^5+\Big(18 t^4-11 u \
t^3+84 u^2 t^2+2 u^3 t+792 u^4\Big) s^4+u \Big(54 t^4+9 u t^3+246 u^2 \
t^2+41 u^3 t+660 u^4\Big) s^3+u^2 \Big(84 t^4+95 u t^3+360 u^2 t^2+147 \
u^3 t+374 u^4\Big) s^2+u^3 \Big(64 t^4+115 u t^3+246 u^2 t^2+140 u^3 \
t+132 u^4\Big) s+22 u^4 \Big(t^2+u t+u^2\Big)^2\Big) H(2,y) H(0,0,z) \
(s+u)^2+18 s^2 t^2 u^2 (t+u)^2 \Big(\Big(25 t^4+64 u t^3+78 u^2 t^2+44 u^3 \
t+11 u^4\Big) s^{14}+2 \Big(92 t^5+278 u t^4+393 u^2 t^3+307 u^3 t^2+133 \
u^4 t+27 u^5\Big) s^{13}+\Big(633 t^6+2332 u t^5+3918 u^2 t^4+3896 u^3 \
t^3+2351 u^4 t^2+816 u^5 t+124 u^6\Big) s^{12}+2 \Big(684 t^7+2970 u \
t^6+5715 u^2 t^5+6827 u^3 t^4+5495 u^4 t^3+2871 u^5 t^2+846 u^6 t+96 \
u^7\Big) s^{11}+\Big(2094 t^8+10548 u t^7+21762 u^2 t^6+27160 u^3 \
t^5+25897 u^4 t^4+19800 u^5 t^3+10356 u^6 t^2+2820 u^7 t+231 u^8\Big) \
s^{10}+2 \Big(1200 t^9+6984 u t^8+15768 u^2 t^7+19041 u^3 t^6+16319 u^4 \
t^5
\end{dmath*}
\begin{dmath*}
{\white=}
+14613 u^5 t^4+13072 u^6 t^3+7593 u^7 t^2+2013 u^8 t+105 u^9\Big) \
s^9+\Big(2094 t^{10}+13968 u t^9+35988 u^2 t^8+47158 u^3 t^7+37273 u^4 \
t^6+27528 u^5 t^5+31140 u^6 t^4+31554 u^7 t^3+17931 u^8 t^2+4336 u^9 t+130 \
u^{10}\Big) s^8+2 \Big(684 t^{11}+5274 u t^{10}+15768 u^2 t^9+23579 u^3 \
t^8+22058 u^4 t^7+21537 u^5 t^6+26874 u^6 t^5+27399 u^7 t^4+18504 u^8 \
t^3+7571 u^9 t^2+1472 u^{10} t+24 u^{11}\Big) s^7+\Big(633 t^{12}+5940 u \
t^{11}+21762 u^2 t^{10}+38082 u^3 t^9+37273 u^4 t^8+43074 u^5 t^7+83288 u^6 \
t^6+111834 u^7 t^5+83094 u^8 t^4+33834 u^9 t^3+7744 u^{10} t^2+1056 u^{11} \
t+8 u^{12}\Big) s^6+2 t \Big(92 t^{12}+1166 u t^{11}+5715 u^2 \
t^{10}+13580 u^3 t^9+16319 u^4 t^8+13764 u^5 t^7+26874 u^6 t^6+55917 u^7 \
t^5+62742 u^8 t^4+36199 u^9 t^3+9748 u^{10} t^2+930 u^{11} t+72 \
u^{12}\Big) s^5+t^2 \Big(25 t^{12}+556 u t^{11}+3918 u^2 t^{10}+13654 u^3 \
t^9+25897 u^4 t^8+29226 u^5 t^7+31140 u^6 t^6+54798 u^7 t^5+83094 u^8 \
t^4+72398 u^9 t^3+32492 u^{10} t^2+6012 u^{11} t+84 u^{12}\Big) s^4+2 t^3 \
u \Big(32 t^{11}+393 u t^{10}+1948 u^2 t^9+5495 u^3 t^8+9900 u^4 t^7+13072 \
u^5 t^6+15777 u^6 t^5+18504 u^7 t^4+16917 u^8 t^3+9748 u^9 t^2+3006 u^{10} \
t+368 u^{11}\Big) s^3+t^4 u^2 (t+u)^2 \Big(78 t^8+458 u t^7+1357 u^2 \
t^6+2570 u^3 t^5+3859 u^4 t^4+4898 u^5 t^3+4276 u^6 t^2+1692 u^7 t+84 \
u^8\Big) s^2+2 t^5 u^3 (t+u)^3 \Big(22 t^6+67 u t^5+141 u^2 t^4+200 u^3 \
t^3+320 u^4 t^2+312 u^5 t+72 u^6\Big) s+t^6 u^4 (t+u)^4 \Big(11 t^4+10 u \
t^3+18 u^2 t^2+16 u^3 t+8 u^4\Big)\Big) H(0,z) H(0,2,y) (s+u)^2
\end{dmath*}

\begin{dmath*}
{\white =}
+3 t^2 \
(s+t) u^2 \Big(\Big(17 t^6+1338 u t^5+1209 u^2 t^4+3688 u^3 t^3+4269 u^4 \
t^2+762 u^5 t+293 u^6\Big) s^{15}+(t+u)^2 \Big(125 t^5+9282 u t^4-1574 \
u^2 t^3+24402 u^3 t^2+2031 u^4 t+1740 u^5\Big) s^{14}+\Big(421 t^8+28668 \
u t^7+79874 u^2 t^6+122578 u^3 t^5+210974 u^4 t^4+216624 u^5 t^3+90402 u^6 \
t^2+18674 u^7 t+4873 u^8\Big) s^{13}+(t+u)^2 \Big(853 t^7+47050 u \
t^6+99979 u^2 t^5+64212 u^3 t^4+244839 u^4 t^3+99258 u^5 t^2+22313 u^6 \
t+8520 u^7\Big) s^{12}+4 \Big(288 t^{10}+12551 u t^9+71057 u^2 t^8+139212 \
u^3 t^7+179702 u^4 t^6+266064 u^5 t^5+276429 u^6 t^4+139626 u^7 t^3+41347 \
u^8 t^2+13917 u^9 t+2547 u^{10}\Big) s^{11}+2 (t+u)^2 \Big(546 t^9+13510 \
u t^8+92097 u^2 t^7+124154 u^3 t^6+79909 u^4 t^5+282798 u^5 t^4+136349 u^6 \
t^3+25806 u^7 t^2+17949 u^8 t+4260 u^9\Big) s^{10}+\Big(745 t^{12}+5476 u \
t^{11}+88790 u^2 t^{10}+380158 u^3 t^9+559919 u^4 t^8+624124 u^5 t^7+1280316 \
u^6 t^6+1630684 u^7 t^5+908135 u^8 t^4+264280 u^9 t^3+94614 u^{10} t^2+32670 \
u^{11} t+4873 u^{12}\Big) s^9+(t+u)^2 \Big(361 t^{11}-5942 u t^{10}-13109 \
u^2 t^9+45310 u^3 t^8-83270 u^4 t^7-225424 u^5 t^6+359776 u^6 t^5+341850 u^7 \
t^4+131133 u^8 t^3+49098 u^9 t^2+8565 u^{10} t+1740 u^{11}\Big) \
s^8+\Big(113 t^{14}-5422 u t^{13}-49927 u^2 t^{12}-155092 u^3 t^{11}-318038 \
u^4 t^{10}-778482 u^5 t^9-1327656 u^6 t^8-974758 u^7 t^7-19040 u^8 \
t^6+432306 u^9 t^5+380872 u^{10} t^4
\end{dmath*}
\begin{dmath*}
{\white=}
+185882 u^{11} t^3+39939 u^{12} t^2+2122 \
u^{13} t+293 u^{14}\Big) s^7+t (t+u)^2 \Big(17 t^{12}-2466 u t^{11}-25660 \
u^2 t^{10}-88158 u^3 t^9-136236 u^4 t^8-237708 u^5 t^7-370210 u^6 t^6-209898 \
u^7 t^5-117978 u^8 t^4-9304 u^9 t^3+43600 u^{10} t^2+14262 u^{11} t+49 \
u^{12}\Big) s^6-t^2 u \Big(424 t^{13}+9046 u t^{12}+64296 u^2 \
t^{11}+209881 u^3 t^{10}+368916 u^4 t^9+405336 u^5 t^8+306108 u^6 t^7+185900 \
u^7 t^6+201510 u^8 t^5+255266 u^9 t^4+158896 u^{10} t^3+31479 u^{11} \
t^2-5494 u^{12} t-2252 u^{13}\Big) s^5-t^3 u^2 (t+u)^2 \Big(1038 \
t^{10}+11826 u t^9+41953 u^2 t^8+50370 u^3 t^7-10747 u^4 t^6-92766 u^5 \
t^5-113853 u^6 t^4-35658 u^7 t^3+15137 u^8 t^2+11676 u^9 t+2344 \
u^{10}\Big) s^4-t^4 u^3 (t+u)^3 \Big(1108 t^8+5303 u t^7+2515 u^2 \
t^6-23685 u^3 t^5-48783 u^4 t^4-46581 u^5 t^3-17785 u^6 t^2-2753 u^7 t-67 \
u^8\Big) s^3-t^5 u^4 (t+u)^4 \Big(133 t^6-1574 u t^5-8533 u^2 t^4-14102 \
u^3 t^3-10237 u^4 t^2-2876 u^5 t-167 u^6\Big) s^2+30 t^6 u^5 (t+u)^6 \
\Big(10 t^3+37 u t^2+37 u^2 t+14 u^3\Big) s+120 t^7 u^6 (t+u)^6 \
\Big(t^2+u t+u^2\Big)\Big) H(1,0,z) (s+u)^2
\end{dmath*}

\intertext{}
\begin{dmath*}
{\white =}
+18 s^2 t^2 (s+t)^2 u^2 \
\Big(\Big(11 t^6+162 u t^5+129 u^2 t^4+796 u^3 t^3+129 u^4 t^2+162 u^5 \
t+11 u^6\Big) s^{12}+6 \Big(11 t^7+193 u t^6+349 u^2 t^5+1047 u^3 \
t^4+1027 u^4 t^3+313 u^5 t^2+173 u^6 t+7 u^7\Big) s^{11}+\Big(187 \
t^8+3280 u t^7+8506 u^2 t^6+20204 u^3 t^5+32372 u^4 t^4+18608 u^5 t^3+6910 \
u^6 t^2+2596 u^7 t+73 u^8\Big) s^{10}+2 \Big(165 t^9+2522 u t^8+8324 u^2 \
t^7+17309 u^3 t^6+35002 u^4 t^5+33469 u^5 t^4+14236 u^6 t^3+6119 u^7 \
t^2+1745 u^8 t+53 u^9\Big) s^9+\Big(396 t^{10}+4920 u t^9+19227 u^2 \
t^8+34792 u^3 t^7+71793 u^4 t^6+105726 u^5 t^5+63533 u^6 t^4+25164 u^7 \
t^3+13635 u^8 t^2+3190 u^9 t+168 u^{10}\Big) s^8+2 \Big(165 t^{11}+1725 u \
t^{10}+7356 u^2 t^9+10992 u^3 t^8+12447 u^4 t^7+31129 u^5 t^6+30585 u^6 \
t^5+10947 u^7 t^4+9212 u^8 t^3+6186 u^9 t^2+1323 u^{10} t+109 u^{11}\Big) \
s^7+\Big(187 t^{12}+1872 u t^{11}+8220 u^2 t^{10}+10632 u^3 t^9-15915 u^4 \
t^8-29212 u^5 t^7-9538 u^6 t^6-13152 u^7 t^5+1543 u^8 t^4+19604 u^9 \
t^3+10518 u^{10} t^2+2096 u^{11} t+185 u^{12}\Big) s^6+2 \Big(33 \
t^{13}+372 u t^{12}+1782 u^2 t^{11}+2908 u^3 t^{10}-8202 u^4 t^9-31332 u^5 \
t^8-35999 u^6 t^7-26337 u^7 t^6-11010 u^8 t^5+8183 u^9 t^4+10293 u^{10} \
t^3+3681 u^{11} t^2+623 u^{12} t+45 u^{13}\Big) s^5+\Big(11 t^{14}+182 u \
t^{13}+1142 u^2 t^{12}+3266 u^3 t^{11}-2041 u^4 t^{10}-30594 u^5 t^9-56568 \
u^6 t^8-48084 u^7 t^7-20480 u^8 t^6+13148 u^9 t^5
\end{dmath*}

\begin{dmath*}
{\white =}
+25835 u^{10} t^4+13918 \
u^{11} t^3+3530 u^{12} t^2+452 u^{13} t+19 u^{14}\Big) s^4+2 t u \Big(10 \
t^{13}+115 u t^{12}+604 u^2 t^{11}+1366 u^3 t^{10}-633 u^4 t^9-5650 u^5 \
t^8-5310 u^6 t^7+1442 u^7 t^6+9166 u^8 t^5+12427 u^9 t^4+8168 u^{10} \
t^3+2766 u^{11} t^2+499 u^{12} t+38 u^{13}\Big) s^3+t^2 u^2 (t+u)^2 \
\Big(24 t^{10}+190 u t^9+841 u^2 t^8+1482 u^3 t^7+1601 u^4 t^6+4172 u^5 \
t^5+4881 u^6 t^4+5362 u^7 t^3+3151 u^8 t^2+930 u^9 t+126 u^{10}\Big) s^2+2 \
t^3 u^3 (t+u)^3 \Big(10 t^8+70 u t^7+248 u^2 t^6+399 u^3 t^5+669 u^4 \
t^4+647 u^5 t^3+545 u^6 t^2+236 u^7 t+48 u^8\Big) s+t^4 u^4 (t+u)^6 \
\Big(11 t^4+22 u t^3+39 u^2 t^2+28 u^3 t+33 u^4\Big)\Big) H(0,y) \
H(1,0,z) (s+u)^2+18 s^2 t^2 (s+t)^2 u^2 \Big(4 \Big(2 t^6+36 u t^5+21 u^2 \
t^4+184 u^3 t^3+21 u^4 t^2+36 u^5 t+2 u^6\Big) s^{12}+12 \Big(4 t^7+88 u \
t^6+155 u^2 t^5+501 u^3 t^4+501 u^4 t^3+155 u^5 t^2+88 u^6 t+4 u^7\Big) \
s^{11}+2 \Big(65 t^8+1472 u t^7+3872 u^2 t^6+9748 u^3 t^5+16246 u^4 \
t^4+9748 u^5 t^3+3872 u^6 t^2+1472 u^7 t+65 u^8\Big) s^{10}+2 \Big(105 \
t^9+2168 u t^8+7571 u^2 t^7+16917 u^3 t^6+36199 u^4 t^5+36199 u^5 t^4+16917 \
u^6 t^3+7571 u^7 t^2+2168 u^8 t+105 u^9\Big) s^9+3 \Big(77 t^{10}+1342 u \
t^9+5977 u^2 t^8+12336 u^3 t^7+27698 u^4 t^6+41828 u^5 t^5+27698 u^6 \
t^4+12336 u^7 t^3+5977 u^8 t^2+1342 u^9 t+77 u^{10}\Big) s^8+6 \Big(32 \
t^{11}+470 u t^{10}+2531 u^2 t^9+5259 u^3 t^8+9133 u^4 t^7+18639 u^5 \
t^6+18639 u^6 t^5+9133 u^7 t^4
\end{dmath*}
\begin{dmath*}
{\white=}
+5259 u^8 t^3+2531 u^9 t^2+470 u^{10} t+32 \
u^{11}\Big) s^7+4 \Big(31 t^{12}+423 u t^{11}+2589 u^2 t^{10}+6536 u^3 \
t^9+7785 u^4 t^8+13437 u^5 t^7+20822 u^6 t^6+13437 u^7 t^5+7785 u^8 t^4+6536 \
u^9 t^3+2589 u^{10} t^2+423 u^{11} t+31 u^{12}\Big) s^6+6 \Big(9 \
t^{13}+136 u t^{12}+957 u^2 t^{11}+3300 u^3 t^{10}+4871 u^4 t^9+4588 u^5 \
t^8+7179 u^6 t^7+7179 u^7 t^6+4588 u^8 t^5+4871 u^9 t^4+3300 u^{10} t^3+957 \
u^{11} t^2+136 u^{12} t+9 u^{13}\Big) s^5+\Big(11 t^{14}+266 u \
t^{13}+2351 u^2 t^{12}+10990 u^3 t^{11}+25897 u^4 t^{10}+32638 u^5 t^9+37273 \
u^6 t^8+44116 u^7 t^7+37273 u^8 t^6+32638 u^9 t^5+25897 u^{10} t^4+10990 \
u^{11} t^3+2351 u^{12} t^2+266 u^{13} t+11 u^{14}\Big) s^4+2 t u \Big(22 \
t^{13}+307 u t^{12}+1948 u^2 t^{11}+6827 u^3 t^{10}+13580 u^4 t^9+19041 u^5 \
t^8+23579 u^6 t^7+23579 u^7 t^6+19041 u^8 t^5+13580 u^9 t^4+6827 u^{10} \
t^3+1948 u^{11} t^2+307 u^{12} t+22 u^{13}\Big) s^3+6 t^2 u^2 (t+u)^2 \
\Big(13 t^{10}+105 u t^9+430 u^2 t^8+940 u^3 t^7+1317 u^4 t^6+1682 u^5 \
t^5+1317 u^6 t^4+940 u^7 t^3+430 u^8 t^2+105 u^9 t+13 u^{10}\Big) s^2+4 \
t^3 u^3 (t+u)^3 \Big(16 t^8+91 u t^7+262 u^2 t^6+410 u^3 t^5+530 u^4 \
t^4+410 u^5 t^3+262 u^6 t^2+91 u^7 t+16 u^8\Big) s+t^4 u^4 (t+u)^6 \
\Big(25 t^4+34 u t^3+54 u^2 t^2+34 u^3 t+25 u^4\Big)\Big) H(0,z) \
H(2,0,y) (s+u)^2
\end{dmath*}

\begin{dmath*}
{\white =}
+18 s^2 t^2 u^2 \Big(6 \Big(11 t^6+46 u t^5+156 u^2 \
t^4+102 u^3 t^3+156 u^4 t^2+46 u^5 t+11 u^6\Big) s^{14}+2 \Big(270 \
t^7+1259 u t^6+4389 u^2 t^5+4796 u^3 t^4+5120 u^4 t^3+3729 u^5 t^2+1049 u^6 \
t+204 u^7\Big) s^{13}+\Big(2055 t^8+10770 u t^7+38032 u^2 t^6+57306 u^3 \
t^5+58498 u^4 t^4+54214 u^5 t^3+27450 u^6 t^2+7450 u^7 t+1185 u^8\Big) \
s^{12}+2 \Big(2410 t^9+14457 u t^8+51810 u^2 t^7+97505 u^3 t^6+109719 u^4 \
t^5+108252 u^5 t^4+77806 u^6 t^3+32031 u^7 t^2+8199 u^8 t+1059 u^9\Big) \
s^{11}+\Big(7744 t^{10}+53588 u t^9+198480 u^2 t^8+439066 u^3 t^7+579501 \
u^4 t^6+598332 u^5 t^5+517783 u^6 t^4+294682 u^7 t^3+105495 u^8 t^2+24196 \
u^9 t+2541 u^{10}\Big) s^{10}+2 \Big(4458 t^{11}+35621 u t^{10}+138966 \
u^2 t^9+344982 u^3 t^8+538700 u^4 t^7+609874 u^5 t^6+590760 u^6 t^5+426236 \
u^7 t^4+200386 u^8 t^3+62613 u^9 t^2+12146 u^{10} t+1050 u^{11}\Big) \
s^9+\Big(7450 t^{12}+68928 u t^{11}+290076 u^2 t^{10}+782276 u^3 \
t^9+1406634 u^4 t^8+1787400 u^5 t^7+1901521 u^6 t^6+1654302 u^7 t^5+999621 \
u^8 t^4+402156 u^9 t^3+106839 u^{10} t^2+16578 u^{11} t+1179 u^{12}\Big) \
s^8+2 \Big(2230 t^{13}+24137 u t^{12}+113268 u^2 t^{11}+330723 u^3 \
t^{10}+666867 u^4 t^9+948111 u^5 t^8+1082558 u^6 t^7+1072690 u^7 t^6
\end{dmath*}
\begin{dmath*}
{\white=}
+809304 \
u^8 t^5+421377 u^9 t^4+148590 u^{10} t^3+32739 u^{11} t^2+3807 u^{12} t+207 \
u^{13}\Big) s^7+\Big(1830 t^{14}+23864 u t^{13}+129744 u^2 t^{12}+419622 \
u^3 t^{11}+946629 u^4 t^{10}+1526944 u^5 t^9+1879479 u^6 t^8+1986148 u^7 \
t^7+1729402 u^8 t^6+1108500 u^9 t^5+506934 u^{10} t^4+157722 u^{11} \
t^3+28413 u^{12} t^2+2220 u^{13} t+69 u^{14}\Big) s^6+2 t \Big(232 \
t^{14}+3972 u t^{13}+26409 u^2 t^{12}+97770 u^3 t^{11}+247866 u^4 \
t^{10}+463269 u^5 t^9+652628 u^6 t^8+749862 u^7 t^7+712998 u^8 t^6+514895 \
u^9 t^5+272466 u^{10} t^4+106176 u^{11} t^3+28022 u^{12} t^2+3996 u^{13} \
t+159 u^{14}\Big) s^5+t^2 \Big(55 t^{14}+1614 u t^{13}+14748 u^2 \
t^{12}+66434 u^3 t^{11}+190555 u^4 t^{10}+406308 u^5 t^9+671989 u^6 \
t^8+896290 u^7 t^7+978039 u^8 t^6+811640 u^9 t^5+480198 u^{10} t^4+201792 \
u^{11} t^3+59561 u^{12} t^2+11310 u^{13} t+1083 u^{14}\Big) s^4+2 t^3 u \
\Big(74 t^{13}+1262 u t^{12}+7944 u^2 t^{11}+27682 u^3 t^{10}+67049 u^4 \
t^9+125496 u^5 t^8+189647 u^6 t^7+237429 u^7 t^6+234672 u^8 t^5+169330 u^9 \
t^4+85057 u^{10} t^3+28209 u^{11} t^2+5389 u^{12} t+424 u^{13}\Big) \
s^3+t^4 u^2 (t+u)^2 \Big(192 t^{10}+1868 u t^9+7244 u^2 t^8+17304 u^3 \
t^7+30738 u^4 t^6+40806 u^5 t^5+45573 u^6 t^4+36880 u^7 t^3+20764 u^8 \
t^2+7278 u^9 t+1137 u^{10}\Big) s^2+2 t^5 u^3 (t+u)^3 \Big(64 t^8+420 u \
t^7+1164 u^2 t^6+2189 u^3 t^5+2787 u^4 t^4+2772 u^5 t^3+1870 u^6 t^2+855 u^7 \
t+183 u^8\Big) s+t^6 u^4 (t+u)^6 \Big(47 t^4+64 u t^3+117 u^2 t^2+92 u^3 \
t+83 u^4\Big)\Big) H(0,1,0,z) (s+u)^2
\end{dmath*}

\intertext{}
\begin{dmath*}
{\white =}
+36 s^2 t^2 (s+t)^6 u^2 (t+u)^6 \
\Big(22 s^8+44 (t+3 u) s^7+2 \Big(33 t^2+70 u t+187 u^2\Big) \
s^6+\Big(44 t^3+246 u t^2+147 u^2 t+660 u^3\Big) s^5+\Big(22 t^4+115 u \
t^3+360 u^2 t^2+41 u^3 t+792 u^4\Big) s^4+u \Big(64 t^4+95 u t^3+246 u^2 \
t^2+2 u^3 t+660 u^4\Big) s^3+u^2 \Big(84 t^4+9 u t^3+84 u^2 t^2+54 u^3 \
t+374 u^4\Big) s^2+u^3 \Big(54 t^4-11 u t^3+36 u^2 t^2+63 u^3 t+132 \
u^4\Big) s+u^4 \Big(18 t^4+4 u t^3+18 u^2 t^2+21 u^3 t+22 \
u^4\Big)\Big) H(1,0,0,z) (s+u)^2+36 s^2 t^2 u^3 \Big(\Big(48 t^5-6 u \
t^4+332 u^2 t^3+45 u^3 t^2+90 u^4 t+11 u^5\Big) s^{14}+\Big(420 t^6+282 u \
t^5+2684 u^2 t^4+2292 u^3 t^3+810 u^4 t^2+542 u^5 t+42 u^6\Big) \
s^{13}+\Big(1386 t^7+1149 u t^6+7038 u^2 t^5+12044 u^3 t^4+4972 u^4 \
t^3+2592 u^5 t^2+1202 u^6 t+73 u^7\Big) s^{12}+\Big(2181 t^8-6 u t^7+3 \
u^2 t^6+15414 u^3 t^5+7299 u^4 t^4-896 u^5 t^3+2931 u^6 t^2+1368 u^7 t+106 \
u^8\Big) s^{11}+\Big(985 t^9-9867 u t^8-44480 u^2 t^7-48559 u^3 t^6-45186 \
u^4 t^5-50331 u^5 t^4-19580 u^6 t^3+585 u^7 t^2+1097 u^8 t+168 u^9\Big) \
s^{10}-\Big(2469 t^{10}+31352 u t^9+135204 u^2 t^8+241022 u^3 t^7+253354 \
u^4 t^6+222072 u^5 t^5+127806 u^6 t^4+30846 u^7 t^3+657 u^8 t^2-980 u^9 \
t-218 u^{10}\Big) s^9-\Big(5517 t^{11}+52233 u t^{10}+228083 u^2 \
t^9+506619 u^3 t^8+645063 u^4 t^7+575952 u^5 t^6+368997 u^6 t^5+131304 u^7 \
t^4+16832 u^8 t^3-1527 u^9 t^2-936 u^{10} t-185 u^{11}\Big) s^8-\Big(5797 \
t^{12}+55314 u t^{11}+250929 u^2 t^{10}+646444 u^3 t^9+990366 u^4 \
t^8+1001980 u^5 t^7+717884 u^6 t^6+319380 u^7 t^5+58843 u^8 t^4-6082 u^9 \
t^3-4017 u^{10} t^2-684 u^{11} t-90 u^{12}\Big) s^7-\Big(3849 \
t^{13}+40039 u t^{12}+191778 u^2 t^{11}
\end{dmath*}

\begin{dmath*}
{\white =}
+546879 u^3 t^{10}+969791 u^4 \
t^9+1126182 u^5 t^8+922078 u^6 t^7+512534 u^7 t^6+148569 u^8 t^5-1403 u^9 \
t^4-14004 u^{10} t^3-3726 u^{11} t^2-315 u^{12} t-19 u^{13}\Big) s^6-t \
\Big(1647 t^{13}+19818 u t^{12}+104248 u^2 t^{11}+323862 u^3 t^{10}+638973 \
u^4 t^9+819288 u^5 t^8+723606 u^6 t^7+455700 u^7 t^6+180305 u^8 t^5+24144 \
u^9 t^4-13260 u^{10} t^3-8070 u^{11} t^2-1683 u^{12} t-66 u^{13}\Big) \
s^5-t^2 \Big(413 t^{13}+6240 u t^{12}+38435 u^2 t^{11}+133574 u^3 \
t^{10}+291906 u^4 t^9+405133 u^5 t^8+357472 u^6 t^7+200055 u^7 t^6+58195 u^8 \
t^5-7313 u^9 t^4-13779 u^{10} t^3-6238 u^{11} t^2-1920 u^{12} t-309 \
u^{13}\Big) s^4-t^3 \Big(48 t^{13}+1112 u t^{12}+8730 u^2 t^{11}+35276 \
u^3 t^{10}+86245 u^4 t^9+130602 u^5 t^8+113599 u^6 t^7+38248 u^7 t^6-30777 \
u^8 t^5-50450 u^9 t^4-31999 u^{10} t^3-10602 u^{11} t^2-1774 u^{12} t-114 \
u^{13}\Big) s^3-t^4 u (t+u)^2 \Big(84 t^{10}+898 u t^9+3400 u^2 t^8+6309 \
u^3 t^7+5217 u^4 t^6-2505 u^5 t^5-8514 u^6 t^4-9343 u^7 t^3-5743 u^8 \
t^2-2007 u^9 t-336 u^{10}\Big) s^2-t^5 u^2 (t+u)^3 \Big(48 t^8+200 u \
t^7+135 u^2 t^6-571 u^3 t^5-1726 u^4 t^4-1854 u^5 t^3-1219 u^6 t^2-459 u^7 \
t-90 u^8\Big) s+t^6 u^4 (t+u)^6 \Big(23 t^3+54 u t^2+46 u^2 t+26 \
u^3\Big)\Big) H(1,1,0,z) (s+u)^2+9 (s+t)^2 u^2 (t+u) \Big(2 \Big(92 \
t^7+224 u t^6+807 u^2 t^5+41 u^3 t^4+227 u^4 t^3+71 u^5 t^2+70 u^6 t+20 \
u^7\Big) s^{15}+2 \Big(548 t^8+1800 u t^7+6507 u^2 t^6+5556 u^3 t^5+2001 \
u^4 t^4+2149 u^5 t^3+864 u^6 t^2+535 u^7 t+120 u^8\Big) s^{14}+\Big(3080 \
t^9+12672 u t^8+43909 u^2 t^7+68677 u^3 t^6+39172 u^4 t^5+27176 u^5 \
t^4+19411 u^6 t^3+8427 u^7 t^2+3620 u^8 t+640 u^9\Big) s^{13}+\Big(5400 \
t^{10}+26796 u t^9+85362 u^2 t^8+182471 u^3 t^7+168303 u^4 t^6+105098 u^5 \
t^5+92562 u^6 t^4+53097 u^7 t^3+21885 u^8 t^2+7050 u^9 t+1000 u^{10}\Big) \
s^{12}+\Big(6464 t^{11}+38196 u t^{10}+114563 u^2 t^9+283713 u^3 t^8+385450 \
u^4 t^7+275704 u^5 t^6+227431 u^6 t^5
\end{dmath*}
\begin{dmath*}
{\white=}
+183473 u^7 t^4+90148 u^8 t^3+33770 u^9 \
t^2+8600 u^{10} t+1000 u^{11}\Big) s^{11}+\Big(5400 t^{12}+38196 u \
t^{11}+123420 u^2 t^{10}+320669 u^3 t^9+573417 u^4 t^8+513000 u^5 t^7+343086 \
u^6 t^6+321375 u^7 t^5+218509 u^8 t^4+96742 u^9 t^3+32200 u^{10} t^2+6690 \
u^{11} t+640 u^{12}\Big) s^{10}+\Big(3080 t^{13}+26796 u t^{12}+114563 \
u^2 t^{11}+320669 u^3 t^{10}+641380 u^4 t^9+698714 u^5 t^8+386310 u^6 \
t^7+282028 u^7 t^6+279284 u^8 t^5+162550 u^9 t^4+65139 u^{10} t^3+18779 \
u^{11} t^2+3220 u^{12} t+240 u^{13}\Big) s^9+\Big(1096 t^{14}+12672 u \
t^{13}+85362 u^2 t^{12}+283713 u^3 t^{11}+573417 u^4 t^{10}+698714 u^5 \
t^9+392272 u^6 t^8+129314 u^7 t^7+178058 u^8 t^6+158492 u^9 t^5+72298 u^{10} \
t^4+25025 u^{11} t^3+6257 u^{12} t^2+870 u^{13} t+40 u^{14}\Big) s^8+t \
\Big(184 t^{14}+3600 u t^{13}+43909 u^2 t^{12}+182471 u^3 t^{11}+385450 u^4 \
t^{10}+513000 u^5 t^9+386310 u^6 t^8+129314 u^7 t^7+92160 u^8 t^6+110380 u^9 \
t^5+48211 u^{10} t^4+11865 u^{11} t^3+3700 u^{12} t^2+994 u^{13} t+100 \
u^{14}\Big) s^7+t^2 u \Big(448 t^{13}+13014 u t^{12}+68677 u^2 \
t^{11}+168303 u^3 t^{10}+275704 u^4 t^9+343086 u^5 t^8+282028 u^6 t^7+178058 \
u^7 t^6+110380 u^8 t^5+33784 u^9 t^4-4267 u^{10} t^3-3791 u^{11} t^2-474 \
u^{12} t+42 u^{13}\Big) s^6+t^3 u^2 \Big(1614 t^{12}+11112 u t^{11}+39172 \
u^2 t^{10}+105098 u^3 t^9+227431 u^4 t^8+321375 u^5 t^7+279284 u^6 \
t^6+158492 u^7 t^5+48211 u^8 t^4-4267 u^9 t^3-7188 u^{10} t^2-1770 u^{11} \
t-164 u^{12}\Big) s^5+t^4 u^3 \Big(82 t^{11}+4002 u t^{10}+27176 u^2 \
t^9+92562 u^3 t^8+183473 u^4 t^7+218509 u^5 t^6+162550 u^6 t^5+72298 u^7 \
t^4+11865 u^8 t^3-3791 u^9 t^2-1770 u^{10} t-204 u^{11}\Big) s^4+t^5 u^4 \
(t+u)^2 \Big(454 t^8+3390 u t^7+12177 u^2 t^6+25353 u^3 t^5+27265 u^4 \
t^4+16859 u^5 t^3+4156 u^6 t^2-146 u^7 t-164 u^8\Big) s^3+t^6 u^5 (t+u)^3 \
\Big(142 t^6+1302 u t^5+4095 u^2 t^4+5552 u^3 t^3+3527 u^4 t^2+868 u^5 t+42 \
u^6\Big) s^2+10 t^7 u^6 (t+u)^5 \Big(14 t^3+37 u t^2+37 u^2 t+10 \
u^3\Big) s+40 t^8 u^7 (t+u)^5 \Big(t^2+u t+u^2\Big)\Big) H(1,0,y) \
(s+u)
\end{dmath*}

\begin{dmath*}
{\white =}
-3 s^2 (s+t)^2 u^2 (t+u)^2 \Big(\Big(259 t^6+616 u t^5+426 u^2 t^4-68 \
u^3 t^3+109 u^4 t^2+300 u^5 t+120 u^6\Big) s^{14}+2 \Big(774 t^7+2870 u \
t^6+3872 u^2 t^5+1903 u^3 t^4+425 u^4 t^3+1339 u^5 t^2+1395 u^6 t+420 \
u^7\Big) s^{13}+\Big(4367 t^8+21396 u t^7+40759 u^2 t^6+37858 u^3 \
t^5+19215 u^4 t^4+14984 u^5 t^3+18489 u^6 t^2+11430 u^7 t+2640 u^8\Big) \
s^{12}+2 \Big(3840 t^9+23121 u t^8+56244 u^2 t^7+72977 u^3 t^6+57956 u^4 \
t^5+40878 u^5 t^4+39401 u^6 t^3+31286 u^7 t^2+13545 u^8 t+2460 u^9\Big) \
s^{11}+2 \Big(4602 t^{10}+33792 u t^9+102876 u^2 t^8+170653 u^3 t^7+175620 \
u^4 t^6+137297 u^5 t^5+118565 u^6 t^4+104385 u^7 t^3+62005 u^8 t^2+20475 u^9 \
t+3000 u^{10}\Big) s^{10}+2 \Big(3840 t^{11}+35847 u t^{10}+140082 u^2 \
t^9+290657 u^3 t^8+356593 u^4 t^7+294815 u^5 t^6+230739 u^6 t^5+215383 u^7 \
t^4+164395 u^8 t^3+77358 u^9 t^2+20475 u^{10} t+2460 u^{11}\Big) \
s^9+\Big(4367 t^{12}+55904 u t^{11}+291138 u^2 t^{10}+756156 u^3 \
t^9+1095009 u^4 t^8+963078 u^5 t^7+649134 u^6 t^6+554262 u^7 t^5+519114 u^8 \
t^4+328790 u^9 t^3+124010 u^{10} t^2+27090 u^{11} t+2640 u^{12}\Big) s^8+2 \
\Big(774 t^{13}+15265 u t^{12}+110242 u^2 t^{11}+363648 u^3 t^{10}+642258 \
u^4 t^9+661734 u^5 t^8+447563 u^6 t^7+300068 u^7 t^6+277131 u^8 t^5+215383 \
u^9 t^4+104385 u^{10} t^3+31286 u^{11} t^2+5715 u^{12} t+420 u^{13}\Big) \
s^7+\Big(259 t^{14}+10260 u t^{13}+111068 u^2 t^{12}
\end{dmath*}

\intertext{}
\begin{dmath*}
{\white =}
+486996 u^3 \
t^{11}+1107162 u^4 t^{10}+1479900 u^5 t^9+1291792 u^6 t^8+895126 u^7 \
t^7+649134 u^8 t^6+461478 u^9 t^5+237130 u^{10} t^4+78802 u^{11} t^3+18489 \
u^{12} t^2+2790 u^{13} t+120 u^{14}\Big) s^6+2 t u \Big(777 t^{13}+16074 \
u t^{12}+104819 u^2 t^{11}+330308 u^3 t^{10}+608514 u^4 t^9+739950 u^5 \
t^8+661734 u^6 t^7+481539 u^7 t^6+294815 u^8 t^5+137297 u^9 t^4+40878 u^{10} \
t^3+7492 u^{11} t^2+1339 u^{12} t+150 u^{13}\Big) s^5+t^2 u^2 \Big(3885 \
t^{12}+50940 u t^{11}+249466 u^2 t^{10}+660616 u^3 t^9+1107162 u^4 \
t^8+1284516 u^5 t^7+1095009 u^6 t^6+713186 u^7 t^5+351240 u^8 t^4+115912 u^9 \
t^3+19215 u^{10} t^2+850 u^{11} t+109 u^{12}\Big) s^4+2 t^3 u^3 \Big(2590 \
t^{11}+25470 u t^{10}+104819 u^2 t^9+243498 u^3 t^8+363648 u^4 t^7+378078 \
u^5 t^6+290657 u^6 t^5+170653 u^7 t^4+72977 u^8 t^3+18929 u^9 t^2+1903 \
u^{10} t-34 u^{11}\Big) s^3+t^4 u^4 (t+u)^2 \Big(3885 t^8+24378 u \
t^7+58427 u^2 t^6+79252 u^3 t^5+74207 u^4 t^4+52498 u^5 t^3+26549 u^6 \
t^2+6892 u^7 t+426 u^8\Big) s^2+2 t^5 u^5 (t+u)^3 \Big(777 t^6+2799 u \
t^5+4537 u^2 t^4+5167 u^3 t^3+3936 u^4 t^2+1946 u^5 t+308 u^6\Big) s+t^6 \
u^6 (t+u)^4 \Big(259 t^4+512 u t^3+765 u^2 t^2+512 u^3 t+259 \
u^4\Big)\Big) H(0,y) H(1,z)+9 s^2 (t+u)^2 \Big(2 \Big(20 t^8+50 u \
t^7+21 u^2 t^6-82 u^3 t^5-102 u^4 t^4-82 u^5 t^3+21 u^6 t^2+50 u^7 t+20 \
u^8\Big) s^{16}+2 \Big(140 t^9+505 u t^8+568 u^2 t^7-298 u^3 t^6-1069 u^4 \
t^5-1069 u^5 t^4-298 u^6 t^3+568 u^7 t^2+505 u^8 t+140 u^9\Big) \
s^{15}+\Big(880 t^{10}+4370 u t^9+8221 u^2 t^8+4262 u^3 t^7-6199 u^4 \
t^6-10932 u^5 t^5-6199 u^6 t^4+4262 u^7 t^3+8221 u^8 t^2+4370 u^9 t+880 \
u^{10}\Big) s^{14}+2 \Big(820 t^{11}+5395 u t^{10}+14563 u^2 t^9+17988 \
u^3 t^8+5650 u^4 t^7-8508 u^5 t^6
\end{dmath*}

\begin{dmath*}
{\white =}
-8508 u^6 t^5+5650 u^7 t^4+17988 u^8 \
t^3+14563 u^9 t^2+5395 u^{10} t+820 u^{11}\Big) s^{13}+\Big(2000 \
t^{12}+16930 u t^{11}+60889 u^2 t^{10}+115200 u^3 t^9+112888 u^4 t^8+52018 \
u^5 t^7+18062 u^6 t^6+52018 u^7 t^5+112888 u^8 t^4+115200 u^9 t^3+60889 \
u^{10} t^2+16930 u^{11} t+2000 u^{12}\Big) s^{12}+2 \Big(820 t^{13}+8825 \
u t^{12}+40630 u^2 t^{11}+106430 u^3 t^{10}+162506 u^4 t^9+145433 u^5 \
t^8+94054 u^6 t^7+94054 u^7 t^6+145433 u^8 t^5+162506 u^9 t^4+106430 u^{10} \
t^3+40630 u^{11} t^2+8825 u^{12} t+820 u^{13}\Big) s^{11}+\Big(880 \
t^{14}+12310 u t^{13}+71305 u^2 t^{12}+252860 u^3 t^{11}+542940 u^4 \
t^{10}+672624 u^5 t^9+495141 u^6 t^8+346704 u^7 t^7+495141 u^8 t^6+672624 \
u^9 t^5+542940 u^{10} t^4+252860 u^{11} t^3+71305 u^{12} t^2+12310 u^{13} \
t+880 u^{14}\Big) s^{10}+2 \Big(140 t^{15}+2785 u t^{14}+20491 u^2 \
t^{13}+99450 u^3 t^{12}+294436 u^4 t^{11}+490859 u^5 t^{10}+448931 u^6 \
t^9+254956 u^7 t^8+254956 u^8 t^7+448931 u^9 t^6+490859 u^{10} t^5+294436 \
u^{11} t^4+99450 u^{12} t^3+20491 u^{13} t^2+2785 u^{14} t+140 u^{15}\Big) \
s^9+\Big(40 t^{16}+1490 u t^{15}+14845 u^2 t^{14}+102820 u^3 t^{13}+419280 \
u^4 t^{12}+950788 u^5 t^{11}+1225773 u^6 t^{10}+975710 u^7 t^9+743060 u^8 \
t^8+975710 u^9 t^7+1225773 u^{10} t^6+950788 u^{11} t^5+419280 u^{12} \
t^4+102820 u^{13} t^3+14845 u^{14} t^2+1490 u^{15} t+40 u^{16}\Big) s^8+2 \
t u \Big(90 t^{15}+1540 u t^{14}+16932 u^2 t^{13}+96123 u^3 t^{12}+304282 \
u^4 t^{11}+583798 u^5 t^{10}+762212 u^6 t^9+803305 u^7 t^8+803305 u^8 \
t^7+762212 u^9 t^6+583798 u^{10} t^5+304282 u^{11} t^4+96123 u^{12} \
t^3+16932 u^{13} t^2+1540 u^{14} t+90 u^{15}\Big) s^7+t^2 u^2 \Big(282 \
t^{14}+6622 u t^{13}+54887 u^2 t^{12}+264008 u^3 t^{11}+776536 u^4 \
t^{10}+1517240 u^5 t^9+2171441 u^6 t^8+2431080 u^7 t^7+2171441 u^8 \
t^6+1517240 u^9 t^5+776536 u^{10} t^4+264008 u^{11} t^3+54887 u^{12} \
t^2+6622 u^{13} t+282 u^{14}\Big) s^6
\end{dmath*}
\begin{dmath*}
{\white=}
+2 t^3 u^3 \Big(298 t^{13}+4418 u \
t^{12}+40731 u^2 t^{11}+190625 u^3 t^{10}+505964 u^4 t^9+877790 u^5 \
t^8+1117090 u^6 t^7+1117090 u^7 t^6+877790 u^8 t^5+505964 u^9 t^4+190625 \
u^{10} t^3+40731 u^{11} t^2+4418 u^{12} t+298 u^{13}\Big) s^5+t^4 u^4 \
\Big(536 t^{12}+16810 u t^{11}+131975 u^2 t^{10}+463360 u^3 t^9+936996 u^4 \
t^8+1292362 u^5 t^7+1406138 u^6 t^6+1292362 u^7 t^5+936996 u^8 t^4+463360 \
u^9 t^3+131975 u^{10} t^2+16810 u^{11} t+536 u^{12}\Big) s^4+2 t^5 u^5 \
\Big(848 t^{11}+13094 u t^{10}+64600 u^2 t^9+162207 u^3 t^8+255217 u^4 \
t^7+298424 u^5 t^6+298424 u^6 t^5+255217 u^7 t^4+162207 u^8 t^3+64600 u^9 \
t^2+13094 u^{10} t+848 u^{11}\Big) s^3+t^6 u^6 (t+u)^2 \Big(2062 \
t^8+13122 u t^7+32971 u^2 t^6+48846 u^3 t^5+54292 u^4 t^4+48846 u^5 \
t^3+32971 u^6 t^2+13122 u^7 t+2062 u^8\Big) s^2+4 t^7 u^7 (t+u)^3 \
\Big(158 t^6+746 u t^5+1500 u^2 t^4+1923 u^3 t^3+1500 u^4 t^2+746 u^5 t+158 \
u^6\Big) s+8 t^8 u^8 (t+u)^4 \Big(23 t^4+45 u t^3+67 u^2 t^2+45 u^3 t+23 \
u^4\Big)\Big) H(1,z) H(3,y)+3 s^2 (s+t) u^2 (t+u)^2 \Big(\Big(17 \
t^6-424 u t^5-1038 u^2 t^4-1108 u^3 t^3-133 u^4 t^2+300 u^5 t+120 u^6\Big) \
s^{15}+\Big(113 t^7-2432 u t^6-9046 u^2 t^5-13902 u^3 t^4-8627 u^4 t^3+1042 \
u^5 t^2+2910 u^6 t+840 u^7\Big) s^{14}+\Big(361 t^8-5422 u t^7-30575 u^2 \
t^6-64296 u^3 t^5-66643 u^4 t^4-21748 u^5 t^3+14031 u^6 t^2+12270 u^7 t+2640 \
u^8\Big) s^{13}+\Big(745 t^9-5220 u t^8-49927 u^2 t^7-141944 u^3 \
t^6-209881 u^4 t^5-146102 u^5 t^4-877 u^6 t^3+57146 u^7 t^2+29730 u^8 t+4920 \
u^9\Big) s^{12}+2 \Big(546 t^{10}+2738 u t^9-12316 u^2 t^8-77546 u^3 \
t^7-169106 u^4 t^6-184458 u^5 t^5-65973 u^6 t^4+53495 u^7 t^3+62003 u^8 \
t^2+22935 u^9 t+3000 u^{10}\Big) s^{11}+2 \Big(576 t^{11}+14602 u \
t^{10}
\end{dmath*}

\begin{dmath*}
{\white =}
+44395 u^2 t^9+6575 u^3 t^8-159019 u^4 t^7-299169 u^5 t^6-202668 u^6 \
t^5+31945 u^7 t^4+130735 u^8 t^3+82071 u^9 t^2+23475 u^{10} t+2460 \
u^{11}\Big) s^{10}+\Big(853 t^{12}+50204 u t^{11}+239326 u^2 \
t^{10}+380158 u^3 t^9-5759 u^4 t^8-778482 u^5 t^7-981862 u^6 t^6-306108 u^7 \
t^5+310132 u^8 t^4+327562 u^9 t^3+138034 u^{10} t^2+32010 u^{11} t+2640 \
u^{12}\Big) s^9+\Big(421 t^{13}+48756 u t^{12}+284228 u^2 t^{11}+643716 \
u^3 t^{10}+559919 u^4 t^9-346654 u^5 t^8-1327656 u^6 t^7-1188026 u^7 \
t^6-185900 u^8 t^5+356130 u^9 t^4+244634 u^{10} t^3+72974 u^{11} t^2+14070 \
u^{12} t+840 u^{13}\Big) s^8+\Big(125 t^{14}+28668 u t^{13}+194932 u^2 \
t^{12}+556848 u^3 t^{11}+840628 u^4 t^{10}+624124 u^5 t^9-174342 u^6 \
t^8-974758 u^7 t^7-907984 u^8 t^6-201510 u^9 t^5+170032 u^{10} t^4+108262 \
u^{11} t^3+22743 u^{12} t^2+3630 u^{13} t+120 u^{14}\Big) s^7+t \Big(17 \
t^{14}+9532 u t^{13}+79874 u^2 t^{12}+311220 u^3 t^{11}+718808 u^4 \
t^{10}+1133540 u^5 t^9+1280316 u^6 t^8+835978 u^7 t^7-19040 u^8 t^6-455158 \
u^9 t^5-255266 u^{10} t^4-6292 u^{11} t^3+26245 u^{12} t^2+3544 u^{13} t+420 \
u^{14}\Big) s^6+t^2 u \Big(1338 t^{13}+17115 u t^{12}+122578 u^2 \
t^{11}+473242 u^3 t^{10}+1064256 u^4 t^9+1563708 u^5 t^8+1630684 u^6 \
t^7+1174609 u^7 t^6+432306 u^8 t^5-92986 u^9 t^4-158896 u^{10} t^3-40833 \
u^{11} t^2+2954 u^{12} t+167 u^{13}\Big) s^5+t^3 u^2 \Big(1209 \
t^{12}+30536 u t^{11}+210974 u^2 t^{10}+653148 u^3 t^9+1105716 u^4 \
t^8+1162604 u^5 t^7+908135 u^6 t^6+653214 u^7 t^5+380872 u^8 t^4+92158 u^9 \
t^3-31479 u^{10} t^2-16364 u^{11} t+67 u^{12}\Big) s^4+t^4 u^3 \Big(3688 \
t^{11}+49261 u t^{10}+216624 u^2 t^9+465668 u^3 t^8+558504 u^4 t^7+411820 \
u^5 t^6+264280 u^6 t^5
\end{dmath*}

\intertext{}
\begin{dmath*}
{\white =}
+237894 u^7 t^4+185882 u^8 t^3+72173 u^9 t^2+5494 \
u^{10} t-2344 u^{11}\Big) s^3+t^5 u^4 (t+u)^2 \Big(4269 t^8+21666 u \
t^7+42801 u^2 t^6+45136 u^3 t^5+32315 u^4 t^4+22162 u^5 t^3+17975 u^6 \
t^2+9856 u^7 t+2252 u^8\Big) s^2+t^6 u^5 (t+u)^3 \Big(762 t^6+3225 u \
t^5+6713 u^2 t^4+8777 u^3 t^3+5973 u^4 t^2+1975 u^5 t+49 u^6\Big) s+t^7 \
u^6 (t+u)^4 \Big(293 t^4+568 u t^3+843 u^2 t^2+568 u^3 t+293 \
u^4\Big)\Big) H(0,1,z)+18 s^2 t^2 (s+t)^2 u^2 (t+u)^2 \Big(\Big(11 \
t^4+20 u t^3+24 u^2 t^2+20 u^3 t+11 u^4\Big) s^{14}+2 \Big(33 t^5+91 u \
t^4+115 u^2 t^3+119 u^3 t^2+100 u^4 t+44 u^5\Big) s^{13}+\Big(187 t^6+744 \
u t^5+1142 u^2 t^4+1208 u^3 t^3+1245 u^4 t^2+976 u^5 t+336 u^6\Big) \
s^{12}+2 \Big(165 t^7+936 u t^6+1782 u^2 t^5+1633 u^3 t^4+1366 u^4 t^3+1677 \
u^5 t^2+1363 u^6 t+406 u^7\Big) s^{11}+\Big(396 t^8+3450 u t^7+8220 u^2 \
t^6+5816 u^3 t^5-2041 u^4 t^4-1266 u^5 t^3+5406 u^6 t^2+5360 u^7 t+1391 \
u^8\Big) s^{10}+2 \Big(165 t^9+2460 u t^8+7356 u^2 t^7+5316 u^3 t^6-8202 \
u^4 t^5-15297 u^5 t^4-5650 u^6 t^3+4428 u^7 t^2+4099 u^8 t+897 u^9\Big) \
s^9+\Big(187 t^{10}+5044 u t^9+19227 u^2 t^8+21984 u^3 t^7-15915 u^4 \
t^6-62664 u^5 t^5-56568 u^6 t^4-10620 u^7 t^3+14826 u^8 t^2+9784 u^9 t+1783 \
u^{10}\Big) s^8+2 \Big(33 t^{11}+1640 u t^{10}+8324 u^2 t^9+17396 u^3 \
t^8+12447 u^4 t^7-14606 u^5 t^6-35999 u^6 t^5-24042 u^7 t^4+1442 u^8 \
t^3+9648 u^9 t^2+4481 u^{10} t+668 u^{11}\Big) s^7+\Big(11 t^{12}+1158 u \
t^{11}+8506 u^2 t^{10}+34618 u^3 t^9+71793 u^4 t^8+62258 u^5 t^7-9538 u^6 \
t^6-52674 u^7 t^5-20480 u^8 t^4+18332 u^9 t^3+18756 u^{10} t^2+6076 u^{11} \
t+702 u^{12}\Big) s^6+2 u \Big(81 t^{12}+1047 u t^{11}+10102 u^2 \
t^{10}+35002 u^3 t^9+52863 u^4 t^8+30585 u^5 t^7-6576 u^6 t^6
\end{dmath*}

\begin{dmath*}
{\white =}
-11010 u^7 \
t^5+6574 u^8 t^4+12427 u^9 t^3+6297 u^{10} t^2+1397 u^{11} t+113 \
u^{12}\Big) s^5+u^2 \Big(129 t^{12}+6282 u t^{11}+32372 u^2 t^{10}+66938 \
u^3 t^9+63533 u^4 t^8+21894 u^5 t^7+1543 u^6 t^6+16366 u^7 t^5+25835 u^8 \
t^4+16336 u^9 t^3+5137 u^{10} t^2+760 u^{11} t+33 u^{12}\Big) s^4+2 t u^3 \
\Big(398 t^{11}+3081 u t^{10}+9304 u^2 t^9+14236 u^3 t^8+12582 u^4 t^7+9212 \
u^5 t^6+9802 u^6 t^5+10293 u^7 t^4+6959 u^8 t^3+2766 u^9 t^2+591 u^{10} t+48 \
u^{11}\Big) s^3+t^2 u^4 (t+u)^2 \Big(129 t^8+1620 u t^7+3541 u^2 t^6+3536 \
u^3 t^5+3022 u^4 t^4+2792 u^5 t^3+1912 u^6 t^2+746 u^7 t+126 u^8\Big) \
s^2+2 t^3 u^5 (t+u)^3 \Big(81 t^6+276 u t^5+227 u^2 t^4+155 u^3 t^3+173 u^4 \
t^2+112 u^5 t+38 u^6\Big) s+t^4 u^6 (t+u)^4 \Big(11 t^4-2 u t^3+15 u^2 \
t^2+14 u^3 t+19 u^4\Big)\Big) H(0,y) H(0,1,z)-18 s^2 t^2 u^2 (t+u)^2 \
\Big(\Big(47 t^4+116 u t^3+144 u^2 t^2+96 u^3 t+39 u^4\Big) s^{16}+2 \
\Big(211 t^5+697 u t^4+1032 u^2 t^3+886 u^3 t^2+487 u^4 t+147 u^5\Big) \
s^{15}+3 \Big(587 t^6+2466 u t^5+4371 u^2 t^4+4476 u^3 t^3+3059 u^4 \
t^2+1466 u^5 t+355 u^6\Big) s^{14}+2 \Big(2246 t^7+11795 u t^6+25344 u^2 \
t^5+30555 u^3 t^4+24384 u^4 t^3+14649 u^5 t^2+6280 u^6 t+1255 u^7\Big) \
s^{13}+\Big(7761 t^8+49924 u t^7+132409 u^2 t^6+195056 u^3 t^5+184034 u^4 \
t^4+125820 u^5 t^3+68009 u^6 t^2+25468 u^7 t+4325 u^8\Big) s^{12}+6 \
\Big(1587 t^9+12289 u t^8+40330 u^2 t^7+74208 u^3 t^6+86445 u^4 t^5+69435 \
u^5 t^4+42172 u^6 t^3+20076 u^7 t^2+6418 u^8 t+944 u^9\Big) \
s^{11}+\Big(8447 t^{10}+78224 u t^9+319917 u^2 t^8+731786 u^3 t^7+1046888 \
u^4 t^6+1017372 u^5 t^5+727368 u^6 t^4
\end{dmath*}
\begin{dmath*}
{\white=}
+404718 u^7 t^3+167685 u^8 t^2+44164 \
u^9 t+5599 u^{10}\Big) s^{10}+2 \Big(2688 t^{11}+30091 u t^{10}+156379 \
u^2 t^9+443281 u^3 t^8+763361 u^4 t^7+878974 u^5 t^6+745371 u^6 t^5+500624 \
u^7 t^4+259751 u^8 t^3+91571 u^9 t^2+19006 u^{10} t+2015 u^{11}\Big) s^9+3 \
\Big(784 t^{12}+11080 u t^{11}+75305 u^2 t^{10}+263492 u^3 t^9+545731 u^4 \
t^8+746656 u^5 t^7+740518 u^6 t^6+577912 u^7 t^5+364335 u^8 t^4+170508 u^9 \
t^3+50345 u^{10} t^2+7940 u^{11} t+660 u^{12}\Big) s^8+2 \Big(318 \
t^{13}+6360 u t^{12}+59340 u^2 t^{11}+252696 u^3 t^{10}+631038 u^4 \
t^9+1061526 u^5 t^8+1285210 u^6 t^7+1158562 u^7 t^6+806991 u^8 t^5+444563 \
u^9 t^4+181058 u^{10} t^3+45120 u^{11} t^2+5167 u^{12} t+295 u^{13}\Big) \
s^7+\Big(80 t^{14}+3066 u t^{13}+43714 u^2 t^{12}+225298 u^3 t^{11}+680647 \
u^4 t^{10}+1441224 u^5 t^9+2222801 u^6 t^8+2456276 u^7 t^7+1932681 u^8 \
t^6+1122364 u^9 t^5+510799 u^{10} t^4+177102 u^{11} t^3+37210 u^{12} \
t^2+2790 u^{13} t+80 u^{14}\Big) s^6+6 t u \Big(58 t^{13}+1689 u \
t^{12}+11116 u^2 t^{11}+42847 u^3 t^{10}+116103 u^4 t^9+225254 u^5 \
t^8+308029 u^6 t^7+293932 u^7 t^6+196649 u^8 t^5+94795 u^9 t^4+34537 u^{10} \
t^3+9491 u^{11} t^2+1574 u^{12} t+58 u^{13}\Big) s^5+t^2 u^2 \Big(1092 \
t^{12}+11428 u t^{11}+67756 u^2 t^{10}+245052 u^3 t^9+575825 u^4 t^8+929600 \
u^5 t^7+1064520 u^6 t^6+875584 u^7 t^5+515905 u^8 t^4+213096 u^9 t^3+59096 \
u^{10} t^2+10508 u^{11} t+1092 u^{12}\Big) s^4+2 t^3 u^3 \Big(394 \
t^{11}+5820 u t^{10}+29696 u^2 t^9+83756 u^3 t^8+156154 u^4 t^7+208169 u^5 \
t^6+204213 u^6 t^5+148024 u^7 t^4+77656 u^8 t^3+27425 u^9 t^2+5475 u^{10} \
t+394 u^{11}\Big) s^3+6 t^4 u^4 (t+u)^2 \Big(173 t^8+987 u t^7+2643 u^2 \
t^6+4684 u^3 t^5+5690 u^4 t^4+4568 u^5 t^3+2519 u^6 t^2+941 u^7 t+173 \
u^8\Big) s^2+2 t^5 u^5 (t+u)^3 \Big(150 t^6+629 u t^5+1361 u^2 t^4+1851 \
u^3 t^3+1353 u^4 t^2+606 u^5 t+150 u^6\Big) s+66 t^6 u^6 (t+u)^4 \
\Big(t^2+u t+u^2\Big)^2\Big) H(2,y) H(0,1,z)-3 s^2 (s+t)^2 u^2 (t+u)^2 \
\Big(\Big(259 t^6+616 u t^5+426 u^2 t^4-68 u^3 t^3+109 u^4 t^2+300 u^5 \
t+120 u^6\Big) s^{14}+2 \Big(774 t^7+2870 u t^6+3872 u^2 t^5+1903 u^3 \
t^4+425 u^4 t^3+1339 u^5 t^2+1395 u^6 t+420 u^7\Big) s^{13}+\Big(4367 \
t^8+21396 u t^7
\end{dmath*}

\begin{dmath*}
{\white =}
+40759 u^2 t^6+37858 u^3 t^5+19215 u^4 t^4+14984 u^5 \
t^3+18489 u^6 t^2+11430 u^7 t+2640 u^8\Big) s^{12}+2 \Big(3840 t^9+23121 \
u t^8+56244 u^2 t^7+72977 u^3 t^6+57956 u^4 t^5+40878 u^5 t^4+39401 u^6 \
t^3+31286 u^7 t^2+13545 u^8 t+2460 u^9\Big) s^{11}+2 \Big(4602 \
t^{10}+33792 u t^9+102876 u^2 t^8+170653 u^3 t^7+175620 u^4 t^6+137297 u^5 \
t^5+118565 u^6 t^4+104385 u^7 t^3+62005 u^8 t^2+20475 u^9 t+3000 \
u^{10}\Big) s^{10}+2 \Big(3840 t^{11}+35847 u t^{10}+140082 u^2 \
t^9+290657 u^3 t^8+356593 u^4 t^7+294815 u^5 t^6+230739 u^6 t^5+215383 u^7 \
t^4+164395 u^8 t^3+77358 u^9 t^2+20475 u^{10} t+2460 u^{11}\Big) \
s^9+\Big(4367 t^{12}+55904 u t^{11}+291138 u^2 t^{10}+756156 u^3 \
t^9+1095009 u^4 t^8+963078 u^5 t^7+649134 u^6 t^6+554262 u^7 t^5+519114 u^8 \
t^4+328790 u^9 t^3+124010 u^{10} t^2+27090 u^{11} t+2640 u^{12}\Big) s^8+2 \
\Big(774 t^{13}+15265 u t^{12}+110242 u^2 t^{11}+363648 u^3 t^{10}+642258 \
u^4 t^9+661734 u^5 t^8+447563 u^6 t^7+300068 u^7 t^6+277131 u^8 t^5+215383 \
u^9 t^4+104385 u^{10} t^3+31286 u^{11} t^2+5715 u^{12} t+420 u^{13}\Big) \
s^7+\Big(259 t^{14}+10260 u t^{13}+111068 u^2 t^{12}+486996 u^3 \
t^{11}+1107162 u^4 t^{10}+1479900 u^5 t^9+1291792 u^6 t^8+895126 u^7 \
t^7+649134 u^8 t^6+461478 u^9 t^5+237130 u^{10} t^4+78802 u^{11} t^3+18489 \
u^{12} t^2+2790 u^{13} t+120 u^{14}\Big) s^6+2 t u \Big(777 t^{13}+16074 \
u t^{12}+104819 u^2 t^{11}+330308 u^3 t^{10}+608514 u^4 t^9+739950 u^5 \
t^8+661734 u^6 t^7+481539 u^7 t^6+294815 u^8 t^5+137297 u^9 t^4
\end{dmath*}

\intertext{}
\begin{dmath*}
{\white =}
+40878 u^{10} \
t^3+7492 u^{11} t^2+1339 u^{12} t+150 u^{13}\Big) s^5+t^2 u^2 \Big(3885 \
t^{12}+50940 u t^{11}+249466 u^2 t^{10}+660616 u^3 t^9+1107162 u^4 \
t^8+1284516 u^5 t^7+1095009 u^6 t^6+713186 u^7 t^5+351240 u^8 t^4+115912 u^9 \
t^3+19215 u^{10} t^2+850 u^{11} t+109 u^{12}\Big) s^4+2 t^3 u^3 \Big(2590 \
t^{11}+25470 u t^{10}+104819 u^2 t^9+243498 u^3 t^8+363648 u^4 t^7+378078 \
u^5 t^6+290657 u^6 t^5+170653 u^7 t^4+72977 u^8 t^3+18929 u^9 t^2+1903 \
u^{10} t-34 u^{11}\Big) s^3+t^4 u^4 (t+u)^2 \Big(3885 t^8+24378 u \
t^7+58427 u^2 t^6+79252 u^3 t^5+74207 u^4 t^4+52498 u^5 t^3+26549 u^6 \
t^2+6892 u^7 t+426 u^8\Big) s^2+2 t^5 u^5 (t+u)^3 \Big(777 t^6+2799 u \
t^5+4537 u^2 t^4+5167 u^3 t^3+3936 u^4 t^2+1946 u^5 t+308 u^6\Big) s+t^6 \
u^6 (t+u)^4 \Big(259 t^4+512 u t^3+765 u^2 t^2+512 u^3 t+259 \
u^4\Big)\Big) H(0,2,y)-18 s^2 t^2 u^2 (t+u)^3 \Big(2 t \Big(7 t^2+15 u \
t+12 u^2\Big) s^{16}+12 \Big(8 t^4+24 u t^3+29 u^2 t^2+12 u^3 \
t-u^4\Big) s^{15}+3 \Big(101 t^5+423 u t^4+720 u^2 t^3+556 u^3 t^2+111 \
u^4 t-31 u^5\Big) s^{14}+\Big(598 t^6+3066 u t^5+7134 u^2 t^4+8992 u^3 \
t^3+5154 u^4 t^2+510 u^5 t-318 u^6\Big) s^{13}+\Big(851 t^7+4495 u \
t^6+13290 u^2 t^5+25506 u^3 t^4+26949 u^4 t^3+12309 u^5 t^2+688 u^6 t-652 \
u^7\Big) s^{12}+6 \Big(158 t^8+586 u t^7+2628 u^2 t^6+7708 u^3 t^5+11649 \
u^4 t^4+9092 u^5 t^3+3324 u^6 t^2+218 u^7 t-155 u^8\Big) s^{11}+\Big(851 \
t^9-417 u t^8+9564 u^2 t^7+62150 u^3 t^6+123813 u^4 t^5+121269 u^5 t^4+67982 \
u^6 t^3+23208 u^7 t^2+2838 u^8 t-1002 u^9\Big) s^{10}+2 \Big(299 \
t^{10}-2075 u t^9-1896 u^2 t^8+23751 u^3 t^7+72903 u^4 t^6+97515 u^5 \
t^5+71423 u^6 t^4+31963 u^7 t^3+10800 u^8 t^2+1966 u^9 t-409 u^{10}\Big) \
s^9+3 \Big(101 t^{11}-1463 u t^{10}-3896 u^2 t^9+208 u^3 t^8+20429 u^4 \
t^7
\end{dmath*}

\begin{dmath*}
{\white =}
+58669 u^5 t^6+80268 u^6 t^5+55172 u^7 t^4+19400 u^8 t^3+4832 u^9 \
t^2+1000 u^{10} t-156 u^{11}\Big) s^8+2 \Big(48 t^{12}-1128 u t^{11}-4410 \
u^2 t^{10}-17090 u^3 t^9-31983 u^4 t^8+6084 u^5 t^7+100874 u^6 t^6+143160 \
u^7 t^5+88338 u^8 t^4+23800 u^9 t^3+2622 u^{10} t^2+582 u^{11} t-81 \
u^{12}\Big) s^7+\Big(14 t^{13}-572 u t^{12}-3036 u^2 t^{11}-35818 u^3 \
t^{10}-106376 u^4 t^9-101688 u^5 t^8+48822 u^6 t^7+213630 u^7 t^6+239397 u^8 \
t^5+134809 u^9 t^4+31274 u^{10} t^3-36 u^{11} t^2+157 u^{12} t-25 \
u^{13}\Big) s^6-6 t u \Big(8 t^{12}+37 u t^{11}+3213 u^2 t^{10}+11984 u^3 \
t^9+14592 u^4 t^8+1987 u^5 t^7-12914 u^6 t^6-20563 u^7 t^5-20499 u^8 \
t^4-11612 u^9 t^3-2423 u^{10} t^2+147 u^{11} t+3 u^{12}\Big) s^5+t^2 u^2 \
\Big(102 t^{11}-5234 u t^{10}-25530 u^2 t^9-39126 u^3 t^8-12408 u^4 \
t^7+24648 u^5 t^6+35505 u^6 t^5+40505 u^7 t^4+40476 u^8 t^3+21240 u^9 \
t^2+3629 u^{10} t-267 u^{11}\Big) s^4+2 t^3 u^3 \Big(-266 t^{10}-1804 u \
t^9-4752 u^2 t^8-4830 u^3 t^7+378 u^4 t^6+5301 u^5 t^5+6806 u^6 t^4+6178 u^7 \
t^3+3642 u^8 t^2+1119 u^9 t+156 u^{10}\Big) s^3+3 t^4 u^4 (t+u)^2 \Big(16 \
t^7-382 u t^6-532 u^2 t^5+8 u^3 t^4+467 u^4 t^3+637 u^5 t^2+265 u^6 t-71 \
u^7\Big) s^2-2 t^5 u^5 (t+u)^3 \Big(48 t^5+151 u t^4-84 u^2 t^3+18 u^3 \
t^2-160 u^4 t-15 u^5\Big) s+t^6 u^7 (t+u)^3 \Big(12 t^3+3 u t^2+2 u^2 \
t-11 u^3\Big)\Big) H(1,z) H(0,3,y)-3 s^2 (s+t)^2 u^2 (t+u)^2 \
\Big(\Big(259 t^6+616 u t^5+426 u^2 t^4-68 u^3 t^3+109 u^4 t^2+300 u^5 \
t+120 u^6\Big) s^{14}+2 \Big(774 t^7+2870 u t^6+3872 u^2 t^5+1903 u^3 \
t^4+425 u^4 t^3+1339 u^5 t^2+1395 u^6 t+420 u^7\Big) s^{13}+\Big(4367 \
t^8+21396 u t^7+40759 u^2 t^6+37858 u^3 t^5+19215 u^4 t^4+14984 u^5 \
t^3+18489 u^6 t^2+11430 u^7 t+2640 u^8\Big) s^{12}+2 \Big(3840 t^9+23121 \
u t^8+56244 u^2 t^7+72977 u^3 t^6+57956 u^4 t^5+40878 u^5 t^4+39401 u^6 \
t^3+31286 u^7 t^2+13545 u^8 t+2460 u^9\Big) s^{11}+2 \Big(4602 \
t^{10}+33792 u t^9+102876 u^2 t^8+170653 u^3 t^7+175620 u^4 t^6+137297 u^5 \
t^5+118565 u^6 t^4+104385 u^7 t^3+62005 u^8 t^2+20475 u^9 t+3000 \
u^{10}\Big) s^{10}
\end{dmath*}
\begin{dmath*}
{\white=}
+2 \Big(3840 t^{11}+35847 u t^{10}+140082 u^2 \
t^9+290657 u^3 t^8+356593 u^4 t^7+294815 u^5 t^6+230739 u^6 t^5+215383 u^7 \
t^4+164395 u^8 t^3+77358 u^9 t^2+20475 u^{10} t+2460 u^{11}\Big) \
s^9+\Big(4367 t^{12}+55904 u t^{11}+291138 u^2 t^{10}+756156 u^3 \
t^9+1095009 u^4 t^8+963078 u^5 t^7+649134 u^6 t^6+554262 u^7 t^5+519114 u^8 \
t^4+328790 u^9 t^3+124010 u^{10} t^2+27090 u^{11} t+2640 u^{12}\Big) s^8+2 \
\Big(774 t^{13}+15265 u t^{12}+110242 u^2 t^{11}+363648 u^3 t^{10}+642258 \
u^4 t^9+661734 u^5 t^8+447563 u^6 t^7+300068 u^7 t^6+277131 u^8 t^5+215383 \
u^9 t^4+104385 u^{10} t^3+31286 u^{11} t^2+5715 u^{12} t+420 u^{13}\Big) \
s^7+\Big(259 t^{14}+10260 u t^{13}+111068 u^2 t^{12}+486996 u^3 \
t^{11}+1107162 u^4 t^{10}+1479900 u^5 t^9+1291792 u^6 t^8+895126 u^7 \
t^7+649134 u^8 t^6+461478 u^9 t^5+237130 u^{10} t^4+78802 u^{11} t^3+18489 \
u^{12} t^2+2790 u^{13} t+120 u^{14}\Big) s^6+2 t u \Big(777 t^{13}+16074 \
u t^{12}+104819 u^2 t^{11}+330308 u^3 t^{10}+608514 u^4 t^9+739950 u^5 \
t^8+661734 u^6 t^7+481539 u^7 t^6+294815 u^8 t^5+137297 u^9 t^4+40878 u^{10} \
t^3+7492 u^{11} t^2+1339 u^{12} t+150 u^{13}\Big) s^5+t^2 u^2 \Big(3885 \
t^{12}+50940 u t^{11}+249466 u^2 t^{10}+660616 u^3 t^9+1107162 u^4 \
t^8+1284516 u^5 t^7+1095009 u^6 t^6+713186 u^7 t^5+351240 u^8 t^4+115912 u^9 \
t^3+19215 u^{10} t^2+850 u^{11} t+109 u^{12}\Big) s^4+2 t^3 u^3 \Big(2590 \
t^{11}+25470 u t^{10}+104819 u^2 t^9+243498 u^3 t^8+363648 u^4 t^7+378078 \
u^5 t^6+290657 u^6 t^5+170653 u^7 t^4+72977 u^8 t^3
\end{dmath*}

\begin{dmath*}
{\white =}
+18929 u^9 t^2+1903 \
u^{10} t-34 u^{11}\Big) s^3+t^4 u^4 (t+u)^2 \Big(3885 t^8+24378 u \
t^7+58427 u^2 t^6+79252 u^3 t^5+74207 u^4 t^4+52498 u^5 t^3+26549 u^6 \
t^2+6892 u^7 t+426 u^8\Big) s^2+2 t^5 u^5 (t+u)^3 \Big(777 t^6+2799 u \
t^5+4537 u^2 t^4+5167 u^3 t^3+3936 u^4 t^2+1946 u^5 t+308 u^6\Big) s+t^6 \
u^6 (t+u)^4 \Big(259 t^4+512 u t^3+765 u^2 t^2+512 u^3 t+259 \
u^4\Big)\Big) H(2,0,y)-18 s^2 t^2 u^2 (t+u)^2 \Big(\Big(83 t^4+224 u \
t^3+312 u^2 t^2+224 u^3 t+83 u^4\Big) s^{16}+\Big(646 t^5+2236 u t^4+3738 \
u^2 t^3+3738 u^3 t^2+2236 u^4 t+646 u^5\Big) s^{15}+3 \Big(795 t^6+3434 u \
t^5+6735 u^2 t^4+8252 u^3 t^3+6735 u^4 t^2+3434 u^5 t+795 u^6\Big) \
s^{14}+\Big(5590 t^7+29722 u t^6+68154 u^2 t^5+96362 u^3 t^4+96362 u^4 \
t^3+68154 u^5 t^2+29722 u^6 t+5590 u^7\Big) s^{13}+\Big(9297 t^8+59792 u \
t^7+162893 u^2 t^6+265856 u^3 t^5+307366 u^4 t^4+265856 u^5 t^3+162893 u^6 \
t^2+59792 u^7 t+9297 u^8\Big) s^{12}+6 \Big(1912 t^9+14697 u t^8+48146 \
u^2 t^7+92627 u^3 t^6+123322 u^4 t^5+123322 u^5 t^4+92627 u^6 t^3+48146 u^7 \
t^2+14697 u^8 t+1912 u^9\Big) s^{11}+\Big(10571 t^{10}+97460 u t^9+390525 \
u^2 t^8+895950 u^3 t^7+1372200 u^4 t^6+1560372 u^5 t^5+1372200 u^6 \
t^4+895950 u^7 t^3+390525 u^8 t^2+97460 u^9 t+10571 u^{10}\Big) s^{10}+2 \
\Big(3555 t^{11}+40253 u t^{10}+202415 u^2 t^9+561210 u^3 t^8+989305 u^4 \
t^7+1257298 u^5 t^6+1257298 u^6 t^5+989305 u^7 t^4+561210 u^8 t^3+202415 u^9 \
t^2+40253 u^{10} t+3555 u^{11}\Big) s^9+3 \Big(1100 t^{12}+16144 u \
t^{11}+105097 u^2 t^{10}+356156 u^3 t^9+736699 u^4 t^8+1061804 u^5 \
t^7+1182566 u^6 t^6+1061804 u^7 t^5+736699 u^8 t^4+356156 u^9 t^3+105097 \
u^{10} t^2+16144 u^{11} t+1100 u^{12}\Big) s^8+2 \Big(471 t^{13}+10059 u \
t^{12}
\end{dmath*}

\intertext{}
\begin{dmath*}
{\white =}
+89007 u^2 t^{11}+370856 u^3 t^{10}+918809 u^4 t^9+1568172 u^5 \
t^8+2010976 u^6 t^7+2010976 u^7 t^6+1568172 u^8 t^5+918809 u^9 t^4+370856 \
u^{10} t^3+89007 u^{11} t^2+10059 u^{12} t+471 u^{13}\Big) s^7+\Big(124 \
t^{14}+5166 u t^{13}+68942 u^2 t^{12}+360694 u^3 t^{11}+1093775 u^4 \
t^{10}+2281660 u^5 t^9+3515021 u^6 t^8+4060864 u^7 t^7+3515021 u^8 \
t^6+2281660 u^9 t^5+1093775 u^{10} t^4+360694 u^{11} t^3+68942 u^{12} \
t^2+5166 u^{13} t+124 u^{14}\Big) s^6+6 t u \Big(102 t^{13}+2718 u \
t^{12}+19299 u^2 t^{11}+75952 u^3 t^{10}+198691 u^4 t^9+370135 u^5 \
t^8+503605 u^6 t^7+503605 u^7 t^6+370135 u^8 t^5+198691 u^9 t^4+75952 u^{10} \
t^3+19299 u^{11} t^2+2718 u^{12} t+102 u^{13}\Big) s^5+t^2 u^2 \Big(1752 \
t^{12}+21508 u t^{11}+127396 u^2 t^{10}+438996 u^3 t^9+986209 u^4 \
t^8+1559032 u^5 t^7+1808388 u^6 t^6+1559032 u^7 t^5+986209 u^8 t^4+438996 \
u^9 t^3+127396 u^{10} t^2+21508 u^{11} t+1752 u^{12}\Big) s^4+2 t^3 u^3 \
\Big(834 t^{11}+10755 u t^{10}+52845 u^2 t^9+146657 u^3 t^8+271322 u^4 \
t^7+363015 u^5 t^6+363015 u^6 t^5+271322 u^7 t^4+146657 u^8 t^3+52845 u^9 \
t^2+10755 u^{10} t+834 u^{11}\Big) s^3+6 t^4 u^4 (t+u)^2 \Big(283 \
t^8+1733 u t^7+4779 u^2 t^6+8360 u^3 t^5+10092 u^4 t^4+8360 u^5 t^3+4779 u^6 \
t^2+1733 u^7 t+283 u^8\Big) s^2+2 t^5 u^5 (t+u)^3 \Big(282 t^6+1178 u \
t^5+2417 u^2 t^4+3162 u^3 t^3+2417 u^4 t^2+1178 u^5 t+282 u^6\Big) s+110 \
t^6 u^6 (t+u)^4 \Big(t^2+u t+u^2\Big)^2\Big) H(1,z) H(2,3,y)+9 s^2 \
(t+u)^2 \Big(2 \Big(20 t^8+50 u t^7+21 u^2 t^6-82 u^3 t^5-102 u^4 t^4-82 \
u^5 t^3+21 u^6 t^2+50 u^7 t+20 u^8\Big) s^{16}+2 \Big(140 t^9+505 u \
t^8+568 u^2 t^7-298 u^3 t^6-1069 u^4 t^5-1069 u^5 t^4-298 u^6 t^3+568 u^7 \
t^2+505 u^8 t+140 u^9\Big) s^{15}+\Big(880 t^{10}+4370 u t^9
\end{dmath*}

\begin{dmath*}
{\white =}
+8221 u^2 \
t^8+4262 u^3 t^7-6199 u^4 t^6-10932 u^5 t^5-6199 u^6 t^4+4262 u^7 t^3+8221 \
u^8 t^2+4370 u^9 t+880 u^{10}\Big) s^{14}+2 \Big(820 t^{11}+5395 u \
t^{10}+14563 u^2 t^9+17988 u^3 t^8+5650 u^4 t^7-8508 u^5 t^6-8508 u^6 \
t^5+5650 u^7 t^4+17988 u^8 t^3+14563 u^9 t^2+5395 u^{10} t+820 u^{11}\Big) \
s^{13}+\Big(2000 t^{12}+16930 u t^{11}+60889 u^2 t^{10}+115200 u^3 \
t^9+112888 u^4 t^8+52018 u^5 t^7+18062 u^6 t^6+52018 u^7 t^5+112888 u^8 \
t^4+115200 u^9 t^3+60889 u^{10} t^2+16930 u^{11} t+2000 u^{12}\Big) \
s^{12}+2 \Big(820 t^{13}+8825 u t^{12}+40630 u^2 t^{11}+106430 u^3 \
t^{10}+162506 u^4 t^9+145433 u^5 t^8+94054 u^6 t^7+94054 u^7 t^6+145433 u^8 \
t^5+162506 u^9 t^4+106430 u^{10} t^3+40630 u^{11} t^2+8825 u^{12} t+820 \
u^{13}\Big) s^{11}+\Big(880 t^{14}+12310 u t^{13}+71305 u^2 t^{12}+252860 \
u^3 t^{11}+542940 u^4 t^{10}+672624 u^5 t^9+495141 u^6 t^8+346704 u^7 \
t^7+495141 u^8 t^6+672624 u^9 t^5+542940 u^{10} t^4+252860 u^{11} t^3+71305 \
u^{12} t^2+12310 u^{13} t+880 u^{14}\Big) s^{10}+2 \Big(140 t^{15}+2785 u \
t^{14}+20491 u^2 t^{13}+99450 u^3 t^{12}+294436 u^4 t^{11}+490859 u^5 \
t^{10}+448931 u^6 t^9+254956 u^7 t^8+254956 u^8 t^7+448931 u^9 t^6+490859 \
u^{10} t^5+294436 u^{11} t^4+99450 u^{12} t^3+20491 u^{13} t^2+2785 u^{14} \
t+140 u^{15}\Big) s^9+\Big(40 t^{16}+1490 u t^{15}+14845 u^2 \
t^{14}+102820 u^3 t^{13}+419280 u^4 t^{12}+950788 u^5 t^{11}+1225773 u^6 \
t^{10}+975710 u^7 t^9+743060 u^8 t^8+975710 u^9 t^7+1225773 u^{10} \
t^6+950788 u^{11} t^5+419280 u^{12} t^4+102820 u^{13} t^3+14845 u^{14} \
t^2+1490 u^{15} t+40 u^{16}\Big) s^8+2 t u \Big(90 t^{15}+1540 u \
t^{14}+16932 u^2 t^{13}+96123 u^3 t^{12}+304282 u^4 t^{11}+583798 u^5 \
t^{10}+762212 u^6 t^9+803305 u^7 t^8+803305 u^8 t^7+762212 u^9 t^6+583798 \
u^{10} t^5+304282 u^{11} t^4+96123 u^{12} t^3+16932 u^{13} t^2+1540 u^{14} \
t+90 u^{15}\Big) s^7+t^2 u^2 \Big(282 t^{14}
\end{dmath*}
\begin{dmath*}
{\white=}
+6622 u t^{13}+54887 u^2 \
t^{12}+264008 u^3 t^{11}+776536 u^4 t^{10}+1517240 u^5 t^9+2171441 u^6 \
t^8+2431080 u^7 t^7+2171441 u^8 t^6+1517240 u^9 t^5+776536 u^{10} t^4+264008 \
u^{11} t^3+54887 u^{12} t^2+6622 u^{13} t+282 u^{14}\Big) s^6+2 t^3 u^3 \
\Big(298 t^{13}+4418 u t^{12}+40731 u^2 t^{11}+190625 u^3 t^{10}+505964 u^4 \
t^9+877790 u^5 t^8+1117090 u^6 t^7+1117090 u^7 t^6+877790 u^8 t^5+505964 u^9 \
t^4+190625 u^{10} t^3+40731 u^{11} t^2+4418 u^{12} t+298 u^{13}\Big) \
s^5+t^4 u^4 \Big(536 t^{12}+16810 u t^{11}+131975 u^2 t^{10}+463360 u^3 \
t^9+936996 u^4 t^8+1292362 u^5 t^7+1406138 u^6 t^6+1292362 u^7 t^5+936996 \
u^8 t^4+463360 u^9 t^3+131975 u^{10} t^2+16810 u^{11} t+536 u^{12}\Big) \
s^4+2 t^5 u^5 \Big(848 t^{11}+13094 u t^{10}+64600 u^2 t^9+162207 u^3 \
t^8+255217 u^4 t^7+298424 u^5 t^6+298424 u^6 t^5+255217 u^7 t^4+162207 u^8 \
t^3+64600 u^9 t^2+13094 u^{10} t+848 u^{11}\Big) s^3+t^6 u^6 (t+u)^2 \
\Big(2062 t^8+13122 u t^7+32971 u^2 t^6+48846 u^3 t^5+54292 u^4 t^4+48846 \
u^5 t^3+32971 u^6 t^2+13122 u^7 t+2062 u^8\Big) s^2+4 t^7 u^7 (t+u)^3 \
\Big(158 t^6+746 u t^5+1500 u^2 t^4+1923 u^3 t^3+1500 u^4 t^2+746 u^5 t+158 \
u^6\Big) s+8 t^8 u^8 (t+u)^4 \Big(23 t^4+45 u t^3+67 u^2 t^2+45 u^3 t+23 \
u^4\Big)\Big) H(3,2,y)
\end{dmath*}

\begin{dmath*}
{\white =}
-36 s^2 t^2 u^3 (t+u)^2 \Big(-12 t \Big(4 t^2+7 \
u t+4 u^2\Big) s^{16}-\Big(413 t^4+1112 u t^3+1066 u^2 t^2+344 u^3 t-23 \
u^4\Big) s^{15}-3 \Big(549 t^5+2080 u t^4+2910 u^2 t^3+1760 u^3 t^2+293 \
u^4 t-64 u^5\Big) s^{14}-\Big(3849 t^6+19818 u t^5+38435 u^2 t^4+35276 \
u^3 t^3+14007 u^4 t^2+482 u^5 t-715 u^6\Big) s^{13}-\Big(5797 t^7+40039 u \
t^6+104248 u^2 t^5+133574 u^3 t^4+86245 u^4 t^3+21235 u^5 t^2-2834 u^6 \
t-1572 u^7\Big) s^{12}-3 \Big(1839 t^8+18438 u t^7+63926 u^2 t^6+107954 \
u^3 t^5+97302 u^4 t^4+43534 u^5 t^3+4746 u^6 t^2-2870 u^7 t-757 u^8\Big) \
s^{11}-\Big(2469 t^9+52233 u t^8+250929 u^2 t^7+546879 u^3 t^6+638973 u^4 \
t^5+405133 u^5 t^4+113599 u^6 t^3-8307 u^7 t^2-12530 u^8 t-2258 u^9\Big) \
s^{10}+\Big(985 t^{10}-31352 u t^9-228083 u^2 t^8-646444 u^3 t^7-969791 u^4 \
t^6-819288 u^5 t^5-357472 u^6 t^4-38248 u^7 t^3+28876 u^8 t^2+11404 u^9 \
t+1557 u^{10}\Big) s^9+3 \Big(727 t^{11}-3289 u t^{10}-45068 u^2 \
t^9-168873 u^3 t^8-330122 u^4 t^7-375394 u^5 t^6-241202 u^6 t^5-66685 u^7 \
t^4+10259 u^8 t^3+10981 u^9 t^2+2326 u^{10} t+240 u^{11}\Big) \
s^8+\Big(1386 t^{12}-6 u t^{11}-44480 u^2 t^{10}-241022 u^3 t^9-645063 u^4 \
t^8-1001980 u^5 t^7-922078 u^6 t^6-455700 u^7 t^5-58195 u^8 t^4+50450 u^9 \
t^3+22836 u^{10} t^2+2866 u^{11} t+202 u^{12}\Big) s^7+\Big(420 \
t^{13}+1149 u t^{12}+3 u^2 t^{11}-48559 u^3 t^{10}-253354 u^4 t^9-575952 u^5 \
t^8-717884 u^6 t^7-512534 u^7 t^6-180305 u^8 t^5+7313 u^9 t^4+31999 u^{10} \
t^3+10093 u^{11} t^2+729 u^{12} t+26 u^{13}\Big) s^6+3 t \Big(16 \
t^{13}+94 u t^{12}+2346 u^2 t^{11}+5138 u^3 t^{10}-15062 u^4 t^9-74024 u^5 \
t^8-122999 u^6 t^7-106460 u^7 t^6-49523 u^8 t^5-8048 u^9 t^4+4593 u^{10} \
t^3+3534 u^{11} t^2+893 u^{12} t+30 u^{13}\Big) s^5+t^2 u \Big(-6 \
t^{12}+2684 u t^{11}+12044 u^2 t^{10}+7299 u^3 t^9-50331 u^4 t^8-127806 u^5 \
t^7-131304 u^6 t^6-58843 u^7 t^5+1403 u^8 t^4+13260 u^9 t^3+6238 u^{10} \
t^2+1774 u^{11} t+336 u^{12}\Big) s^4+2 t^3 u^2 \Big(166 t^{11}+1146 u \
t^{10}+2486 u^2 t^9-448 u^3 t^8-9790 u^4 t^7-15423 u^5 t^6-8416 u^6 t^5+3041 \
u^7 t^4
\end{dmath*}

\intertext{}
\begin{dmath*}
{\white =}
+7002 u^8 t^3+4035 u^9 t^2+960 u^{10} t+57 u^{11}\Big) s^3+3 t^4 \
u^3 (t+u)^2 \Big(15 t^8+240 u t^7+369 u^2 t^6-u^3 t^5-172 u^4 t^4+126 u^5 \
t^3+429 u^6 t^2+355 u^7 t+103 u^8\Big) s^2+t^5 u^4 (t+u)^3 \Big(90 \
t^6+272 u t^5+116 u^2 t^4+114 u^3 t^3+135 u^4 t^2+117 u^5 t+66 u^6\Big) \
s+t^6 u^5 (t+u)^4 \Big(11 t^4-2 u t^3+15 u^2 t^2+14 u^3 t+19 \
u^4\Big)\Big) H(0,0,1,z)+18 s^2 t^2 (s+t)^2 u^2 \Big(2 \Big(55 t^6+282 \
u t^5+849 u^2 t^4+834 u^3 t^3+876 u^4 t^2+306 u^5 t+62 u^6\Big) s^{14}+2 \
\Big(330 t^7+2024 u t^6+6897 u^2 t^5+10755 u^3 t^4+10754 u^4 t^3+8154 u^5 \
t^2+2583 u^6 t+471 u^7\Big) s^{13}+2 \Big(935 t^8+6797 u t^7+25584 u^2 \
t^6+52845 u^3 t^5+63698 u^4 t^4+57897 u^5 t^3+34471 u^6 t^2+10059 u^7 t+1650 \
u^8\Big) s^{12}+2 \Big(1650 t^9+14229 u t^8+58953 u^2 t^7+146657 u^3 \
t^6+219498 u^4 t^5+227856 u^5 t^4+180347 u^6 t^3+89007 u^7 t^2+24216 u^8 \
t+3555 u^9\Big) s^{11}+\Big(3960 t^{10}+40664 u t^9+189546 u^2 t^8+542644 \
u^3 t^7+986209 u^4 t^6+1192146 u^5 t^5+1093775 u^6 t^4+741712 u^7 t^3+315291 \
u^8 t^2+80506 u^9 t+10571 u^{10}\Big) s^{10}+2 \Big(1650 t^{11}+20332 u \
t^{10}+110712 u^2 t^9+363015 u^3 t^8+779516 u^4 t^7+1110405 u^5 t^6+1140830 \
u^6 t^5+918809 u^7 t^4+534234 u^8 t^3+202415 u^9 t^2+48730 u^{10} t+5736 \
u^{11}\Big) s^9+\Big(1870 t^{12}+28458 u t^{11}+189546 u^2 t^{10}+726030 \
u^3 t^9+1808388 u^4 t^8+3021630 u^5 t^7+3515021 u^6 t^6+3136344 u^7 \
t^5+2210097 u^8 t^4+1122420 u^9 t^3+390525 u^{10} t^2+88182 u^{11} t+9297 \
u^{12}\Big) s^8+2 \Big(330 t^{13}+6797 u t^{12}+58953 u^2 t^{11}+271322 \
u^3 t^{10}+779516 u^4 t^9+1510815 u^5 t^8+2030432 u^6 t^7+2010976 u^7 \
t^6+1592706 u^8 t^5+989305 u^9 t^4+447975 u^{10} t^3+144438 u^{11} t^2+29896 \
u^{12} t+2795 u^{13}\Big) s^7+\Big(110 t^{14}+4048 u t^{13}+51168 u^2 \
t^{12}+293314 u^3 t^{11}+986209 u^4 t^{10}+2220810 u^5 t^9+3515021 u^6 \
t^8+4021952 u^7 t^7
\end{dmath*}

\begin{dmath*}
{\white =}
+3547698 u^8 t^6+2514596 u^9 t^5+1372200 u^{10} \
t^4+555762 u^{11} t^3+162893 u^{12} t^2+29722 u^{13} t+2385 u^{14}\Big) \
s^6+2 u \Big(282 t^{14}+6897 u t^{13}+52845 u^2 t^{12}+219498 u^3 \
t^{11}+596073 u^4 t^{10}+1140830 u^5 t^9+1568172 u^6 t^8+1592706 u^7 \
t^7+1257298 u^8 t^6+780186 u^9 t^5+369966 u^{10} t^4+132928 u^{11} t^3+34077 \
u^{12} t^2+5151 u^{13} t+323 u^{14}\Big) s^5+u^2 \Big(1698 t^{14}+21510 u \
t^{13}+127396 u^2 t^{12}+455712 u^3 t^{11}+1093775 u^4 t^{10}+1837618 u^5 \
t^9+2210097 u^6 t^8+1978610 u^7 t^7+1372200 u^8 t^6+739932 u^9 t^5+307366 \
u^{10} t^4+96362 u^{11} t^3+20205 u^{12} t^2+2236 u^{13} t+83 u^{14}\Big) \
s^4+2 t u^3 \Big(834 t^{13}+10754 u t^{12}+57897 u^2 t^{11}+180347 u^3 \
t^{10}+370856 u^4 t^9+534234 u^5 t^8+561210 u^6 t^7+447975 u^7 t^6+277881 \
u^8 t^5+132928 u^9 t^4+48181 u^{10} t^3+12378 u^{11} t^2+1869 u^{12} t+112 \
u^{13}\Big) s^3+t^2 u^4 (t+u)^2 \Big(1752 t^{10}+12804 u t^9+41582 u^2 \
t^8+82046 u^3 t^7+109617 u^4 t^6+103550 u^5 t^5+73808 u^6 t^4+37710 u^7 \
t^3+13665 u^8 t^2+3114 u^9 t+312 u^{10}\Big) s^2+2 t^3 u^5 (t+u)^3 \
\Big(306 t^8+1665 u t^7+4146 u^2 t^6+6477 u^3 t^5+6719 u^4 t^4+4996 u^5 \
t^3+2469 u^6 t^2+782 u^7 t+112 u^8\Big) s+t^4 u^6 (t+u)^6 \Big(124 \
t^4+198 u t^3+252 u^2 t^2+148 u^3 t+83 u^4\Big)\Big) H(0,1,0,y)-18 s^2 \
t^2 u^2 (t+u)^3 \Big(2 t \Big(7 t^2+15 u t+12 u^2\Big) s^{16}+12 \Big(8 \
t^4+24 u t^3+29 u^2 t^2+12 u^3 t-u^4\Big) s^{15}+3 \Big(101 t^5+423 u \
t^4+720 u^2 t^3+556 u^3 t^2+111 u^4 t-31 u^5\Big) s^{14}+\Big(598 \
t^6+3066 u t^5+7134 u^2 t^4+8992 u^3 t^3+5154 u^4 t^2+510 u^5 t-318 \
u^6\Big) s^{13}+\Big(851 t^7+4495 u t^6+13290 u^2 t^5+25506 u^3 t^4+26949 \
u^4 t^3+12309 u^5 t^2+688 u^6 t-652 u^7\Big) s^{12}+6 \Big(158 t^8+586 u \
t^7+2628 u^2 t^6+7708 u^3 t^5+11649 u^4 t^4+9092 u^5 t^3+3324 u^6 t^2+218 \
u^7 t-155 u^8\Big) s^{11}+\Big(851 t^9-417 u t^8+9564 u^2 t^7+62150 u^3 \
t^6
\end{dmath*}
\begin{dmath*}
{\white=}
+123813 u^4 t^5+121269 u^5 t^4+67982 u^6 t^3+23208 u^7 t^2+2838 u^8 \
t-1002 u^9\Big) s^{10}+2 \Big(299 t^{10}-2075 u t^9-1896 u^2 t^8+23751 \
u^3 t^7+72903 u^4 t^6+97515 u^5 t^5+71423 u^6 t^4+31963 u^7 t^3+10800 u^8 \
t^2+1966 u^9 t-409 u^{10}\Big) s^9+3 \Big(101 t^{11}-1463 u t^{10}-3896 \
u^2 t^9+208 u^3 t^8+20429 u^4 t^7+58669 u^5 t^6+80268 u^6 t^5+55172 u^7 \
t^4+19400 u^8 t^3+4832 u^9 t^2+1000 u^{10} t-156 u^{11}\Big) s^8+2 \
\Big(48 t^{12}-1128 u t^{11}-4410 u^2 t^{10}-17090 u^3 t^9-31983 u^4 \
t^8+6084 u^5 t^7+100874 u^6 t^6+143160 u^7 t^5+88338 u^8 t^4+23800 u^9 \
t^3+2622 u^{10} t^2+582 u^{11} t-81 u^{12}\Big) s^7+\Big(14 t^{13}-572 u \
t^{12}-3036 u^2 t^{11}-35818 u^3 t^{10}-106376 u^4 t^9-101688 u^5 t^8+48822 \
u^6 t^7+213630 u^7 t^6+239397 u^8 t^5+134809 u^9 t^4+31274 u^{10} t^3-36 \
u^{11} t^2+157 u^{12} t-25 u^{13}\Big) s^6-6 t u \Big(8 t^{12}+37 u \
t^{11}+3213 u^2 t^{10}+11984 u^3 t^9+14592 u^4 t^8+1987 u^5 t^7-12914 u^6 \
t^6-20563 u^7 t^5-20499 u^8 t^4-11612 u^9 t^3-2423 u^{10} t^2+147 u^{11} t+3 \
u^{12}\Big) s^5+t^2 u^2 \Big(102 t^{11}-5234 u t^{10}-25530 u^2 t^9-39126 \
u^3 t^8-12408 u^4 t^7+24648 u^5 t^6+35505 u^6 t^5+40505 u^7 t^4+40476 u^8 \
t^3+21240 u^9 t^2+3629 u^{10} t-267 u^{11}\Big) s^4+2 t^3 u^3 \Big(-266 \
t^{10}-1804 u t^9-4752 u^2 t^8-4830 u^3 t^7+378 u^4 t^6+5301 u^5 t^5+6806 \
u^6 t^4+6178 u^7 t^3+3642 u^8 t^2+1119 u^9 t+156 u^{10}\Big) s^3+3 t^4 u^4 \
(t+u)^2 \Big(16 t^7-382 u t^6-532 u^2 t^5+8 u^3 t^4+467 u^4 t^3+637 u^5 \
t^2+265 u^6 t-71 u^7\Big) s^2-2 t^5 u^5 (t+u)^3 \Big(48 t^5+151 u t^4-84 \
u^2 t^3+18 u^3 t^2-160 u^4 t-15 u^5\Big) s+t^6 u^7 (t+u)^3 \Big(12 t^3+3 \
u t^2+2 u^2 t-11 u^3\Big)\Big) H(0,3,2,y)-18 s^2 t^2 u^2 (t+u)^2 \
\Big(\Big(55 t^4
\end{dmath*}

\begin{dmath*}
{\white =}
+148 u t^3+192 u^2 t^2+128 u^3 t+47 u^4\Big) s^{16}+2 \
\Big(232 t^5+807 u t^4+1262 u^2 t^3+1126 u^3 t^2+612 u^4 t+173 u^5\Big) \
s^{15}+6 \Big(305 t^6+1324 u t^5+2458 u^2 t^4+2648 u^3 t^3+1862 u^4 t^2+872 \
u^5 t+201 u^6\Big) s^{14}+2 \Big(2230 t^7+11932 u t^6+26409 u^2 t^5+33217 \
u^3 t^4+27682 u^4 t^3+16830 u^5 t^2+7005 u^6 t+1347 u^7\Big) \
s^{13}+\Big(7450 t^8+48274 u t^7+129744 u^2 t^6+195540 u^3 t^5+190555 u^4 \
t^4+134098 u^5 t^3+72590 u^6 t^2+26532 u^7 t+4375 u^8\Big) s^{12}+6 \
\Big(1486 t^9+11488 u t^8+37756 u^2 t^7+69937 u^3 t^6+82622 u^4 t^5+67718 \
u^5 t^4+41832 u^6 t^3+19931 u^7 t^2+6288 u^8 t+910 u^9\Big) \
s^{11}+\Big(7744 t^{10}+71242 u t^9+290076 u^2 t^8+661446 u^3 t^7+946629 \
u^4 t^6+926538 u^5 t^5+671989 u^6 t^4+379294 u^7 t^3+157923 u^8 t^2+41472 \
u^9 t+5271 u^{10}\Big) s^{10}+2 \Big(2410 t^{11}+26794 u t^{10}+138966 \
u^2 t^9+391138 u^3 t^8+666867 u^4 t^7+763472 u^5 t^6+652628 u^6 t^5+448145 \
u^7 t^4+237429 u^8 t^3+84416 u^9 t^2+17568 u^{10} t+1903 u^{11}\Big) s^9+3 \
\Big(685 t^{12}+9638 u t^{11}+66160 u^2 t^{10}+229988 u^3 t^9+468878 u^4 \
t^8+632074 u^5 t^7+626493 u^6 t^6+499908 u^7 t^5+326013 u^8 t^4+156448 u^9 \
t^3+46699 u^{10} t^2+7420 u^{11} t+638 u^{12}\Big) s^8+2 \Big(270 \
t^{13}+5385 u t^{12}+51810 u^2 t^{11}+219533 u^3 t^{10}+538700 u^4 \
t^9+893700 u^5 t^8+1082558 u^6 t^7+993074 u^7 t^6+712998 u^8 t^5+405820 u^9 \
t^4+169330 u^{10} t^3+42843 u^{11} t^2+4984 u^{12} t+295 u^{13}\Big) \
s^7+\Big(66 t^{14}+2518 u t^{13}+38032 u^2 t^{12}+195010 u^3 t^{11}+579501 \
u^4 t^{10}
\end{dmath*}

\intertext{}
\begin{dmath*}
{\white =}
+1219748 u^5 t^9+1901521 u^6 t^8+2145380 u^7 t^7+1729402 u^8 \
t^6+1029790 u^9 t^5+480198 u^{10} t^4+170114 u^{11} t^3+36457 u^{12} \
t^2+2808 u^{13} t+83 u^{14}\Big) s^6+6 t u \Big(46 t^{13}+1463 u \
t^{12}+9551 u^2 t^{11}+36573 u^3 t^{10}+99722 u^4 t^9+196920 u^5 t^8+275717 \
u^6 t^7+269768 u^7 t^6+184750 u^8 t^5+90822 u^9 t^4+33632 u^{10} t^3+9403 \
u^{11} t^2+1592 u^{12} t+61 u^{13}\Big) s^5+t^2 u^2 \Big(936 t^{12}+9592 \
u t^{11}+58498 u^2 t^{10}+216504 u^3 t^9+517783 u^4 t^8+852472 u^5 \
t^7+999621 u^6 t^6+842754 u^7 t^5+506934 u^8 t^4+212352 u^9 t^3+59561 u^{10} \
t^2+10778 u^{11} t+1137 u^{12}\Big) s^4+2 t^3 u^3 \Big(306 t^{11}+5120 u \
t^{10}+27107 u^2 t^9+77806 u^3 t^8+147341 u^4 t^7+200386 u^5 t^6+201078 u^6 \
t^5+148590 u^7 t^4+78861 u^8 t^3+28022 u^9 t^2+5655 u^{10} t+424 \
u^{11}\Big) s^3+3 t^4 u^4 (t+u)^2 \Big(312 t^8+1862 u t^7+5114 u^2 \
t^6+9264 u^3 t^5+11523 u^4 t^4+9432 u^5 t^3+5226 u^6 t^2+1942 u^7 t+361 \
u^8\Big) s^2+2 t^5 u^5 (t+u)^3 \Big(138 t^6+635 u t^5+1406 u^2 t^4+1938 \
u^3 t^3+1431 u^4 t^2+633 u^5 t+159 u^6\Big) s+3 t^6 u^6 (t+u)^4 \Big(22 \
t^4+48 u t^3+71 u^2 t^2+46 u^3 t+23 u^4\Big)\Big) H(1,0,1,z)-18 s^2 t^2 \
(s+t)^3 u^2 \Big(\Big(11 t^6-30 u t^5+213 u^2 t^4-312 u^3 t^3+267 u^4 \
t^2+18 u^5 t+25 u^6\Big) s^{13}+\Big(31 t^7-410 u t^6-369 u^2 t^5-2238 \
u^3 t^4-3629 u^4 t^3+882 u^5 t^2-157 u^6 t+162 u^7\Big) s^{12}+6 \Big(4 \
t^8-169 u t^7-548 u^2 t^6-1214 u^3 t^5-3540 u^4 t^4-2423 u^5 t^3+6 u^6 \
t^2-194 u^7 t+78 u^8\Big) s^{11}-2 \Big(8 t^9+525 u t^8+3009 u^2 t^7+6178 \
u^3 t^6+20238 u^4 t^5+34836 u^5 t^4+15637 u^6 t^3+2622 u^7 t^2+1500 u^8 \
t-409 u^9\Big) s^{10}-\Big(47 t^{10}+414 u t^9+4737 u^2 t^8+13612 u^3 \
t^7+40505 u^4 t^6+122994 u^5 t^5+134809 u^6 t^4+47600 u^7 t^3+14496 u^8 \
t^2+3932 u^9 t-1002 u^{10}\Big) s^9-3 \Big(13 t^{11}-178 u t^{10}-49 u^2 \
t^9+3534 u^3 t^8+11835 u^4 t^7+41126 u^5 t^6+79799 u^6 t^5+58892 u^7 \
t^4+19400 u^8 t^3+7200 u^9 t^2
\end{dmath*}

\begin{dmath*}
{\white =}
+946 u^{10} t-310 u^{11}\Big) s^8-2 \Big(6 \
t^{12}-513 u t^{11}-2157 u^2 t^{10}+378 u^3 t^9+12324 u^4 t^8+38742 u^5 \
t^7+106815 u^6 t^6+143160 u^7 t^5+82758 u^8 t^4+31963 u^9 t^3+11604 u^{10} \
t^2+654 u^{11} t-326 u^{12}\Big) s^7+2 u \Big(295 t^{12}+1920 u \
t^{11}+4830 u^2 t^{10}+6204 u^3 t^9+5961 u^4 t^8-24411 u^5 t^7-100874 u^6 \
t^6-120402 u^7 t^5-71423 u^8 t^4-33991 u^9 t^3-9972 u^{10} t^2-344 u^{11} \
t+159 u^{12}\Big) s^6+3 u \Big(32 t^{13}+350 u t^{12}+3168 u^2 \
t^{11}+13042 u^3 t^{10}+29184 u^4 t^9+33896 u^5 t^8-4056 u^6 t^7-58669 u^7 \
t^6-65010 u^8 t^5-40423 u^9 t^4-18184 u^{10} t^3-4103 u^{11} t^2-170 u^{12} \
t+31 u^{13}\Big) s^5+u^2 \Big(-48 t^{13}+3608 u t^{12}+25530 u^2 \
t^{11}+71904 u^3 t^{10}+106376 u^4 t^9+63966 u^5 t^8-61287 u^6 t^7-145806 \
u^7 t^6-123813 u^8 t^5-69894 u^9 t^4-26949 u^{10} t^3-5154 u^{11} t^2-333 \
u^{12} t+12 u^{13}\Big) s^4+2 t u^3 \Big(266 t^{12}+2617 u t^{11}+9639 \
u^2 t^{10}+17909 u^3 t^9+17090 u^4 t^8-312 u^5 t^7-23751 u^6 t^6-31075 u^7 \
t^5-23124 u^8 t^4-12753 u^9 t^3-4496 u^{10} t^2-834 u^{11} t-72 \
u^{12}\Big) s^3-6 t u^4 (t+u)^2 \Big(17 t^{10}-71 u t^9-381 u^2 t^8-637 \
u^3 t^7-293 u^4 t^6+591 u^5 t^5+705 u^6 t^4+627 u^7 t^3+256 u^8 t^2+50 u^9 \
t+4 u^{10}\Big) s^2+t^2 u^5 (t+u)^3 \Big(48 t^8+428 u t^7+828 u^2 t^6+573 \
u^3 t^5-481 u^4 t^4-687 u^5 t^3-585 u^6 t^2-198 u^7 t-30 u^8\Big) s-t^3 \
u^6 (t+u)^6 \Big(14 t^4+12 u t^3+21 u^2 t^2+12 u^3 t+14 u^4\Big)\Big) \
H(2,1,0,y)-18 s^2 t^2 u^2 (t+u)^2 \Big(\Big(83 t^4+224 u t^3+312 u^2 \
t^2+224 u^3 t+83 u^4\Big) s^{16}+\Big(646 t^5+2236 u t^4+3738 u^2 \
t^3+3738 u^3 t^2+2236 u^4 t+646 u^5\Big) s^{15}+3 \Big(795 t^6+3434 u \
t^5+6735 u^2 t^4+8252 u^3 t^3+6735 u^4 t^2+3434 u^5 t+795 u^6\Big) \
s^{14}+\Big(5590 t^7+29722 u t^6+68154 u^2 t^5+96362 u^3 t^4+96362 u^4 \
t^3+68154 u^5 t^2+29722 u^6 t+5590 u^7\Big) s^{13}+\Big(9297 t^8
\end{dmath*}
\begin{dmath*}
{\white=}
+59792 u \
t^7+162893 u^2 t^6+265856 u^3 t^5+307366 u^4 t^4+265856 u^5 t^3+162893 u^6 \
t^2+59792 u^7 t+9297 u^8\Big) s^{12}+6 \Big(1912 t^9+14697 u t^8+48146 \
u^2 t^7+92627 u^3 t^6+123322 u^4 t^5+123322 u^5 t^4+92627 u^6 t^3+48146 u^7 \
t^2+14697 u^8 t+1912 u^9\Big) s^{11}+\Big(10571 t^{10}+97460 u t^9+390525 \
u^2 t^8+895950 u^3 t^7+1372200 u^4 t^6+1560372 u^5 t^5+1372200 u^6 \
t^4+895950 u^7 t^3+390525 u^8 t^2+97460 u^9 t+10571 u^{10}\Big) s^{10}+2 \
\Big(3555 t^{11}+40253 u t^{10}+202415 u^2 t^9+561210 u^3 t^8+989305 u^4 \
t^7+1257298 u^5 t^6+1257298 u^6 t^5+989305 u^7 t^4+561210 u^8 t^3+202415 u^9 \
t^2+40253 u^{10} t+3555 u^{11}\Big) s^9+3 \Big(1100 t^{12}+16144 u \
t^{11}+105097 u^2 t^{10}+356156 u^3 t^9+736699 u^4 t^8+1061804 u^5 \
t^7+1182566 u^6 t^6+1061804 u^7 t^5+736699 u^8 t^4+356156 u^9 t^3+105097 \
u^{10} t^2+16144 u^{11} t+1100 u^{12}\Big) s^8+2 \Big(471 t^{13}+10059 u \
t^{12}+89007 u^2 t^{11}+370856 u^3 t^{10}+918809 u^4 t^9+1568172 u^5 \
t^8+2010976 u^6 t^7+2010976 u^7 t^6+1568172 u^8 t^5+918809 u^9 t^4+370856 \
u^{10} t^3+89007 u^{11} t^2+10059 u^{12} t+471 u^{13}\Big) s^7+\Big(124 \
t^{14}+5166 u t^{13}+68942 u^2 t^{12}+360694 u^3 t^{11}+1093775 u^4 \
t^{10}+2281660 u^5 t^9+3515021 u^6 t^8+4060864 u^7 t^7+3515021 u^8 \
t^6+2281660 u^9 t^5+1093775 u^{10} t^4+360694 u^{11} t^3
\end{dmath*}
\begin{dmath*}
{\white=}
+68942 u^{12} \
t^2+5166 u^{13} t+124 u^{14}\Big) s^6+6 t u \Big(102 t^{13}+2718 u \
t^{12}+19299 u^2 t^{11}+75952 u^3 t^{10}+198691 u^4 t^9+370135 u^5 \
t^8+503605 u^6 t^7+503605 u^7 t^6+370135 u^8 t^5+198691 u^9 t^4+75952 u^{10} \
t^3+19299 u^{11} t^2+2718 u^{12} t+102 u^{13}\Big) s^5+t^2 u^2 \Big(1752 \
t^{12}+21508 u t^{11}+127396 u^2 t^{10}+438996 u^3 t^9+986209 u^4 \
t^8+1559032 u^5 t^7+1808388 u^6 t^6+1559032 u^7 t^5+986209 u^8 t^4+438996 \
u^9 t^3+127396 u^{10} t^2+21508 u^{11} t+1752 u^{12}\Big) s^4+2 t^3 u^3 \
\Big(834 t^{11}+10755 u t^{10}+52845 u^2 t^9+146657 u^3 t^8+271322 u^4 \
t^7+363015 u^5 t^6+363015 u^6 t^5+271322 u^7 t^4+146657 u^8 t^3+52845 u^9 \
t^2+10755 u^{10} t+834 u^{11}\Big) s^3+6 t^4 u^4 (t+u)^2 \Big(283 \
t^8+1733 u t^7+4779 u^2 t^6+8360 u^3 t^5+10092 u^4 t^4+8360 u^5 t^3+4779 u^6 \
t^2+1733 u^7 t+283 u^8\Big) s^2+2 t^5 u^5 (t+u)^3 \Big(282 t^6+1178 u \
t^5+2417 u^2 t^4+3162 u^3 t^3+2417 u^4 t^2+1178 u^5 t+282 u^6\Big) s+110 \
t^6 u^6 (t+u)^4 \Big(t^2+u t+u^2\Big)^2\Big) H(2,3,2,y)  \Bigg\}
\Big/ \Big(  {27 s^3 t^3 u^3 (s+t)^6 (s+u)^6 (t+u)^6} \Big)
\end{dmath*}

\intertext{}
\end{dgroup*}
 }


\end{document}